\documentclass{aa}  
\makeatletter

\renewcommand*\aa@pageof{, page \thepage{} of \pageref*{LastPage}}

\usepackage{natbib}
\bibpunct{(}{)}{;}{a}{}{,}%
\def\bibfont{\aa@bibliographyfont}%
\makeatother

\usepackage[T1]{fontenc}
\usepackage{amsmath}	
\DeclareMathOperator{\erf}{erf}
\usepackage{gensymb}
\usepackage{bm}	
\usepackage{graphicx}
\usepackage{newtxtext}
\usepackage[smallerops,cmbraces]{newtxmath}
\usepackage{xcolor}
\usepackage{colortbl}
\usepackage{nicematrix}
\definecolor{xlinkcolor}{cmyk}{1,1,0,0}
\usepackage[
  bookmarks=true,         
  pdfnewwindow=true,      
  colorlinks=true,        
  linkcolor=xlinkcolor,   
  citecolor=xlinkcolor,   
  filecolor=xlinkcolor,   
  urlcolor=xlinkcolor,    
  final=true,
]{hyperref}
\usepackage{soul}
\usepackage{xspace} \xspaceaddexceptions{]\}}
\usepackage{overpic}
\usepackage[normalem]{ulem} 
\usepackage{array}

\usepackage{bm} 
\usepackage{textgreek} 

\usepackage[capitalize]{cleveref} 
\crefname{section}{Sect.}{Sects.}
\creflabelformat{equation}{#2#1#3} 
\crefname{enumi}{item}{items} 

\usepackage{siunitx} 
\DeclareSIUnit[number-unit-product = ]\percent{\char`\%} 

\usepackage{csquotes} 

\definecolor{blackberry}{HTML}{8D1D75}
\definecolor{lightblue}{rgb}{0.1,0.5,0.89}

\newcommand{\txt}[1]{\mathrm{#1}} 

\begin{document}

\def\textfraction{0}
\def\topfraction{1}
\def\bottomfraction{1}
\def\dbltopfraction{1}
\def\floatpagefraction{.9}
\def\dblfloatpagefraction{.9}
\def\labelitemi{$\bullet$}

\definecolor{blackberry}{HTML}{8D1D75}
\definecolor{lightblue}{rgb}{0.1,0.5,0.89}
\definecolor{teal}{HTML}{20B2AA}
\definecolor{mblue}{rgb}{0.07, 0.50, 0.99} 
\definecolor{pinky}{rgb}{0.8,0.04,0.8}
\definecolor{mr}{rgb}{0.95,0.87,0.67}
\definecolor{hr}{rgb}{0.92,0.69,0.94}
\definecolor{uhr}{rgb}{0.52,0.81,098}
\definecolor{xhr}{rgb}{0.85,0.85,0.83}
\definecolor{uhrl}{rgb}{0.65,0.91,0.90}
\newcommand{\magpath}{\textsc{Magneticum Pathfinder}\xspace}
\newcommand{\Bz}{\textit{Box0}\xspace}
\newcommand{\Bzmr}{\textit{Box0/mr}\xspace}
\newcommand{\Bo}{\textit{Box1a}\xspace}
\newcommand{\Bomr}{\textit{Box1a/mr}\xspace}
\newcommand{\Bt}{\textit{Box2}\xspace}
\newcommand{\Bthr}{\textit{Box2/hr}\xspace}
\newcommand{\Btb}{\textit{Box2b}\xspace}
\newcommand{\Btbhr}{\textit{Box2b/hr}\xspace}
\newcommand{\Bth}{\textit{Box3}\xspace}
\newcommand{\Bthmr}{\textit{Box3/mr}\xspace}
\newcommand{\Bthhr}{\textit{Box3/hr}\xspace}
\newcommand{\Bthuhr}{\textit{Box3/uhr}\xspace}
\newcommand{\Bf}{\textit{Box4}\xspace}
\newcommand{\Bfuhr}{\textit{Box4/uhr}\xspace}
\newcommand{\Bfi}{\textit{Box5}\xspace}
\newcommand{\Bfixhr}{\textit{Box5/xhr}\xspace}
\newcommand{\xhr}{\textit{xhr}\xspace}
\newcommand{\uhr}{\textit{uhr}\xspace}
\newcommand{\hr}{\textit{hr}\xspace}
\newcommand{\mr}{\textit{mr}\xspace}
\newcommand{\subfind}{\textsc{SubFind}\xspace}
\newcommand{\Msun}{M_\odot}
\newcommand{\tb}{\textbf}
\newcolumntype{C}[1]{>{\centering\let\newline\\\arraybackslash\hspace{0pt}}m{#1}}
\def\H4{$H_4$\xspace}
\def\h4{$\langle h_4\rangle$\xspace}
\def\rh{$r_{1/2}$}
\def\th{$\theta$\xspace}
\def\lr{$\lambda_{R_{1/2}}$\xspace}
\def\bH4{$\bm{H_4}$\xspace}
\def\bh4{$\bm{\langle h_4\rangle}$\xspace}
\def\ion#1#2{#1\,\textsc{#2}\xspace}
\def\HI{\ion{H}{i}}
\def\MgII{\ion{Mg}{ii}}
\def\CII{\ion{C}{ii}}
\def\CIV{\ion{C}{iv}}
\def\NII{\ion{N}{ii}}
\def\OII{\ion{O}{ii}}
\def\OIII{\ion{O}{iii}}

\title{Encyclopedia \textsc{Magneticum}: \\ Scaling Relations from Cosmic Dawn to Present Day}
\titlerunning{Magneticum Scaling Relations}

\author{
    Klaus Dolag\thanks{dolag@usm.lmu.de}\inst{\ref{inst:usm},\ref{inst:mpa}}
    \and
    Rhea-Silvia Remus\inst{\ref{inst:usm}}
    \and
    Lucas M. Valenzuela\inst{\ref{inst:usm}}
    \and
    Lucas C. Kimmig\inst{\ref{inst:usm}}
    \and
    Benjamin Seidel\inst{\ref{inst:usm}}
    \and\\
    Silvio Fortun\'e\inst{\ref{inst:usm},\ref{inst:origins}}
    \and
    Johannes Stoiber\inst{\ref{inst:usm}}
    \and
    Anna Ivleva\inst{\ref{inst:usm}}
    \and
    Tadziu Hoffmann\inst{\ref{inst:usm}}
    \and
    Veronica Biffi\inst{\ref{inst:inaf},\ref{inst:ifpu}}
    \and\\
    Ilaria Marini\inst{\ref{inst:eso}}
    \and
    Paola Popesso\inst{\ref{inst:eso},\ref{inst:origins}}
    \and
    Stephan Vladutescu-Zopp\inst{\ref{inst:usm}}
    }
\authorrunning{K. Dolag et al.}

\institute{
    Universit\"ats-Sternwarte, Fakult\"at f\"ur Physik, Ludwig-Maximilians-Universit\"at München, Scheinerstr.~1, 81679 M\"unchen, Germany\label{inst:usm}
    \and
    Max-Planck-Institut f\"ur Astrophysik, Karl-Schwarzschild-Str.~1, 85748 Garching, Germany\label{inst:mpa}
    \and
    INAF – Osservatorio Astronomico di Trieste, Via Tiepolo 11, 34143 Trieste, Italy\label{inst:inaf}
    \and 
    IFPU – Institute for Fundamental Physics of the Universe, Via Beirut 2, 34014 Trieste, Italy\label{inst:ifpu}
    \and
    European Southern Observatory, Karl-Schwarzschild-Str.~2, 85748, Garching bei M\"unchen, Germany\label{inst:eso}
    \and
    Excellence Cluster ORIGINS, Boltzmannstr.~2, 85748 Garching bei M\"unchen, Germany\label{inst:origins}
    }

\date{Received XXX / Accepted YYY}

\abstract
{Galaxy and halo scaling relations, connecting a broad range of parameters, are well established from observations. The origin of many of these relations and their scatter is still a matter of debate. It remains a sizable challenge for models to simultaneously and self-consistently reproduce as many scaling relations as possible.}
{We introduce the \magpath hydrodynamical cosmological simulation suite, to date the suite that self-consistently covers the largest range in box volumes and resolutions. It is the only cosmological simulation suite that is tuned on the hot gas content of galaxy clusters instead of the stellar mass function. By assessing the successes and shortcomings of tuning to the hot gas component of galaxy clusters, we aim to further our understanding of the physical processes shaping the Universe. We analyze the importance of the hot and cold gas components for galaxy and structure evolution.}
{We analyze 28 scaling relations, covering large-scale global parameters as well as internal properties for halos ranging from massive galaxy clusters down to galaxies, and show their predicted evolution from $z=4$ to $z=0$ in comparison with observations. These include the halo-to-stellar-mass and Kennicutt--Schmidt relations, the cosmic star formation rate density as well as the Fundamental Plane.}
{\magpath matches a remarkable number of the observed scaling relations from $z=4$ to $z=0$, including challenging relations like the number density of quiescent galaxies at cosmic dawn, the mass--size evolution, the mass--metallicity relation, the Magorrian relation, and the temperature--mass relation. We compile our data to allow for straightforward future comparisons.}
{Galaxy properties and scaling relations arise naturally from feedback implementations that capture the evolution of the hot gas component down to $z=0$. Similarly, the large scatter in observables at high redshift is crucial to distinguish the various galaxy formation models that reproduce the same $z=0$ relations.}

\keywords{methods: data analysis -- methods: numerical -- galaxies: clusters: general -- galaxies: evolution -- galaxies: general -- galaxies: halos }


\maketitle

\section{Introduction}

In the field of structure formation and evolution from individual galaxies up to galaxy clusters, a multitude of relations, either scaling relations between global or internal properties, or, the evolution of them across cosmic time have been found in the last 60 years, and their existence has been established observationally with large surveys, systematically mapping the sky at lower redshifts. Such scaling relations provide insights into how structures have grown through cosmic time, and as such are an invaluable tool to study galaxy formation from dwarf galaxies to galaxy clusters (see \citealt{donofrio:2021} for a review on scaling relations).
At higher redshifts, observational evidence becomes more scarce; however, with the newest measurements using JWST, more of these relations have been probed up to redshifts $z=4$ or are currently under investigation, verifying or disproving predictions from models and challenging our understanding of how galaxies form and evolve.

These scaling relations cover a large range of properties. Among those that are studied most extensively are two different types. First, there are those that target the properties of entire halos, for example
the halo \citep[e.g.][]{sheth:2001a,tinker:2008,bocquet:2016,Driver2022_HMF}, stellar \citep[e.g.][]{panter:2004,marchesini:2009,muzzin:2013,Driver2022_GSMF,weaver:2023}, gas \citep[e.g.][]{berta:2013,saintonge:2017}, and black hole (BH) mass functions \citep[e.g.][]{shankar:2004,marconi:2004,shankar:2013},
the stellar-mass--halo-mass relation \citep[e.g.][]{mandelbaum:2006,gonzalez:2013,vanderburg:2014,hudson:2015,kravtsov:2018},
the number densities of all and of quiescent galaxies \cite[e.g.][]{brammer:2011,muzzin:2013,weaver:2022,chworowsky:2024},
the cosmic gas fractions \citep[e.g.][]{lagana:2011,lovisari:2015,lovisari:2020, gonzalez:2013},
the temperature--mass relation \citep[e.g.][]{finoguenov:2001,lagana:2013,giacintucci:2019}, the X-ray luminosity--mass relation \citep[e.g.][]{Pratt2009,lovisari:2015,anderson:2015,lovisari:2020}, the integrated Compton Y --mass relation \citep[e.g.][]{Planck2013_SZ}, the X-ray luminosity--temperature relation \citep[e.g.][]{Pratt2009,Sun2009,lovisari:2015,lovisari:2020}, the entropy--temperature relation \citep[e.g.][]{Sun2009,bahar:2024},
and the cosmic star formation rate \citep[e.g.][]{madau:2014}. Many of the above relations are established benchmarks in the formation and evolution of large-scale structure.
Second, there are those more focussed on galaxies, among which are
the galaxy star formation main sequence \cite[e.g.][]{pearson:2018,santini:2017,leslie:2020,popesso:2019},
the Kennicutt--Schmidt relation \citep[e.g.][]{schmidt:1959,kennicutt:1989,kennicutt:1998},
the color--mass relation \citep[e.g.][]{kauffmann:2004},
the mass--metallicity relation \citep[e.g.][]{lequeux:1979,tremonti:2004,gallazzi:2005,kudritzki:2021},
the mass--age relation \citep[e.g.][]{gallazzi:2005,gallazzi:2014,neumann:2021,saracco:2023},
the stellar-mass--angular-momentum relation \citep[e.g.][]{fall:1983,romanowsky:2012,fall:2013,obreschkow:2015},
the galaxy mass--size relation \citep[e.g.][]{shen:2003,vanderwel:2014,lange:2015,vanderwel:2024},
the galaxy Fundamental Plane \citep[e.g.][]{hyde:2009,bezanson:2013,bezanson:2015,zahid:2016},
and the kinematics--shape relation \citep[e.g.][]{emsellem:2007,emsellem:2011,cappellari:2011,vandesande:2019}.
In addition, there are scaling relations known to exist that connect the properties of supermassive black holes (SMBHs) to their host galaxies, namely
the Magorrian relation \citep[e.g.][]{magorrian:1998,haering:2004,harikane:2023,maiolino:2024},
the black hole--$\sigma$ relation \citep[e.g.][]{ferrarese:2000,salviander:2007},
and the correlation between the star formation rate and the black hole mass \citep[e.g.][]{piotrowska:2022}. 
This is by no means a complete list of all scaling relations that have been found in the field of structure formation, but those listed here are an important subset.

As demonstrated by this list, scaling relations connect a multitude of different galaxy and halo properties to one another, illustrating the complexity of the physical processes that drive structure formation from cosmic dawn to the present day. They prove that the different components of gas, stars, and dark matter are involved in an intricate dance, interacting with one another in sometimes subtle and sometimes clearly visible ways. Disentangling these processes is among the most difficult puzzles that our Universe presents us with. However, on top of the intrinsic challenges of disentangling these processes, the technical and observational methods used to derive quantities vary strongly, especially with respect to the definition of structural parameters \citep[e.g.][]{courteau:1996,courteau:1997}, the fitting algorithms or morphologies \citep[e.g.][]{courteau:2007}, or the environment \citep[e.g.][]{cappelari:2013}, adding biases and additional challenges in interpreting the observations. This is important to keep in mind whenever simulations or models are compared to observed scaling relations, as such biases can strongly impact the results \citep[see review by ][]{donofrio:2021}.

\begin{figure*}
    \centering
    \includegraphics[width=0.85\textwidth]{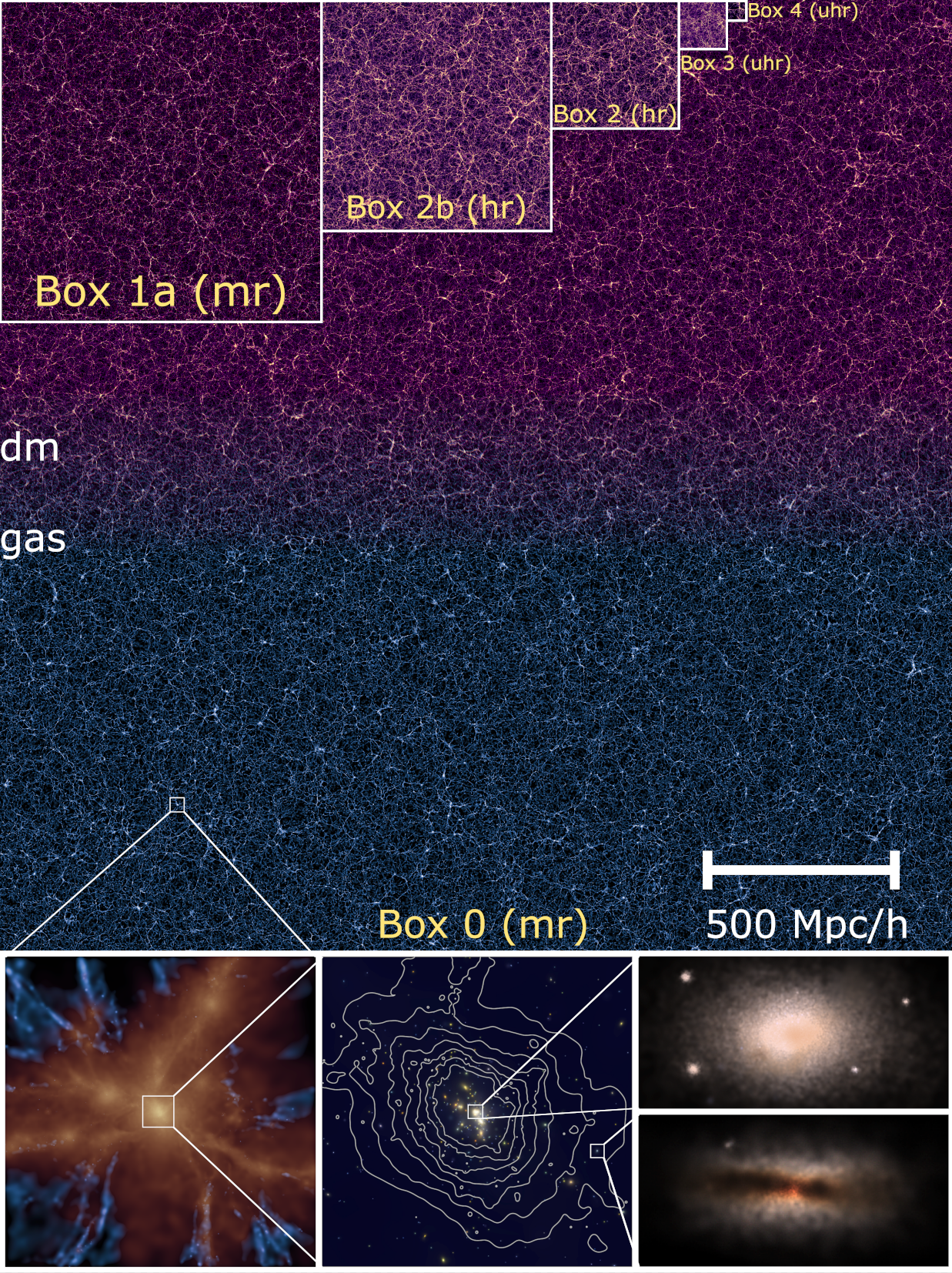}
    \caption{Visualization of the different boxes of \magpath. Zooming from the largest scales (Gpc scales) into galaxy clusters (Mpc scales) and further down onto individual galaxies (tens of kpc scales).
    \textit{Top:} The simulation volumes from \magpath to scale. The inlay panels at the top show the dark matter density. The large background panel shows the dark matter density transitioning into the gas density from top to bottom for \Bz.
    \textit{Bottom left:} A massive galaxy cluster from \Bt and its large-scale environment. Shown is the gas density split into a hot (red-yellow) and a cold (blue) component, with the temperature splitting the two phases at $T_\mathrm{cut}=10^4\mathrm{K}$.
    \textit{Bottom center:} Zoom in on the same galaxy cluster from \Bt, where the stellar component is visualized using Splotch \citep{dolag:2008} and the gas surface density contours are overlaid in white. The image has a side length of 5.59~cMpc.
    \textit{Bottom right:} Two galaxies taken from \Bf, where the top shows a massive elliptical galaxy and the bottom an edge-on disk galaxy of MW-mass. Both images were created using the dust radiative transfer code {\textsc{Skirt}} \citep{baes:2011, camps:2020}.}
    \label{fig:sim}
\end{figure*}

\def\mystrut{\rule{0pt}{1.1em}}
\begin{table*}
    \centering
    \begin{NiceTabular}{l| c c c c c c c } 
                                         & \Bz     & \Bo     & \Btb   & \Bt    & \Bth    & \Bf   & \Bfi   \\
    \hline\hline
    Volume [$\mathrm{Gpc}^3$]\mystrut    & 56.35    & 2.06 & 0.75 & 0.125 & 0.006 & 0.00032 & 0.000017 \\
    Boxlength [$\mathrm{Mpc}$]\mystrut   & 3820     & 1300     & 910     & 500     & 180     & 68     & 26     \\
    Boxlength [$h^{-1}\mathrm{Mpc}$]\mystrut & 2688     & 896      & 640     & 352     & 128     & 48     & 18     \\[0.2em]
    \hline
    \cellcolor{mr}{{\mr}\mystrut}   &
      \cellcolor{mr}{$2\times 4536^3$} &
      \cellcolor{mr}{$2\times 1512^3$} & &
      \cellcolor{mr}{$2\times  594^3$} &
      \cellcolor{mr}{$2\times  216^3$} &
      \cellcolor{mr}{$2\times   81^3$} &  \\
    \cellcolor{hr}{{\hr}\mystrut}   &  &  &
      \Block[l,draw=hr,  line-width=1.5pt]{}{$2\times 2880^3$} &
      \cellcolor{hr}{$2\times 1564^3$} &
      \cellcolor{hr}{$2\times  576^3$} &
      \cellcolor{hr}{$2\times  216^3$} &
      \cellcolor{hr}{$2\times   81^3$} \\
    \cellcolor{uhr}{{\uhr}\mystrut} &  &  &  &  &
      \Block[l,draw=uhrl,line-width=1.5pt]{}{$2\times 1536^3$} &
      \cellcolor{uhr}{$2\times  576^3$} &
      \cellcolor{uhr}{$2\times  216^3$} \\
    \cellcolor{xhr}{{\xhr}\mystrut} & & & & & & &
      \cellcolor{xhr}{$2\times  576^3$}\\
    \end{NiceTabular}
    \vskip\VSpaceAfterTabCaption
    \caption{Volumes and initial particle numbers for the different simulations of the \magpath simulation suite. All simulations that are available at $z=0$ are highlighted in solid colors. All simulations are set up with the same number of DM and gas particles per box and resolution, and that number is constant for the DM particles throughout the simulations. The simulations \Btbhr and \Bthuhr have been run until $z=0.25$ and $z=1.9$, respectively, and are thus marked by framed boxes instead of solid colors. \Bthuhr has been performed with an updated BH model, as described in the text, and is thus not marked in the \uhr resolution color blue but rather in teal color. An \hr version of that simulation is also available but not explicitly listed here.}
    \label{tab:Boxes}
\end{table*}

One process that is known to be extremely efficient in establishing scaling relations between different quantities in structure formation is the merging of structures. Structures in our Universe grow hierarchically \citep[e.g.][]{peebles:1965,white:1978,blumenthal:1985}; especially after redshifts of $z=2$, structure growth through dry (i.e. non-gas-dominated) merger events is one of the dominant channels of structure formation \citep[e.g.][]{naab:2009,oser:2010}. Such (dry) hierarchical growth is being discussed as cause for the formation and evolution of the present-day Magorrian relation \citep[e.g.][]{hirschmann:2010}, the Fundamental Plane \citep{donofrio:2025}, the correlation between the number of globular clusters in a galaxy and the halo mass of the galaxy \citep{valenzuela:2021}, or even the star formation main sequence \citep{kelson:2014}. This part of galaxy formation can be well described by the central limit theorem \citep[see][for a review]{fischer:2010}. 
Self similar models are able to describe global quantities -- especially at large masses -- and give insights into the evolution of structures and the importance of additional baryonic physics on their formation processes \citep[e.g.][]{boehringer:2012}. In addition, on large scales, mergers can be incorporated into self similar models \citep[e.g.][]{2021MNRAS.506..839Z}. 

However, hierarchical growth cannot explain the observed scatter in these relations, as collisionless growth according to the central limit theorem will only tighten a relation over time \citep[e.g.][]{hirschmann:2010}.
Nevertheless, it is extremely important to not simply reproduce a tight scaling relation when studying the physics behind that relation, but also to understand the origin of the scatter found for the relation. For example, \citet{valenzuela:2021} showed that the scatter in the globular-cluster--halo-mass relation encodes the amount of smooth accretion compared to merging events. On the effect of mergers on scaling relations of galaxy clusters, it was shown that scatter can lead to insights into the interplay between the long term and short term evolutionary processes \citep[e.g.][]{2011ApJ...729...45R,2012MNRAS.419.1766K}. For many other scaling relations, unfortunately, the scatter is yet not well understood, especially at higher redshifts and the early Universe when hierarchical dry merging playes only a minor role. Nevertheless, this is a growing field of research that will prove invaluable for understanding galaxy and structure formation \citep[e.g.][for the stellar mass Tully--Fisher relation]{ristea:2024}.

While scaling relations are not suited as tools for generating models to explain galaxy or structure formation \citep{donofrio:2021}, they can be used to inform us about how accurately a given model describes the underlying physics important for galaxy formation, by comparing the predictions from the model to the observed scaling relations. They can furthermore inform our models about how to adjust free parameters. Some of the most successful models that helped to improve our understanding of galaxy and structure formation in recent years are fully hydrodynamical cosmological simulations, where galaxies, groups, and clusters of galaxies evolve self-consistently through cosmic time from initial conditions that mimic the density distributions in the early universe as inferred from measurements of the cosmic microwave background with probes such as WMAP \citep{komatsu:2011} or PLANCK \citep{Planck:2016}.

The field of fully hydrodynamical cosmological simulations has been developing with increasing speed in the last decade, with one of the first such simulations on the market being \Bt~mr and \Bt~hr of the \magpath simulation suite \citep{biffi:2013,hirschmann:2014}. Further simulation suites have been developed since then, building on the successes and failures of the previous simulation attempts. 
Among them,
Eagle \citep{schaye:2015},
HorizonAGN \citep{dubois:2014},
Illustris \citep{vogelsberger:2014},
Illustris-TNG \citep{springel:2018},
Simba \citep{dave:2019},
FIREbox \citep{feldmann:2023},
MilleniumTNG \citep{pakmor:2023},
and Flamingo \citep{schaye:2023},
which are employing different underlying hydrodynamical schemes from AMR (e.g. HorizonAGN) to SPH (e.g. \magpath, Eagle, Flamingo, Simba, FIREbox) and hybrid codes like Arepo (e.g. Illustris, Illustris-TNG, Millenium-TNG), but also including different treatments of the physics in form of subgrid models. However, all simulations require a fine-tuning of the free parameters of their subgrid-models \citep[see recent review][]{2025arXiv250206954V}, and most of these suites of simulations have been fine-tuned to reproduce the stellar mass functions and other stellar-mass based properties at $z=0$. Reproducing those scaling relations is thus not self-consistent from the included physics, but rather a result of the fine-tuning. It is those scaling relations beyond the ones used to inform the free parameters of the subgrid models that are the real successes of the simulations if reproduced, but from the combination of both or more importantly the failure in reproducing a given scaling relation, we ultimately learn to decipher which physics govern the formation of structures at different redshifts. Thus, the most important test for quantifying our understanding of structure formation through such simulations is the comparison to observed scaling relations, targeting not just those that are used for tuning but rather targeting all of them.

As shown by \citet{popesso:2024}, those simulations that are tuned to reproduce the stellar mass functions struggle in recovering gas scaling relations, but also kinematic properties \citep[e.g.][]{vandesande:2019}. The \magpath simulation suite, however, is the only simulation suite from the cosmological hydrodynamical simulations on the market that is not tuned to the stellar mass functions but rather to generate hot atmospheres in galaxy clusters at $z=0$, which they do successfully as shown by \citet{popesso:2024}. Therefore, testing the ability of this suite of simulations in reproducing a broad range of galaxy and structure scaling relations is highly informative and complementary to all the other simulation suites.

The \magpath\footnote{www.magneticum.org} simulations are to date still the suite of simulations spanning the largest range of box volumes and resolutions with the same physics, and as such can self-consistently test the full range of structure formation, from the physics and evolution of galaxies to the evolution of the cosmic web, the formation of galaxy clusters through cosmic time, and even cosmology. Different simulations of this suite have been analyzed in previous studies, starting already more than 10 years ago with X-ray properties of galaxy clusters by \citep{biffi:2013},
black hole properties by \citet{hirschmann:2014},
galaxy angular momentum properties by \citet{teklu:2015},
impact on baryons on halo mass function by \citet{bocquet:2016},
thermal SZ signal and pressure profiles by \citep{2016MNRAS.463.1797D,2017MNRAS.469.3069G},
dark matter--baryon interactions in galaxies by \citet{remus:2017},
metal content of clusters and galaxies by \citet{dolag:2017},
galaxy cluster properties by \citet{lotz:2019,lotz:2021},
and many others.

In this work we will, for the first time, present the scaling relations obtained from these simulations covering 7 orders of magnitude in mass and four orders of magnitude in resolution, at redshifts from $z=4$ to $z=0$. It should also replace the legendary Dolag et al. 2012 (in prep) and all variations of years for this citation which meanwhile exist in the literature. In Section~\ref{sec:sim} we introduce the full set of simulations that are part of the \magpath simulation suite. The details of the physics included in these simulations as well as the definitions of the quantities used throughout this work are described in Section~\ref{sec:gadget}. Section~\ref{sec:mf} introduces the halo, stellar, gas and BH mass functions for the \magpath simulations in comparison to observations. Section~\ref{sec:scale0} is dedicated to overall quantity evolutions with redshift, including the cosmic star formation rate density,
the depletion timescale evolution, the overall and quenched number densities of galaxies through cosmic time, and the maximum halo mass evolution.
In Section~\ref{sec:scale1}, we present the scaling relations of global halo quantities, that is,
the stellar-mass--halo-mass relation,
the baryon conversion efficiency,
the Compton-Y--halo-mass relation,
the gas-mass--halo-mass relation,
the temperature--mass relation,
the X-ray-luminosity--mass relation,
and the gas-metallicity--mass relation.
Section~\ref{sec:scale2} is dedicated to the galaxy scaling relations, including
the star formation main sequence,
the Kennicutt--Schmidt relation,
the color--mass relation,
the mass--metallicity relation,
the age--mass relation,
the mass--angular-momentum relation,
the mass--size relation,
the Fundamental Plane,
and the kinematics--shape relation ($\lambda_\mathrm{R}$--$\epsilon$).
Finally, the black hole scaling relations are presented in Section~\ref{sec:scale3}, including
the Magorrian relation,
the BH-mass--$\sigma$ relation, 
the distribution of black hole masses between quiescent and star forming populations,
and the connection between the galaxy star formation rate and black hole mass.
Section~\ref{sec:DataAvail} provides an overview over the data availability of the \magpath simulations, presenting a short review on the available flat and fully sky lightcones and the cosmological web portal.
In Section~\ref{sec:end} we summarize our results and present our conclusions.

Given the extensive nature of this suite of simulations, releasing the full, underlying simulation data to the public is very challenging and practically not feasible. However, data from individual projects are directly available in the data section of the project webpage\footnote{http://www.magneticum.org/data.html} and a significant subset of the simulation data can be processed or extracted directly through the cosmological web portal\footnote{https://c2papcosmosim.uc.lrz.de/} as described by \citet{ragagnin:2017webportal}, and full access to the data can be given upon request. 

\begin{figure*}
    \includegraphics[width=0.99\textwidth]{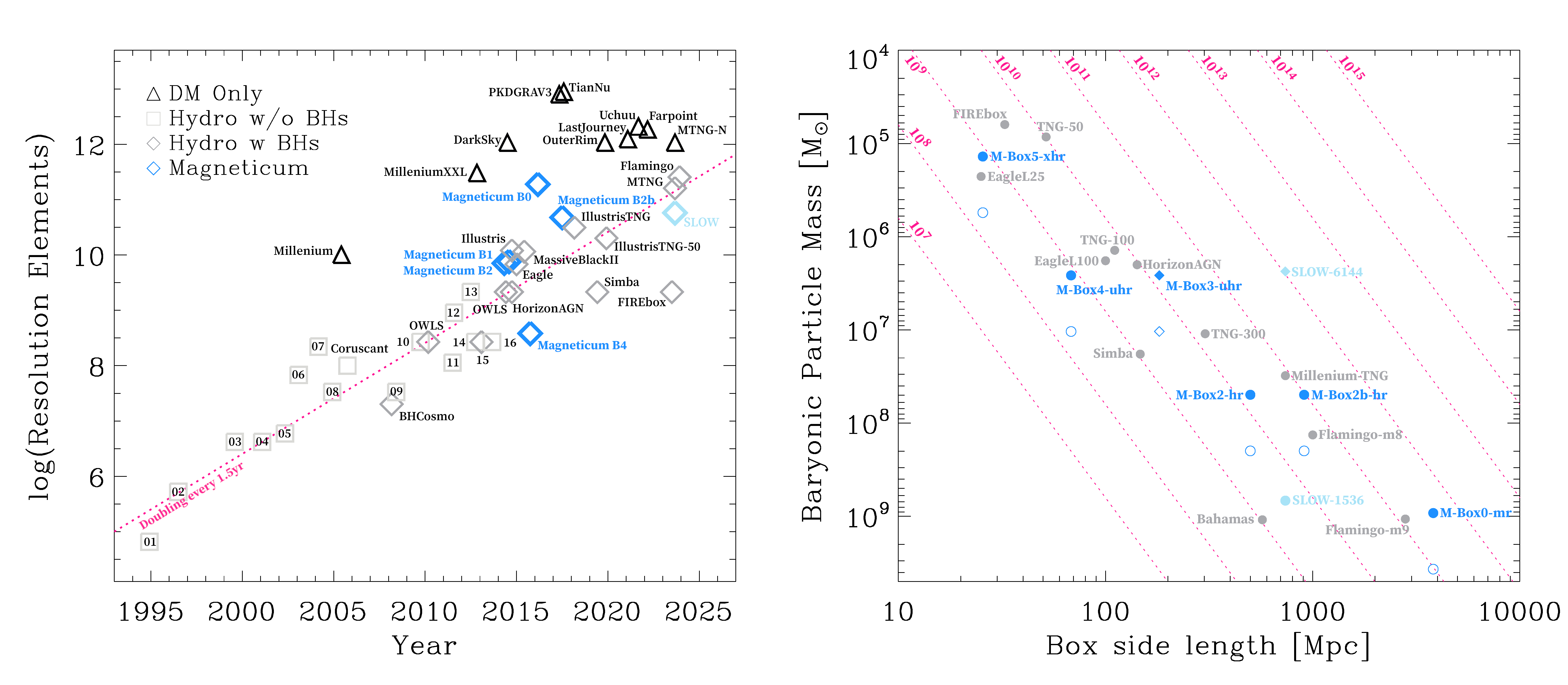}
    \caption{Placing simulations in resolution context.
    \textit{Left panel:} Evolution of the number of resolution elements used in (hydrodynamical) cosmological simulations over the last thirty years (inspired by \citet{genel:2014}). The \magpath simulations are marked in blue \citep[presented by][]{hirschmann:2014,teklu:2015,saro:2014,bocquet:2016,ragagnin:2017webportal}, and the \magpath local Universe spinoff SLOW is shown in light blue \citep{dolag:2023}.
    Light gray squares mark hydrodynamical simulations without BH treatment
    (01: \citet{metzler:1994};
     02: \citet{katz:1996};
     03: \citet{pearce:1999};
     04: \citet{dave:2001};
     05: \citet{murali:2002};
     06: \citet{springel:2003_cosmosim};
     07: \citet{borgani:2004};
     08: \citet{kay:2004};
     Coruscant: \citet{dolag:2005};
     09: \citet{oppenheimer:2008};
     10: \citet{planelles:2009};
     11: \citet{dave:2011};
     12: \citet{deboni:2011};
     13: \citet{cui:2012};
     14: \citet{vogelsberger:2012};
     16: \citet{dave:2013});
    dark gray diamonds mark hydrodynamical simulations including BH treatment
    (BHCosmo: \citet{dimatteo:2008};
     OWLS: \citet{schaye:2010};
     15: \citet{puchwein:2013};
     OWLS: \citet{vanDaalen:2014};
     HorizonAGN: \citet{dubois:2014};
     Illustris: \citet{vogelsberger:2014};
     Eagle: \citet{schaye:2015};
     MassiveBlackII: \citet{khandai:2015};
     IllustrisTNG: \citet{springel:2018};
     Simba: \citet{dave:2019};
     IllustrisTNG-50: \citet{nelson:2019};
     FIREbox: \citet{feldmann:2023};
     MTNG: \citet{pakmor:2023};
     Flamingo: \citet{schaye:2023});
    and black triangles mark dark matter only simulations
    (Millenium: \citet{springel:2005_mil};
     MilleniumXXL: \citet{angulo:2012};
     DarkSky: \citet{skillman:2014};
     PKDGRAV3: \citet{potter:2017};
     TianNu: \citet{emberson:2017};
     OuterRim: \citet{heitmann:2019};
     LastJourney: \citet{heitmann:2021};
     Uchuu: \citet{ishiyama:2021};
     Farpoint: \citet{frontiere:2022};
     MTNG-N: \citet[][simulation with only Neutrinos]{hernandezAguayo:2023}).
    \textit{Right Panel:} Baryonic particle mass versus box length for the subset of hydrodynamical simulations including BH treatment. The number of resolution elements is indicated by the pink dotted lines. For the \magpath simulations where one gas particles can spawn up to four stellar particles, two symbols are shown: the open circles mark the gas resolution before spawning, solid circles mark the average stellar particle masses. In addition to the simulations shown in the left panel, we also include the Bahamas simulations \citep{mccarthy:2017}.}
    \label{fig:sim_comp}
\end{figure*}


\section{The \magpath Simulations}
\label{sec:sim}

\begin{figure}
    \includegraphics[width=0.5\textwidth]{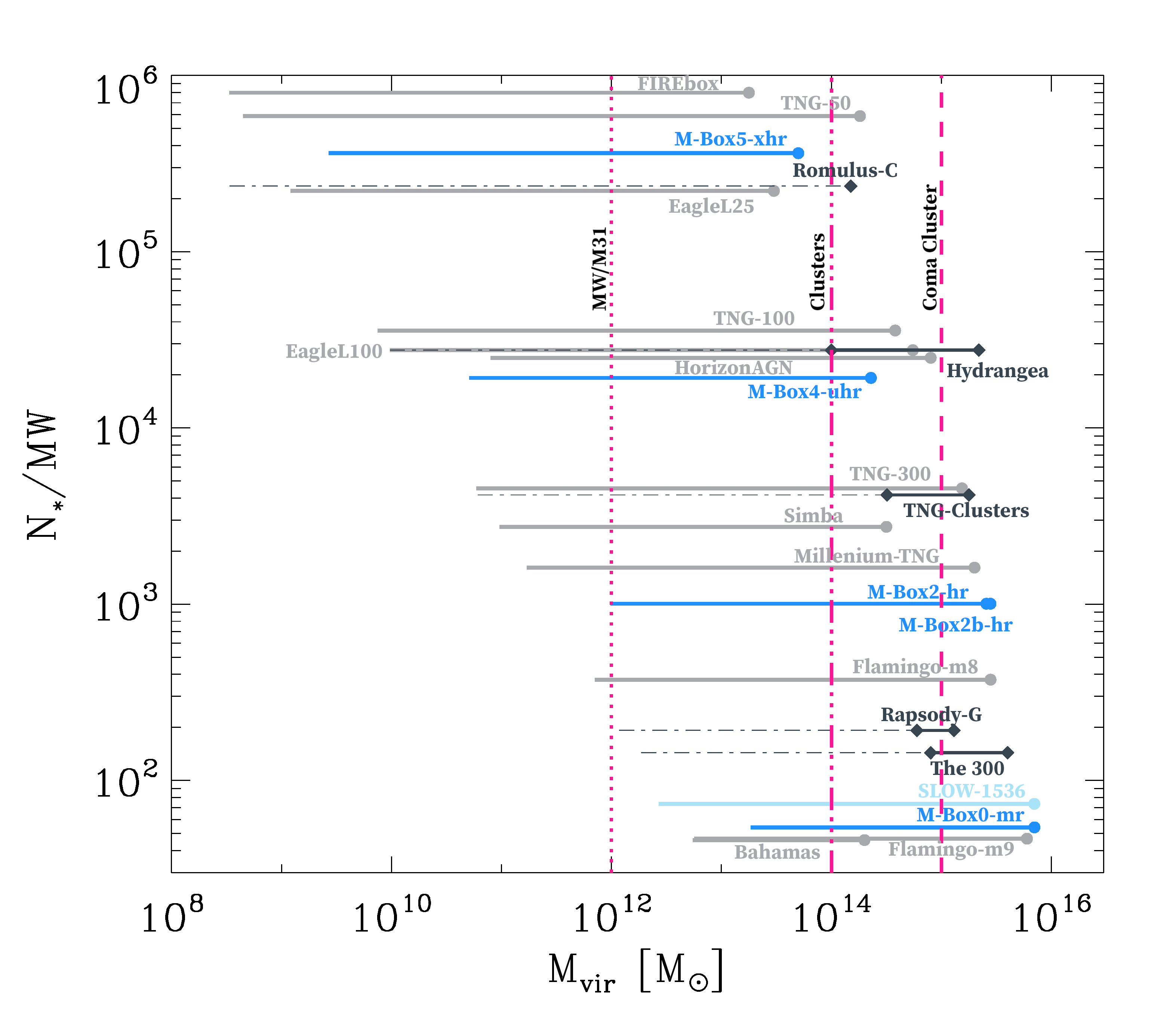}
    \caption{Stellar particle resolution for Milky-Way mass galaxies versus the stellar mass ranges encompassed by different simulation suites including both hydrodynamical cosmological simulations (round circles and solid lines) and hydrodynamical zoom-in simulations (diamonds and dash-dotted lines). The \magpath simulations are shown in blue, with the SLOW simulation \citep{dolag:2023} marked in light blue. The rightmost symbol on each line marks the virial mass of the most massive structure in that simulation, while the length of the lines indicates a mass limit of 1000 DM particles in a halo. For Zoom simulations, the leftmost diamond marks the virial mass of the smallest zoom object in the given suite. Pink vertical lines mark the virial mass of the Milky Way (dotted), the threshold splitting galaxy groups from clusters ($M_\mathrm{vir}=10^{14}\,M_\odot$, dash-dot-dot-dotted), and the virial mass of the Coma cluster (dashed).
    Included for comparison are
    FIREbox \citep{feldmann:2023},
    TNG-50 \citep{nelson:2019},
    Eagle \citep{schaye:2015},
    Romulus-C with a single cluster \citep{tremmel:2019},
    Illustris\-TNG-100 and Illustris\-TNG-300 \citep{springel:2018,pillepich:2018},
    Horizon\-AGN \citep{dubois:2014},
    Hydrangea \citep[Eagle Clusters Zooms;][]{bahe:2017},
    TNG-Clusters zooms \citep{nelson:2024},
    Simba \citep{dave:2019},
    MilleniumTNG \citep{pakmor:2023},
    Flamingo \citep{schaye:2023},
    \mbox{Rhapsody-G} Cluster Zooms \citep{hahn:2017},
    \mbox{The-300} cluster zooms \citep{cui:2018},
    and the Bahamas simulations \citep{mccarthy:2017}.}
    \label{fig:sim_res}
\end{figure}

The \magpath simulations are a set of cosmological, hydrodynamical simulations following the formation of cosmological structures in a hitherto unreached range of resolutions and volumes. They consist of a set of large-scale and high-resolution simulations, following more than $10^{10}$ particles and beyond while taking into account many physical processes to allow detailed comparison to a variety of multi-wavelength observational data. The simulations cover 7 different box volumes, numbered from largest to smallest, with \Bz being the largest box of $(3820~\mathrm{Mpc})^3$, \Bo with $(1300~\mathrm{Mpc)^3}$, \Btb with $(910~\mathrm{Mpc)^3}$, \Bt with $(500~\mathrm{Mpc})^3$, \Bth~with $(180~\mathrm{Mpc})^3$, \Bf with $(68~\mathrm{Mpc})^3$, down to \Bfi with $(26~\mathrm{Mpc)}^3$. The relative sizes of these simulations are shown in comparison in the upper big panel of Fig.~\ref{fig:sim}, with the smallest box, \Bfi, not plotted, as it would be minuscule at that scale. Shown are the dark-matter cosmic webs for all simulations, and for the largest box, \Bz, we show in the lower half the cosmic web as seen from the gas, clearly showing that both gas and DM trace the cosmic web nicely. All boxes are simulated with DM-only and with full hydrodynamics to allow to study the influence of the baryonic components on the forming structures in detail.\looseness-1

Four different resolutions exist for the \magpath suite of simulations: the medium resolution, \mr; the high-resolution, \hr; the ultra-high resolution, \uhr; and the extra-high resolution, \xhr. Not all boxes exist in all resolutions, as the highest resolutions at the largest box sizes would require a number of particles too large to handle for the current generation of supercomputers. Table~\ref{tab:Boxes} gives an overview of the different cosmological boxes simulated within the \magpath project.

The two largest volumes, \Bz and \Bo, only exist for the \mr resolution, with \Bz including an extremely high amount of resolution elements with initially $2\times4536^3$ particles. Fig.~\ref{fig:sim_comp} places this into context in a comparison to other simulations. The left panel of Fig.~\ref{fig:sim_comp} shows the number of resolution elements of the simulations against the time of their first appearance in the literature. The \magpath boxes are marked in blue, with the local Universe \magpath spinoff SLOW shown in light blue \citep{dolag:2023}. Light gray squares mark hydrodynamical simulation volumes without black holes (BHs), and dark gray diamonds mark those hydrodynamical cosmological simulations that include BHs and a treatment of their feedback. For comparison, some of the large and well-known dark-matter-only simulations are shown as black triangles. Overall, the number of resolution elements used in hydrodynamical simulations is seen to roughly double every 1.5 years, as indicated by the pink dotted line. The figure also makes evident that \magpath \Bz was an extraordinary outlier at the time of its first appearance in 2016 in terms of technological achievement, as it took further 8~years for another hydrodynamical simulation with a similar number of resolution elements to appear on the market (Flamingo, \citealp{schaye:2023}). \Bzmr was run on 131072~cores on SuperMUC at the Leibniz-Rechenzentrum in Garching, utilizing a total of 160~TB of main memory. The simulation showed an excellent scaling behavior compared to the smaller simulations. It was the first, hydrodynamical simulation which allowed to study the effect of baryonic physics onto the halo mass function, covering it to very massive galaxy clusters \citep{bocquet:2016} and allowed to construct very large and deep light-cones of kinetic and thermal SZ effect \citep{2018MNRAS.478.5320S}.

From the right panel of Fig.~\ref{fig:sim_comp} it can be seen that \Bz is still the largest simulation volume to date, although the number of initial resolution elements for the Flamingo-m9 run is slightly larger, as indicated by the pink dotted lines. This figure shows the baryonic particle mass of the different simulations versus the box side length for the fully hydrodynamical simulations with BH physics shown in the left panel. Again, the \magpath simulations are shown in blue, and two simulations of the \magpath local Universe spin off SLOW are shown in light blue. This figure also highlights one specialty of the \magpath simulations that will be discussed in detail in Section~\ref{sec:starform}: every gas particle can spawn up to four stellar particles, which then have approximately 1/4th of the mass of the gas particle. The stellar resolution is thus a factor of 4 higher than the gas resolution, which is why two different symbols are shown for every \magpath simulation. Open blue circles indicate the gas particle mass of the simulation, filled blue circles mark the stellar particle mass. This feature accounts for the fact that usually molecular clouds do not convert all their gas into stars (see \citealt{chevance:2023} for a review on molecular clouds and star formation efficiencies).

The highest resolution available for box volumes \Bt and \Btb is \hr. These two simulations are on the one hand large enough to harbor statistically representative amounts of galaxy clusters, while on the other hand their \hr resolution level is high enough to resolve galaxy clusters in detail. These two simulations have therefore been used already to study galaxy clusters in detail, for example the X-ray scaling relations \citep{biffi:2013}, the quenching and anisotropies of galaxies in clusters \citep{lotz:2019}, the behavior of post-starburst galaxies in clusters \citep{lotz:2021}, the properties of proto-clusters at $z\approx4$ and their evolution \citep{remus:2023}, and the use of the ICL and BCG fractions as dynamical clocks \citep{kimmig:2025b}. \Btbhr is the most computationally expensive simulation of them, covering a cosmic volume of almost 1~Gpc$^3$. 
This simulation was performed using the full capacity of the SuperMUC supercomputer for an entire week; it did not reach $z=0$ by the end of this week and therefore only exists up to $z=0.25$. The fact that \Bomr was computationally less expensive than \Btbhr already shows that for hydrodynamical simulations the scaling does not only depend on the number of resolution elements, but that in fact an increasing resolution adds many more computations in the hydrodynamics part of the code, especially through shorter time steps between particle wake-up calls, and explains why reaching the upper right corner of the right panel of Fig.~\ref{fig:sim_comp} is especially difficult.
Nevertheless, \Btb has proven extremely valuable in producing theoretical counterparts for interpreting data from large surveys or instruments, such as PLANCK, SPT, DES, and eROSITA. In fact, this simulation has already been used to compare cluster pressure profiles to detailed SZ observations by PLANCK \citep{pc:2013} and SPT \citep{mcdonald:2014}, and to explain some of the most peculiar massive structures observed at redshifts of $z\approx4$ that require large volumes in order to be found in simulations, due to cosmic variance \citep{remus:2023}. The lower left and central panels of Fig.~\ref{fig:sim} show an example cluster selected from \Bt. The left panel shows the gas of the large-scale environment, with hot gas ($T>10^5~\mathrm{K}$) marked in red and cold gas ($T<10^5~\mathrm{K}$) marked in blue. The filamentary structure of the surroundings can be seen well. The middle panel shows a zoom-in on the center of the cluster, with the contours indicating the gas and the colors showing the stellar component.

The large \magpath simulation boxes are supplemented by simulations of smaller volumes with ultra-high resolution to follow the evolution of galaxies and, especially, the AGN population in detail. This includes \Bfuhr, which has run down to $z\approx0$ with the fiducial BH model as used for all the other simulations, and the larger volume \Bthuhr, which has been run down to $z=1.9$. \Bthuhr was performed with an advanced BH feedback model introduced by \citet{steinborn:2015} and is the youngest in the immediate \magpath simulation family. It was explicitly performed to study the impact of BH feedback at early redshifts in extremely high resolution, as comparisons between the fiducial BH model and the new model performed on \Bthhr level by \citet{steinborn:2015} have shown that the new model especially performs better at high redshifts. This simulation has demonstrated its capability to reproduce the challenging number densities observed with JWST, setting it apart from all other simulations currently on the market as shown by \citet{kimmig:2025} and \citet{remus:2025}. We will therefore include this simulation in the study presented here despite its results only being available to $z=2$, as this is where the most crucial differences between the models appear.
The \uhr resolution has been used for several comparisons between galaxy properties from the simulation and observations, especially in the field of angular momentum properties \citep{teklu:2015}, galaxy kinematics \citep{schulze:2018,bellstedt:2018,vandesande:2019,schulze:2020,vandesande:2021} and the interactions between dark matter and baryons \citep{remus:2017,harris:2020,derkenne:2021,derkenne:2023}. Two examples of galaxies from \Bfuhr are shown in the lower right panels of Fig.~\ref{fig:sim}, with the upper panel depicting an elliptical galaxy and the lower depicting the edge-on view of a disk galaxy, calculated with the dust radiative transfer code {\textsc{Skirt}} \citep{baes:2011,camps:2020}.

Finally, the highest resolution in the \magpath family is the \xhr level, which has so far only been used for \Bfi. This simulation volume, however, is too small to be used for statistically representative studies and thus has only been used for calibrations so far.
The softening parameters and the particle masses for DM, gas, and stars for the different resolution levels of the \magpath simulation suite are given in Table~\ref{tab:Resolution}.
For our simulations we chose the cosmological parameters corresponding to WMAP-7 data \citep{komatsu:2011}. Thus, the background cosmology is described by $\Omega_M=0.272$, $\Omega_B=0.0456$, $\Omega_\Lambda=0.728$ and $h=0.704$. The initial power-spectra follow an index of $n=0.963$ and are normalized to $\sigma_8=0.809$.

\begin{figure}
    \includegraphics[width=0.5\textwidth]{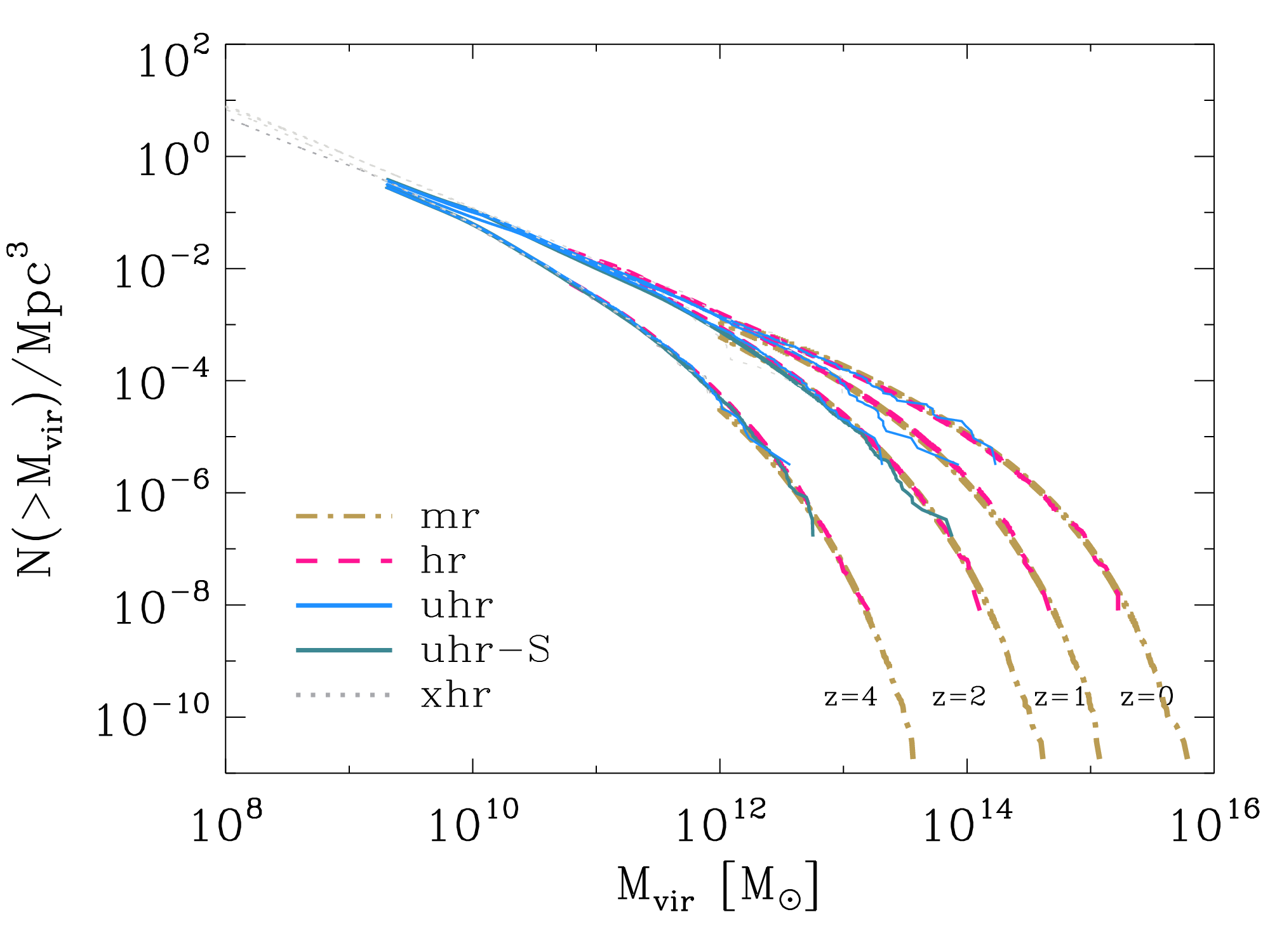}
    \caption{The halo mass function obtained from the \magpath simulations,
    at the redshifts of $z=4$, $2$, $1$, and $0$.
    Shown are:
    \Bfixhr   (gray,      $z=4$,~$2$),
    \Bfuhr    (blue,      $z=4$,~$2$,~$1$,~$0$),
    \Bthuhr   (turquoise, $z=4$,~$2$),
    \Bthr     (pink,      $z=4$,~$2$,~$1$,~$0$),
    \Btbhr    (pink,      $z=4$,~$2$,~$1$),
    and\Bzmr (gold,      $z=4$,~$2$,~$1$,~$0$).}
    \label{fig:halo_mass}
\end{figure}

The large range of simulation volumes and resolution levels in the \magpath suite allowed, for the first time in 2016, to self-consistently study the full range of structures from galaxy clusters and groups to individual galaxies and AGNs in a statistical manner. Fig.~\ref{fig:sim_res} demonstrates the full capacity of the \magpath simulation suite. On the $y$-axis it shows the number of stellar particles that a Milky-Way (MW) mass galaxy of $M_\mathrm{MW}^*=5\times10^{10}\,M_\odot$ would consist of at the given simulation's stellar resolution. On the $x$-axis, the maximum virial mass and resolution extent for a given simulation are shown. For the hydrodynamic cosmological simulations, that is, \magpath in blue and all others in light gray, the filled circle marks the most massive structure in the given simulation, while the solid line marks the full range of virial masses that is resolved with more than 1000 DM particles. For zoom-in simulation suites, marked by dark gray lines and diamonds, the right-most diamond marks the most massive object in the zoom suite, and the left-most diamond marks the least massive object in the zoom suite. The dash-dotted dark gray lines mark the 1000 DM particle resolution limit for the zoom suite. As can be seen, the \magpath simulations range from obtaining hundreds of Coma-like clusters in the largest box suited for cosmological studies, to thousands of clusters with galaxies down to MW mass resolved well enough for counting and simple statistics in the flagship-runs \Bt and \Btb, to the small volume barely large enough to host four clusters with masses above $M_\mathrm{vir}>1\times10^{14}\,M_\odot$ but well enough resolved to study MW-like galaxies in detail. However, this figure also cautions against using the wrong boxes for studying a given question: while the high-resolution simulations on the market are well suited for studying galaxies in detail, none of them contains a massive cluster with $M_\mathrm{vir}>1\times10^{15}\,M_\odot$, and thus probing for the progenitors of such massive clusters in comparison to observations cannot yield a positive result, as such nodes are not contained in the simulation volume. This is important to keep in mind especially when comparing to the extreme outlier objects observed at high redshifts, as none of the high-resolution box volumes is large enough to contain such outliers in abundance, if at all. Here, \Bth is the sole outlier, and is thus to date still the best-suited simulation for studying galaxies at cosmic dawn.

\begin{table}
    \centering
    \begin{tabular}{l | c@{~~}c@{~~}c@{~~~~~}c@{~~~~~}c} 
     & $\epsilon_\mathrm{gas/DM}$ & $\epsilon_\mathrm{*}$ & $m_\mathrm{DM}$ & $m_\mathrm{gas}$ & $m_\mathrm{*}$ \\
     & [$h^{-1}\mathrm{kpc}$]         & [$h^{-1}\mathrm{kpc}$]    & [$h^{-1}M_\odot$]   & [$h^{-1}M_\odot$]    & [$h^{-1}M_\odot$]  \\[0.2em]
    \hline
    \cellcolor{mr}{\mr\mystrut}   &             10\hphantom{.00} & 5\hphantom{.00} & $1.3\times10^{10}$            & $2.6\times10^{9}$ & $6.5\times10^{8}$ \\
    \cellcolor{hr}{\hr\mystrut}   &  \hphantom{0}3.75            & 2\hphantom{.00} & $6.9\times10^{8\hphantom{0}}$ & $1.4\times10^{8}$ & $3.5\times10^{7}$ \\
    \cellcolor{uhr}{\uhr\mystrut} &  \hphantom{0}1.4\hphantom{0} & 0.7\hphantom{0} & $3.6\times10^{7\hphantom{0}}$ & $7.3\times10^{6}$ & $1.8\times10^{6}$ \\
    \cellcolor{xhr}{\xhr\mystrut} &  \hphantom{0}0.45            & 0.25            & $1.9\times10^{6\hphantom{0}}$ & $3.9\times10^{5}$ & $1.0\times10^{5}$ \\
    \end{tabular}
    \vskip\VSpaceAfterTabCaption
    \caption{Gravitational softening and particle mass resolution for the simulations. Note that every gas particle can form up to four stellar particles, with their mass depending on the gas density as described in Sec.~\ref{sec:starform}, and as such the gas particle mass given here is only the gas particles initial mass, and the star particle mass is a quarter of the gas particle mass.
    }
    \label{tab:Resolution}
\end{table}

The combination of the different boxes allows studying the formation of objects covering almost eight orders of magnitude in mass, as can be seen from Fig.~\ref{fig:halo_mass}, which shows the evolution of the halo mass function obtained from the hydrodynamical runs at four different redshifts of $z=4$, $2$, $1$, and $0$. As can be seen directly, they agree well with one another over all orders of magnitude, and also reflect the results found for the DM-only runs marked by the light gray lines. For a detailed study on the effect of baryon physics onto the mass function, see \citet{bocquet:2016}. We here already introduce the color scheme that we will use throughout this work and that is also shown in Tables~\ref{tab:Boxes} and~\ref{tab:Resolution}: The \mr resolution simulation results are shown in gold, and are obtained solely from\Bzmr. The \hr resolution simulation results are shown in pink, and are usually obtained from both \Btb and \Bt for global scaling relations, but in case of galaxy properties in some cases only from \Bt if the calculation from \Btb is too expensive and we do not expect extreme outliers to be important. The \uhr resolution simulation \Bthuhr with the fiducial BH model as introduced by \citet{hirschmann:2014} and \citet{teklu:2015} is shown in blue (\uhr), while the \Bthuhr simulation with the advanced BH model as introduced by \citet{steinborn:2015} is shown in turquoise (\uhr-S).

These are the simulations that constitute the current status of the \magpath simulation suite. They are complemented by a set of 15 additional simulations of \Bomr with varying cosmologies using different combinations of $\sigma_8$, $\Omega_0$, $H_0$, and $\Omega_b$, as presented by \citet{singh:2020}. These simulations are mostly used to study the impact of cosmology on large-scale structure, but have been extended to proto-cluster studies by \citet{remus:2023}. They are run with the same physics as the \magpath simulation suite. However, we do not include them here as that would go beyond the scope of the current work and has already partially been done by \citet{singh:2020}.

\subsection{Initial Conditions}
\label{sec:ini}

The initial conditions for all simulations were generated using the \textsc{N-GenIC} code \citep{springel:2005_mil,angulo:2012}, an initial conditions generation code that takes an initial particle grid with a given resolution (usually a glass file,but here we used a regular grid to be able to obtain the particle numbers matching exactly the same mass for the different resolutions and box sizes).  The power spectrum for the \magpath simulation set was constructed using analytical Transfer Function from \citet{1998ApJ...496..605E,eisenstein:1999}, taking baryonic effects into account. This allows a better comparison to theoretical predictions. All simulation boxes used the same random number generator seed and the same Boxes at different resolution start form the largest random field to obtain the phases and amplitudes of the perturbation field across the different resolutions. The displacement field is then advanced to the desired starting redshift using the Zel'dovich approximation \citep{zeldovich:1970b}.


\section{The Simulation Models}
\label{sec:gadget}

The simulations have been carried out with an advanced version of the \textsc{Gadget} \citep{springel:2005} code, which uses an entropy-conserving formulation of smoothed particle hydrodynamics (SPH) \citep{springel:2002}. Additionally, it includes prescriptions for radiative cooling, UV background heating, and star formation as well as related feedback processes. The latter are followed with a sub-resolution model for the multi-phase structure of the interstellar medium \citep{springel:2003}. The \magpath simulations were performed using a massively improved version of the code (\textsc{P-Gadget3-XXL}), which includes a hybrid OpenMP/MPI parallelization, efficient memory optimizations, and handling of large amounts of snapshot data. In the following, we will provide an overview of these improvements and the implemented models.

\subsection{Hydrodynamics of the gas}

We use the SPH method to follow the hydrodynamical evolution of the gas \citep[for an excellent review see, e.g.][]{price:2012}. Furthermore, we implemented several of the most modern techniques to improve the accuracy, stability and reliability of our SPH simulations \citep{beck:2015}. Thereby, we follow the hydrodynamics of the gas with a time-dependent low-viscosity scheme \citep{cullen:2010}, which we couple together with a higher-order \citep{hu:2014} Balsara switch \citep{balsara:1995}.  Additionally, we use a time-dependent variant of an artificial conductivity scheme \citep{price:2012,tricco:2013} and a time-step limiting particle wake-up scheme as proposed by \cite{saitoh:2009} and first used in the \textsc{Gadget} code by \cite{pakmor:2012}. For physical heat transport we use isotropic thermal conduction with a conduction coefficient of $\kappa=1/20$ in regard to the classical Spitzer value \citep[see][ but also \citealp{arth:2014} for more details on that choice]{dolag:2004}. Finally, we use the Wendland $C^{6}$ kernel functions \citep{dehnen:2012} with 295 neighbors in three dimensions. Our entire hydrodynamical toolbox is presented and its performance analyzed in detail by \citet{beck:2015}. We only give a short example of these  improvements in the following.

\begin{figure}
    \begin{center}
    \includegraphics[width=0.475\textwidth]{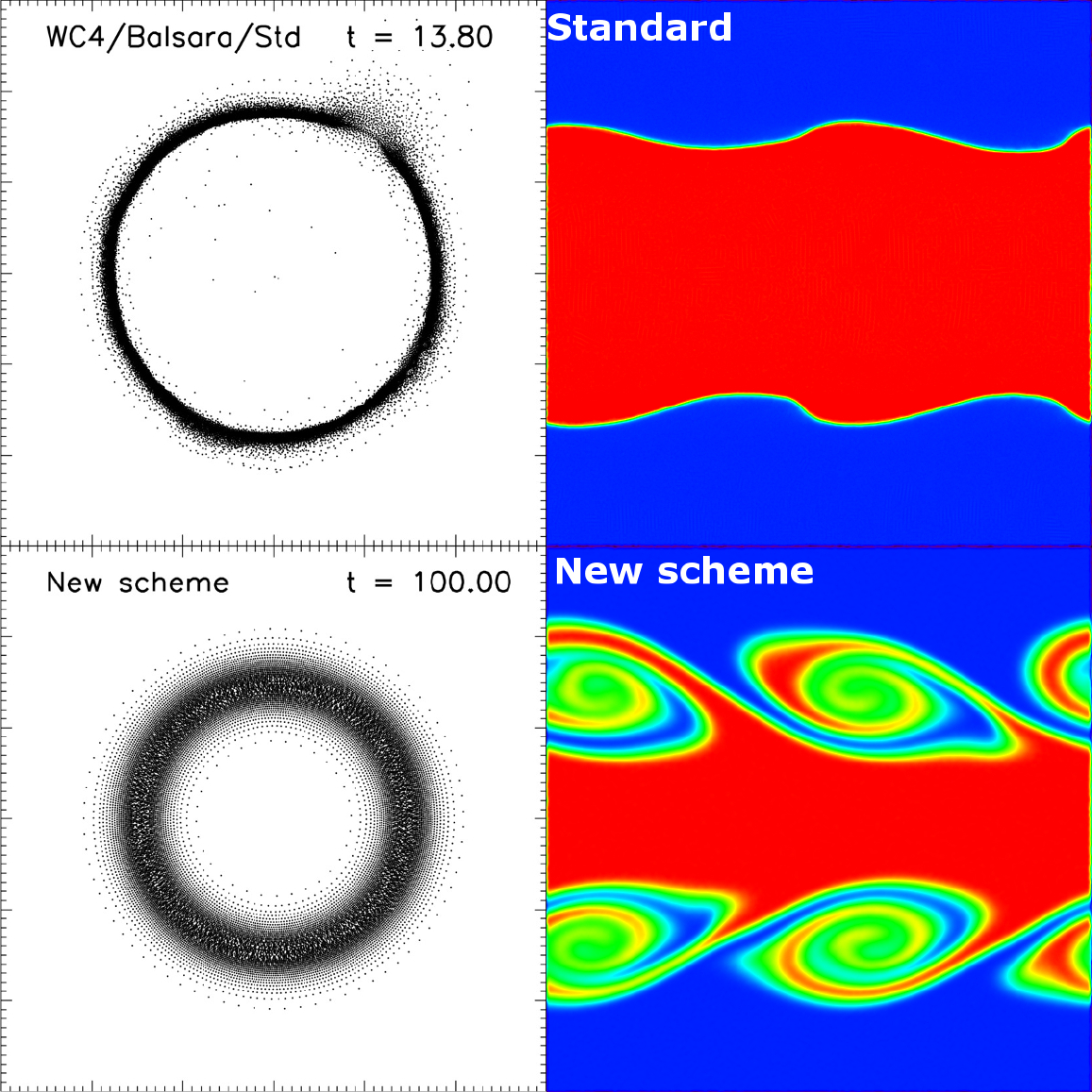}
    \caption{Results of the Keplerian Ring test (\textit{left column}) and the Kelvin-Helmholtz instability test (\textit{right column}). The \textit{upper row} shows the performance of the old SPH scheme, while the performance of the new scheme presented by \citet{beck:2015}, which was used to carry out the \magpath simulations, is shown in the \textit{lower row}. In the Keplerian Ring test particles are orbiting a central point mass, and it is a useful test for investigations of galactic disk rotation and stability. Historic SPH (top left panel) contains too much artificial viscosity and the ring becomes unstable after two dynamical times ($T=2\pi$). In contrast, our updated SPH scheme (bottom left panel) is able to preserve the stability of the ring for long times. Additionally, the improved prescriptions of SPH allow the Kelvin-Helmholtz instability to develop and form prominent roll-ups (bottom right panel), to initiate perturbations between two shearing layers. Historic SPH again is unable to perform fluid phase mixing and development of the instability (top right panel).}
    \label{fig:magneticum_hydro}
    \end{center}
\end{figure}

This strongly improves the the Keplerian Ring \citep{cartwright:2009} test, with the results shown in the left column of Fig.~\ref{fig:magneticum_hydro}, to illustrate the performance of the improved SPH prescriptions. 20 000 particles of equal masses are set up sampling a two-dimensional ring with a Gaussian surface density profile, with a peak at radius $R=15.0~\mathrm{kpc}$ and a standard deviation of $\sigma=2.0~\mathrm{kpc}$. For numerical reasons we initialize the distribution in concentric shifted circles and not in a random fashion. The particles are set on Keplerian orbits with a rotation period of $T=2\pi$ around a central $10^9\,M_\odot$ point mass. We choose the sound speed orders of magnitudes smaller than the orbital velocity to ensure thermal stability of the ring. The upper panel shows the results of the old standard SPH scheme, where the amount of artificial viscosity is too large. This results in a numerically induced transport of angular momentum and the development of a run-away instability, which causes a break-up of the entire ring structure (upper right corner of the ring in this case). The results from the new improved SPH scheme are shown in the lower panel. No break up of the ring structure appears, as the amount of artificial viscosity is significantly reduced and only applied where it is necessary to capture shocks. Thus, shear flows or rotating objects are stable for more dynamical times, which is for example important for the stability and rotation of galactic disks. 

In addition, we test the performance of the improved SPH prescriptions in the Kelvin--Helmholtz instability test \citep{read:2010}. For this, we set up 1\,548\,288 particles of equal masses using a cubic lattice in a three-dimensional periodic box with dimensions $\Delta{x}=256~\mathrm{kpc}$, $\Delta{y}=256~\mathrm{kpc}$, and $\Delta{z}=16~\mathrm{kpc}$, which is centered around $0,0,0$. In the central half of the box ($|y|<64~\mathrm{kpc}$) we initialize 512\,000 particles with a density of $\rho_1=6.26\times10^{3}\,M_\odot/\mathrm{kpc}^3$, a temperature of $T_1=5\times10^{6}~\mathrm{K}$, and a velocity in $x$-direction of $v_1=-40~\mathrm{km}/\mathrm{s}$. In the outer half of the box ($|y|>64$) we initialize 1\,036\,288 particles with a density of $\rho_2=3.13\times10^{3}\,M_\odot/\mathrm{kpc}^3$, a temperature of $T_2=10^{7}~\mathrm{K}$, and a velocity in $x$-direction of $v_2=+40~\mathrm{km}/\mathrm{s}$. To trigger the instability, we perturb the velocity in $y$-direction with an exponentially damped sine mode of amplitude $4~\mathrm{km}/\mathrm{s}$ and wavelength $128~\mathrm{kpc}$. The results of the test are shown in the right column of Fig.~\ref{fig:magneticum_hydro}. The top panel again shows the results from the old standard SPH scheme, where the mixing between fluid phases and the development of the prominent Kelvin--Helmholtz roll-ups is suppressed, giving a physically incorrect numerical solution. The lower panel of Fig.~\ref{fig:magneticum_hydro} shows the results from the new improved SPH scheme, where the conduction scheme and the low-viscosity scheme promote fluid phase mixing, leading to the formation of the Kelvin--Helmholtz roll-ups.

The formation and evolution of galactic disks in the \magpath simulations is a direct result of the improved hydrodynamical method. A proper description of fluid phase mixing prevents the formation of artificial cold blobs of gas, which can be falsely interpreted as galactic structures, but are completely of a numerical origin. A proper description of artificial viscosity promotes the formation and subsequent stability of galactic disks and also allows to track turbulent motions in the large-scale structure such as galaxy clusters.

\subsection{Cooling, Star Formation, Stellar Feedback, and Chemical Enrichment}
\label{sec:starform}

Cooling, star formation, and the chemical enrichment through feedback from the stars are important ingredients in simulating galaxy formation. In our simulations, radiative cooling rates are computed by following the procedure presented by \citet{wiersma:2009}. We account for the presence of the cosmic microwave background (CMB) and of ultraviolet (UV)/X-ray background radiation from quasars and galaxies, as computed by \citet{haardt:2001}. The contributions to cooling from each one of 11 elements (H, He, C, N, O, Ne, Mg, Si, S, Ca, Fe) have been pre-computed using the publicly available CLOUDY photo-ionization code \citep{ferland:1998} for an optically thin gas in (photo-)ionization equilibrium.  

In the multiphase model for star formation \citep{springel:2003}, the ISM is treated as a two-phase medium where clouds of cold gas form through cooling from the hot gas whenever gas particles are above a given threshold density. The star-formation within the sub-grid model is controlled by a star-formation timescale of 1.5~Gyr at the threshold.  
Different than in other simulations, each of our gas particles can spawn up to four star particles. This spawning allows a more smooth description of the otherwise stochastic star-formation treatment and only gradually lowers the gas particle mass and resulting in star and gas particles having a large variety in masses. This splitting traces slightly better the circumstance that molecular clouds in reality convert only about 10\%, but never more than 60\% of their mass into stars \citep[e.g.][]{chevance:2023}. Consequently, a single gas particle containing the molecular cloud should not be entirely consumed by a single star forming event, but instead remnants of its particular metal distribution should remain within the ISM -- which we account for by allow every gas particle to spawn up to four generations of stars. This also improves the ability of the underlying description to follow multiple enrichment events.

Each of our star particles represents an ensemble of stars with a shared age and metallicity, and the internal masses of the stellar population of our star particles are distributed according to a Chabrier initial mass function (IMF; \citealp{chabrier:2003}). For all stars, lifetime functions according to \citet{padovani:1993} are assumed, and the stellar particle suffers mass loss according to these lifetime functions as the more massive stars die first. We assume a binary fraction of 7\% following \citet{greggio:1983} and \citet{matteucci:1986} for stars between 3 and $16\,M_\odot$ \citep[see][]{tornatore:2007} mass loss returned to neighboring gas particles from AGB and SNIa, while SNII are still treated in the instant recycling approximation and only affect the according star-forming particles. Note that this also leads to decreasing mass of the stellar tracer particles, which in case of a Chabrier IMF can account to almost 50\% of the initial mass of the stellar population formed at early times.

Gas within the multiphase model is heated by feedback from SNII and which is calculated directly from the fraction of massive stars for the given IMF. This feedback can evaporate the cold clouds and therefore can regulate the star-formation within the multi-phase treatment of star-forming particles. The stellar particles continuously evolve their stellar population and intermediate and low-mass stars on the asymptotic giant branch (AGB) distribute continuously mass and metals to neighboring gas particles. The released energy by SNII ($10^{51}\,\mathrm{erg}$) is modeled to trigger galactic winds with a mass loading rate being proportional to the star formation rate (SFR) to obtain a resulting wind velocity of $v_{\mathrm{wind}} = 350~\mathrm{km}/\mathrm{s}$. In addition, supernovae type I (SNI) are modeled following \citet{matteucci:2003} and \citet{tornatore:2007}. 

The detailed  prescription of chemical evolution follows \citet{tornatore:2007}, where metals are produced by SNII, SNIa, and AGB stars. Metal and energy release are coupled to the initial metalicity of the stellar populations and follow the different masses of the progenitor stars to properly account for mass-dependent lifetimes, with a lifetime function according to \citealp{padovani:1993}. Metallicity-dependent stellar yields are incorporated following \citet{woosley:1995} for SNII, \citet{vandenHoek:1997} for AGB stars, and \citet{thielemann:2003} for SNIa. Note that throughout the performance of the simulation suite, tables were extended for more metals, as described by \citet{dolag:2017}.

\subsection{Black Hole physics}

\subsubsection{The Fiducial \magpath BH Model}
\label{sec:bhmodel}

Most importantly, our simulations also include a prescription for BH growth and for feedback from active galactic nuclei (AGN) based on the model presented by \citet{springel:2005feedback} and \citet{diMatteo:2005} including the same modifications as the study of \citet{fabjan:2010} and some new, minor changes. As for star formation, the accretion onto BHs and the associated feedback adopts a sub-resolution model. BHs are represented by collision-less ``sink particles'' that can grow in mass by accreting gas from their environments, or by merging with other BHs.  

The gas accretion rate $\dot{M}_\bullet$ is estimated by using the Bondi-Hoyle-Lyttleton approximation \citep{hoyle:1939,bondi:1944,bondi:1952}: 
\begin{equation}\label{Bondi}
    \dot{M}_\bullet = \frac{4 \pi G^2 M_\bullet^2 \alpha   \rho}{(c_s^2 + v^2)^{3/2}}, 
\end{equation}
where $\rho$ and $c_s$ are the density and the sound speed of the surrounding (ISM) gas, respectively, $v$ is the velocity of the black hole relative to the surrounding gas and $\alpha$ is a boost factor for the density, typically set to $100$ following \citet{springel:2005feedback}, unless a more detailed description as introduced by \citet{booth:2009} is used. This boost factor accounts for the fact that in cosmological simulations we can not resolve the intra-cluster medium (ICM) properties within the vicinity of the BH, where higher resolution would result in greater turbulence and cooling funneling gas more efficiently toward the center \citep{gaspari:2013}. The BH accretion is always limited to the Eddington rate (maximum possible accretion for balance between inward directed gravitational force and outward directed radiation pressure): 
\begin{equation}
    \dot{M}_\bullet = \min(\dot{M}_\bullet, \dot{M}_{\mathrm{edd}}).
\end{equation}
Note that the detailed accretion flows onto the BHs are unresolved, we can only capture BH growth due to the larger scale gas distribution, which is resolved. 

Once the accretion rate is computed for each BH particle, its mass continuously grows. To model the loss of this accreted mass from the gas particles, a stochastic criterion is used to select the surrounding gas particles from which the mass is drawn. Unlike in \citet{springel:2005feedback}, in which a selected gas particle contributes to accretion with \textit{all} its mass, we include the possibility for a gas particle to accrete \textit{only with a slice of its mass}, which corresponds to 1/4 of its original mass. This way, each gas particle can contribute with up to four generations of BH accretion events, thus providing a more continuous description of the accretion process, while also retaining a smoother evolution (metal enrichment, heating) of the gas around the black hole. 

The total released energy $\dot{E}$ is related to the BH accretion rate by
\begin{equation}\label{Lrad}
    L_{\mathrm{rad}} = \epsilon_{\mathrm{r}} \dot{M}_\bullet c^2,  
\end{equation} 
where $\epsilon_{\mathrm{r}}$ is the radiative efficiency, for which we adopt a fixed value of 0.2. Here we are using a slightly larger  value than the one commonly assumed (=0.1) for a radiatively efficient accretion disk onto a non-rapidly spinning BH according to \citet{shakura:1973} (see also \citealp{springel:2005, diMatteo:2005}). Instead, given the resolution of the underlying simulation, we follow the observations for the high-mass end of BHs by \citet{davis:2011}.

We assume that a fraction $\epsilon_{\mathrm{f}}$ of this energy is thermally coupled to the surrounding gas so that 
\begin{equation}
    \dot{E}_{\mathrm{f}} = \epsilon_{\mathrm{r}}\epsilon_{\mathrm{f}}\dot{M}_\bullet c^2 
\end{equation} 
is the rate of the energy feedback. $\epsilon_{\mathrm{f}}$ is a free parameter and typically set to $0.15$ (as usually done in simulations that follow the metal depending cooling function, see for example \citealp{booth:2011}). The energy is distributed kernel weighted to the surrounding gas particles in an SPH like manner. 

\begin{figure}
    \includegraphics[width=1.0\columnwidth]{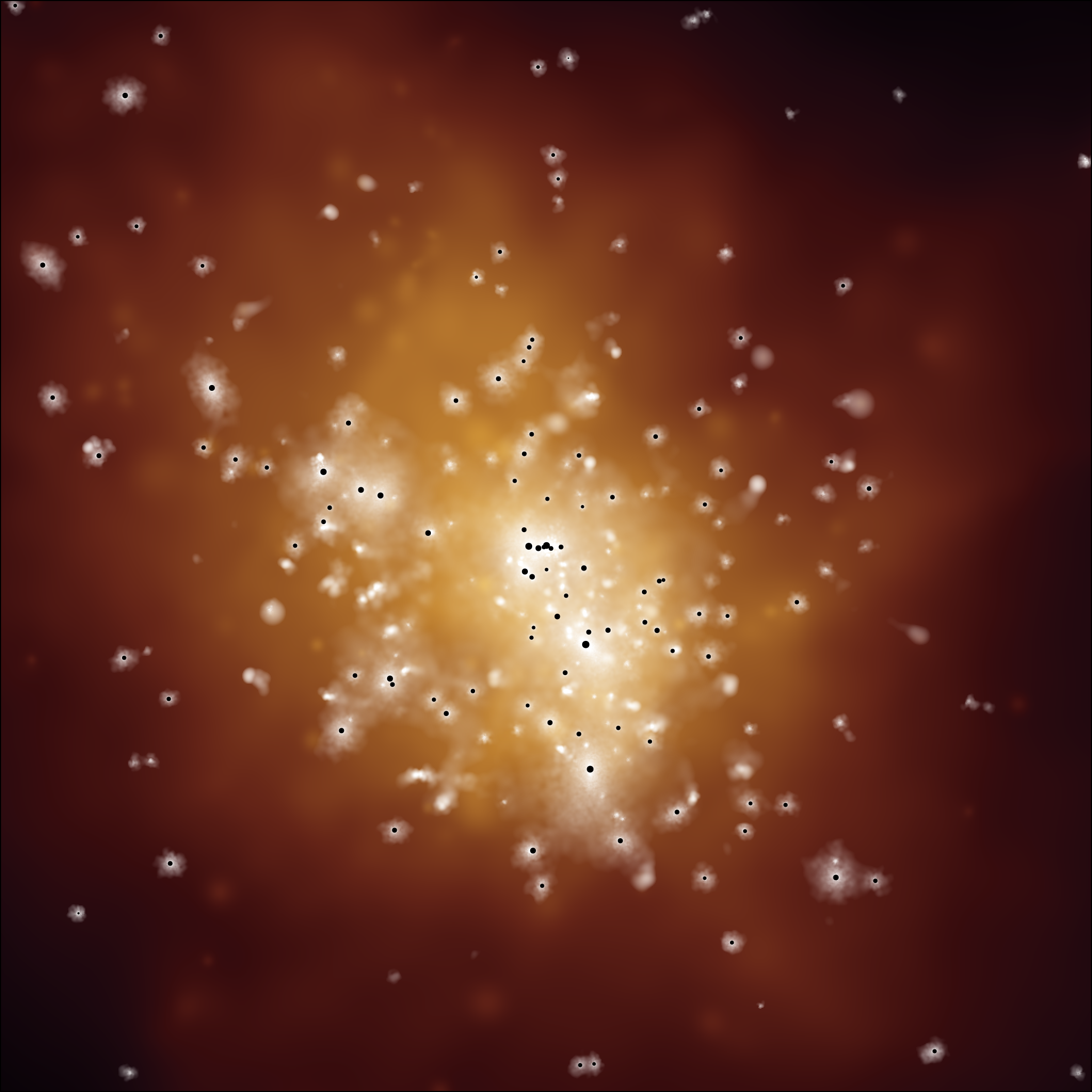}
    \caption{The 19th most massive galaxy cluster from the \Bthr simulation at $z=0$. The gas mass is visualized with the background orange colors and the stellar component is overplotted according to its magnitude in the r band filter. All BHs are shown as the black markers, where the marker radii are scaled logarithmically with the BH mass. The depicted region is a cube with side length 1.86~cMpc, which roughly corresponds to a third of the virial diameter. As can clearly be seen, despite the fact that the black holes are not pinned in \magpath, they reside well within the galaxies even in such a strongly interacting cluster.}
    \label{fig:bh_in_sub}
\end{figure}

Additionally, we incorporated the feedback prescription according to \citet{fabjan:2010}: we account for a transition from a quasar- to a radio-mode feedback (see also \citealp{sijacki:2007}) whenever the accretion rate falls below an Eddington-ratio of 
\begin{equation}
    f_{\mathrm{edd}} := \dot{M}_{\mathrm{r}}/ \dot{M}_{\mathrm{edd}} < 10^{-2}.
\end{equation}
During this radio-mode feedback we assume a 4 times larger feedback efficiency than in the quasar mode. This way, we want to account for massive BHs, which are radiatively inefficient (having low accretion rates), but which are efficient in heating the ICM by inflating hot bubbles in correspondence with the termination of AGN jets. The total efficiency thereby is basically 0.1 (as $\epsilon_{\mathrm{r}}\cdot\epsilon_{\mathrm{f}}\cdot4=0.2\cdot0.15\cdot4=0.12$), as suggested by
\citet{churazov:2005}.

Note that we also, in contrast to \citet{springel:2005feedback}, modify the mass growth of the black hole by taking into account the feedback, i.e.
\begin{equation}
    \Delta M_\bullet = (1-\epsilon_{r})\dot{M}_\bullet \Delta t.
\end{equation}
Furthermore, we introduced some additional, technical modifications of the original implementation, which we will now summarize:  
\begin{enumerate}
    \item One difference with respect to the original implementation by \citet{springel:2005feedback} concerns the seeding of BH particles. In the implementation by \citet{springel:2005feedback}, BH particles are seeded in a halo whenever it first reaches a minimum (total) friends-of-friends (FoF) halo mass, where the FoF is performed on the dark matter particles only. In order to guarantee that BHs are seeded only in halos representing clearly resolved galaxies, where sufficient star formation took place, our implementation performs a FoF algorithm on star particles, grouping them with a linking length of about 0.05 times the mean separation of the DM particles.\footnote{Note that this linking length is thus much smaller than the usually used values of 0.15--0.20 to identify virialized halos.}

    In the simulations presented here, a total stellar mass of roughly $10^{10}\,h^{-1}M_\odot$ is needed (corresponding to a couple of hundreds of star particles) for a halo to be seeded with a BH particle (starting with a seed mass of $2\times10^5\,h^{-1}M_\odot$). While the BH then grows very fast until it reaches the stellar-mass--BH-mass relation, this recovers the BH feedback within the galaxies that would have been present if the resolution had allowed to seed BHs earlier. This also avoids imprinting any stellar-mass--BH-mass relation from the beginning. Finally, we choose the seeded BHs at the position of the star particle with the largest binding energy within the FoF group, instead of at the dark matter particle with the maximum density, as originally implemented.

    \item In the original implementation by \citet{springel:2005feedback}, black holes are forced to remain within the host galaxy by pinning them to the position of the particle found having the minimum value of the potential among all the particles lying within the SPH smoothing length computed at the BH position. Within a cosmological context an aside effect of this criterion is that, due to the relatively large values of SPH smoothing lengths, a BH can be removed from the host galaxy whenever it becomes a satellite, and is spuriously merged into the BH hosted by the central halo galaxy. We have relaxed this criterion and do not apply any pinning of the BH particles to the minimum potential within the smoothing length. 

    To avoid that the BH particles wander away from the center of galaxies by numerical effects, we take several measures in addition to the original implementation of the BH treatment: first, we enforce a stricter momentum conservation within the implementation of gas accretion by forcing momentum conservation for the smooth accretion of the gas and then do not model any momentum transfer when swallowing gas.\footnote{Note that otherwise one would statistically account for the momentum transfer of accreted gas twice.} Additionally, we implemented the conservation of momentum and center of mass when two BH particles merge.\footnote{Note that in the original scheme the merged BH have had the position and velocity of the BH with the smaller particle ID.}
\end{enumerate}

Moreover, in contrast to the original implementation, we have included explicitly a dynamical friction force, which is switched on unless the underlying simulation has a high enough resolution so that the cosmological simulations can numerically resolve dynamical friction reasonably well. To estimate the typical friction force induced onto a BH particle, we use the following approximation of the Chandrasekhar formula \citep{chandrasekhar:1943}:
\begin{equation}
   F_\mathrm{df} = -4 \pi \left(\frac{GM_\bullet}{v}\right)^2\rho\ln(\Lambda) \left(\erf(x)-\frac{2x}{\sqrt{\pi}}e^{-x^2}   \right)\frac{\vec{v}}{v}, 
\end{equation}
where $G$ is the gravitational constant and $M_\bullet$ is the mass of the BH. The local density $\rho$ in the vicinity of the black hole as well as for the relative velocity $\vec{v}$ is calculated using only the stellar and the dark matter components around the black hole. The Coulomb logarithm is calculated as
\begin{equation}
    \ln(\Lambda) = \ln\left(\frac{R v}{G M_\bullet}\right)
\end{equation}
and $x=v\sqrt{2}/\sigma$, where we estimate $\sigma$ as one third of the maximum circular velocity of the hosting sub-halo and for $R$ (as typical size of the system) we use the half-mass radius of the sub-halo hosting the BH. The parameters of the hosting sub-halo for each BH particle are updated every time \subfind is executed on-the-fly.   

This way a BH particle remains within the host galaxy, even if it becomes a satellite of a larger halo and, compared to the original scheme, we are able to track BHs also in satellite galaxies in cluster environments. When the BHs are not placed artificially at the minimum of the potential, of course, there is no guarantee (due to numerical noise, 2 body scattering or when two BHs are merging) that black hole particles always stay exactly at the local potential minimum. But due to the above handling of the dynamical friction, with evolving time during the simulation, BHs sink towards the minimum potential and typical displacements from the true potential minimum are smaller than the effective gravitational softening and therefore, orders of magnitude smaller than the typical smoothing radius used for estimating the parameters in the accretion model or for distributing the feedback energy. Therefore, they do not play any significant role for the behavior of the model, aside from preventing spurious merging of a satellite black hole onto the host.

Fig.~\ref{fig:bh_in_sub} shows a visualization of the 19th most massive galaxy cluster from the \Bthr run at $z=0$ from a cube with 1.86~cMpc side length, focusing on the most massive cluster forming. The gas density is shown in the background in orange colors, and the stellar component in the r-band is overplotted in white. The position of all BHs within this extracted region are marked as black points with sizes according to their mass, which nicely reflect the ability of our implementation to keep the BHs at the center of both the central and satellite galaxies. Note that further improvements on the handling of dynamical friction lead to even better results on the BH particle positions \citep[see discussion by][]{2024A&A...692A..81D}.

\subsubsection{Advanced Black Hole Model}
\label{sss:abhm}

For a subset of the simulations, namely a version of \Bthhr and the \Bthuhr run, an advanced BH model was used. That particular model and the resulting differences between the standard model and the new model using \Bthhr are described in detail by \citet{steinborn:2015}, and we here only give a short summary of that model and refer the reader to \citet{steinborn:2015,steinborn:2016,steinborn:2018} for more details. 

The main differences concern black hole mass growth as well as feedback. First, the accretion following Eq.~\ref{Bondi} is performed separately for hot and cold gas (split by a temperature cut of $T=5\times10^{5}\,\mathrm{K}$), which allows for differing boost factors of $\alpha=10$ and $\alpha=100$, respectively. The hot phase better matches the assumptions of Bondi-Hoyle-Lyttleton accretion (adiabatic, isotropic sphere of gas), while \citet{gaspari:2013} show that cold streams feed the black hole much more efficiently and thus require a higher boost factor when they are unresolved in cosmological simulations. This split has the added benefit of inherently distinguishing a cold, rapidly rotating gas disk in star-forming galaxies from the hot, diffuse medium of a massive elliptical galaxy, instead of smoothing over both phases and applying a common boost factor. 

Second, the transition between quasar and radio mode feedback is smoothed (instead of being a step function of the Eddington-ratio, as is typical -- see Sec.~\ref{sec:bhmodel}), and coupled to both the black hole mass as well as the mass accretion rate following observational findings \citep{steinborn:2015}. Combined, these changes result in a much more rapid initial growth of the black holes at high redshifts \citep[see also Sec.~\ref{sec:bhmf}]{steinborn:2015}, and, as shown by \citet{kimmig:2025} and \citet{remus:2025}, this model is extremely successful in reproducing the quenched fractions at redshifts around $z=4$, an issue that other simulations consistently fail at \citep[see][for details]{remus:2025}. Therefore, in this work we also employ the highest resolution simulation performed using the model by \citet{steinborn:2015}, namely \Bthuhr, and thus it is marked by the teal colors instead of blue and named \uhr-S throughout this work. As the high-redshift Universe turns out to be a rather crucial testbed for the employed physics in simulations, especially the AGN prescriptions, the differences between this model and the fiducial runs in terms of the scaling relations are especially of interest and will be highlighted throughout this work (see e.g. Sec.~\ref{sec:numbdens} and Sec.~\ref{sec:scale3}). They demonstrate the importance of testing scaling relations against observations over a broad range of redshifts to refine our understanding of the physical processes still missing in the simulations to fully describe and reproduce the Universe.

\subsubsection{Black Hole Model limitations}
\label{sss:bhml}

Despite the simplicity and the various limitations of such a description of the evolution of black holes and the associated feedback, this basic model in general reflects a reasonably fair description of an effective feedback model for the ICM. Therefore, many of the current cosmological simulations are using a similar basic description, however with various modifications of the details. Our choices of the detailed realization and implementation of the AGN model and the choice of parameters are based on various previous simulations \citep{2008ApJ...687L..53P,fabjan:2010,2010MNRAS.406..936P,2013MNRAS.431.1487P,2015ApJ...813L..17R,2016A&A...596A.101P} which investigated the impact of AGN feedback on various properties of galaxy clusters. However, the different choices of parameters or the various modifications of the detailed descriptions of the underlying physical processes can lead to large differences in the properties of galaxies, CGM, ICM, and the BH and AGN population. While a detailed description of the impact of the different models is outside the scope of this paper, interested readers can find a detailed summary of the different AGN implementations currently used in \citet{2025arXiv250206954V}, and an extensive discussion of their impact on the predicted AGN population in \citep{2022MNRAS.509.3015H}. Additionally, the CAMELS project \citep{2021ApJ...915...71V} offers extensive comparisons between the different models.

It has also to be emphasized, that AGN feedback operates over a very wide range. This is in both, how real AGN operates in large values spawned in accretion rates, often then simple distinguished as quasar or radio mode, as well as the large spread how and where the feedback energy is deposited. Later challenge all attempts of modeling these processes in large scale, cosmological simulations, especially when jets are involved. Here, the models are often adjusted at the given resolution to just effectively work. This sometimes means that depositing the energy in a simulation at the wrong scale is effectively compensated by the different amount of energy deposited. A detailed discussion what this means for the \magpath AGN model can be found in \citet{gonzalez_villalba:2025}. 

\subsection{Post-Processing}
\label{sec:subfind}

We use \subfind \citep{springel:2001,dolag:2009} to define halo and sub-halo properties. \subfind identifies substructures as locally overdense, gravitationally bound groups of particles. Starting with a halo identified through the Friends-of-Friends algorithm with a linking length of $b=0.16$, a local density at the position of every particle is estimated by summing up the individual density fields of all particle species estimated by the standard SPH kernel method. Then, starting from isolated density peaks, additional particles are added in sequence of decreasing density. Whenever a saddle point in the global density field is reached that connects two disjoint overdense regions, the smaller structure is treated as a substructure candidate, followed by merging the two regions. All substructure candidates are subjected to an iterative unbinding procedure with a tree-based calculation of the potential. These structures can then be associated with galaxies and their integrated properties (like stellar mass or star-formation rate) can be calculated, as we will discuss in the following.

For each halo, \subfind finds one most massive sub-halo, which defines the particles allocated to the host halo itself. We will, in the following, call wherever required these most massive sub-halos the \textit{main} of a given halo, and the stellar body within that \textit{main} the \textit{main galaxy}. Note that we neglect any splitting of the brightest cluster galaxies (BCGs) from their intra-cluster light (ICL) through this procedure here, and that such splits will only occur in the post-processing as discussed in Sec.~\ref{sec:localprops}.
All other sub-halos within a halo, that is, all sub-halos that are not defined as the \textit{main}, are defined as satellite galaxies of that particular halo, and we will refer to those sub-halos, if necessary, as \textit{satellites}. Note that due to the nature of how \subfind cuts the satellites from the mains, the satellite galaxies contain less dark matter or gas than they might originally have had. As such, throughout this paper we will show scaling relations involving gas properties or dark matter properties always only for the mains and not for the satellites if not stated otherwise.

\subsubsection{Global properties}
\label{sec:globalprops}

These global halo properties are calculated on-the-fly from the \subfind algorithm directly.
For every halo we determine its mass and size using different spherical overdensity criteria,
corresponding to the most commonly used values as listed in Table~\ref{tab:over}.
The virial radius and mass are based on the spherical top-hat collapse model,
according to which a collapsed overdensity in an Einstein--de-Sitter Universe
will virialize at a mean density of $1+\Delta=179$ times the critical density
at the time of the collapse of the structure \citep[e.g.][]{eke:1996}. For our case with a $\Lambda$CDM cosmology, we follow the approximation of the collapsing top-hat overdensity by \citet{bryan:1998} as 
\begin{equation}
    \Delta_\mathrm{vir}(z)=(18\pi^2 + 82 x - 39 x^2)/\Omega_m(z),
\end{equation}
with $x=x(z)=\Omega_m(z)-1$, and $\Omega_m(z)$ the total matter density at the given redshift. The factor $\Delta_\mathrm{vir}(z)$ goes from a value of around $177$~at $z=10$ to $97$~at $z=0$. 

Meanwhile, the other criteria are variants using different overdensity thresholds that are fixed across cosmic time. In all cases, corresponding radii and (total) masses are determined simultaneously
such that the mean enclosed density $\bar{\rho}_\mathrm{halo}$ satisfies
the given threshold criterion. For each criterion, in addition to the total mass we also extract the gas mass $M^\mathrm{gas}$ and the stellar mass $M^{*}$, as well as the mean gas temperature $T$ of the gas enclosed within the halo radius from the \subfind output, where we use the following in this work for $R^{500}_\mathrm{crit}$:
\begin{itemize} 
    \item[$\bullet$] $\mathbf{M_\mathrm{500}^*}$, the full stellar mass within $R^{500}_\mathrm{crit}$ including the main galaxy and the satellites,
    \item[$\bullet$] $\mathbf{M_\mathrm{500}^\mathrm{gas}}$, the total gas mass within $R^{500}_\mathrm{crit}$,
    \item[$\bullet$] $\mathbf{T_\mathrm{500}}$, the mean gas temperature within $R^{500}_\mathrm{crit}$,
    \item[$\bullet$] $\mathbf{Y_\mathrm{500}}$, the dimensionless SZ signal within $R^{500}_\mathrm{crit}$,
\end{itemize}

Other global gas properties presented in this work have been computed through direct post-processing of the simulation data:
\begin{itemize} 
    \item[$\bullet$] $\mathbf{L_\mathrm{500}^\mathrm{SXR}}$, the soft X-ray luminosity of gas within $R^{500}_\mathrm{crit}$,
    \item[$\bullet$] $\mathbf{Z_\mathrm{Fe,mw}}$, the mass-weighted gas iron abundance within $R^{500}_\mathrm{crit}$.
\end{itemize}

For the X-ray luminosity, the simulation output was first processed using the {\sc Phox} algorithm \citep{biffi:2012, biffi:2013, vladutescu:2023} to produce discrete X-ray photons from accurate spectral models reflecting the thermodynamic state of a gas particle in the simulation. The rest-frame luminosity in the $0.5-2\,\unit{\keV}$ band (SXR) is then calculated for each halo by selecting photons within a sphere of $R^{500}_\mathrm{crit}$ from the halo center obtained from \subfind. The mass-weighted iron abundance is the ratio of the total iron mass to the total hydrogen mass of the non-star-forming hot (with temperature $T>10^5\,\mathbf{K}$) gas located within a sphere of $R^{500}_\mathrm{crit}$ from the halo center. 

\begin{table}
    \caption{{\normalsize\rule{0pt}{1em}}Commonly used density thresholds for specifying halo radii. The corresponding masses are defined as the total mass contained within that radius, centered around the deepest point of the halo potential.}
    \centering\def\arraystretch{1.5}
    \begin{tabular}{l@{ }l | l}
    \multicolumn{2}{c|}{Definition} & \multicolumn{1}{c}{mean enclosed density}\\\hline
    $R_\mathrm{vir}$,        & $M_\mathrm{vir}$        & $\bar{\rho}_\mathrm{halo}=\Delta_\mathrm{vir}(z)\,\rho_\mathrm{crit}$ \\
    $R^{200}_\mathrm{crit}$, & $M^{200}_\mathrm{crit}$ & $\bar{\rho}_\mathrm{halo}=200\,\rho_\mathrm{crit}$ \\
    $R^{500}_\mathrm{crit}$, & $M^{500}_\mathrm{crit}$ & $\bar{\rho}_\mathrm{halo}=500\,\rho_\mathrm{crit}$ \\
    \end{tabular}
    \label{tab:over}
\end{table}

\subsubsection{Local properties}
\label{sec:localprops}

Local properties of halos are those that are not given directly on the fly by \subfind, but are calculated in post-processing from the particle data, using the \subfind output to locate the halo and its associated stars, gas particles, and BHs as a basis. Here, particles associated with satellites inside a larger halo are excluded, such that the quantities always belong to the object itself, independent of whether they are a main halo or a satellite as introduced above.
For these, we calculate three-dimensional quantities inside spheres, but also projected quantities in 2D. Such two-dimensional quantities are for the most part calculated along the line-of-sight $z$-axis of the box, as this should resemble random projections and is a completely arbitrary choice. Only for the $\lambda_R$ and ellipticity measurements is the edge-on projection of the individual galaxies used.

For all the following local properties, the galaxies are shifted in real and velocity space according to the stellar component to properly center the galaxy since the halo and subhalo positions obtained from \subfind are the position of the particle at the deepest potential point. The centering is performed following \citet{valenzuela:2024}: the spatial center is determined using a shrinking sphere method \citep{power:2003} and the velocity frame of reference is obtained from the mass-weighted mean velocity of all stellar particles except those with the 10\% highest absolute velocities relative to the median of all the considered particles.

For the three-dimensional quantities we use four different radial cuts throughout this work:
\begin{itemize}
    \item[$\bullet$] $\mathbf{R_{0.1}}$, the radius that describes 10\% of the virial radius $R_\mathrm{vir}$ as defined in Sec.~\ref{sec:globalprops}. This radius scales with redshift for a fixed mass.
    
    \item[$\bullet$] $\mathbf{R_{1/2}}$, the radius that contains half of the stellar mass of a halo, calculated using the total stellar mass within 10\% of $R_\mathrm{vir}$, i.e. $R_{0.1}$. This is a more characteristic quantity for galaxies. Note that we use the $R_\mathrm{vir}$ of the main halo also for the satellites, as this is usually much larger, but we only use those particles associated with the satellite by \subfind. This is done to ensure a consistent treatment.
    
    \item[$\bullet$] $\mathbf{3R_{1/2}}$ is three times $R_{1/2}$.
    
    \item[$\bullet$] $\mathbf{R_{30}}$ or $\mathbf{R_{100}}$ is a fixed aperture of $30~\mathrm{kpc}$ or $100~\mathrm{kpc}$, with the former used for galaxies and the latter for central galaxies of galaxy clusters. This is a radius measure that does not adapt with redshift.
\end{itemize}

Using these radial cuts, we calculate the following quantities from the particles assigned by \subfind to the respective main or satellite galaxy:
\begin{itemize}
    \item[$\bullet$] $\mathbf{M^*_{R_{0.1}}}$, $\mathbf{M^*_{3R_{1/2}}}$, $\mathbf{M^*_{30\,kpc}}$, and $\mathbf{M^*_{100\,kpc}}$, the total stellar mass of all stellar particles within the designated radius.
    
    \item[$\bullet$] $\mathbf{M^{cg}_{R_{0.1}}}$ and $\mathbf{M^{cg}_{3R_{1/2}}}$, the total mass of all cold gas particles inside the designated radius, with cold defined here as gas particles with a temperature below $10^5~\mathrm{K}$ or with a non-zero star formation rate as commonly done in simulations that do not resolve cooling through molecular lines \citep[see e.g.][]{katz:1996,maio:2007}.
 
    \item[$\bullet$] $\mathbf{SFR_{R_{0.1}}}$ and $\mathbf{SFR_{3R_{1/2}}}$, the star formation rate of a galaxy inside $R_{0.1}$ or $3R_{1/2}$, respectively, calculated in $M_\odot/\mathrm{yr}$. Here, the current star formation rates of all gas particles within the respective radius are summed up.

    \item[$\bullet$] $\mathbf{M_g}$ and $\mathbf{M_r}$, the absolute AB magnitudes in the SDSS g and r band filters, obtained from the summed fluxes of all the stellar particles within $3R_{1/2}$, where the fluxes are calculated from the stellar population synthesis code by \citet{bruzual:2003}, using their CB07 models assuming a \citet{chabrier:2003} IMF.

    \item[$\bullet$] $\mathbf{j^*_{R_{0.1}}}$ and $\mathbf{j^*_{3R_{1/2}}}$, the stellar specific angular momentum calculated from all stars within the designated radius. We calculate 
    \begin{equation}
       j^* = \frac{\sum_i m_i r_i \times v_i}{\sum_i m_i},
    \end{equation}
    where the sums run over the stellar particles, with the particle mass $m_i$, 3D radial distance $r_i$, and velocity $v_i$, following the description by \citet{teklu:2015}.

    \item[$\bullet$] $\mathbf{Z^*_{R_{0.1}}}$ and $\mathbf{Z^*_{3R_{1/2}}}$, the stellar metallicity calculated from all stars within the designated radius. We calculate the mass-weighted metallicity
    \begin{equation}
       Z^* = \frac{\sum_i m_i Z_i}{\sum_i m_i}, 
    \end{equation}
    where the sums run over the stellar particles, with the metal mass fraction $Z_i$ being the mass of metals above helium divided by the total stellar particle mass for each stellar particle~$i$.
    \item[$\bullet$] $\mathbf{t^*_{R_{0.1}}}$ and $\mathbf{t^*_{3R_{1/2}}}$, the stellar age calculated from all stars within the designated radius. We calculate the mass-weighted age
    \begin{equation}
       t^* = \frac{\sum_i m_i t_i}{\sum_i m_i}, 
    \end{equation}
    where the sums run over the stellar particles, with the age $t_i$ of each stellar particle $i$ being the lookback time at which the stellar particle was formed.
    \item[$\bullet$] $\mathbf{M_{bh}}$ is defined as the mass of the most massive black hole within $3$~stellar half-mass radii. 
\end{itemize}

\begin{figure*}
\begin{overpic}[width=1.0\textwidth]{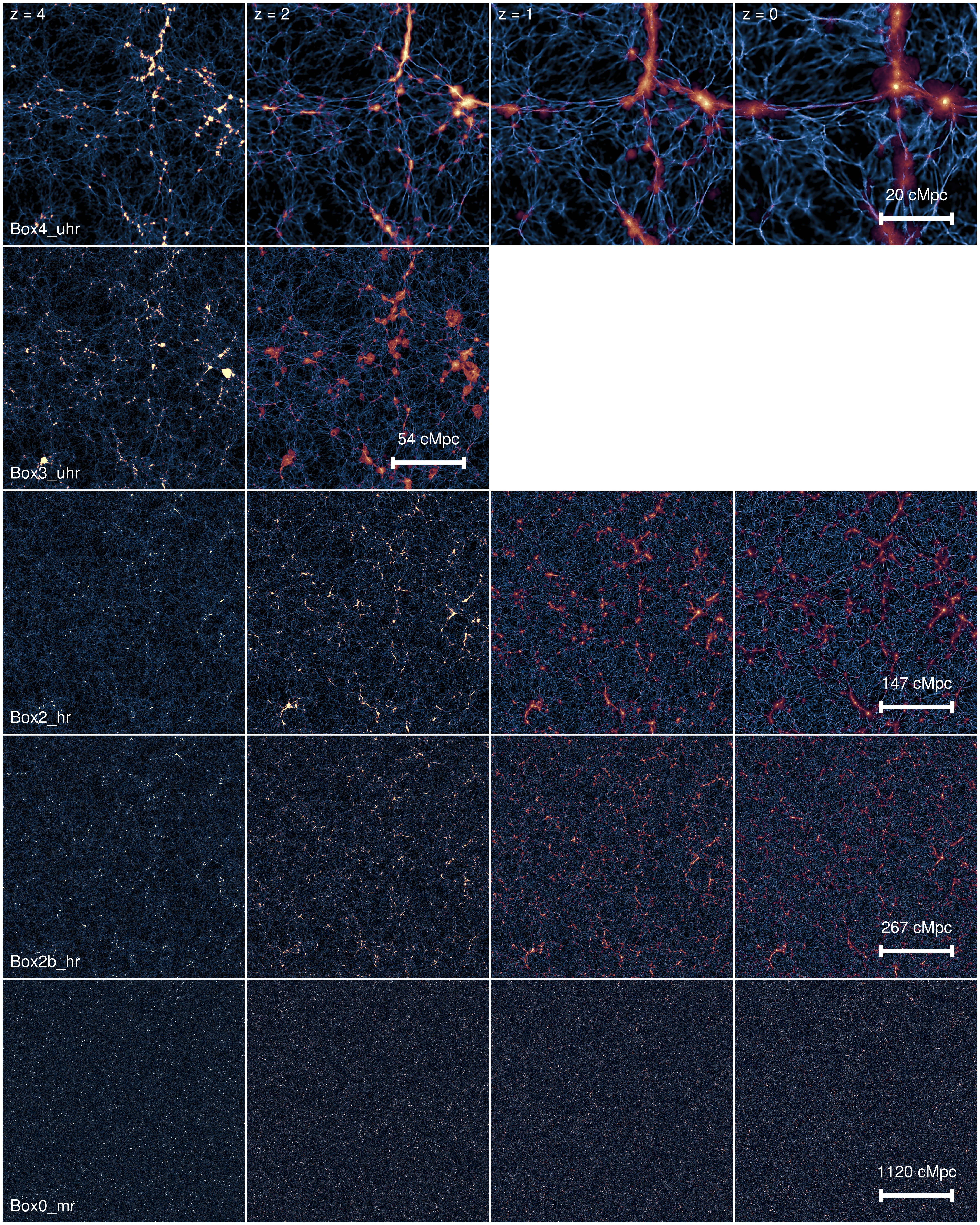}
\put(40.1,69.6){\begin{tabular}{|C{15mm}|C{10mm}|C{10mm}|C{10mm}|C{10mm}|C{10mm}|}
\hline
    & \Bf\newline\uhr & \Bt\newline\hr & \Bt\newline\hr & \Btb\newline\hr & \Bz\newline\mr \\ \hline
$\mathrm{N}_\mathrm{part}$ & $2\times$ \newline $576^3$ & $2\times$\newline $576^3$ & $2\times$\newline$ 1564^3$ & $2\times$\newline$2880^3$ & $2\times$\newline$4536^3$ \\ \hline
$\mathrm{L}_\mathrm{box}$\newline$[h^{-1}\mathrm{Mpc}]$ & 48 & 128 & 352 & 640 & 2688 \\ \hline
$m_\mathrm{*}$\newline$[h^{-1}M_\odot]$ & $1.8$\newline$\times10^{6}$ & $1.8$\newline$\times10^{6}$ & $3.5\times$\newline$10^{7}$ & $3.5\times$\newline$10^{7}$ & $6.5\times$ \newline$10^{8}$ \\ \hline
$\epsilon_\mathrm{*}$\newline$[h^{-1}\mathrm{kpc}]$ & 0.7\hphantom{0} & 0.7\hphantom{0} & 2\hphantom{.00} & 2\hphantom{.00} & 5\hphantom{.00} \\ \hline
\end{tabular}}
\end{overpic}
\caption{Evolution of the large-scale structure in \magpath. Shown is the gas overdensity of the hot gas component (magma colorscale) and the cold gas component (black to blue colorscale) for \Bfuhr (upper most row) down to \Bzmr (lower most row), at four different redshifts $z=4$, $2$, $1$, and $0$, from left to right. We define cold gas as all gas particles with a temperature $T\leq10^4\,\mathrm{K}$, and additionally all particles actively undergoing star formation. Hot gas particles are all particles not belonging to the former group. The thickness of the slices are 1~cMpc for both boxes.}
\label{fig:boxevo}
\end{figure*}

\begin{figure*}
\includegraphics[width=1.0\textwidth]{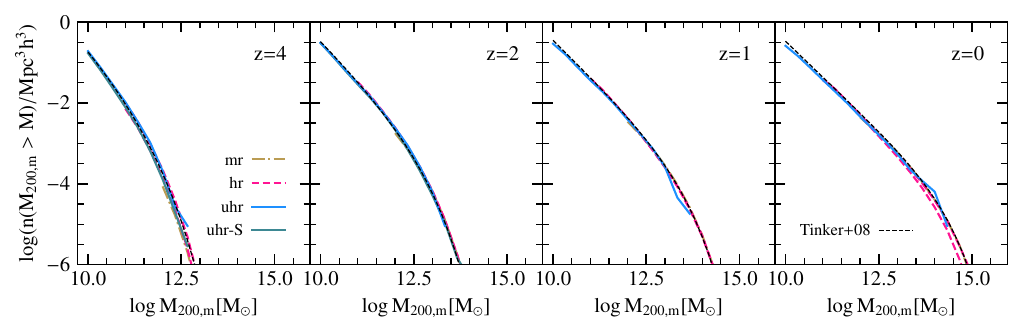}
\caption{The halo mass function in the different \magpath simulations and resolutions, at $z=4$, $2$, $1$, and $0$, from left to right. The virial mass $M_\mathrm{vir}$ is used as the halo mass. Simulations on the \mr level are shown in gold dash-dotted lines (\Bz), simulations on the \hr level are shown in pink dashed lines (\Bt and \Btb), and the fiducial \uhr level simulation is shown in blue solid lines (\Bf). The \uhr level simulation with the advanced BH model is shown as turquoise solid lines (\Bth), labeled as \uhr-S. The black dashed line shows the prediction from the model by \citet{tinker:2008}.}
\label{fig:hmf}
\end{figure*}

For the two-dimensional quantities we use two different radial cuts throughout this work, each determined in the $x$-$y$-plane of the simulation box, where only the particles with 
$z$-coordinates within $R_{0.1}$ along the line-of-sight axis are taken into account (i.e. filtering particles within a cylinder with depth $2R_{0.1}$):
\begin{itemize}
    \item[$\bullet$] $\mathbf{R_{1/2}^{2D}}$, the 2D radius that contains half of the stellar mass of a halo, calculated using the total stellar mass within $R_{0.1}$ (i.e. a cylinder with depth $2R_{0.1}$ and radius $R_{0.1}$). As for the 3D $R_{1/2}$, we use $R_\mathrm{vir}$ of the main halo also for the satellites.
    
    \item[$\bullet$] $\mathbf{3R_{1/2}^{2D}}$ is three times $R_{1/2}^\mathrm{2D}$.
\end{itemize}

Using these 2D radial cuts, we calculate the following quantities from the particles assigned by \subfind to the respective main or satellite galaxy from the $x$-$y$-plane, again only considering particles with $z$-coordinates within $R_{0.1}$ along the line-of-sight:
\begin{itemize}
    \item[$\bullet$] $\mathbf{M^*_{3R_{1/2}^{2D}}}$, $\mathbf{M^*_{R_{1/2}^{2D}}}$, the total stellar mass of all stellar particles within the designated 2D radius.
    
    \item[$\bullet$] $\mathbf{M^{cg}_{3R_{1/2}^{2D}}}$, $\mathbf{M^{cg}_{R_{1/2}^{2D}}}$, the total mass of all cold gas particles inside the designated 2D radius, with cold defined as above for the 3D cold gas masses.
    
    \item[$\bullet$] $\mathbf{\bm{\sigma}^*_{3R_{1/2}^{2D}}}$, $\mathbf{\bm{\sigma}^*_{R_{1/2}^{2D}}}$, the stellar velocity dispersion within the designated 2D radius, calculated as the standard deviation of the $z$-axis line-of-sight velocities of the stellar particles.
\end{itemize}

Finally, for the resolved stellar kinematics parameters, we determined the light-weighted two-dimensional quantities from the edge-on projection of the individual galaxies. For this, we used the method by \citet{valenzuela:2024} to find the best-fitting ellipsoid and rotate the galaxy into the frame of reference given by its shape tensor. For this we calculated the light-weighted 3D shape at $3R_{1/2}$ with the unweighted iterative shape determination method keeping the ellipsoidal volume constant, modified to be weighted by the r-band fluxes of the stellar particles instead of their masses. In the edge-on projection, we compute the following properties:
\begin{itemize}
    \item[$\bullet$] $\mathbf{R_e^{2D,edge}}$, the 2D edge-on effective radius that contains half of the stellar light of a halo, calculated using all the stellar flux in the r-band within $R_{0.1}$ (i.e. a cylinder with depth $2R_{0.1}$ and radius $R_{0.1}$). As for the 2D and 3D $R_{1/2}$, we use $R_\mathrm{vir}$ of the main halo also for the satellites.

    \item[$\bullet$] $\mathbf{\bm{\epsilon}_e}$, the ellipticity $1 - b/a$ given from the ellipse axis ratios at $R_e^\mathrm{2D,edge}$, obtained from the r-band light-weighted 2D shape determined with the same shape method as described above, but applied to only two dimensions (unweighted iterative shape determination method keeping the surface area constant).
    
    \item[$\bullet$] $\mathbf{\bm{\lambda}_{R_e}}$, the 2D projected quantification of how rotation or dispersion dominated a galaxy is within the ellipse of the same surface area as a circle with radius $R_e^\mathrm{2D,edge}$ \citep{emsellem:2007, emsellem:2011}. We calculate $\lambda_{R_e}$ following the approach from \citet{jesseit:2009} and \citet{schulze:2018} by binning the stellar particles in 2D with the centroidal Voronoi tesselation algorithm from \citet{cappellari:2003}, targeting a minimum of 100~stellar particles per cell, and using the following equation:
    \begin{equation}
        \lambda_R = \frac{\sum_k F_k R_k |\overline{V}_k|}{\sum_k F_k R_k \sqrt{\overline{V}_k^2 + \sigma_k^2}},
    \end{equation}
    where the sums run over the Voronoi cells, with the total r-band flux $F_k$, the cell 2D radial distance $R_k$, the mean line-of-sight velocity within the cell $|\overline{V}_k|$, and the line-of-sight velocity dispersion within the cell $\sigma_k$.
\end{itemize}


\section{Scaling Relations Through Cosmic Time I: Mass Functions}
\label{sec:mf}

Structure formation constantly assembles matter into structures of varying size and mass.  Fig.~\ref{fig:boxevo} shows this as the evolution of the cosmic web, from $z=4$ on the left down to $z=0$ on the right, for \Bfuhr in the upper row down to \Bzmr in the lower row. The gas is colored according to its temperature, with yellow to red colors showing hot gas, and blue to white colors showing cold gas, in increasing temperature for both. Here, the cold gas is all gas with temperatures below $10^4\,\mathrm{K}$, while the hot gas has temperatures above $10^4\,\mathrm{K}$. As can be seen for both boxes, filamentary structure of the cosmic web is already visible at $z=4$ and becomes increasingly more pronounced with lower redshifts. Simultaneously, the gas temperature in the filaments slowly increases, but especially in the most massive nodes that evolve the fastest the temperature heats up already at $z=4$, with increasing temperature and extent of hot atmospheres towards low redshifts. This can be seen for all box volumes, despite the different length scales depicted, clearly demonstrating the self-similar nature of our Universe. 

A powerful tool to understand this assemblage and the interplay of the physical processes involved are mass functions. However, mass functions can be obtained from different points of view. In the theoretical structure formation studies, the halo masses play the most crucial role as they indeed map the growth of structures globally through cosmic time the most accurately. On the other hand, these are impossible to observe, and thus observationally the stellar mass functions are the most accessible mass functions. Unfortunately, they are hampered by the fact that stars are made from gas, and only the baryonic mass function should mirror the halo mass function as it is because the stellar mass function is convoluted by the information about the star formation efficiency at different masses and redshifts. Thus, the gas mass function complements these two mass functions, however, it is nearly impossible to observed as this includes gas of different temperatures covering ten orders of magnitude, and thus observing the gas mass function requires completeness over several observational bands. Finally, we can also look at the BH mass function, however, this is notoriously difficult to observe as we can only observe BHs indirectly anyway. We will show the results usually for four different redshifts, $z=4$, $2$, $1$, and $0$
with the \magpath simulations shown in color and the observations and in some cases models shown in black and gray. All quantities presented here are calculated as described in Sec.~\ref{sec:globalprops}. In the following, we will show all four mass functions for the \magpath simulations and, wherever possible, compare to observations or models.

\begin{figure*}
\includegraphics[width=1.0\textwidth]{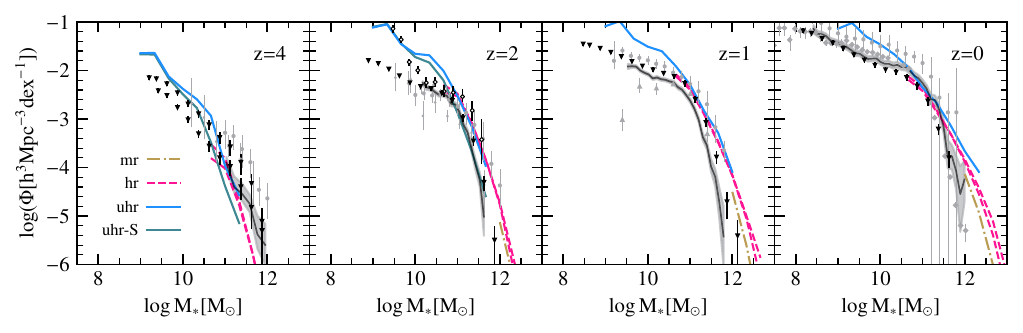}
\caption{The stellar mass function in the \magpath simulations for redshifts $z=4$, $2$, $1$, and $0$. The line colors indicate the different simulations with their respective resolutions as indicated in the legend. The black lines with the grey shaded errors are taken from \citep{muzzin:2013}. Observations from the COSMOS survey \citep{weaver:2022,weaver:2023} are shown with the black downwards pointing triangles. Additional observational data from \citet{perez-gonzalez:2008} (circles), \citet{panter:2004} (diamonds), \citet{marchesini:2009} (tri-down), \citet{bundy:2005} (upwards pointing triangles) are shown with grey symbols.  For z=2 we include measurements from \citep{santini:2012} (blue diamonds)}
\label{fig:smf}
\end{figure*}

\subsection{The Halo Mass Function}
\label{sec:hmf}

Among the most well established scaling relations used to assess structure formation in the Universe is the halo mass function (HMF). This function captures the purely gravitational part of the assembly history of the Universe and is theoretically well understood with analytical models dating back to the Press-Schechter function first presented in the seminal work by \citet{press:1974}. According to this theoretical prediction the shape of the HMF is sensitive only to the underlying cosmology, which is why it is mainly employed in constraining cosmological parameters and deviations from the standard cosmology. Modern cosmological simulations capture the full non-linear evolution of these halos, and more advanced analytical models \citep[e.g.][]{sheth:2001a,tinker:2008} have since improved the precision of our understanding of the gravitational assembly of structures further. While the shape of the low-mass end of the HMF is still not fully understood from a theoretical perspective (being dominated by non-linear gravitational and possibly even baryonic effects), this function can be used as a baseline for numerical simulations. The HMF including baryonic effects have been pioneered in \citet{bocquet:2016}, where updates set of fitting parameters for the Tinker mass function are given for different overdensities and accounting for the baryonic effects. For an even more in-depth analysis of the baryonic effects on the HMF, see \citet{2021MNRAS.500.2316C}, while \citet{2021MNRAS.500.5056R} presents the cosmology dependence from the multi-cosmology set of simulations.

Fig.~\ref{fig:hmf} shows the HMF for the \magpath simulations \Bzmr (gold dash-dotted lines), \Btb and \Bthr (pink dashed lines), and \Bfuhr fiducial BH (blue solid lines) as well as \Bthuhr advanced BH (turquoise solid lines), in comparisons with the fitted model by \citet[][black dashed lines]{tinker:2008}. We used the fitting parameters for $\Delta=200$ in combination with computing the HMF from the simulations with the $M_{200,m}$ masses. As can be seen directly, the HMF is consistent between different resolutions at all redshifts and all box volumes and resolutions, and in excellent agreement with theoretical predictions.

\subsection{The Stellar Mass Function}
\label{sec:smf}

While galaxies assemble their mass and form stars, they are governed by a complex interplay of numerous physical processes, like gas accretion, feedback and stripping processes. These processes also shape how stellar mass is distributed among galaxies. Therefore, the stellar mass function (SMF) is a key metric for understanding and modeling galaxy evolution accurately. Since early measurements of the local SMF in the infrared and near infrared based on the 2MASS survey \citep{cole:2001}, significant progress has been made with regards to sample size, accuracy, and redshift coverage enabling detailed comparisons between models and observations.

Despite the fact that the observationally inferred stellar mass function suffers from systematics in the luminosity and stellar mass determinations, as well as flux limits \citep[see][]{2013MNRAS.436..697B}, it was already realized in early theoretical investigations \citep[e.g.][]{2003ApJ...599...38B} that different feedback processes shape the low- and high-mass ends of the luminosity function differently. Such cosmological simulations use different descriptions of stellar and AGN feedback, leading to different levels of agreement with the observations. While here we concentrate on a qualitative comparison with our implementation, a detailed description of the differences in stellar and AGN feedback descriptions in cosmological simulations, as well as references to various quantitative comparisons between such implementations, can be found in \citet{2025arXiv250206954V}.

Fig.~\ref{fig:smf} shows the comparison of the \magpath SMF obtained from the different resolutions and volumes with observations, at $z=4$, $2$, $1$, and $0$, from left to right. The simulations are shown in colors as indicated in the legend and Fig.~\ref{fig:hmf}. New COSMOS observations by \citet{weaver:2022} and \citet{weaver:2023} are marked as black downward triangles, while older data by \citet[][circles]{perez-gonzalez:2008}, \citet[][diamonds]{panter:2004}, \citet[][tri-down]{marchesini:2009}, and \citet[][upwards triangles]{bundy:2005} are marked in gray. 

\begin{figure*}
\includegraphics[width=1.0\textwidth]{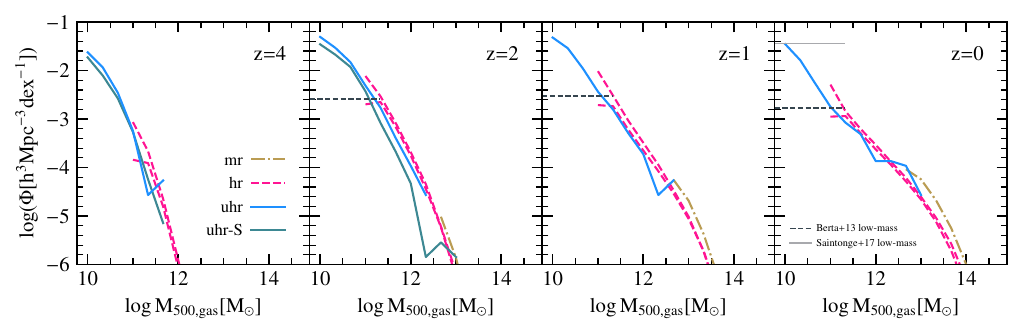}
\caption{The total gas mass function in the \magpath simulations for redshifts $z=4$, $2$, $1$, and $0$, as contained within $R^{500}_\mathrm{crit}$. The line colors indicate the different simulations with their respective resolutions as indicated in the legend. We chose the gas mass within $R^{500}_\mathrm{crit}$ as this is what can be observed with X-ray observations of hot gas, although this is not what is usually observed for galaxies that are dominated by cold gas which is much less extended in its majority than the hot gas component. Note that we do not differentiate between hot and cold gas in this plot. The horizontal lines are from the low-mass end of observational cold gas studies \citep{berta:2013, saintonge:2017} and can only be used to compare the order of magnitude of our low-gas mass halos with respect to the gas mass function.}
\label{fig:gmf}
\end{figure*}

First of all, we find the SMF to be largely consistent across different resolutions and simulation volumes, indicating the subgrid prescriptions to be robust with regards to resolution. The strongest deviations are found for \Bthuhr with the advanced BH model, which leads to slightly less massive galaxies at the high-mass end at around $z=2$ and overall at $z=4$, in better agreement with the observations. It was already shown by \citet{steinborn:2015} that the advanced BH model is slightly better in reproducing the stellar mass functions at high redshifts than the fiducial model. The major difference between the two models, however, becomes evident with respect to the number of quiescent galaxies relative to the star-forming galaxies at high redshift: the advanced BH model simulations are much better in reproducing quiescent galaxies as was shown by \citet{steinborn:2015}, especially at high redshift as shown by \citet{lustig:2023} for $z=2$ and \citet{kimmig:2025} at $z>2$, and which we will discuss in Sec.~\ref{sec:numbdens} in more detail.

Overall, we find agreement with current observational constraints throughout the cosmic evolution, with deviations mainly for the very low-mass end at all redshifts and at low redshifts also for very high stellar masses. While the agreement at the high-mass end around $z=2$ and $z=1$ between simulations and observations is excellent, the simulations show much larger values at $z=0$ than what is observed. Interestingly, the observed stellar mass function at the high-mass end does not seem to evolve strongly between $z=1$ and $z=0$, while the simulations show a continuous growth of the high-mass end. Interestingly, \citet{muzzin:2013} find larger numbers at the high-mass end around $z=0.2-0.5$ than \citet{panter:2004} find at $z=0$. This could be due to our immediate local Universe not harboring many relaxed clusters for which the BCGs had time to assemble more stellar mass from disrupting satellites \citep{kimmig:2025b}, while the large simulation volumes contain plenty of such relaxed galaxy clusters, which could also be the reason for the larger values found by \citet{muzzin:2013} at higher redshifts. Unfortunately, the newer measurements by \citet{weaver:2022} and \citet{weaver:2023} do not reach to values below $10^4\,h^3~\mathrm{Mpc}^{-3}\,\mathrm{dex}^{-1}$ to shed more light on this issue. Thus, it is possible that the overshooting of the simulation at the high-mass end is not as dramatic as it might seem if compared only to \citet{panter:2004}.
Furthermore, observations by \citet{gonzalez:2013} show much larger values overall than observations by \citet{panter:2004} or \citet{weaver:2022} and \citet{weaver:2023}, which clearly indicates that there are systematic differences between different measurements even at $z=0$, adding additional difficulties to the comparison at the high-mass end.

At $z=4$, the simulations interestingly show slightly too few massive galaxies compared to the observations by \citet{weaver:2023}, but this could also be due to the largest simulation volumes not having enough resolution to resolve the most massive galaxies at $z=4$, and the better resolved simulations not being large enough to capture the cosmic variance. This would agree with the findings of the \magpath galaxy number densities to be in excellent agreement with observations, as discussed by \citet{kimmig:2025} and \citet{remus:2025} for all redshift ranges from $z=8$ to $z=2$, and further shown in Sec.~\ref{sec:numbdens}.

At the low-mass end, the \magpath simulations at all redshifts find too many galaxies, with the strongest deviations for the highest resolved simulations of the \uhr level. Only at around $z=4$, this deviation is small, at all other redshifts the deviation increases. While at high redshift, variations in the CDM model could potentially have a significant effect on the stellar mass function, either through the properties of the dark matter \citep{2018ApJ...854....1M} or dark energy \citep{2020ApJ...900..108M}, at low redshift differences more likely root in the details of the feedback description. Here it is most likely that our galaxies at the low-mass end form too many stars from the gas due to overcooling, having insufficient stellar feedback, while BHs do not contribute at this mass range. Our stellar feedback description neither depends on the mass of the halo, nor is it as strong as what has been found by the NIHAO simulations to be needed to reproduce proper dwarf galaxies \citep[e.g.][]{wang:2015,tollet:2019}. As soon as the AGN feedback from the BH acts on a galaxy, this overcooling is stopped and thus the stellar mass function can be reproduced successfully. This is most important for the higher-resolution simulations and clearly shows that an updated stellar feedback model will be needed in the future for even higher-resolution simulations to capture the dwarf galaxy formation properly.

\begin{figure*}
\includegraphics[width=1.0\textwidth]{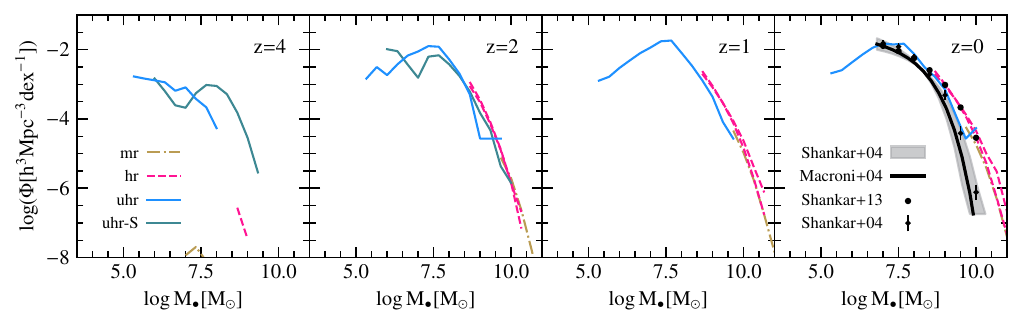}
\caption{The total BH mass function in the \magpath simulations for redshifts $z=4$, $2$, $1$, and $0$. The line colors indicate the different simulations with their respective resolutions as indicated in the legend. Additionally we show observations from \citet{shankar:2004} (black diamonds and grey shaded area), \citet{marconi:2004} (black line) and \citep{shankar:2013}(black dots).}
\label{fig:bhmf}
\end{figure*}

\subsection{The Gas Mass Function}
\label{sec:gmf}

As stars are formed from the gas that has been assembled in a similar fashion as the dark matter \citep[e.g.][]{seidel:2025}, the gas mass function also provides important insight into the complex interplay between star formation, feedback, and assembly history. Unfortunately, it is nearly impossible to measure a gas mass function from observations over a large range of masses. This is due to the fact that most of the gas in galaxies is present in form of cold, often molecular gas, while in galaxy groups and galaxy clusters the gas is present in form of hot ionized gas. Cold gas is usually measured through \HI or molecular gas measurements using tracers like CO, while hot gas is detected through its bremsstrahlung emission in X-ray. The hot gas is also usually far more extended than the cold gas already at high redshifts \citep[e.g.][]{chen:2024}, adding additional problems to detection of a gas mass function. To further add to the complications, the stage in between hot gas and cold gas, the warm gas component, is virtually impossible to measure and is thought to account for the missing mass in the baryon budget \citep[e.g.][]{eckert:2015}.

However, from simulations we are not hampered by this issue and therefore we can measure a complete gas mass function over a large mass range without deviating between cold and hot gas. Fig.~\ref{fig:gmf} shows the gas mass function found for the different resolutions and volumes of the \magpath simulations from $z=4$ to $z=0$ from left to right. As can be seen, the simulations show overall agreement between resolutions and volumes, and the shape of the gas mass function broadly resembles a Schechter function \citep{schechter:1976}. As the hot gas component is usually measured within $R^{500}_\mathrm{crit}$, we use the gas mass within $R^{500}_\mathrm{crit}$ here, assuming that the contribution of hot gas for the smaller masses is negligible and thus the dominant contribution originates from the much more concentrated cold gas. Thus, for gas masses below $M_\mathrm{500,gas}<10^{11}\,M_\odot$ this will be dominated by cold gas, for $10^{11}\,M_\odot<M_\mathrm{500,gas}<10^{12}\,M_\odot$ we see a mix, and for $M_\mathrm{500,gas}>10^{12}\,M_\odot$ most of the gas is in a hot gas phase. For a detailed study of the evolution of the cold gas within the simulated galaxies and its connection to the star-formation as well as the chemical enrichment, we refer the reader to \citet{kudritzki:2021, teklu:2023}.

For the cold gas, measuring a gas mass function from observations has been done by measuring the molecular gas for example by \citet{saintonge:2017} from the xCOLD GASS survey using CO as tracers. They find a luminosity function that resembles a Schechter function, slightly steeper at the low luminosity end than what was reported by \citet{keres:2003} who find a plateau at a value of $\Phi \approx 10^{-1.4}\,h^3~\mathrm{Mpc}^{-3}\,\mathrm{dex}^{-1}$ 
Similarly, \citet{berta:2013} measured the molecular gas masses from the PEP/GOODS-Herschel surveys, from $z=0$ to $z=2$, and also find a Schechter-like behavior with a plateau around $\Phi \approx 10^{-3.3}\,h^3~\mathrm{Mpc}^{-3}\,\mathrm{dex}^{-1}$, slightly lower at lower redshifts than at higher redshifts. While we cannot compare these values directly to our simulation output of the total gas, at the low-mass end galaxies should be dominated by the cold gas with negligible hot gas, which is why we include the average low-mass end of the relations from \citet{saintonge:2017} and \citet{berta:2013} in Fig.~\ref{fig:gmf} for an expectation value from the cold gas. We see that our highest resolution volume that actually resolves galaxies properly reaches towards these values at the low-mass end. However, we do not find a plateau like \citet{berta:2013} and \citet{keres:2003}, but rather a continuing increase for lower masses like reported by \citet{saintonge:2017}.

Most interestingly, the molecular mass function reaches up to $2\times10^{11}\,M_\odot$ at $z=1\dots 2$, but the molecular mass function only reaches maximum masses of $6\times10^{10}\,M_\odot$ at $z=0$, showing that the relative amount of cold molecular gas decreases with time. Similarly, for the molecular gas \citet{darvish:2018} showed from ALMA observations that the molecular gas fractions are larger at higher redshifts, with an overall decrease towards lower redshift, but for the simulations it was shown by \citet{teklu:2023} that this does not occur from a lack of gas, but rather from the distribution of the gas being more extended and thus less dense, leading to less molecular gas as a consequence. Similar results have been reported by \citet{bera:2023}, who also do not find a lack of gas to be responsible for decreasing star formation but rather an increasing inefficiency in converting atomic gas to molecular gas. This is an additional problem when discussing the gas mass function, as for the cold gas the different molecular stages need to be considered which are not considered in the simulation separately. 
Therefore, the gas mass function as shown here is a prediction from the simulations and cannot be properly compared to observations beyond what was discussed above.

\subsection{The BH Mass Function}
\label{sec:bhmf}

The final mass function shown here is the black hole mass function. We present the BH mass function for the \magpath simulation volumes and resolutions in Fig.~\ref{fig:bhmf} in color, for $z=4$, $2$, $1$, and $0$, from the left to right panel. For comparison, observations by \citet{marconi:2004}, \citet{shankar:2004}, \citet{shankar:2009}, and \citet{shankar:2013} are shown at $z=0$. We find an overall agreement between the simulations of different resolutions and volumes from $z=2$ to $z=0$ once the BHs have reached the Magorrian relation after seeding (see Sec.~\ref{sec:maggo} for more details on this). Since the BH seeding masses are different for the different resolutions, this convergence starts at different BH masses, and below this the function declines. This decline was already seen and discussed by \citet{hirschmann:2014} as a side-effect of our BHs not being seeded on the Magorrian.
At $z=4$, the different resolutions are not yet converged due to the resolution not being sufficient for all but the \uhr resolutions to have established clear relations yet. Here, the differences between the box volumes also become very apparent, as the larger volume of \Bthuhr has more massive structures already at $z=4$ than the smaller volume of \Bf, and thus also contains more massive BHs yet. In addition, we also see the effect of the advanced BH model as discussed already by \citet{steinborn:2015}, that the BHs in the advanced model grow faster than in the fiducial model. As demonstrated by \citet{steinborn:2015} for the \hr level simulations, this actually does not continue to low redshifts. In fact, the advanced model BHs grow less strongly at lower redshifts and are in better agreement with observations than the fiducial model.

As discussed already by \citet{hirschmann:2014} for \Bthr and \Bfuhr, the BH mass functions from the \magpath simulations are in good agreement with the observations albeit overshooting the high-mass end, similar to what was seen for the stellar mass function for the fiducial BH model runs, while only the advanced BH model runs do not overshoot at the high-mass end. However, as also shown by \citet{steinborn:2015}, even our fiducial model agrees with the observations by \citet{shankar:2013}, who find larger BH masses overall. Unfortunately, the BH mass function is extremely difficult to measure especially at higher redshifts, where instead usually AGN luminosity functions are measured. The agreement between the AGN luminosity function from the fiducial BH model simulations of the \magpath simulations from $z=4$ to $z=0$ have been shown and discussed in detail by \citet{hirschmann:2014}, finding again a slight overshooting at the high-mass end at $z=0$ and overall good agreement at higher redshifts, with \citet{steinborn:2015} showing that the advanced BH model performs better at the high-mass end again. 

The detailed properties of the AGN in our model are discussed in \citep{hirschmann:2014}, while for the AGN luminosity functions we refer the reader to \citet{hirschmann:2014}, \citet{steinborn:2015}, and \citet{2018MNRAS.481.2213B}. In addition, a detailed discussion of the halo occupation number can be found in Chapter~6 of \citet{ediss22616}, while a detailed discussion of the occurrence of dual AGN can be found in \cite{steinborn:2016}. A detailed discussion of the fueling mechanism and some example light curves of the AGN in the simulations are presented in \citet{steinborn:2018}. The environmental dependence of AGN activity is discussed in detail in \citet{2024A&A...683A..57R}, while in \citet{gonzalez_villalba:2025} a detailed evaluation of the BH feedback description compared to observations on cluster and group scales can be found.


\section{Scaling Relations Through Cosmic Time II:
Global Redshift Evolutions}
\label{sec:scale0}

\begin{figure*}
	\centering
    \includegraphics[width=.49\textwidth]{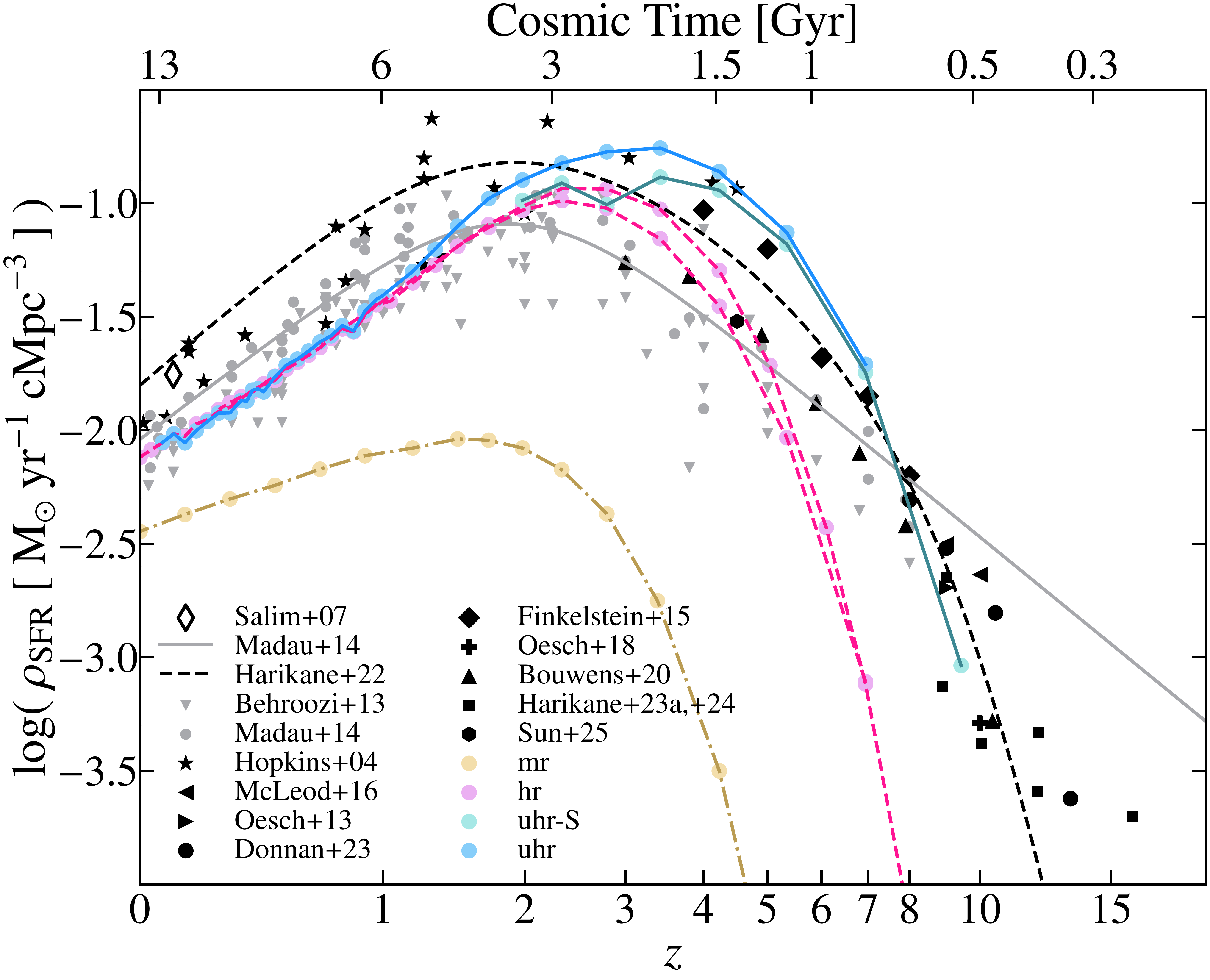}
    \includegraphics[width=.49\textwidth]{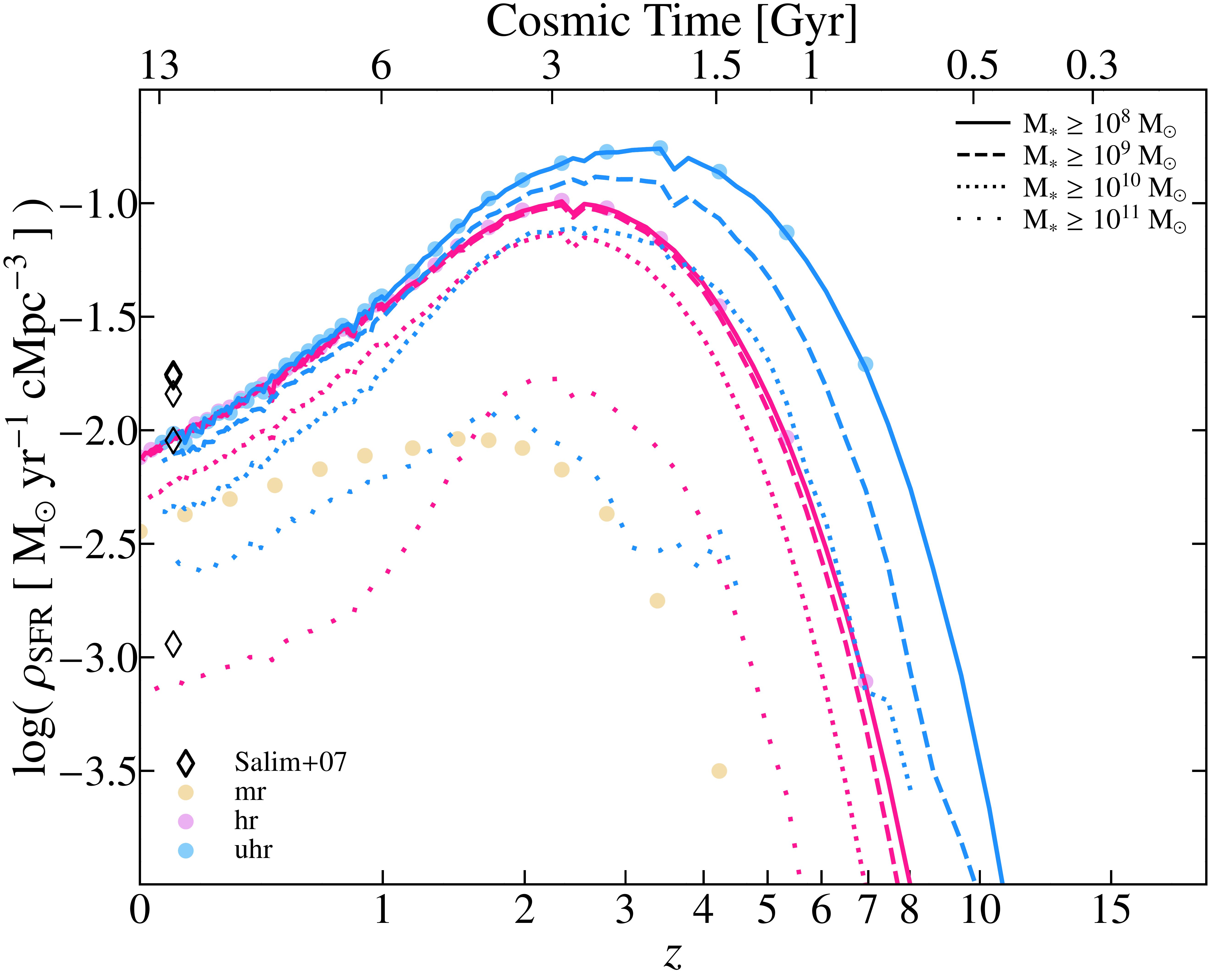}
	\caption{
    Evolution of the cosmic star formation rate density.
    Observations are shown in gray and black colors, simulation data are shown in colors.
    \textit{Left:} The total SFRD of each \magpath box is shown as points for each snapshot and connected by lines for comparison to the fitting results by \citet{harikane:2022} and \citet{madau:2014}.
    We have included the compiled data sets by \citet{hopkins:2004}, \citet{madau:2014}, and \citet{behroozi:2013}, which include data from (far-/mid-)UV, (far-/mid-)IR, and ugriz observations. 
    In black color we have added additional observational data. 
    The work by \citet{hopkins:2004, hopkins:2007} includes measurements from radio, far-IR, H$\alpha$, and UV wavelengths and covers the lower redshift range until $z \sim 5$.
    On the high-redshift end, we include UV data by \citet{finkelstein:2015} and \citet{oesch:2018} as well as ALMA data by \citet{bouwens:2020} and Hubble Ultra Deep Field IR results by \citet{oesch:2013} with the latter combined with Hubble Frontier Fields and CLASH pointings by \citet{mcleod:2016}.
    \citet{sun:2025} combine ALMA with spectroscopic JWST data to consider the contribution of dust-obscured galaxies.
    Spectroscopic JWST data is also used for the measurements by \citet{donnan:2023} and \citet{harikane:2023a} in addition to photometric JWST data by \citet{harikane:2024}.
    \citet{mcleod:2016, oesch:2013}. \textit{Right panel:} Cosmic star formation rate density calculated with different lower stellar mass cuts of $M_{*,\txt{min}}\ \in\ [10^{8},\ 10^{9},\ 10^{10},\ 10^{11}]\,M_\odot$, shown as solid, dashed, narrow-dotted, and dotted lines, respectively, for \Bfuhr in blue and \Bthr in pink. Values for \Bzmr are only possible for the highest mass cut due to the resolution, shown in gold. Colored circles mark the SFRD obtained from all particles as in the left panel. For comparison, observations at $z=0$ from \citet{salim:2007} are included as black diamonds, split for the same mass cuts, with lowest to highest cut from top to bottom.
	}
    \label{fig:sfrd}
\end{figure*}

With mass functions properly in place at all redshifts, this now invites questions on how the four different components of gas, stars, dark matter and BHs interact to establish such relations, as the \magpath simulations were not tuned to reproduce these relations and as such these quantities grew naturally. 
Thus, it is now time to look at the star formation that takes place from the gas reservoir in halos, producing the stellar component that, although being the smallest generally, is also the most visible and therefore observable. Consequently, our first scaling law to investigate is the cosmic star formation rate density with redshift, the depletion timescales, and the onset of quenching with redshift.

\subsection{Cosmic Star Formation Rate Density}
\label{sec:sfrdens}
Observations across cosmic time in different wavelengths show that cosmic star formation evolves from high to low redshifts.
Combined observations across wavelengths reveal a peaking distribution with the highest star formation rate densities (SFRD) $\rho_\txt{SFR}(z)$ at around $z\approx2$ \citep[e.g.][]{madau:2014}.

In Fig.~\ref{fig:sfrd}, we compare the evolution of the cosmic star formation rate density in the \magpath simulations to a compilation of observations, with colors indicating the different resolutions and volumes of the simulation, over the full redshift range from $z=20$ to present day, in both panels. In the left panel, we also include observations in black and gray. As several observations assume the IMF by \citet{salpeter:1955}, while the simulation and other data assume a Chabrier IMF \citep{chabrier:2003}, we divide by a conversion factor of $1.64$ following \citet{madau:2014} to convert the respective values for a proper comparison to the simulation data.
As shown in the left panel of Fig.~\ref{fig:sfrd}, all simulation volumes and resolutions display peaking SFRD distributions, with the \uhr resolution simulations predicting the peak of star formation slightly earlier ($z\approx3$) than the \hr simulations ($z\approx2.5$).\footnote{While the values from particle data are shown here, using values from \subfind output is indistinguishable, but since the particle data is more precise we use the particle data.} \ For the \mr resolution, the peak is even below $z=2$ and at much lower SFR densities, but this is a consequence of the lower mass cut of $M_*>10^{11}\,M_\odot$, as we will discuss further down in more detail.

In the left panel of Fig.~\ref{fig:sfrd}, observations from different surveys are included. For redshifts below $z=4$, these are observations from \citet[][black stars]{hopkins:2004}, \citet[][gray circles]{madau:2014}, and \citet[][open diamond]{salim:2007}. In addition, the data compiled by \citet{behroozi:2013} are shown as gray triangles. The gray solid line marks the fit from \citet{madau:2014} to the respective data.
Overall, the higher-resolution simulations are in very good agreement with observations, tending to the lower end of observations towards $z=0$ and higher SFRD at $z>3$, but well within the observed scatter.
With the recent surge in JWST observations extending current data to very high redshifts, the SFRD appears to decline with a steeper slope \citep[e.g.][]{harikane:2022} than the broken power-law fit by \citet{madau:2014} provides, with the model from \citet{harikane:2022} shown as black dashed line. Individual observations at high redshift are included as black symbols, namely from \citet[][right triangles]{oesch:2013}, \citet[][diamonds]{finkelstein:2015}, \citet[][left triangles]{mcleod:2016}, \citet[][pluses]{oesch:2018}, \citet[][upward triangles]{bouwens:2020}, \citet[][circles]{donnan:2023}, \citet[][squares]{harikane:2023a,harikane:2024}, and \citet[][hexagons]{sun:2025}. The deviation from the fit by \citet{madau:2014} is immediately apparent, while the agreement found with the model by \citet{harikane:2022} is in excellent agreement with even the newer observations at high redshifts, where the deviation from the model only occurs at redshifts higher than $z=12$.

This distribution of the observations at high redshifts before $z=4$ is in especially good agreement with the data of the \uhr simulations, as discussed already by \citet{kimmig:2025}, while the \hr simulations lack the necessary resolution to produce the required SFRD at high redshifts. The effect of the larger volume can be seen directly when comparing the two dashed pink lines, with the higher one marking the larger volume of \Btbhr, and the lower one the smaller volume of \Bthr. The effect of the advanced BH model can also be seen when combining the turquoise line marking the advanced model simulation \Bthuhr, compared to the blue line marking the fiducial \Bfuhr simulation: on the one hand, \Bthuhr contains SFRD already at higher redshift as the box volume is larger and thus also more rare nodes are included. On the other hand, the fiducial run produces a higher SFRD peak, while the advanced BH model causes a lower SFRD peak despite the larger volume. This emphasizes again the need for large volumes with high resolutions to reproduce the observations at cosmic dawn and explain the formation of the very first galaxies, and that this epoch is also a crucial testbed for understanding the interactions of the BHs and their galaxies, as discussed in detail by \citet{kimmig:2025}. We refer the reader to that paper for more details.

At redshifts below $z=2$, all simulations but the \mr runs converge nicely, in good agreement with observations. 
By comparing the data from all boxes we find that a decrease in resolution (from blue to red to yellow lines) also comes with a decrease in overall SFRD, as well as a lower peak redshift $z_\txt{peak}$.
We focus on this aspect in more detail in the right panel of Fig.~\ref{fig:sfrd} by setting different minimum stellar mass thresholds for galaxies to be included in calculating the SFRD. We use the thresholds $M_{*,\txt{min}}\ \in\ [10^{8},\ 10^{9},\ 10^{10},\ 10^{11}]\,M_\odot$, indicated as solid, dashed, narrow dotted, and loosely dotted lines, respectively.
The lowest threshold for all simulations agrees well with the full box values obtained from the particle data, marked by the colored circles as in the left panel of Fig.~\ref{fig:sfrd}. For the \uhr level marked in blue, we see that the SFRD already drops and the onset is delayed if the second cut level of $M_{*,\txt{min}} = 10^{9}\,M_\odot$ is used, while there is basically no difference between these two cuts for the \hr simulations, as for those the halos in the first and second cut level are not resolved. For the third cut level, $M_{*,\txt{min}} = 10^{10}\,M_\odot$, indicated by the narrow dotted lines, we see that \uhr and \hr level are much closer at higher redshifts, and agree well at low redshifts where they both now provide values below the SFRD obtained from all particles. For the last mass cut, $M_{*,\txt{min}} = 10^{11}\,M_\odot$, the \uhr level volume generally contains not very many galaxies due to its small size, so the statistics are not as good as for the \hr box. However, both resolutions show a much lower SFRD and a much later peak for this mass cut, in much better agreement with the result from the \mr resolution where the particle resolution resembles the $M_{*,\txt{min}} = 10^{11}\,M_\odot$ mass cut. 

This clearly demonstrates also the difficulty in constructing the SFRD function with redshift as the galaxy masses already play a crucial role between $M_{*,\txt{min}} = 10^{9}\,M_\odot$ and $M_{*,\txt{min}} = 10^{10}\,M_\odot$ and can shift the resulting SFRDs. This is nicely confirmed by local Universe observations from \citet{salim:2007}, shown as black open diamonds in the right panel of Fig.~\ref{fig:sfrd} using approximately the same cuts as in the simulations.
Consistent with their observations we find that the largest drop-offs of cosmic star formation come with mass thresholds of $10^{10}$ and $10^{11}\,M_\odot$.
Thus, for future studies mass cuts are an important quantity to consider when comparing simulations and observations, as also discussed for the FIREbox simulations by \citet{feldmann:2023}. This is also a scaling relation highly sensitive to the implemented physics, and as such an excellent testbed for new updated modules as well as our understanding of galaxy formation at cosmic dawn.

As the overall SFRD is driven by different galaxies over cosmic time, it is also of interest to see how the individual galaxies evolve in the simulations and how this is reflected in the age distribution of the stellar population. Here we want to refer the reader to stellar age comparisons of different cosmological simulations (including \magpath) with the SAMI Galaxy Survey at $z\approx0$ \citep{vandesande:2019}, with BGGs over cosmic time $(z=0.08\text{--}1.30)$ in the COSMOS field \citep{2024A&A...690A.315G}, with massive galaxies at $z\approx2.7$ \citep{lustig:2023}, and with massive and quiescent galaxies at $z\approx3\text{--}4.5$ \citep{nanayakkara:2024}. In addition, the predicted ages in the simulations have been found to be similar to those predicted by galaxy look-back evolution models \citep{kudritzki:2021}, while the amount of in-situ versus accreted stars in the simulated galaxies is discussed in \citep{2022ApJ...935...37R}.

\begin{figure*}
	\centering
    \includegraphics[width=.49\textwidth]{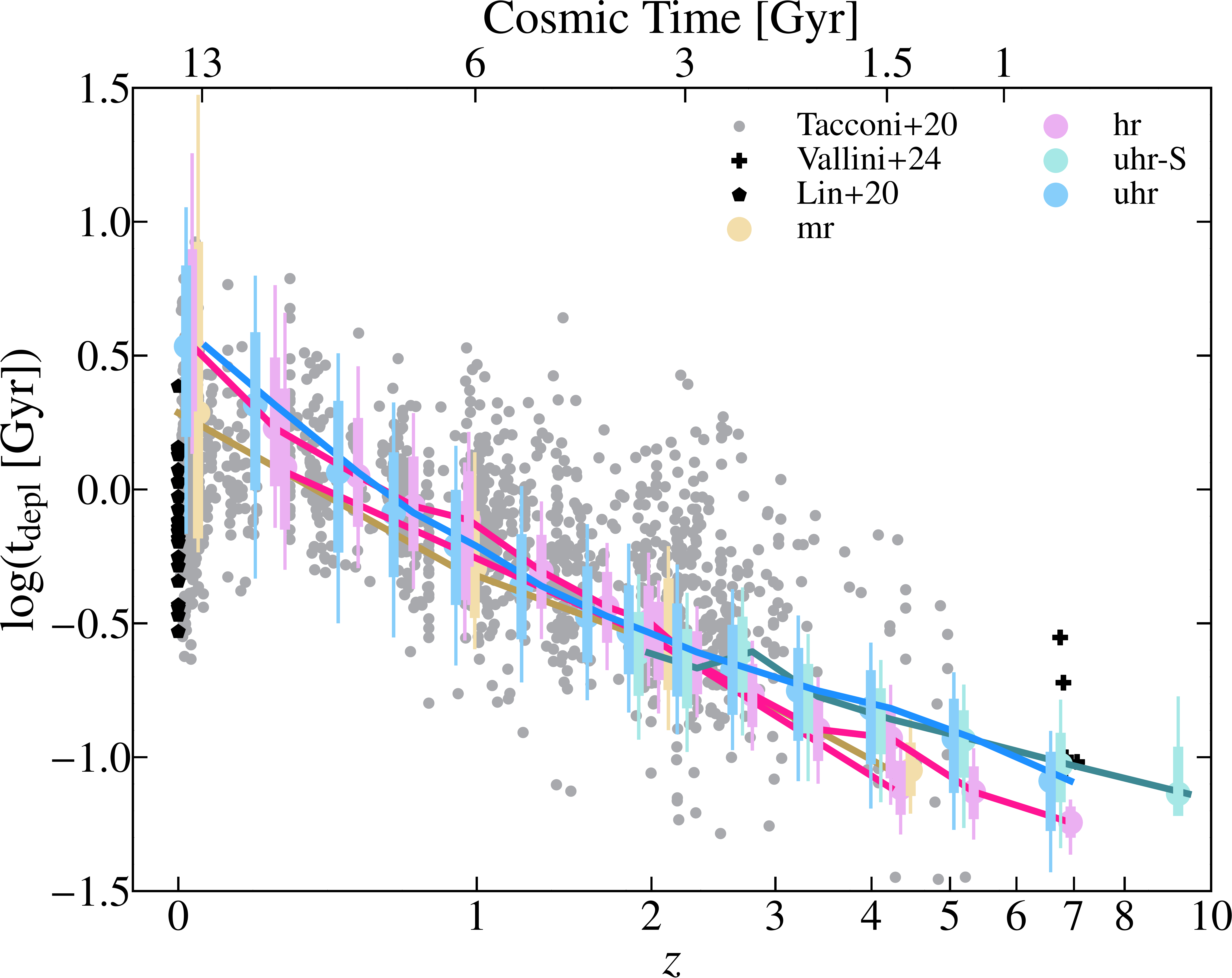}
    \includegraphics[width=.49\textwidth]{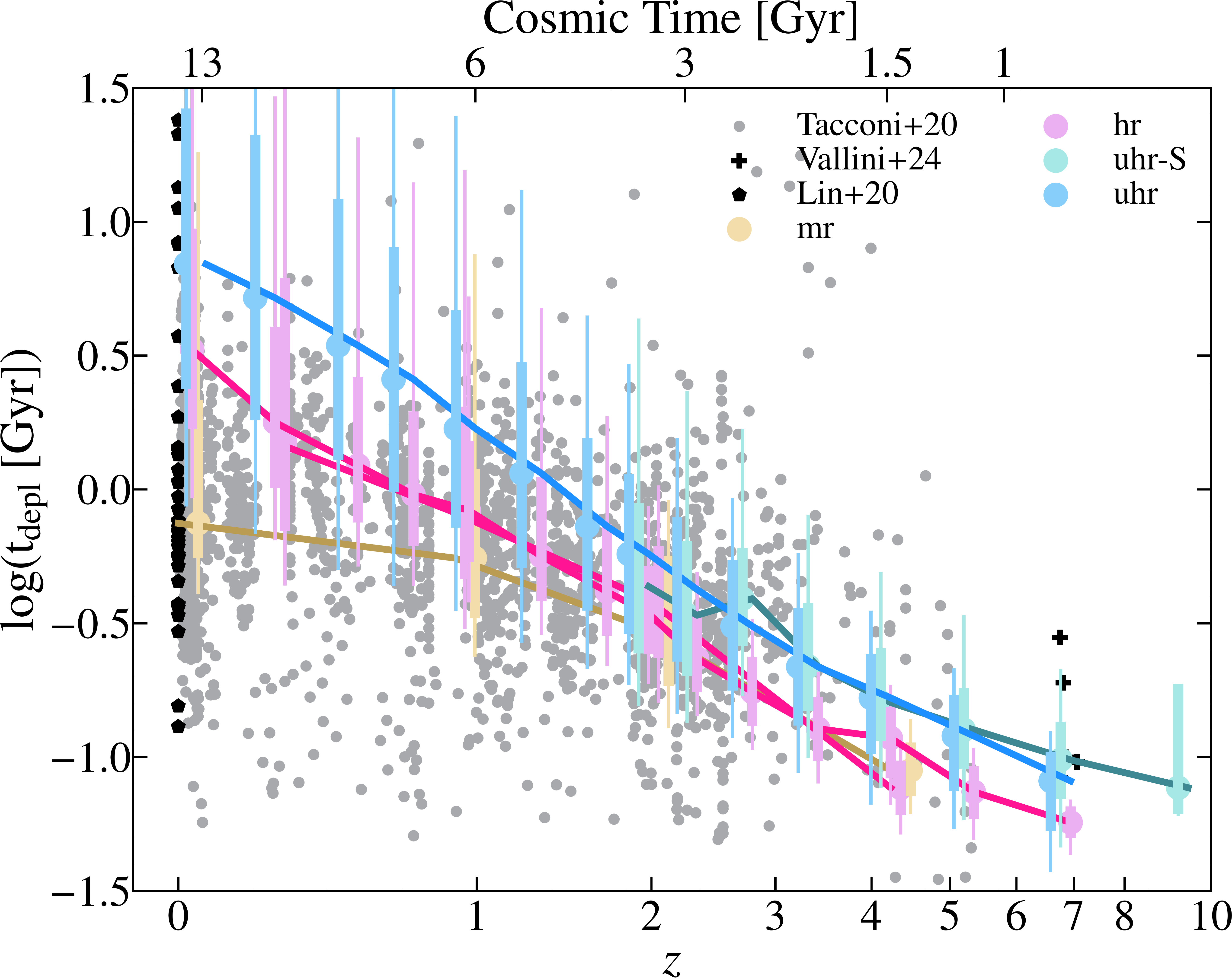}
	\caption{Evolution of depletion time scales in the \magpath simulations (colored according to resolution) and observations (black/gray). For the simulation data, the points represent the median depletion times with a 1- and 2-$\sigma$ scatter as thick and thin vertical bars, respectively.
    The simulation data points have been shifted by $\Delta \txt{log}(1+z) \sim 10^{-2}$ for visualization. 
    The lines highlight the median evolution without an artificial x-axis offset as for the circles.
    The bulk of observational data consists of the data set used by \citet{tacconi:2020} (grey circles). 
    On the high- and low-redshift ends, we include results by \citet{vallini:2024, lin:2020} (black crosses and pentagons) based on ALMA and MaNGA data.
    In the left-hand figure, we select for galaxies that lie within a mass range of $1.07\times 10^9\,M_\odot \leq M_* \leq 1.55\times 10^{12}\,M_\odot$ and $|\Delta \mathrm{MS}| \leq 0.6$ scatter around the main sequence description by \citet{speagle:2014}, following the approach by \citet{tacconi:2020}.
    In the figure on the right, we include all galaxies within the mass range and non-zero star formation rates.
	}
    \label{fig:tdepl}
\end{figure*}

\subsection{Depletion Timescale Evolution}
\label{sec:depl}

From the gas reservoir of a galaxy, the stars are formed within a given time. This time scale, describing the efficiency with which gas is converted into stars, is called the depletion time $t_\txt{depl}$, as described in the review by \citet{tacconi:2020}. The depletion time is calculated as 
\begin{equation}
t_\txt{depl} = \frac{M_\txt{molgas}}{\mathrm{SFR}},
\end{equation}
which gives the timescale on which the reservoir of gas available for star-formation $M_\txt{molgas}$ would be consumed by the current star-formation rate SFR.
\citet{madau:2014} compare the star formation rate density evolution to the mean specific star formation rate evolution, which declines consistently instead of peaking.
This highlights the difference between the cosmic star formation history and the histories of individual galaxies.
Declining specific star formation rates are consistent with observations of increasing depletion times as an inverse measure for gas conversion efficiency \citep[e.g.][]{tacconi:2020}. 

In simulations, we do not have the values for the molecular gas, but can only approximate this from the cold gas mass that a galaxy has at its leverage. We approximated the value as $t_\txt{depl,sim} = M_\txt{cld}/\txt{SFR}$ using cold gas mass with a temperature threshold of $T \leq 10^5\,\mathrm{K}$.
Fig.~\ref{fig:tdepl} compares the values of $t_\txt{depl,sim}$ for the \magpath simulations (colored according to their resolutions) to observations from the compilation by \citet[][gray circles]{tacconi:2020}, and additional observations from \citet[][black pentagons]{lin:2020}, and \citet[][black plus signs]{vallini:2024}. For all simulations and observations, we only consider galaxies with masses between $1.07\times 10^{9}\,\Msun < M_* < 1.55\times 10^{12}\,\Msun$, following \citet{tacconi:2020}. We account for the fact that we trace all cold gas while the observations use molecular gas by using a conversion factor between all cold gas and molecular gas of 0.25 following the results by \citet{valentini:2023}.

The right panel of Fig.~\ref{fig:tdepl} shows the direct comparison as described above. The scatter for both simulations and observations is large, with $1\sigma$ errors shown as thick lines and $2\sigma$ errors indicated by thin lines for the simulations. The overall trends of the observations are recovered well, but there are large differences between the simulations as we do not see good convergence. This changes, however, if we apply the additional criterion commonly used and described by \citet{tacconi:2020}, namely excluding all galaxies that are above or below the star formation main sequence as described by \citet{speagle:2014} by more than $|\Delta \txt{MS}| \leq 0.6$.
Using this, we recover a redshift trend consistent with observations, as shown in the left panel of Fig.~\ref{fig:tdepl}. We also find good agreement now between the simulations of all resolution levels. This clearly shows the importance of the main sequence cut in obtaining the tight relation for the depletion time with redshift, and also shows that this can be nicely recovered by the \magpath simulations. We do not even see an impact of the two different BH models on the depletion times, as the relation found for \Bthuhr and \Bfuhr are nearly identical.

This shows that for both the simulations and the observations, the depletion times become larger with lower redshifts, that is the efficiency with which gas is converted into stars declines. As discussed by \citet{teklu:2023}, this is mostly driven by the spacial distribution of the gas and not by a lack of gas in general, as the conversion of gas into stars depends on the density of the gas, and at lower redshifts the cold gas is less centrally concentrated than at high redshifts. This clearly shows that the star formation properties of galaxies depend on their radial distribution inside a galaxy, and thus this belongs to the family of local scaling relations that we will investigate in Sec.~\ref{sec:scale2}. The star formation main sequence is tightly connected to the depletion timescale. While these relations are Sec.~\ref{sec:mainsequence}, we refer the reader to detailed discussion based on the \magpath simulations for the relation of the star-formation rate to the gas properties of the galaxies in \citet{teklu:2023} as well as their evolution across the main sequence, as extensively discussed in \citet{fortune:2025}. However, before we turn to the local scaling relations, there are the global scaling relation that we want to consider in Sec.~\ref{sec:scale1}, and two more evolution trends that we discuss in the following.

\begin{figure}
    \includegraphics[width=0.99\columnwidth]{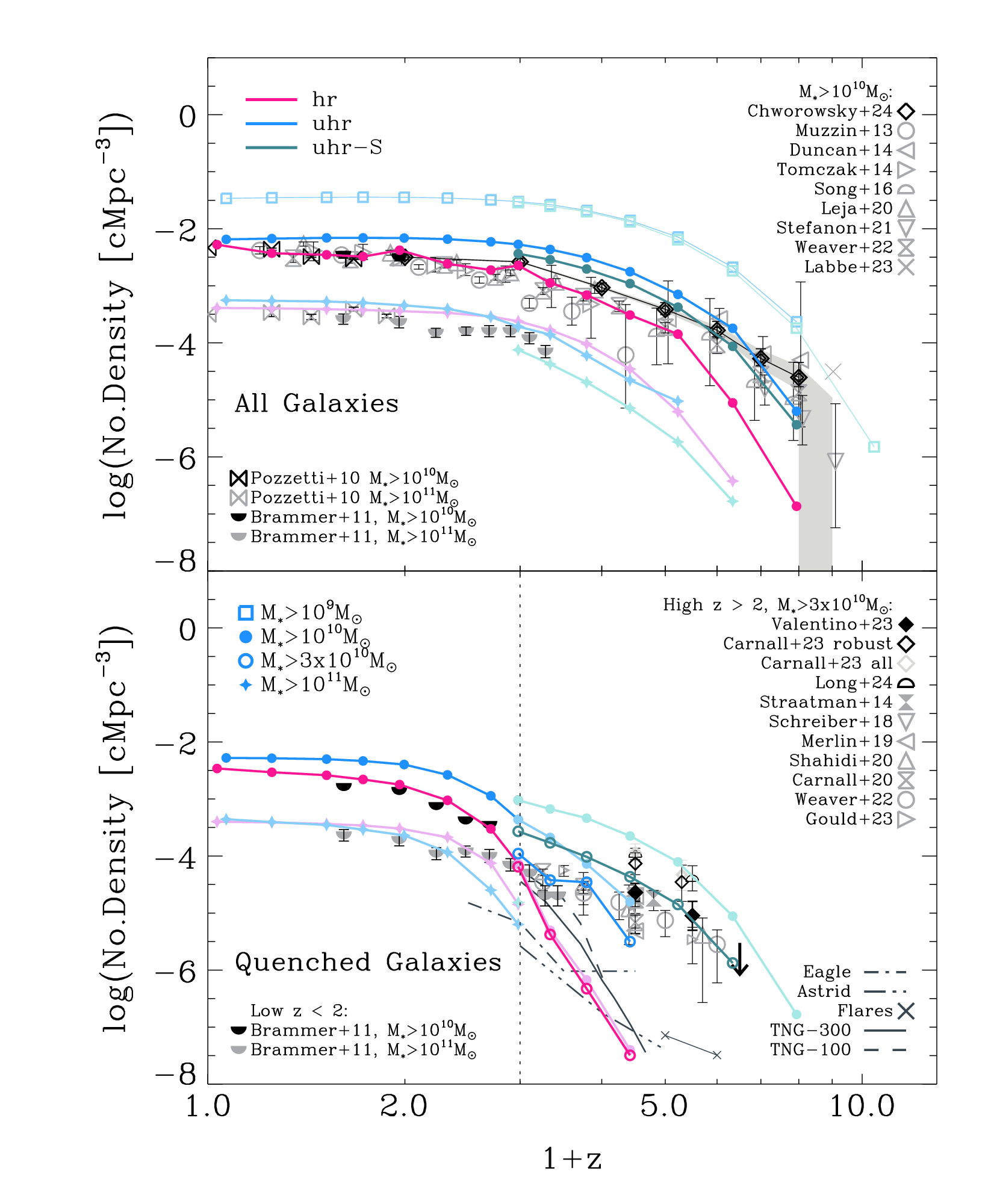}
    \caption{Number densities (\textit{upper panel}) and quenched number densities (\textit{lower panel}) through cosmic time for the \magpath simulations shown in color as indicated by the legend. Three different mass cuts are applied for the simulations, $M_*>10^{9}\,M_\odot$ marked by open squares, $M_*>10^{10}\,M_\odot$ marked by filled circles, and $M_*>10^{11}\,M_\odot$ marked by filled stars. For the quiescent galaxies, an additional mass cut of $M_*>3\times10^{10}\,M_\odot$ is included at $z>2$, as observations at these redshifts are only available with this mass cut. For the total number densities, low redshift observations with two mass cuts of $M_*>10^{10}\,M_\odot$ and $M_*>10^{11}\,M_\odot$ from \citet{pozzetti:2010} and \citet{brammer:2011} are included as black and gray symbols respectively, the latter being also included in the lower panel for the quenched galaxy fractions. For the total number densities, the only JWST measurement is provided by \citet{chworowsky:2024}, with pre-JWST over a large redshift range from \citet{muzzin:2013,tomczak:2014,leja:2020,weaver:2022}, and at high redshift beyond $z=2$ only from \citet{duncan:2014,song:2016,stefanon:2021,labbe:2023}. For quenched fractions above $z=2$, observations are only available for a lower mass cut of $M_*>3\times10^{10}\,M_\odot$, with the mass change in observations indicated by the vertical dotted line. JWST data are included from \citet{valentino:2023,carnall:2023,long:2024} in black, and pre-JWST data from \citet{straatman:2014,schreiber:2018,merlin:2019,shahidi:2020,carnall:2020,weaver:2022,gould:2023} in gray. Simulation comparisons for $M_*>3\times10^{10}\,M_\odot$ (or $M_*>10^{10}\,M_\odot$) for TNG300, TNG100 and Astrid from \citet{weller:2025} is included together with results for Flares from \citet{gould:2023} and for Eagle from \citet{long:2024}.}
    \label{fig:numb_dens}
\end{figure}

\subsection{Number Densities and Quenched Number Densities}
\label{sec:numbdens}

The counterpart to the star-forming galaxies are the quiescent galaxies, and especially with the advent of new JWST observations the number of quenched galaxies has proven to be a challenge for our understanding of galaxy formation and cosmological simulations as such. Therefore, the final scaling relation to investigate here is the number density of galaxies through cosmic time, and the number density of quenched galaxies. These number densities as obtained from the \magpath simulations are shown in color in Fig.~\ref{fig:numb_dens}, in comparison to observations in black and gray. As we have seen in Sec.~\ref{sec:sfrdens}, the \mr resolution does not cover galaxies below a mass of $M_*<10^{11}\,M_\odot$, and thus we only include the \hr and the \uhr resolution simulations in this comparison.

The upper panel of Fig.~\ref{fig:numb_dens} shows the total number density of galaxies for three different mass cuts from the simulations: $M_*>10^{9}\,M_\odot$ in light color marked by squares, $M_*>10^{10}\,M_\odot$ in dark color marked by filled circles, and $M_*>10^{11}\,M_\odot$ in light color marked by stars. The first cut level is only resolved for the \uhr simulations. We clearly see an overall agreement between the different simulations, with the largest deviations at high redshifts before $z\approx2$. Below $z=2$, number densities above all thresholds stay rather constant, with the \hr resolution having generally slightly fewer halos above a given mass threshold than the \uhr resolution simulations. At high redshifts, the \hr resolution has a later onset, as discussed already by \citet{kimmig:2025} and \citet{remus:2025}, while here the \uhr resolution simulations are in good agreement with the advanced BH model simulation having slightly fewer halos above a given mass threshold than the fiducial run.

For comparison, several observations are included. This figure is an extension to low redshifts of the figure presented by \citet{kimmig:2025}, and thus it contains the same observations at $z>2$. For a more detailed discussion on the high redshift number densities between the \magpath simulations and observations and how they are related to the feedback descriptions implemented, we refer the reader to \citet{kimmig:2025} and \citet{nanayakkara:2024}. The included observations at $z>2$ are, as indicated in the legend: new JWST measurements by \citet{chworowsky:2024} shown in black, and pre-JWST data by \citet{muzzin:2013,duncan:2014,tomczak:2014,song:2016,leja:2020,stefanon:2021,weaver:2022,labbe:2023}. All these observations have a lower mass threshold of $M_*>10^{10}\,M_\odot$, comparable directly to the dark colored lines from the simulations. Below $z\approx2$, observations by \citet[][bowties]{pozzetti:2010} and \citet[][halfmoons]{brammer:2011} are included in addition to the data by \citet{weaver:2022}, \citet{muzzin:2013}, and \citet{leja:2020} that extend to low redshifts as well. For the two new observations by \citet{pozzetti:2010} and \citet{brammer:2011}, two mass cuts were available, the first for $M_*>10^{10}\,M_\odot$ comparable to the other samples and shown in black, and a higher mass cut of $M_*>10^{11}\,M_\odot$ shown as gray symbols, comparable to the light-colored simulation values marked by the stars.
As found for the high redshifts by \citet{kimmig:2025}, the number densities of the \magpath simulations are also in excellent agreement with observations at low redshifts, for both mass cuts. 

The lower panel of Fig.~\ref{fig:numb_dens} shows the number density of quenched galaxies through cosmic time. We define a galaxy to be quiescent if their specific star formation rate is below $0.2/t_\mathrm{hub}$, following \citet{carnall:2020}\footnote{The result is nearly identical if instead of using 0.2 a factor if 0.3 is used following \citet{franx:2008}.}. Different than the total number density, the quenched number density has proven to be a real challenge for simulations to reproduce at high redshifts, as shown by \citet{remus:2025}, with only the \magpath simulation advanced BH model being capable of reproducing the observations successfully as demonstrated by \citet{kimmig:2025}. As the observed quenched fractions at redshifts higher than $z\approx2$ are given with a different mass cut of $M_*>3\times10^{10}\,M_\odot$, we include in the lower panel of Fig.~\ref{fig:numb_dens} also that mass cut for the simulations, marked by open symbols. For better comparison, we mark the threshold of $M_*>1\times10^{10}\,M_\odot$ by darker colors for $z<2$, and for $z>2$ the darker colors mark the threshold of $M_*>3\times10^{10}\,M_\odot$.

Again, we see that the advanced BH model simulation shown in turquoise is in very good agreement with the new JWST measurements marked in black at $z>2$ from \citet{valentino:2023,carnall:2023,long:2024}. Interestingly, even the fiducial model shown in blue is in agreement with the pre-JWST observations that give slightly lower quenched galaxy number densities, marked by gray symbols \citep{straatman:2014,schreiber:2018,merlin:2019,shahidi:2020,carnall:2020,weaver:2022,gould:2023}. This is surprising as none of the other simulations are capable of only being near the observations, as shown in the lower panel of Fig.~\ref{fig:numb_dens} as dark gray lines for TNG100, TNG300, and Astrid as presented by \citet{weller:2025}, Flares as shown by \citet{gould:2023}, and Eagle as shown by \citet{long:2024}. This is most likely due to the fact that the BHs in the \magpath simulations are not pinned to the potential minimum, and thus do not merge instantaneously once two galaxies are merged, but have a delayed merging time, as discussed in Sec.~\ref{sec:bhmodel}. However, the advanced BH model is much better at even forming quenched galaxies at redshifts of $z=5$ or higher, as discussed by \citet{kimmig:2025}.

We extend the analysis to low redshifts in this work. We include the measurements by \citet{brammer:2011}, split in two mass bins of $M_*>10^{10}\,M_\odot$ and $M_*>10^{11}\,M_\odot$ marked in black and gray, respectively. A comparison to our simulation again shows excellent agreement for $M_*>10^{11}\,M_\odot$, and for the \hr resolution also for $M_*>10^{10}\,M_\odot$. The fiducial \uhr model shown in blue produces too many quenched galaxies at low redshifts compared to observations with a steep increase between $z=2$ and $z=1$. We do not see indications for that to happen for the advanced BH model, unfortunately we do not have number densities below $z=2$ from this model at \uhr level. However, as discussed by \citet{steinborn:2015} for the \hr level simulations, the advanced model performs better in reproducing quenched galaxies compared to observations at all redshifts.

\begin{figure}
    \includegraphics[width=0.99\columnwidth]{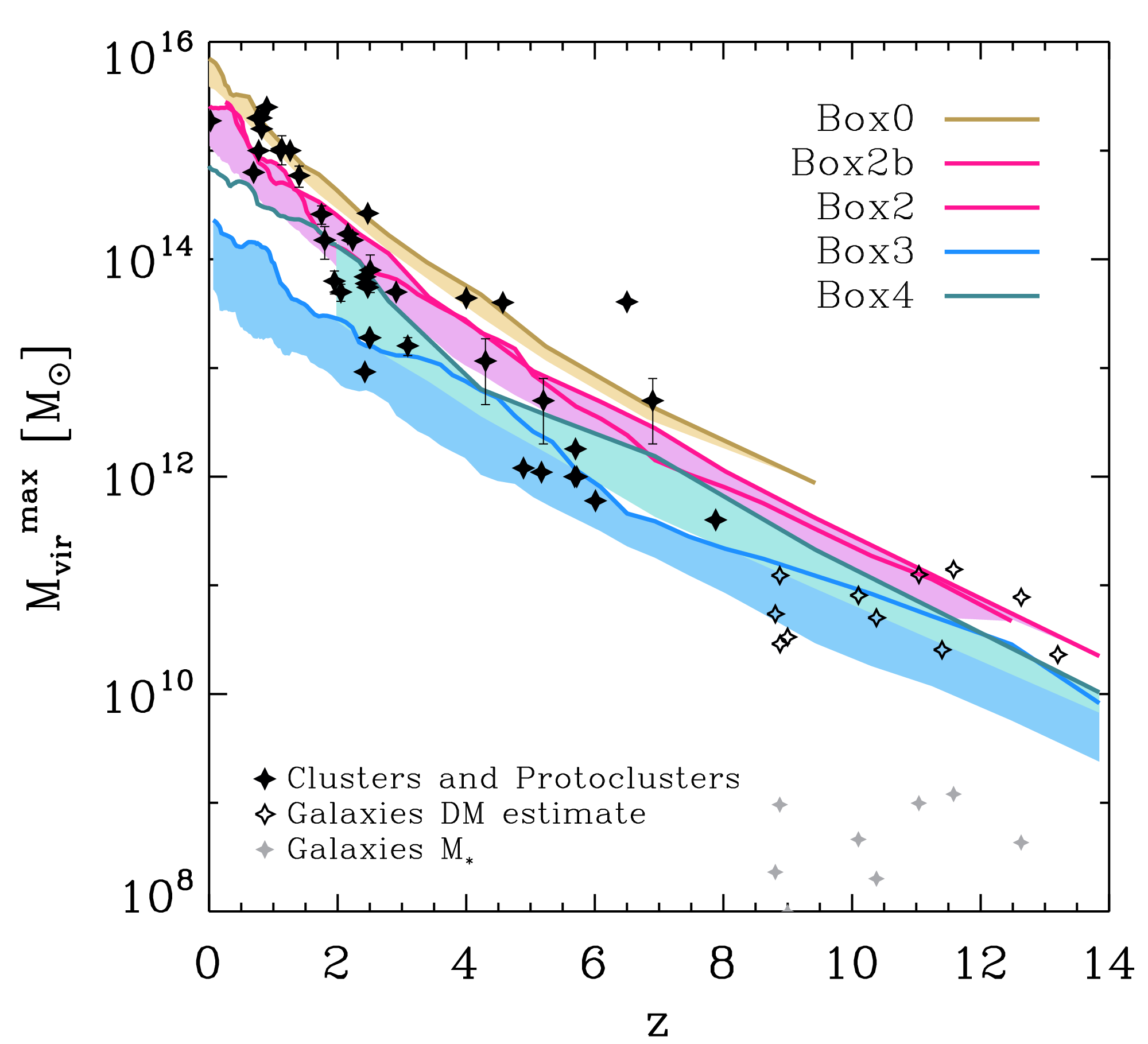}
    \caption{Maximum halo mass at a given redshift present in the different simulation volumes through cosmic time for the \magpath simulations shown in color as indicated by the legend. The solid line marks the most massive halo, while the shaded region shows the range in mass covered by the 10 most massive structures at that redshift. For comparison, observations of clusters and protoclusters as compiled by \citet{remus:2023} are included as filled black stars, with two new proto-clusters added at $z=6.9$ by \citet{arribas:2024} and $z=7.88$ by \citet{morishita:2023}. In addition, at redshift beyond z=8 halo mass estimates for the galaxies compiled by \citet{harikane:2024} are included as open black stars. These estimates are estimated assuming the z=0 stellar-mass--halo-mass relation from observations by \citet{hudson:2015} to calculate the halo mass from the stellar mass. Stellar masses as given by \citet{harikane:2024} are shown as light gray stars.} 
    \label{fig:mass_most_massive}
\end{figure}

Other tests concerning the number of quenched galaxies have been discussed in previous studies using the \magpath simulations. Showing them all is beyond the scope of this work, and here we will only give a short overview. \citet{lustig:2023} investigate stellar mass functions and cumulative number densities as a function of stellar mass for quiescent galaxies at $z\approx2.7$. They find excellent agreement with various observations for the \magpath simulation \Bthuhr, while e.g. IllustrisTNG significantly overpredicts number densities and the stellar mass function for quiescent galaxies at this redshift. \citet{remus:2023} provide a prediction for the quiescent satellite fraction across redshifts for halo masses ranging from group to cluster scale in \Btbhr. They find good agreement with several observational results: for massive clusters \citep{strazzullo:2019}, for groups and low-mass clusters obtained with DETECTIFz from the REFINE survey \citep{sarron:2021}, and even the protocluster studied by \citet{mcconachie:2022}. Lastly, \citet{lotz:2019} showed the radial distribution of quenched galaxy fractions in galaxy clusters for the large samples of \Bthr and \Btbhr, in comparison with galaxy clusters observed at low-redshift, finding very good agreement between simulations and observations. They demonstrated that in such environments star forming galaxies are usually quenched within the first passage in the cluster, which can only be prevented if the galaxy is either very massive and thereby shields its gas from stripping, or is the galaxy is on a very circular orbit where the pressure on the gas is lower. Thus, we conclude that the \magpath simulations are very well suited to reproduce quiescent galaxies through cosmic time over a large range of halo masses, and that a careful treatment of the BH feedback, especially at high redshifts, is key to understanding the origin of the first quiescent galaxies at cosmic dawn.

\subsection{The Most Massive Halos per Redshift}\label{sec:mostmassive}
Finally, we return to the evolution of halo masses of structures through cosmic time, as they are the hosts of the stellar components and as such their evolution and growth limits also the growth of the baryonic observables. As shown by \citet{remus:2023}, the slope of the relation of the most massive collapsed halo at a given redshift is driven by the value of $\sigma_8$, and thus can be used as an independent measure of this elusive quantity. Unfortunately, the most massive structure at a given redshift is difficult to measure observationally, as it requires all sky deep surveys for completeness. In simulation, this can be directly measured and predicted, as shown in Fig.~\ref{fig:mass_most_massive} for the \magpath simulations.

Fig.~\ref{fig:mass_most_massive} shows the maximum virial mass at a given redshift that is reached in the different simulation volumes as solid lines, shown in colors as given by the legend. The shaded areas mark the range in mass of the 10 most massive structures inside the volume at each redshift. This is similar to what has been shown by \citet{remus:2023} but extended to higher redshifts. As expected, the larger boxes include more massive halos at each redshift due to the larger volumes and thus larger modes of growth included in the simulation. However, what can also be seen is that the slopes of the relation are similar for all simulation volumes, just shifted to higher masses for larger box volumes. The resolution limitation can be seen in the fact that \Bzmr does not resolve halos at redshifts higher than $z\approx9$, i.e. below $M_\mathrm{vir}\approx10^{12}M_\odot$. For the \hr resolution, halos down to about $M_\mathrm{vir}\approx3\times10^{10}M_\odot$ are resolved, which appear earlier in the larger box volume of \Btb than \Bt. Finally, the \uhr resolution allows for halos to be found down to $M_\mathrm{vir}\approx10^{9}M_\odot$, which appear even earlier than $z=14$ in both \Bthuhr and \Bfuhr. This clearly demonstrates the issue for high redshift studies, which require both large box volumes but also very high resolutions. Note that we used the most massive halo obtained from \Bthhr for $z<2$ since \Bthuhr only run until $z=2$. However, since the boxes are run with the same initial conditions but simply lower resolutions, we can clearly see that the most massive halo has the same mass in both resolutions, which should be the case and is a nice confirmation that nothing went wrong and that halo crossmatching between the simulations of different resolutions is possible, as also done by \citet{remus:2025} before.

In addition to the simulation output, we included observations of individual massive objects at different redshifts in Fig.~\ref{fig:mass_most_massive}: Black stars mark observations of clusters and protoclusters, as compiled by \citet{remus:2023}, based on the compilation by \citet{miller:2018} with additional points from \citet{kim:2019}, \citet{kubo:2016}, \citet{oteo:2018}, \citet{daddi:2021}, the Spiderweb protocluster from \citet{shimakawa:2014}, the ORELSE cluster by \citet{tomczak:2017}, and the merging group of protoclusters called HYPERION \citep{cucciati:2018}. At redshifts earlier than $z=4$, protoclusters are included from \citet{calvi:2021}, \citet{chanchaiworawit:2019}, and the compilation provided by F. Sinigaglia sampled from observations by \citet{shimasaku:2003}, \citet{venemans:2004}, \citet{ouchi:2005}, \citet{toshikawa:2012,toshikawa:2014,toshikawa:2020}, and \citet{higuchi:2019}. Newly added are the protoclusters by \citet{arribas:2024} and at $z=7.88$ by \citet{morishita:2023}. At such high redshifts, the difference between a protocluster core and individual galaxies in fluent, and usually only characterized by the environment of the galaxy indicating an overdensity in ongoing assembly. 

The figure clearly demonstrates that, given a large enough volume, all these massive structures can also be found in simulations. However, they also clearly are exceptional objects at their given redshifts, at which they are among the most massive assembling nodes of that period. Note, however, that this does not imply for them to be the progenitors of the most massive objects at present-day, for that was shown by \citet{remus:2023} to not be the case, as a structure that assembles early or is an overdensity at a given time does not need to continue growing into the future. Moreover, \citet{seidel:2024} showed that, in fact, this is also true for some of the most massive structures found at present-day when simulated into the future in the framework of the \magpath spinoff simulation SLOW. Note that the outstandingly massive object at $z\approx 6.5$ reported by \citet{chanchaiworawit:2019} is still somewhat of a challenge that needs to be understood in the future in more detail; one of the likely explanations is that this is a protocluster region in assembly and not the measurement of a yet assembled protocluster core, and thus this mass is not yet virialized. That region will, in that case, do so in the future, but it at its given redshift an overestimate of the virialized mass.

Beyond $z=8$, halo mass measurements have not (yet) been provided, however, measurements of galaxy masses are present in the literature, as given for example by the compilation by \citet{harikane:2024}. These stellar masses are shown as small gray stars in Fig.~\ref{fig:mass_most_massive}. Following the findings that the stellar-mass--halo-mass (SMHM) relation does not change significantly with redshift for the simulations between $z=4$ and present day as discussed in Sec.~\ref{sec:smhm}, and a similar result found from observations between $z=10$ and $z=6$ by \citet{stefanon:2021}, we use the SMHM relations found from lensing measurements at $z=0$ by \citet{hudson:2015} to calculate the halo masses from the observed stellar masses reported by \citet{harikane:2024}, which are shown in Fig.~\ref{fig:mass_most_massive} as open black stars. Under these assumptions, the massive galaxies at high redshifts do not pose a problem for $\Lambda$CDM, as they are hosted by halos which are in compliance with halo masses predicted by the simulations to be present already in this early Universe. As discussed by \citet{kimmig:2025}, the stellar resolutions of the \magpath simulations are, unfortunately, not high enough to resolve these galaxies, which is work to be done in the future by the \magpath spinoff MAGNETICUM-DAWN, to prove these hypotheses and understand the formation of the earliest galaxies in the Universe in more detail.

Overall, the \magpath simulations are very successful in reproducing a large range of global scaling relations through cosmic time, despite not being tuned to the stellar mass functions at $z=0$. We will now investigate the scaling relations that are based on more local distributions, and thus not all of them can include the lowest resolution simulation of the \magpath simulations, as the \mr resolution is simply not capable of resolving all quantities studied in the following. However, the fact that the simulations showed excellent convergence through resolutions is promising when connecting both local and global structure growth processes through halo masses and cosmic times, and clearly highlights the strength of this extraordinarily large suite of simulations.

\begin{figure*}
    \includegraphics[width=0.99\textwidth]{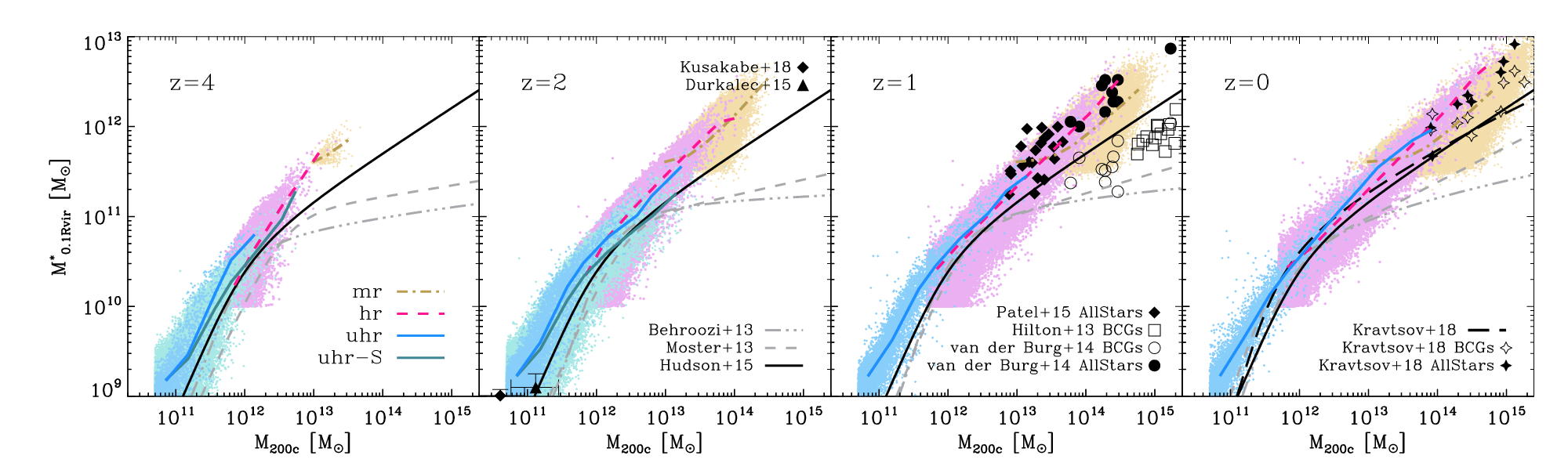}
    \caption{Stellar-mass--halo-mass relations for the \magpath simulations, with colors indicating the different resolutions as indicated in the legend. Here, the halo mass is given as $M_\mathrm{200,c}$, and for the stellar mass all mass within 10\% of $R_\mathrm{vir}$ that is allocated to the central galaxy by \subfind is used, denoted $M_\mathrm{0.1Rvir}^*$. The black solid line shows the SMHM relation found from lensing at $z\approx0$ by \citet{hudson:2015}, as a reference at all redshifts. In addition, different observations are included in black, with solid symbols marking measurements of all stars and open symbols indicating the BCG measurements only. Included observations at $z=0$ are by \citet{kravtsov:2018} as black stars and black dashed line, at $z=1$ observations from \citet{vanderburg:2014} are shown as circles, and observations by \citet{patel:2015} and by \citet{hilton:2013} are shown as diamonds and squares, respectively. These observations are all obtained from X-ray measurements. At $z=2$, two data points from Halo Occupation measurements by \citet[][triangle]{durkalec:2015} and \citet[][diamond]{kusakabe:2018} are included. In addition, predictions from the semi-analytic models by \citet{behroozi:2013} and \citet{moster:2013} are shown as dash-dotted and dashed gray lines, respectively.}
    \label{fig:smhm200}
\end{figure*}

\section{Scaling Relations Through Cosmic Time~III: Global~Scaling~Relations}
\label{sec:scale1}

The first kind of global quantity scaling relations are those that directly correlate halo properties such as halo mass, stellar mass, gas mass, and temperature, or simply count halos with given properties or their evolution with redshift. Thus, these can be analyzed on all resolution levels of the \magpath simulations, and the convergence between resolutions and box volumes can be tested, at all redshifts. Generally, structures form and evolve in time on all resolution levels and box volumes. To illustrate that, Fig.~\ref{fig:boxevo} shows the evolution of the cosmic web, from $z=4$ on the left down to $z=0$ on the right, for \Bfuhr in the upper row and \Bzmr in the lower row. This clearly demonstrates the self-similar nature of our Universe.  Based on the halos that are identified at the different redshifts in these boxes, we will now present the scaling relations for the halos in detail in the following. 

\subsection{Stellar-Mass--Halo-Mass Relation and the Baryon~Conversion Efficiency}
\label{sec:smhm}

As seen from the mass functions, the dark matter follows a simple linear growth theory, while the baryonic components have a more complicated pattern driven by the interplay of gas cooling, star formation, feedback, and heating mechanisms. These baryonic processes inside the dark matter halos are of different importance for different halo masses, and thus the relation between the halo mass and the stellar mass is not a 1:1 relation. For low-mass halos, stellar feedback is thought to be most efficient in suppressing star formation, while at large halo masses the AGN feedback and the hot gas atmosphere of groups and clusters suppress cooling and star formation. Star formation is thus most efficient at around $M_\mathrm{200,c}\approx 10^{12}\,M_\odot$. This is reflected in the so-called stellar-mass--halo-mass (SMHM) relation. 

Unfortunately, there are multiple different definitions of the SMHM relations with respect to both quantities. Halo masses are extremely difficult to measure observationally and are usually either obtained from X-ray measurements assuming that the hot gas traces the dark matter, or from strong or weak lensing where the dark matter can be measured indirectly from its gravitational impact. This also leads to different measurements of both the halo and the stellar masses. One commonly used halo mass measurement is $M_\mathrm{200,c}$, but for observations based on X-ray measurements $M_\mathrm{500,c}$ is more typically used. It is possible to estimate one from the other, as for example described by \citet{vanderburg:2014}, but for these estimates assumptions have to be made. These halo measurements are compared to the stellar mass, for which again different definitions exist: one can either calculate all stellar mass within a given aperture, or just the mass of the central galaxy, which in the case of galaxy clusters is the brightest cluster galaxy (BCG). For galaxies hosted by halos below the group mass threshold of $M_\mathrm{200,c}=10^{13}\,M_\odot$, this is usually simply only the galaxy, and there is no large difference between the stellar mass of the galaxy and the stellar mass within the whole aperture. Above halos of group mass, however, this is a different picture. Here, the question arises what is the content of the BCG, and what is actually in the satellite galaxies that, from observations, are usually masked, which is a different procedure than the potential-based algorithms used to find satellite galaxies in simulations (see Sec.~\ref{sec:subfind}). Thus, comparisons even between observations, but also between simulations and observations, need to be taken with this issue in mind.

Therefore, we present here three different versions of the SMHM relation, including comparisons to observations that match the different definitions. The first compares the halo mass defined as $M_\mathrm{200,c}$ to the stellar mass defined as the stellar mass within 10\%$R_\mathrm{vir}$ that is allocated by \subfind to the central galaxy and not to any satellite. The resulting SMHM is presented in Fig.~\ref{fig:smhm200}, for the redshifts $z=4$, $2$, $1$, and $0$, from left to right, with the different simulation resolutions marked by different colors as given in the legend. As can be seen immediately, we find good convergence between the different resolutions at all redshifts, clearly demonstrating that our subgrid models scale well with resolution.

\begin{figure*}
    \includegraphics[width=0.99\textwidth]{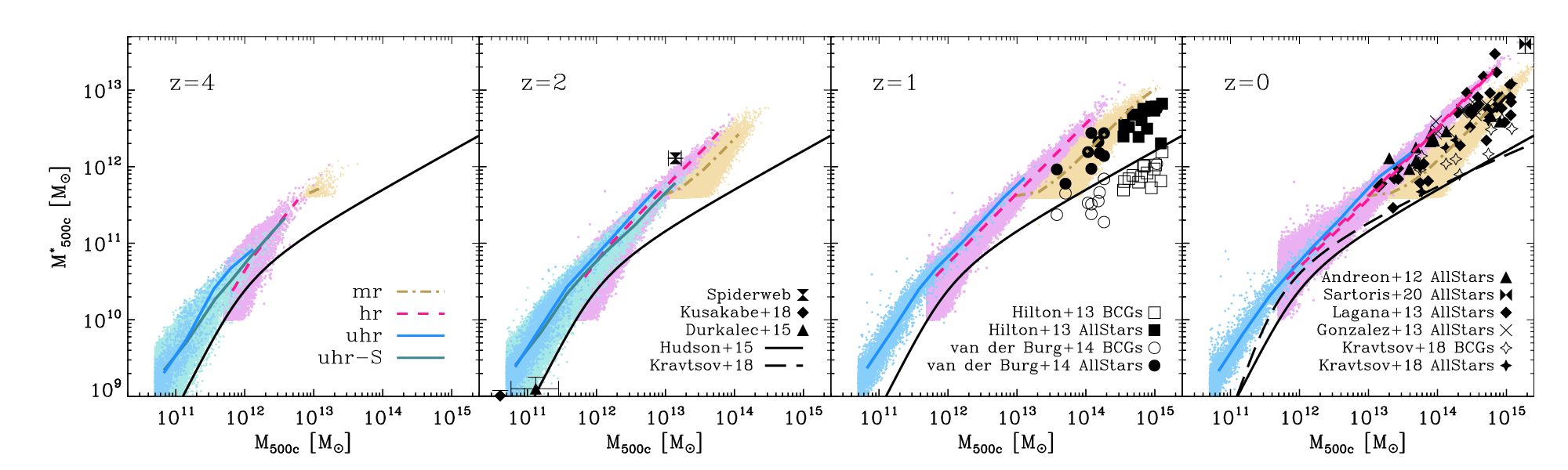}
    \caption{Stellar-mass--halo-mass relations for the \magpath simulations, with colors indicating the different resolutions as indicated in the legend. Here, the halo mass is given as $M_\mathrm{500,c}$, and for the stellar mass all stars within $R_\mathrm{500}$ are used, denoted $M_\mathrm{500c}^*$, independent if they belong to a satellite, the BCG, or the ICL. The black solid line shows the SMHM relation found from lensing at $z\approx0$ by \citet{hudson:2015}, as a reference at all redshifts. In addition, different observations are included in black, with solid symbols marking measurements of all stars and open symbols indicating the BCG measurements only. Included observations at $z=0$ are by \citet{kravtsov:2018} as black stars and black dashed line, from \citet{lagana:2013} as black diamonds, \citet{sartoris:2020} as black bowtie, \citet{andreon:2012} as black triangles, and from \citet{gonzalez:2013} as black x. At $z=1$, observations from \citet{vanderburg:2014} are shown as circles, and observations by \citet{hilton:2013} are shown as squares. These observations are all obtained from X-ray measurements. At $z=2$, two data points from Halo Occupation measurements by \citet[][triangle]{durkalec:2015} and \citet[][diamond]{kusakabe:2018} are included. In addition, the data point for the Spiderweb protocluster is shown at $z=2$, with stellar masses from \citet{seymour:2007} and halo mass measurements from \citet{tozzi:2022}. Note that both values are measured within 100~kpc, which would be only about half of $R_{500}$ for the halo mass reported by \citet{tozzi:2022}.}
    \label{fig:smhm500}
\end{figure*}

This definition of stellar mass is as close to measuring the mass of the BCG as systematically possible from simulations without inspecting every galaxy separately, ignoring the fact that most likely part of the ICL will be added to the stellar mass of the BCG by this method. It is in any case very difficult to spatially split the ICL and BCG \citep{remus:2017_icl}. We thus expect our simulated values to be larger than stellar masses of observed BCGs, which are shown as open symbols in Fig.~\ref{fig:smhm200}, but smaller than what is found from observations if all stars are accounted in a similar aperture from observations, shown as filled black symbols in Fig.~\ref{fig:smhm200} at $z=1$ and $z=0$. We show individual observations compiled from the literature from \citet[][open and filled stars]{kravtsov:2018} using X-ray measurements to infer the halo mass and optical and IR observations to obtain stellar masses at $z=0$, and at $z=1$ from the GCLASS survey by \citet[][open and filled circles]{vanderburg:2014} using SED fitting for the stellar masses and velocity dispersion proxies for the halo masses, from SPT measurements combined with ACT for dynamical masses and Spitzer photometry for the stellar masses as presented by \citet[][open squares]{hilton:2013}, and halo masses obtained from Chandra and XMM-Newton with stellar masses from Spitzer-IMACS as presented by \citet[][filled diamonds]{patel:2015}.
Our simulation is closer to the all stars measurements than the BCG only measurements, which was to be expected from the BCG measuring method as discussed before.

In Fig.~\ref{fig:smhm200} we also show the measured SMHM relation observed from the weak lensing survey CFHTLenS by \citet{hudson:2015} at $z=0$ as a solid black line, and we include this line for reference in all redshift panels. Overall, our simulation is usually slightly above this SMHM at all redshifts, but does not show a strong evolution with redshift. This is different from what was suggested based on semi-analytical models (SAMs) by \citet{moster:2013} and \citet{behroozi:2013}, who find a strong evolution with redshift as indicated by the light gray dashed and dashed-dotted lines, respectively. Interestingly, none of the observations agree with these models, especially at the high-mass end where the models suggest much smaller masses in the BCGs than the observations by \citet{hudson:2015} or any of the other individual measurements. \citet{kravtsov:2018} argue that this could partially be due to the stellar mass function used to build the SAMs. If they use a different SMF instead, one with larger stellar masses present at $z=0$, they obtain the black dashed line shown in the right panel of Fig.~\ref{fig:smhm200} based on the model by \citet{behroozi:2013}. This demonstrates clearly the issue already addressed in the previous section discussing the stellar mass function, as small changes at the high-mass end can lead to strong differences in the obtained SMHM relation.

Note that by using 10\%$R_\mathrm{vir}$ as the aperture cut for obtaining the stellar masses we have implemented an aperture that varies with halo mass. Instead, as discussed for example by \citet{pillepich:2018} and \citet{brough:2024}, a fixed aperture could be used. Following \citet{pillepich:2018}, we chose a fixed aperture of $30~\mathrm{kpc}$ to obtain stellar masses, which leads to much smaller stellar masses at the high-mass end while not affecting the low-mass end, as shown in Fig.~\ref{fig:SMHM_ap30}. If such an aperture is used, the \magpath simulations at all resolutions resemble the observations by \citet{hudson:2015} perfectly at $z=0$, in excellent agreement with the results found for IllustrisTNG by \citet{pillepich:2018} and EAGLE by \citet{matthee:2017}. A comparison of the different simulations was already discussed in length by \citet{remus:2022} and we will not go into details on this here. Note, however, that \citet{brough:2024} showed that $30~\mathrm{kpc}$ is not representative of what is found as cut for BCGs, but instead if a fixed aperture should be used, $100~\mathrm{kpc}$ would be closer to what is observed. In addition, especially at the high-mass end, observational apertures are usually not fixed \citep[e.g.][]{kluge:2020}. We agree with that assessment, which is why we included that particular figure only in the appendix for completeness.

At higher redshifts, measurements of the SMHM relation are rare, and we have included at $z=2$ two measurements, one of Ly-$\alpha$ emitters \citep{kusakabe:2018} and another one using VIMOS ultra deep survey observations \citep{durkalec:2015}, that both use Halo Occupation Distribution clustering matches to determine the halo mass. This method is completely different from the other methods used in this work, and the error bars are huge. However, they provide a guideline for what could be expected at the low-mass end from observations for the position of the SMHM relation. Interestingly, these observations agree well with what is found in our simulation, having larger values than the $z=0$ relation from \citet{hudson:2015} reports.

As discussed before, based on X-ray observations halo masses are usually only obtained within $R_\mathrm{500,c}$, and thus we show in Fig.~\ref{fig:smhm500} the SMHM relation for the stellar and the dark matter mass obtained within $R_\mathrm{500,c}$, dubbed as $M_\mathrm{500,c}$ and $M^*_\mathrm{500,c}$, respectively. Note that $M^*_\mathrm{500,c}$ contains all stars within $R_\mathrm{500,c}$, those still bound in satellites as well as the BGC and the ICL. The different resolutions of the \magpath simulations again show excellent convergence across the different volumes and time with the exception of the medium resolution, which towards redshift zero has typically a factor two less stellar mass within the very massive clusters. Again, the $z=0$ SMHM relation observed by \citet{hudson:2015} from weak lensing is shown as a solid black line at all redshifts for reference. As discussed before, the simulations are above the relation reported by \citet{hudson:2015}, but show the bend at the same halo mass as \citet{hudson:2015}, at about $M_\mathrm{500c}\approx10^{12}\,M_\odot$ at all redshifts. We do not find a change in the position of this bend with redshift, different than what was suggested by the SAM models from \citet{behroozi:2013} and \citet{moster:2013}. 

\begin{figure*}
    \includegraphics[width=0.99\textwidth]{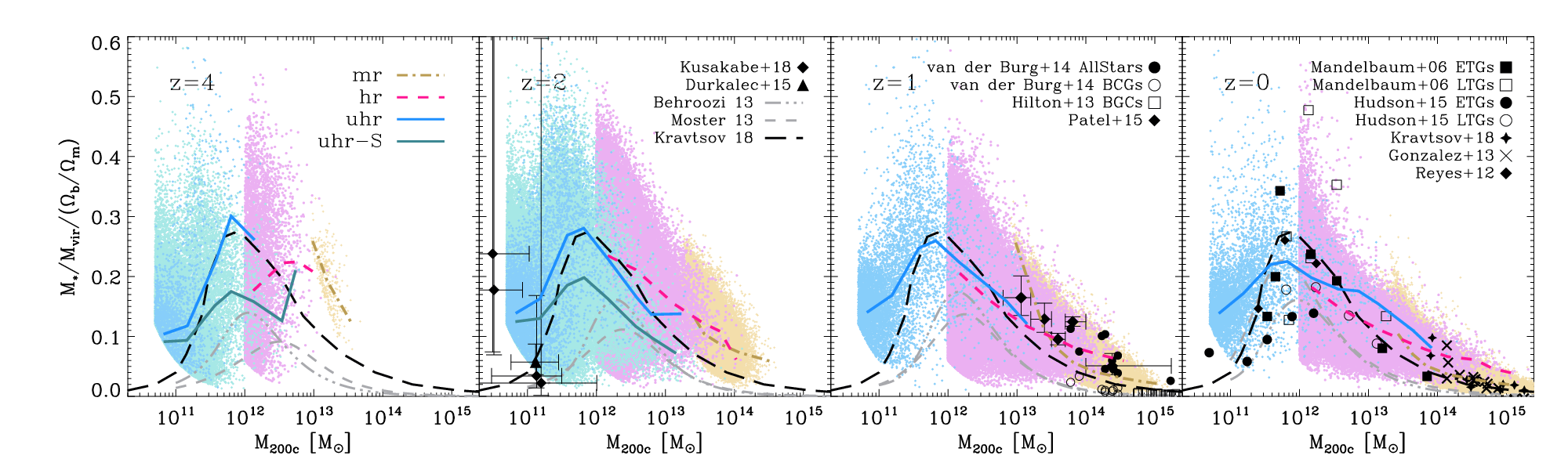}
    \caption{Baryon conversion efficiency versus halo mass given as $M_\mathrm{200,c}$, for all resolutions of the \magpath simulations in color as indicated by the label. At all redshifts, the $z=0$ relation from the model by \citet{kravtsov:2018} is included as black dashed line, as well as the relations from the SAM models by \citet[][gray dashed curve]{moster:2013} and \citet[][gray dash-dotted lines]{behroozi:2013} as predicted for the different redshifts. In addition, at $z=0$ observational data are included from lensing by \citet{mandelbaum:2006} for ETGs (filled squares) and LTGs (open squares), by \citet{hudson:2015} for ETGs (filled circles) and LTGs (open circles), and by \citet[][diamond]{reyes:2012}, and from X-ray measurements by \citet[][black x]{gonzalez:2013} and \citet[][black stars]{kravtsov:2018}. At $z=1$, observations from X-ray measurements by \citet[][circles]{vanderburg:2014}, \citet[][squares]{hilton:2013}, and \citet[][diamonds]{patel:2015} are shown. At $z=4$, measurements from Halo Occupation Distribution methods are shown from \citet[][diamonds]{kusakabe:2018} and \citet[][triangle]{durkalec:2015}.}
    \label{fig:barcon}
\end{figure*}

The \magpath simulations again agree well with the observations based on X-ray measurements by \citet{kravtsov:2018}, \citet{lagana:2013}, \citet{andreon:2012}, and \citet{gonzalez:2013} at $z=0$, and by \citet{vanderburg:2014} and \citet{hilton:2013} at $z=1$, at the group mass scale. For very massive clusters (e.g. $M > 10^{15}M_\odot$) there is a factor $2-3$ difference in the stellar mass between the simulations and the bulk of the observations. This is a long standing problem and it is not clear how much of this difference is caused by possibly missing diffuse light in the observations and how much is caused by not enough suppression of star-formation in massive galaxies due to imperfections of the implemented AGN feedback in the simulations. Note that this is also directly linked to the long standing problem of the observed metal content within the ICM which appears in tension with the observed, total amount of stars as source, the so-called iron conundrum (for a recent reassessment, see Biffi et al. 2025, submitted). However, while the \magpath simulations well reproduce the observed metal content of the ICM \citep[for a recent review, see][]{biffi:2018}, the larger stellar mass in simulated clusters corresponds naturally to what is needed as source for the metals (by construction), and thus lower masses would lead to a lowered metal content. Furthermore, it is worth to notice that some clusters from the \citet{lagana:2013} sample and the \citet{andreon:2012} sample, as well as the very deep observation of a CLASH cluster presented by \citet{sartoris:2020}, report an observed stellar content as massive as found in the simulations, in actually excellent agreement with the \magpath simulations. Thus, more detailed investigations of the observations are of high importance to further inform the simulations on that matter.

We also find no relevant shift between scaling based on $M_\mathrm{500c}$ and $M_\mathrm{200c}$ at the low-mass end, and thus the comparison at $z=2$ to the HOD data from \citet{kusakabe:2018} and \citet{durkalec:2015} is the same as for Fig.~\ref{fig:smhm200}. Interestingly, here we can add another data point at $z=2$ based on X-ray measurements, namely the well-known Spiderweb protocluster at $z=2.156$ \citep{miley:2006}, shown as black hourglass. Stellar masses within $100~\mathrm{kpc}$ were measured for the central galaxy without satellites by \citet{seymour:2007}, and within the same aperture of $100~\mathrm{kpc}$ \citet{tozzi:2022} report a halo mass based on the X-ray measurements, assuming a constant radial distribution of the gas. However, as shown by \citet{lepore:2024}, the Spiderweb protocluster actually has a cool core, and thus the radial profile is not constant. \citet{lepore:2024} report a smaller halo mass from that, as will be discussed later. Given on the halo mass reported by \citet{tozzi:2022}, $100~\mathrm{kpc}$ unfortunately is not close to $R_\mathrm{500,c}$ but rather about half of that, but if the halo mass is indeed a bit smaller as reported by \citet{lepore:2024} it becomes closer to $R_\mathrm{500,c}$. In any case, as the measurement is within the same consistent radii, we included the data point in Fig.~\ref{fig:smhm500}. Interestingly, the point is slightly above the SMHM relation found for the \magpath simulations, different than at lower redshifts, which could be either due to the very peculiar nature of the Spiderweb protocluster, the smaller radial range of the measurement, or a higher-redshift reality. Unfortunately, to disentangle that, more observations at these redshifts are needed and thus it is work for the future. 

The relation between the stellar mass and the halo mass can also be expressed in form of the baryon conversion efficiency, which enhances the deviations from the average values and highlights the differences more clearly and thus is shown in Fig.~\ref{fig:barcon}. Again, the different \magpath simulation resolutions and volumes are indicated by color as stated in the legend. As can be seen, we again see excellent convergence between the different simulations for the fiducial runs at all redshifts from $z=2$ to $z=0$, but see large differences at $z=4$, which are most likely due to the fact that the lower resolution but larger volume simulations do not have many halos yet that are large enough to be properly resolved, and thus they are not very numerous. Interestingly, the differences between the fiducial and the advanced BH model simulations becomes much more evident than in the SMHM relations: while the scatter and the peak of the baryon conversion efficiency are in good agreement between both simulations, the overall baryon conversion efficiency is lower for the advanced model run \Bthuhr than for the fiducial run \Bthhr. Overall, we do not find an evolution in the baryon conversion efficiency with redshift, contrary to the SAM predictions from \citet{moster:2013} and \citet{behroozi:2013}, which are shown as light gray curves in Fig.~\ref{fig:barcon}. In fact, for the \magpath simulations the peak of the baryon conversion efficiency is at a halo mass of $M_\mathrm{200,c}\approx10^{12}\,M_\odot$ for all redshifts from $z=4$ to $z=0$.

\begin{figure*}
    \includegraphics[width=\textwidth]{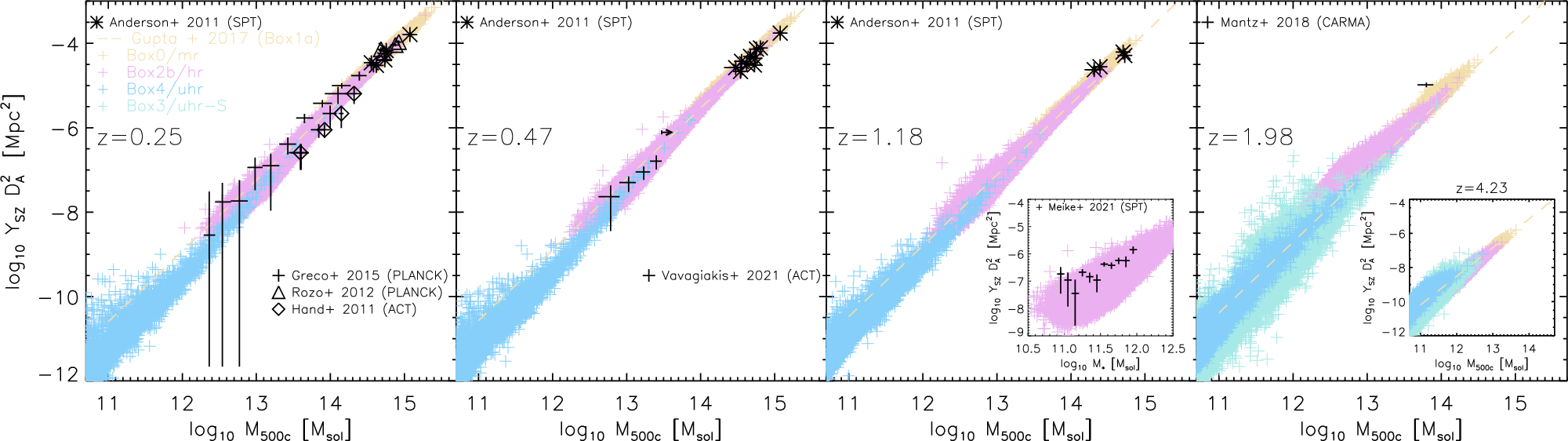}
    \caption{Evolution of the Compton $Y$ inside $R_{500c}$ vs.\ halo mass $M_{500c}$, for redshifts $z\approx2$, $1.2$, $0.5$, and $0.25$. Differently colored points indicate varying resolutions and box sizes, while the data points with error bars are observations as presented by \citet{2011ApJ...736...39H}, \citet{2011ApJ...738...48A}, \citet{2012ApJ...760...67R}, \citet{2018A&A...620A...2M}, \citet{2021PhRvD.104d3503V}, \citet{2021ApJ...913...88M}, and \citet{2015ApJ...808..151G}.
    }
    \label{fig:szmasscaling}
\end{figure*}

The comparison to observation again reveals interesting results, which were already discussed in detail by \citet{teklu:2017} for the \magpath \Bfuhr simulations at $z=0$: we find the scatter in the simulation to be of a similar order of magnitude as that found in observations, with the observations showing the peak of the baryon conversion efficiency at a similar halo mass than the \magpath simulations. In addition, the overall behavior with halo mass is in excellent agreement with observations, showing a lower efficiency at the high-mass end due to the implemented AGN feedback, and a lower efficiency at the low-mass end due to the stellar feedback. 

Overall, the \magpath simulations show a slightly too large efficiency at low halo masses, most likely due to the stellar feedback not being strong enough as already discussed in the context of the stellar mass function low-mass end. At the high-mass end, the simulation agrees well with observations, although the simulation shows larger values than the model predictions or the BCG-only measurements. Especially at $z=1$, the measurements by \citet{patel:2015} find larger baryon conversion efficiencies than even the simulations, but especially than the older models.

At $z=0$, weak lensing observations from the CFHTLenS survey by \citet{hudson:2015}, but also observations from galaxy--galaxy weak lensing from SDSS by \citet{mandelbaum:2006} and \citet{reyes:2012} have split their observations between early-type galaxies (ETGs) and late-type galaxies (LTGs), finding slightly larger efficiencies for LTGs than for ETGs. Using \Bfuhr, \citet{teklu:2017} showed that we find a similar split, albeit not as strong, for this simulation at $z=0$, but since we only have enough resolution to split ETGs from LTGs in the \uhr resolution simulations, we do not attempt this here again and refer the reader to the work by \citet{teklu:2017} for more details on that matter.

To conclude, we find an excellent convergence between the different simulation volumes and resolutions. Furthermore, the \magpath simulations appear to produce too many stars at the low-mass end due to the stellar feedback not being strong enough, and too many stars at the high-mass end due to the AGN feedback not being able to stop cooling from the halo again. Depending on the definition of the stellar and the halo mass, we find good or even excellent agreement with the observations, or an overall too large amount of stars being included in the halos, although this could also originate from the halo finder algorithm being strongly different from observational methods. 
In general, the shape of the SMHM relation and the baryon conversion efficiency relation are in excellent agreement with observations at $z=0$, and show a similar scatter, even if an offset is found, at all redshifts where observations could be compared. We do not find any evolution with redshift since $z=4$, in disagreement with results by \citet{shuntov:2022} who claim to see a shift in peak mass towards higher stellar and dark matter masses with redshift, similar to the model by \citet{moster:2013} and \citet{behroozi:2013}. They compare to the EAGLE, TNG100, and HorizonAGN simulations, claiming that for the cluster-mass end too many satellites are found compared to observations, interpreting that this is due to simulations not quenching the galaxies properly in clusters. However, all these simulations are rather small box volumes and their clusters therefore are low in number and more importantly all of them are non-relaxed due to the box size. Thus, they should contain more satellite galaxies than the average galaxy cluster, as there was no time yet for satellites to be destroyed, as shown by \citet{kimmig:2025b}. Therefore, it is not yet clear how exactly the SMHM relation and the baryon conversion efficiency behave at different redshifts, and it will be an interesting comparison to be done in the future.

\subsection{Compton-Y--Halo-Mass Relation}
\label{sec:szhm}

\begin{figure*}
    \includegraphics[width=\textwidth]{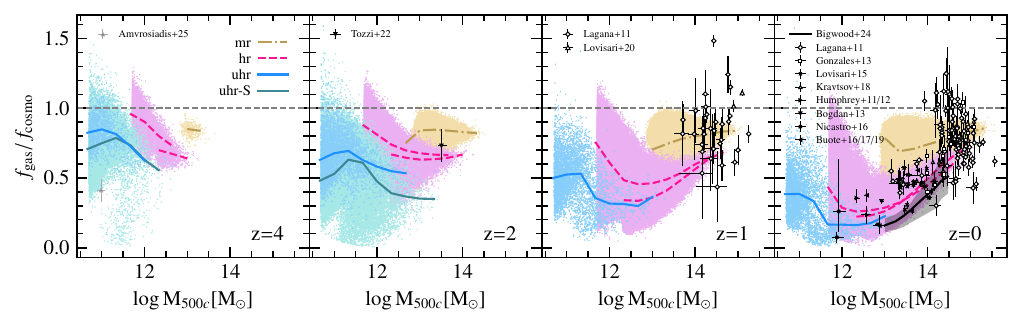}
    \caption{Evolution of the gas mass fraction $f_{\rm gas}/f_{\rm cosmo}$ inside $R_{500c}$ vs.\ halo mass $M_{500c}$, for redshifts $z=4$, $2$, $1$, and $0$. The dashed horizontal line marks the threshold of $f_{\rm gas}/f_{\rm cosmo}=1$. Differently colored points indicate varying resolutions and box sizes, while the respective lines represent their median values. The observed data points are from samples by 
    \citet{bigwood:2024,lagana:2011, gonzalez:2013, kravtsov:2018, humphrey:2011, humphrey:2012, bogdan:2013,  nicastro:2016, buote:2016, buote:2017, buote:2019, lovisari:2020,tozzi:2022}; and \cite{amvrosiadis:2025}.}
    \label{fig:gasmasfraction}
\end{figure*}

The (hot) gas content in halos plays an important role in understanding the physics that govern structure formation through cosmic time, since they are the end-product of both accretion and feedback, and thus any combination of physics that successfully represents the real Universe requires the hot gas content to match.
The chosen feedback description within \magpath leads to predicted pressure profiles for simulated massive galaxy clusters that match the observations extremely well \citep{pc:2013,mcdonald:2014}. Although the bulk of the total Y-signal comes from regions around $R_{500c}$, this relations is not very sensitive to the AGN feedback description (unless very extreme models are used), as soon as once restrict to a more central part (like $R_{2500}$) it gets clear that a AGN feedback like used in \magpath is needed to make the simulations matching the observations \citep[see extensive discussion in][]{10.1093/mnras/stx318}. In a detailed study, \citet{2017MNRAS.469.3069G} even proposed an improvement in the parametrization of the universal pressure profile based on the simulations which overall also fits better to the observations. Detailed analysis of deep light-cones also demonstrated that based on the PLANCK SZ measurement, the mean, the PdF and the SZ power spectrum up to $l\approx1000$ are well reproduced by the signal extracted from the simulated light-cones \citep{2016MNRAS.463.1797D}. In addition, the mean thermal energy density within the universe as extracted from the simulations is in very good agreement with the observational values available up to the redshift of $z\approx1$ and compatible with the upper limits available up to the redshift $z\approx3$ \citep{2021PhRvD.104h3538Y,2024PhRvD.109f3513C}. The predicted value of the pairwise kinetic SZ effect is compatible with the observations within the observed halo mass range \citep{2018MNRAS.478.5320S}. 

Fig.~\ref{fig:szmasscaling} shows a summary of the predicted Compton $Y$ parameter for the \magpath simulations over a wide range of masses and redshifts compared to observations at a similar redshift, with the resolutions given in different colors as indicated by the legend, at $z=0.25,0.47,1.18,1.98$ from left to right, with $z=4.23$ given in the small inlet in the rightmost panel. At all redshifts, we find convergence between the different resolutions and box volumes. For comparison, several observations are included: For the SPT sample of galaxy clusters \citep{2011ApJ...738...48A} shown as black stars, we selected clusters close to the redshift displayed and corrected for the $E_z$ factor between the observed and the displayed redshift. In the $z=0.25$ panel, we also added the ACT clusters as presented by \citet{2011ApJ...738...48A} as black open diamonds. Here we also corrected for the $E_z$ factor between the observed and the displayed redshift and also converted the observationally used $M_{200m}$ to $M_{500c}$ following the scaling reported by \citet{2021MNRAS.500.5056R}, which we also extended to the scaling of the according $Y$ parameter in the simulation.
In addition, we also added data points for A1413, A478, A2204, and RXJ1720.1+2638 from the PLANCK measurements \citep[][black triangles]{2012ApJ...760...67R}, also correcting for the $E_z$ factor between the observed and the displayed redshift. Furthermore, we added the PLANCK data for local brightest galaxies \citep{2015ApJ...808..151G} as black plus. To obtain halo masses from reported stellar masses, we use the scaling by either \citet{moster:2013} or \citet{kravtsov:2018}, finding an order of magnitude difference at the cluster mass scale between the two methods. Therefore, for illustration purposes, we decided to take the mean value between the two methods to add these data points. We note that for the largest mass bin, only a lower bound for the corresponding halo mass has been given. 

In the $z=0.47$ panel, in addition to the SPT galaxy clusters from \citet{2011ApJ...738...48A}, we also added the ACT data from \citet{2021PhRvD.104d3503V} as black crosses, where we converted the quoted $M_{vir}$ to $M_{500c}$ following the scaling reported by \citet{2021MNRAS.500.5056R}. At a redshift around $z\approx1$ we include the SPT galaxy clusters \citep{2011ApJ...738...48A}. As any stellar-to-halo mass conversion is largely uncertain at this high redshift, we placed a sub-panel where we compare SPT data for DES/WISE galaxies \citep{2021ApJ...913...88M} data directly to the scaling with stellar mass from \Bthr. At $z=2$ we added the CARMA data-point from \citet{2018A&A...620A...2M}. The sub-panel shows the scaling relation at $z=4.23$. Both the $z=2$ and the $z=4.23$ panels additionally show data points from the advanced AGN model simulation \Bthuhr. It is clear that, especially at very high redshift, the feedback treatment differs, which results in an improved passive fraction of galaxies as discussed in section \ref{sec:numbdens}. In all panels, the pink dashed line represents the fit based on \Bomr as presented by \citet{2017MNRAS.469.3069G}. Although this fit was obtained from massive clusters, it represents the simulation results quite well over a much larger mass range and across all redshifts displayed, highlighting that the simulations only mildly deviate from a universal scaling at halo masses corresponding to galaxy scales. 

Overall, we find excellent agreement between the different resolutions, and between the simulations and observations, at all inspected redshifts. This is true for both the comparison of the Compton-Y to the halo masses, but also to the stellar masses. This clearly demonstrates that the \magpath simulations are exceptionally well suited for galaxy cluster studies, as they obtain hot atmospheres in groups and clusters from the highest to the lowest redshifts in agreement with observations, which other simulations strongly struggle to achieve \citep[e.g.,][]{popesso:2024,bigwood:2024}.

\begin{figure*}
    \includegraphics[width=0.99\textwidth]{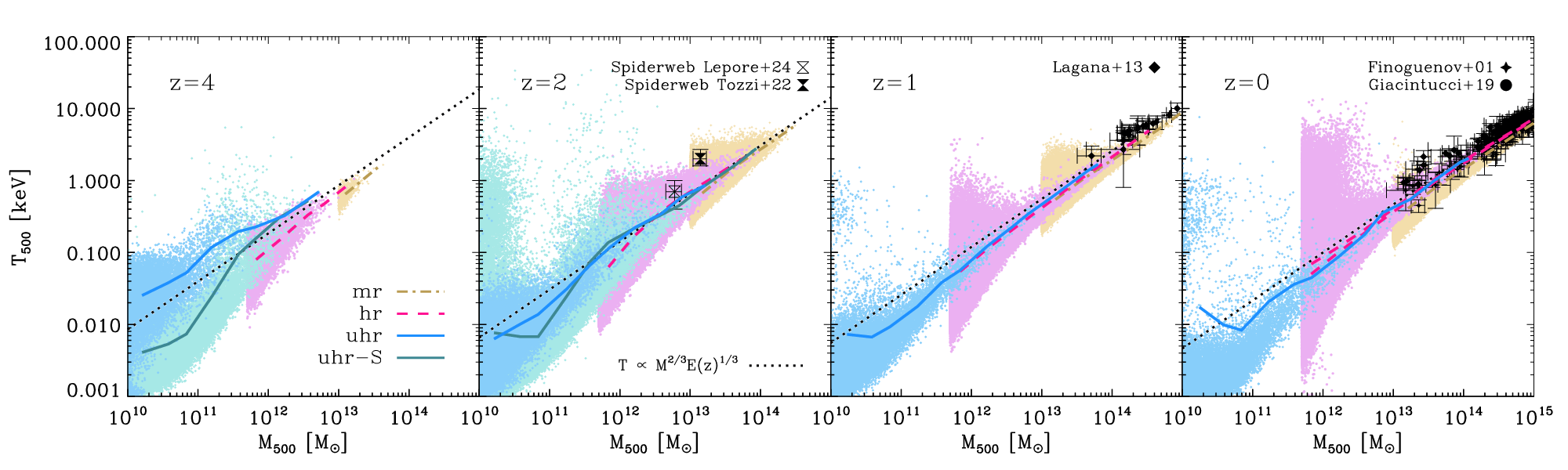}
    \caption{Gas temperature versus halo mass within $R_\mathrm{500,c}$ from $z=4$ to $z=0$ from left to right panel for the \magpath simulations at different resolutions as indicated in the legend. Observations by \citet{finoguenov:2001} and \citet{giacintucci:2019} at $z=0$ are shown as black stars and circles, respectively. At $z=0$ and $z=1$, observations from \citet{lagana:2013} are included as well. At $z=2$, observations for the Spiderweb protocluster are shown according to \citet[][filled hourglass]{tozzi:2022} or \citet[][open hourglass]{lepore:2024}. The dotted line shows the relation found from a self-similarity model from \citet{boehringer:2012}.}
    \label{fig:m_t_evol}
\end{figure*}

\subsection{Gas-Mass--Halo-Mass Relation}
\label{sec:gmhm}

The baryon distribution in cosmic structures is an important indicator for the nature and evolution of halo assembly. Especially, the amount of gas that a structure contains to form stars or to be present as a hot atmosphere is an important quantity for galaxy and galaxy cluster formation. This is also strongly related to the description of the feedback, both for the stellar feedback \citep{2006MNRAS.365.1021E} as well as the AGN feedback description \citep{2013MNRAS.431.1487P}. An in-depth analysis on the baryonic fraction and its evolution in the \magpath can be found in \citet{Angelinelli2022, Angelinelli2023}.  Fig.~\ref{fig:gasmasfraction} shows the gas mass fraction $f_{\rm gas} / f_{\rm cosmo}$ against the total mass $M_{500c}$ inside $R_{500c}$ of all halos from the \magpath simulations, with different resolutions colored as indicated in the legend, above our chosen mass cut for each respective resolution for redshifts $z=4$, $2$, $1$, and $0$. Analogously to other works, we adapt a cosmic baryon fraction $f_{\rm cosmo} = 0.168$ following \citet{ettori:1999}. Halos below the gray horizontal line at $f_{\rm gas}/f_{\rm cosmo}=1$ thus have a lower gas content than the upper limit defined by the cosmic baryon fraction. In each panel, colored solid lines indicate the median values for the simulations in different resolutions, while markers show observed values from different surveys and redshifts. The black solid line in the panel for $z=0$ is  Naturally, the only galactic data is from the Milky Way at $z=0$, indicated by two filled circles, which represent two models by \citet{nicastro:2016}.

For the extragalactic sources, the halo masses $M_{500c}$ were derived with varying methods. \citet{bigwood:2024} derived their halo mass estimates from a joint constraint of weak lensing and kinetic Sunyaev–Zel’dovich effect. \citet{lagana:2011} applied a scaling relation between the gas and total mass, which they inferred from a sample of dynamically relaxed groups and clusters. \citet{gonzalez:2013} used the $T_X$--$M_{500c}$ relation between X-ray temperature and halo mass  calibrated by \citet{vikhlinin:2009} in order to infer total masses. A rather common way is to derive the halo mass under the assumption of hydrostatic equilibrium, as was done by \citet{humphrey:2011,humphrey:2012,lovisari:2015, lovisari:2020, buote:2016,buote:2017} and \citet{buote:2016}, although this conjecture strictly only holds in dynamically relaxed objects. \citet{bogdan:2013}, on the other hand, calculated the total masses using the baryonic Tully-Fisher relation, which associates the halo mass $M_{200c}$ with the maximum rotational velocity of the galaxies $V_{\rm max}$ via $M_{200}\propto V_{\rm max}^{3.23}$. In order to compare this data set with the rest of the observed data points (where the total mass is generally given inside $R_{500c}$), we applied a conversion factor of $0.7$ to scale their values for $M_{200c}$ to estimates of $M_{500c}$. This conversion factor was reported by \citet{2021MNRAS.500.5056R} and is based on the analysis of the multi-cosmology runs of the \magpath simulations. Similarly, we have included scaling parameters based on work by \citet{2021MNRAS.500.5056R} while adding the data by \cite{tozzi:2022}, where the authors inferred the halo mass from self-similar mass-temperature scaling (cf. Sec.~\ref{sec:temp}). 

The gas masses, on the other hand, could be calculated by integrating the measured X-ray profiles for most observations shown in Fig.~\ref{fig:gasmasfraction}, particularly at low redshift. The exception is the data by \citet{bigwood:2024}, where the authors applied a halo mass dependent scaling, which is based on findings by \citet{akino:2022,salcido:2023}. At high redshift, however, robust X-ray measurements to large radial distances become increasingly difficult. Hence the gas mass for the Spiderweb protocluster in the panel for $z=2$ was approximated by assuming constant gas density inside an aperture of 100~kpc by \citet{tozzi:2022}. Although the estimate for the virial radius lies around $R_{500c}=220~\mathrm{kpc}$, this value is generally poorly defined for such a dynamically active object. Hence we decided to include this data in our comparison. An even more tricky observation to compare with is the high-redshift galaxy SPT-2147 in the panel for $z=4$ by \citet{amvrosiadis:2025}. Observing galaxies at such enormous distances is highly challenging, which is why the measurements for the dynamical mass and gas mass for SPT-2147 are within a relatively small aperture of $4~\mathrm{kpc}$. Although that is most probably much smaller than $R_{500c}$, we decided to include this measurement in our figure for completeness, but color the data point in gray to clarify that this is not a rigorous comparison.

Focusing on the panel for $z=0$, we see a trend of decreasing gas mass fractions with halo mass, until a turnover point around $M_\mathrm{500c}\approx10^{12.5}\,M_\odot$. Then the fraction starts to increase again, coming close to cosmic baryon fraction for the most massive clusters. This degree of gas abundance inside galaxies and clusters is mostly determined by the baryon conversion efficiency. As discussed in \cref{sec:smhm}, the baryon conversion efficiency is increasing with halo mass for galaxies below a similar turnover point in mass around $M_\mathrm{200c}\approx10^{12}\,M_\odot$. Therefore, we see the relative decrease in gas mass for objects in this mass range. But when the halo mass further increases, AGN feedback starts to efficiently heat the surrounding gas, causing a drop in conversion efficiency. But since the halo masses are now sufficiently massive, this hot intra-cluster gas will still be bound, regardless of its high temperature and unability to form stars. Hence we see this increase in gas mass fractions again towards galaxy clusters.

\subsection{Temperature--Mass Relations}
\label{sec:temp}

As mentioned before, gas exists in different temperature stages for halos of different mass. Low-mass halos host galaxies where most of the gas is still cold, usually with a large amount in molecular form, building the reservoir for star formation potential of galaxies. The more massive the halo, the better it can also hold a halo of hot gas, gas that has been accreted onto the galaxy and then heated by feedback processes and ejected out into the halo. Alternatively, some part of the hot gas also originates from merger shocks that are very efficient in heating gas as well. The hottest halos are usually found in galaxy clusters, shining bright in X-ray and as such the least problematic to measure of all the gas components. As these hot halos are also extremely efficient in suppressing star formation, obtaining the hot and cold gas phases of halos over several orders of magnitude in mass correctly is extremely crucial to also reproduce realistic galaxies as well as galaxy clusters and their galaxy populations. As shown already by \citet{popesso:2024}, the \magpath simulations capture the X-ray luminosity--halo mass scaling relation from observations extremely well, better than other simulations like Flamingo, Eagle, Simba, or IllustrisTNG. Here, we now focus directly on the temperature instead of the X-ray luminosity, allowing us to extend the study also to Milky-Way mass galaxies and below.

Fig.~\ref{fig:m_t_evol} shows the gas-temperature--halo-mass relation for the \magpath simulations for all resolution levels as indicated in the legend, compared to observations. The median of the distribution found for the different resolutions is shown as dark lines, with the individual halos shown as points. As can be seen directly from the plot, the different resolutions converge extremely well below $z=2$, and for the high-mass end also reasonably at $z=4$, but it is clear that at $z=4$ the halos are still in assembly and the hot halos are still in the process of being built up. Only the most massive nodes at that redshift have already started to build up hot halos, as shown by \citet{remus:2023} in more detail for the protocluster regime. At $z=4$, we also see again the deviation between the fiducial BH model runs of \Bfuhr in blue and the advanced BH model of \Bthuhr in turquoise: below $M_\mathrm{500,c} \approx10^{12}\,M_\odot$, both simulations deviate from one another, with the fiducial model having overall hotter halos than the advanced model. This is particularly interesting as the advanced model is much more efficient in producing massive quenched galaxies already at $z=4$ \citep[e.g.][]{kimmig:2025,remus:2025}, which is commonly assumed to be due to AGN feedback that heats the halo. However, as already discussed by \citet{kimmig:2025}, this picture of too simple, as the AGN at these high redshifts is in an epic dance with gas inflow, finding a complex balance between quenching and rejuvenation as shown by \citet{remus:2025}, with the quenching also depending on the position in the cosmic web at high redshifts. And while there is overall more cold gas available at high redshifts in the advanced BH model simulations, we can see that all simulations produce heated halos at the group mass (and thus protocluster) level at $z=4$, where the most massive nodes of this epoch have already grown enough to hold a hot atmosphere.

\begin{figure*}
    \centering
    \includegraphics[]{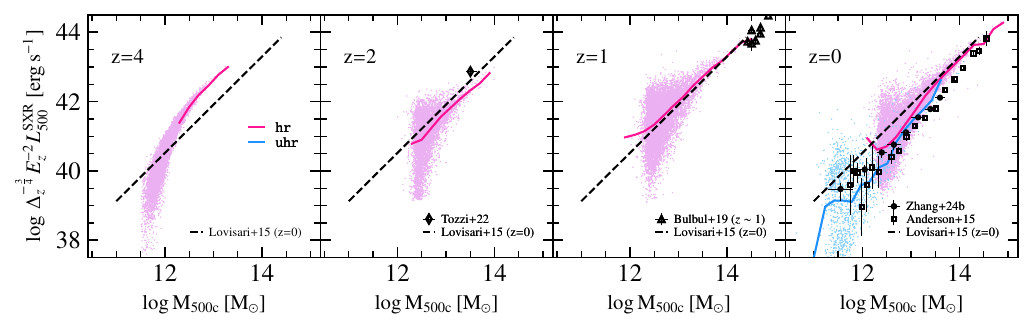}
    \caption{X-ray luminosity as a function of halo mass within $R_\mathrm{500}^\mathrm{crit}$ from $z=4$ to $z=0$ from the left to right panel for the \magpath simulations at various resolutions.
    The black dashed line shows the observed relation for groups and clusters at $z=0$ from \citet{lovisari:2015}. Intrinsic X-ray luminosities have been calculated in the $0.5-2\,\mathrm{keV}$ (SXR) band using the {\sc Phox} algorithm \citep{biffi:2012, biffi:2013}, accounting for redshift-dependent scaling following \citet{boehringer:2012}.
    Observations by \citet{anderson:2015} and \citet{zhang:2024b} at $z=0$ are shown as squares and circles, respectively, and are derived from X-ray stacking experiments on photometric galaxy catalogs.
    Results for the intrinsic X-ray luminosity of \Bfuhr at $z=0$ are published in \citet{vladutescu:2025}.
    }
    \label{fig:LxM500}
\end{figure*}

Overall, at all redshifts we see that the gas generally follows a tight correlation between its temperature and the halo mass, best described by a variation of the self-similar model using a redshift-independent density contrast, shown as dotted line in all panels of Fig.~\ref{fig:m_t_evol}: 
\begin{equation}
    T \propto M^{2/3}\times E(z),
\end{equation}
where $E(z) = H(z)/H_0$ \citep[e.g.][]{hogg:1999}. The original self-similar model proposed a correlation where $ T \propto M^{2/3}\times E(z)^{2/3}$, however, the equation given above is a better fit to the simulations. This was already discussed in detail by \citet{biffi:2014} and \citet{truong:2018}. For a detailed discussion on the expected scaling relations for fixed and redshift-dependent definitions see \citet{boehringer:2012}. While we here obtain the quantity directly from the simulations, as for example shown by \citet{biffi:2014}, it also holds if X-ray mock observations are used thanks to the extremely well developed X-ray mock image generator {\sc Phox} introduced by \citet{biffi:2012} as shown also by \citet{biffi:2013}.

We also compare the simulations to observations in Fig.~\ref{fig:m_t_evol}. At $z=0$, X-ray based observations from \citet{finoguenov:2001} and \citet{giacintucci:2019} are shown as black stars or black circles, respectively. In addition, we show data by the large sample of 123 halos presented by \citet{lagana:2013} as black diamonds, detected with XMM-Newton and Chandra data, ranging from $M_\mathrm{500,c}=10^{13}\,M_\odot$ to $M_\mathrm{500,c}=10^{15}\,M_\odot$. The sample stretches from $z=1$ to $z=0$, and we show clusters with redshifts $z>0.6$ in the panel for $z=1$, and those with $z,0.4$ in the panel with $z=0$. We find excellent agreement between the \magpath simulations and observations, even with a similar scatter at both $z=0$ and $z=1$.

At $z=2$, X-ray detections become increasingly difficult, so only one object could be added to the comparison, namely the Spiderweb protocluster \citep{miley:2006}. As discussed already in previous sections, the data for this protocluster is extremely rich, and it is among the best studied object at $z\approx2$. Two measurements of the temperature and the halo mass can be found in the literature, which both are included in the second panel of Fig.~\ref{fig:m_t_evol}.
Both measurements are performed within a radius of $100~\mathrm{kpc}$, which depending on the assumed mass is somewhere between the value for $R_\mathrm{500}$ or $0.5\,R_\mathrm{500}$. First, \citet{tozzi:2022} present a measurement where they measure the total mass assuming the measured temperature of $2\,\mathrm{keV}$ from \citet{seymour:2007} as a flat temperature profile. More recently, from X-ray spectroscopy \citet{lepore:2024} report a non-flat profile, as they find a radial gradient in the temperature, reaching about $2\,\mathrm{keV}$ only at the largest radii of $100~\mathrm{kpc}$, resulting in an average temperature of $0.7\pm0.3 \,\mathrm{keV}$, leading also to a smaller total mass. The measurement by \citet{tozzi:2022} is shown in Fig.~\ref{fig:m_t_evol} as filled hourglass, while the measurement by \citet{lepore:2024} is shown as open hourglass. Interestingly, the two measurements do not lie within each others' error bars, as those are only the methodological error bars and not those including the discussed differences. Both measurements agree well with the \magpath simulations, with the data by \citet{lepore:2024} bringing the Spiderweb protocluster closer to the average behavior, while for the finding by \citet{tozzi:2022} the Spiderweb protocluster is closer to being an outlier. This is in agreement with what we already found for the SMHM relation, where the measurements by \citet{tozzi:2022} also were closer to being an outlier than the average.

The findings by \citet{lepore:2024} indicate the Spiderweb protocluster to be a cool-core protocluster. This is especially interesting as cool-core clusters have this strong dip in temperature in the centers where gas cools from the halo and rekindles star formation in the BCGs, sometimes even leading to massive starbursts like seen in the Spiderweb protocluster, but also at about $z=0.2$ for the Phoenix cluster \citep{reefe:2025} or other more local cool core clusters \citep{donahue:2015,tremblay:2015}. For the \magpath simulations, \citet{gonzalez_villalba:2025} showed that cool core clusters are most common at halo masses around $M_\mathrm{500,c}\approx10^{14}\,M_\odot$, with the AGN feedback preventing cooling from the halo at lower masses, while mergers destroy cool cores at higher masses extremely efficiently. This is also in agreement with the finding by \citet{kimmig:2025b}, that the most massive clusters tend to be currently more unrelaxed than the spread of low-mass clusters where extremely relaxed clusters commonly occur. The fact that the most massive clusters are currently in highly disturbed phases is also discussed in detail by \citet{kimmig:2023} and for the local Universe by \citet{seidel:2024}, even following the local clusters into the future.
Interestingly, this does not disturb the gas-temperature--halo-mass relation, as usually these cool cores are confined to the centers of the clusters. However, as the Spiderweb protocluster demonstrates, for redshifts of $z=2$ and higher this is important to take into account, since the cool cores can extend out to $100~\mathrm{kpc}$ or more \citep{lepore:2024}. Nevertheless, it will be especially interesting to see if the temperature--mass relation holds in observations to redshifts as high as $z=4$ or even higher, as predicted by the simulations.

\subsection{Lx-Mass relations} \label{sec:lxm}

As outlined in the previous sections, global halo properties such as the total gas mass and average halo temperature can be derived from X-ray observables. 
At typical temperatures and densities in the extended gas atmosphere of halos, the primary cooling processes occur via Bremsstrahlung radiation and metal line cooling, both of which are visible in X-ray.
Thus, the total X-ray luminosity of a halo is directly connected to the gas temperature and gas mass, which are governed by the underlying gravitational potential, i.e. the halo mass.
Especially galaxy clusters are among the most X-ray-luminous objects in the sky, and their X-ray luminosity is tightly coupled to their total mass \citep{vikhlinin:2009, lovisari:2015}.
While some biases stemming from hydrostatic assumptions and spherical symmetry lead to 15\% lower total mass estimates than predicted by simulations, these biases are well understood today \citep{shi:2016_hydro}.

Similar connections between the halo mass and X-ray luminosity also exist for galaxy groups, and they appear to follow scaling relations consistent with cluster measurements \citep[e.g.][]{lovisari:2015, bahar:2022}.
However, recent studies using the \magpath simulations and data of the eROSITA first sky scan suggest that a potential bias towards low entropy groups affects group scaling relations \citep{bahar:2024, Marini2024, Marini2025, Marini2025_assembly}.
\cite{Marini2025_assembly} investigate the physical origins behind the scatter in the $L_X-M$ relation, finding that at the group scales, both AGN feedback and accretion history can significantly impact the halo's hot gas fraction, in turn impacting the X-ray emissivity.
Recent results from eROSITA's 4th all-sky scan \citep{Merloni2024} on galaxies reveal a scaling relation again consistent with groups and clusters \citep{zhang:2024b} with similar slope and slightly lower normalization, confirming earlier results from the ROSAT all-sky survey \citep{anderson:2015}.
This is in contrast to the long-standing assumption that baryonic processes would lead to significantly steeper slopes in the $L_X-M_\mathrm{500}$ relation of galaxies \citep[see e.g.][]{kim:2015}.
A parallel study using eROSITA investigated the X-ray surface brightness distribution and the resulting gas density distributions.
The derived profile slopes for various mass ranges are significantly flatter than predicted by simulations, giving new constraints on the types and strength of possible feedback mechanisms \citep{zhang:2024a}.

Fig.~\ref{fig:LxM500} shows the X-ray luminosity against total mass within $R_\mathrm{500}$ for the \magpath simulation \Bthr as pink points for $z=4,2,1,0$ from left to right. 
The measurements for \Bthr are obtained from the X-ray lightcones generated with \Bt for halos with $z<0.2$ \citep[][see also Sec. \ref{sec:DataAvail}]{Marini2024}.
X-ray luminosities from \magpath are given as intrinsic luminosities in the 0.5-2 \unit{\keV} band.
A comparison study for halos in \magpath \Bfuhr and eROSITA galaxy results has been conducted by \citet{vladutescu:2025}, which includes models for the contribution from stellar X-ray point sources \citep{vladutescu:2023}.
They found very good agreement at $z=0$ for the $L_X-M_{500}$ scaling relations compared to \citet{zhang:2024b}, and their measurements are also shown in the rightmost panel of Fig. \ref{fig:LxM500} as blue points.
In all panels, the black dashed line shows the observational derived scaling law by \citet{lovisari:2015} using their group sample together with the HIFLUCS cluster sample for $z=0$.
The solid colored lines indicate the median of the \magpath simulation data.

At $z=0$, observational data by \citet{anderson:2015} and \citet{zhang:2024b} are shown as black squares and circles, respectively.
Here, the convergence between the different box resolutions and sizes becomes apparent.
Furthermore, the $L_X-M_{500}$ relation closely follows the relation for groups and clusters \citep{lovisari:2015} for \Bthr while the \Bfuhr results are in very good agreement with scaling relations derived for galaxies even displaying a larger scatter towards the low-mass end \citep{anderson:2015, zhang:2024b, vladutescu:2025}. As seen previously, in their overlap region we find excellent convergence between the different resolutions of the \magpath simulations.

\begin{figure*}
    \centering
    \includegraphics[width=0.99\linewidth]{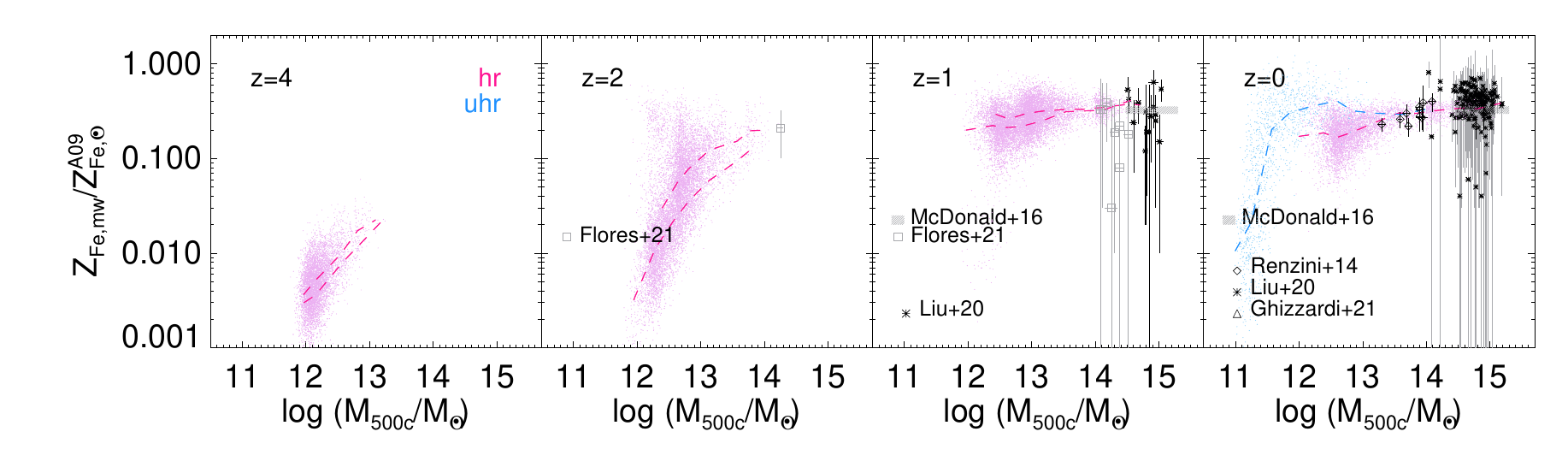}
    \caption{Gas iron abundance as a function of halo mass within $R_\mathrm{500,c}$, from $z=4$ to $z=0$ for the \magpath{} simulations.
    Observational data at $z=0$ are from \cite{renzini:2014} and \cite{ghizzardi:2021} at $z=0$ (black diamonds and triangles, respectively), from \cite{mcdonald:2016} and \cite{liu:2020} at $z=0$ and $z=1$ (grey shaded area and black asterisks), and from \cite{flores:2021} at $z=1$ and $z=2$ (grey squares). 
    All iron abundances are normalized to the solar reference value by \cite{asplund:2009}.
    \label{fig:zfe-mass}}
\end{figure*}

In the other panels of Fig. \ref{fig:LxM500}, we show the $L_X-M_{500}$ relation for the redshifts $z=4$, $2$, and $1$ from left to right. For $z=1$ we include a sub-sample of the SPT cluster sample \citep{bulbul:2019} shown as black triangles.
For $z=2$, we show a measurement of the Spiderweb proto-cluster \citep{tozzi:2022} indicated as a black diamond.
Across the redshift regimes $z<2$, \magpath simulation results are in excellent agreement with observations with respect to the median.
For lower halo masses, the intrinsic scatter generally increases towards low-mass halos but remains consistent with observations.

As the lightcone for \Bthr presented by \citet{Marini2024} does not reach beyond $z\sim2$, we additionally generated a deep exposure X-ray photon simulation of \Bthr at $z=4$ using the {\sc Phox} algorithm \citep{biffi:2012, biffi:2013}, and compute X-ray luminosities from photons extracted within $R_{500}$ of a specified halo.
We additionally rescale luminosities using the self-similar correction given by \citet{boehringer:2012} as indicated on the y-axis label, enabling the comparison to the \citet{lovisari:2015} relation.
While $z=4$ shows a power-law relation for halo masses $M_{500}>10^{12}\,\Msun$, at lower halo masses the intrinsic luminosity drops considerably with respect to standard scaling relations.
As was pointed out in previous sections, the peculiar behavior at this redshift must be attributed to the complex interplay between the halo environment, its evolutionary state, and the activity of the central SMBH \citep{kimmig:2025}.
High amounts of dense cold gas at high redshift have to be treated separately from the hot gas accretion onto the central SMBH to properly account for early growth, as shown by \citet{steinborn:2015,kimmig:2025}, which was not included for \Bthr. Additionally, low mass halos by definition are more prone to resolution based effects.

\subsection{Relation between gas metallicity and total mass}
\label{sec:zfemass}

The amount, distribution and relative abundances of chemical elements in the hot atmosphere of cosmic structures is a key tracer of the interaction between stars and gas via chemical and energetic feedback processes. 
Interestingly, the global level of gas chemical enrichment from groups to clusters of galaxies is found to vary little in time, for $z\lesssim 2$, and with halo mass~\cite[][]{mernier:2018,biffi:2018,gastaldello:2021}, in contrast with other global properties such as temperature or luminosity.
Fig.~\ref{fig:zfe-mass} shows results on the gas chemical enrichment in \magpath{} halos from galaxy to cluster scales. Specifically, we investigate the global mass-weighted iron abundance $Z_\mathrm{Fe,mw}$ of the diffuse non-star-forming gas within $R_\mathrm{500c}$ as a function of halo total mass, $M_\mathrm{500c}$, for halos in the \Bthr, \Btbhr and \Bfuhr runs of the \magpath{} simulations. 
Abundances are normalized to the reference solar iron abundance by~\cite{asplund:2009}, $Z_\mathrm{Fe,\odot}^\mathrm{A09}$.
The four panels refer to the redshifts $z=4,2,1, \mathrm{and}~0$, from left to right, respectively.

Consistently with previous studies~\cite[e.g.][]{dolag:2017,2017MNRAS.468..531B,biffi:2018}, we find that the \magpath{} halos present a very shallow dependency of the gas iron abundance on the total mass of the halo, and negligible variation in the average value for $z<2$. As extensively discussed in \citet{2017MNRAS.468..531B} this is due to the AGN feedback, which expels the gas of galaxies already at high redshift (e.g. $z>2$) before the as is accreted onto the galaxy cluster in the outskirts, leading to a very homogeneous and constant gas metallically in a large fraction of the cluster volume, especially outside the core 
Recently, radial profiles of gas iron abundance have been measured out to large cluster-centric distances for an increasing number of systems from X-ray observations of the diffuse emission of the ICM in the low- and intermediate-redshift Universe~\cite[][]{urban:2017,liu:2020}.
This allowed to probe larger and larger regions of clusters, confirming an essentially homogeneous enrichment level across different systems with variations mostly confined to the innermost regions~\cite[see review by][]{mernier:2018}.
At $z=0$, predictions from the \magpath{} halos show very good agreement with findings by \cite{ghizzardi:2021}, for the X-COP sample, and by \cite{renzini:2014}, both probing the highest mass end at $z<0.1$. 
Data from \cite{mcdonald:2016} for a large sample of mass-selected clusters in the redshift range $0<z<1.5$, observed with the Chandra, XMM-Newton and Suzaku X-ray telescopes, indicate as well an average value of $<Z/Z_\mathrm{Fe,\odot}^\mathrm{A09}> = 0.23 \pm 0.01$ at $z=0.6$ (marked by the shaded area in Fig.~\ref{fig:zfe-mass}). Compared to our simulated abundances in halos of similar mass, this is slightly lower both at $z=0$ and $z=1$.

\begin{figure*}
    \includegraphics[width=0.99\textwidth]{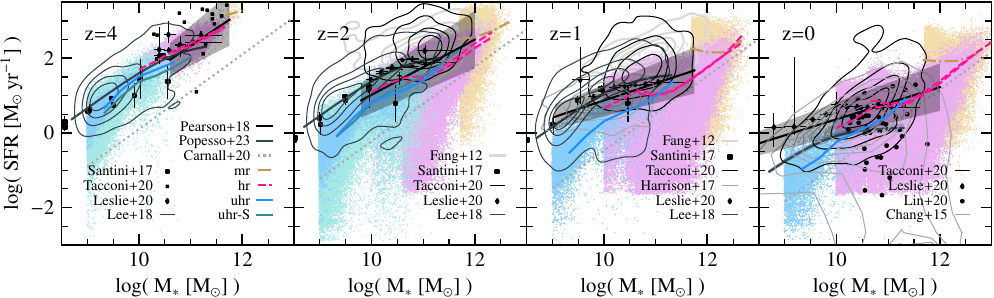}
    \caption{Main sequence of star-forming galaxies at four redshifts $z = 4$, $2$, $1$, and $0$, from left to right. The \magpath simulation data is displayed in colors according to each box resolution as indicated in the legend. Median lines are calculated using active galaxies only, that is galaxies above the threshold by \citet{carnall:2020} shown as a dotted line at each redshift.
    Observations are shown in gray and black. Data points are associated with each redshift by allowing for a deviation of $\Delta z = 0.4$ from the measured redshift. Solid lines and shaded scatter range are two analytical fits covering a large redshift range: \citet{pearson:2018} provide a SFMS fit without turnover mass, while \citet{popesso:2023} compiled data from various observations for a description including a turnover mass. 
    We additionally included observational data covering the mass range across all redshifts.
    The observations included in the work by \citet{tacconi:2020} are displayed as black contour lines except in the subpanel of $z=4$, where only few individual galaxies are present (black circles).
    \citet{leslie:2020} cover all redshift bins until $z\approx 0.4$ with stacked data from radio observations (black diamonds).
    At higher redshift we include the results by \citet{santini:2017} (black squares) with SED fitting to HST and Keck data, and EGS survey results by \citet{fang:2012} (light gray contours) and CANDELS+GOODS (UV \& IR) data by \citet{lee:2018} (dark gray contours).
    Similar to the values for the SFRD in Sec.~\ref{sec:sfrdens}, we calibrated the $M_*$ and SFR values by \citet{santini:2017} to convert from the assumption of the Salpeter IMF to the Chabrier IMF.
    At $z \approx 1$ we include KROSS (near-IR) results by \citet{harrison:2017}.
    \citet{alvarez-marquez:2023} add observations at high redshifts with ALMA data of lensed systems.
    At $z=0$ we include the results by \citet{lin:2020} with a combination of ALMA and MaNGA data, as well as values derived from SED fitting by \citet{chang:2015} to SDSS and WISE photometric data.
    }
    \label{fig:sfms}
\end{figure*}

A thorough study by \cite{liu:2020} of Chandra archival data for 186 clusters with redshift $0.04<z<1.07$ also indicates that the average gas iron abundance within $R_\mathrm{500c}$ does not evolve much across the sample below $z\sim 1$. Despite the larger uncertainties on the individual measurements, the level of enrichment in the observed sample is on average very well in line with the \magpath{} predictions. 
At $z>1$, the simulation data indicate a dependency of gas iron abundance on mass which reaches values in the most-massive halos that are in line with those at $z=0$. Compared to the recent study by ~\cite{flores:2021}, on 10 massive clusters at redshifts $1.05 < z < 1.71$ observed with Chandra and XMM-Newton, the global iron abundance within $R_\mathrm{500c}$ of the \magpath{} clusters agrees well. The \hr runs predict an average iron abundance very close to the value estimated in their highest-redshift cluster (at $z=1.71$).
At the highest redshift, $z=4$, we only show results from the largest boxes at high resolution and notice a much lower abundance level, reaching $Z_\mathrm{Fe,mw}\sim0.01$--$0.03Z_\mathrm{Fe,\odot}^\mathrm{A09}$. The dependency of gas global enrichment on halo mass is nonetheless as shallow as we find at $z\lesssim2$. Such lower values could be partially due to resolution effects, but we do expect the majority of the hot diffuse gas in halos at this epoch to be pristine, with a lower mass fraction of hot highly-enriched gas. Nevertheless, we already see that the most massive protoclusters at $z=4$ start to enrich the hot atmosphere with metals, as shown by \citet{remus:2023}.

We note that while the absolute abundance values can be influenced by modeling assumptions, mainly on the IMF and stellar yields \cite[e.g. studies by][]{tornatore:2007}, the trends with mass and redshifts must be mostly related to the impact of physical processes driving the baryonic evolution. Stellar and AGN feedback and star formation, in particular, shape the distribution of stars, gas and hence chemical elements within cosmic structures.


\section{Scaling Relations Through Cosmic Time~IV: Local~Scaling~Relations}
\label{sec:scale2}
We now leave the global scaling relations behind and turn our attention to the smaller scales of galaxies and their properties. A multitude of scaling relations are known for galaxies to exist, and reproducing them all from simulations has been a long standing challenge. Given that the \magpath simulations have not been tuned to reproduce the stellar components of galaxies but rather the hot gas component of galaxy clusters, galaxies appear self-consistently in the simulation, and both the successes and shortcomings of the simulations in reproducing galaxy properties are highly important to inform our understanding of the complex processes that shape the diverse galaxy populations that we observe through cosmic time. In the following, we will present 9 of the core scaling relations for galaxies, selected to be the most important relations according to the literature.

\subsection{Galaxy Star Formation Main Sequence}
\label{sec:mainsequence}

The main sequence of star-forming galaxies is an empirical relation between their stellar masses and their star formation rates. When comparing this relation at high and low redshifts, observations indicate that the main sequence evolves across cosmic time \citep[e.g.][]{speagle:2014, pearson:2018, leslie:2020, popesso:2023}, whereas the evolution at redshifts of $z>1$ appears to be less apparent \citep{santini:2017}. Despite being studied in great detail, the differences between the relations inferred from different observations are large, and not even for the slope of the relation has agreement been found, with some observations favoring a turn-over mass \citep[e.g.][]{popesso:2023}, while others do not \citep[e.g.][]{speagle:2014,pearson:2018}.

In Fig.~\ref{fig:sfms}, we compare the star formation main sequence from the \magpath simulations to observations at redshifts $z=4$, $2$, $1$, and $0$, with different resolutions and volumes shown in color. 
We allow for a redshift difference of $\Delta z = 0.4$ to match the redshift ranges from observational data and the four values of $z=4$, $2$, $1$, $0$ selected for the simulation data. We show all galaxies from the simulations to illustrate the full range of SFR observed at a given stellar mass. However, the main sequence is usually calculated from observations for star-forming galaxies only, which is why we apply the criterion requiring $\txt{sSFR} \geq 0.2 / t_\txt{H}$ with the Hubble time $t_\txt{H}$ as used by \citet{carnall:2020} to distinguish between star-forming and quiescent galaxies, a slightly lower criterion than implemented by \citet{franx:2008}. The median values shown in darker lines in Fig.~\ref{fig:sfms} are calculated only from the star-forming galaxies in order to be more comparable to observational criteria, but we verified that the relations barely change if all galaxies are accounted for. An extended discussion on that matter is provided by \citet{fortune:2025}, and we refer the reader to this work for more details on how the main sequence changes with different cuts.

\begin{figure*}
    \includegraphics[width=0.99\textwidth]{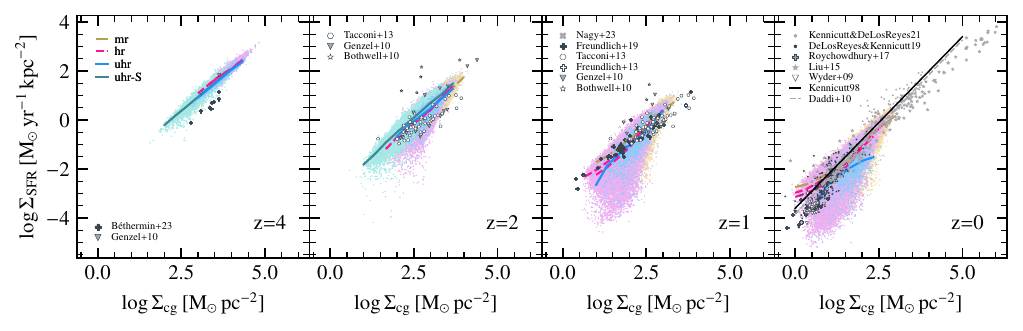}
    \caption{
    Kennicutt--Schmidt relation of the simulated galaxies compared to observations at four selected redshifts. The simulated galaxies are selected by a lower stellar mass cut of $M^*_{R_{0.1}} \geq 5\times10^9\,M_\odot$ for \uhr resolution, $5\times10^{10}\,M_\odot$ for \hr resolution, and $5\times10^{11}\,M_\odot$ for \mr resolution, as well as a lower cold gas mass cut ($M^\mathrm{cg}_{R_{0.1}}$) of half those values. The colored lines trace the median relation for the respective simulations. The observations are taken from \citet{kennicutt:1998}, \citet{wyder:2009}, \citet{bothwell:2010}, \citet{daddi:2010}, \citet{genzel:2010}, \citet{freundlich:2013}, \citet{tacconi:2013}, \citet{liu:2015}, \citet{roychowdhury:2017}, \citet{de_los_reyes:2019}, \citet{freundlich:2019}, \citet{kennicutt:2021}, and \citet{nagy:2023}. For the higher-redshifts, we plotted galaxies at redshifts $z=0.55$--1.4 in the $z=1$ panel, $z=1.6$--2.7 in the $z=2$ panel, and $z=3.3$--4.7 in the $z=4$ panel.
    }
    \label{fig:ks-relation}
\end{figure*}

Again, we find convergence between the different resolutions and volumes of the \magpath simulations, also between the fiducial and the advanced BH model runs. Furthermore, we find good agreement with observations at all redshifts within the large range of measurements provided by observations. We include the following observations: solid lines and shaded scatter mark the analytical fits from \citet{pearson:2018} without turn-over mass in dark gray, and from \citet{popesso:2023} including a turn-over mass in medium gray, at all redshifts.
At $z=2$, $1$, and $0$, data by \citet{tacconi:2020} are displayed as black contour lines, and at $z=4$ individual galaxies from that survey are shown due to low numbers. At all redshifts, stacked radio data from \citet{leslie:2020} are shown as black diamonds. Data by \citet{santini:2017} at high redshifts are included as black squares, results from \citet{fang:2012} are shown as light gray contours, and CANDELS+GOODS (UV \& IR) data by \citet{lee:2018} are shown as dark gray contours. At $z=1$ we include near-IR results from \citet{harrison:2017}, and at $z=0$ combined ALMA and MaNGA data from \citet{lin:2020} as black pentagons, as well as values derived from SED fitting by \citet{chang:2015} to SDSS and WISE photometric data in gray contours.

Overall we find a correlation between star formation rates and stellar masses in the \magpath simulations that match observations especially well at high redshifts. 
Towards lower redshifts, the star formation rates in the simulation tend to the lower part of the observational scatter.
However, the data distribution as well as the analytical function forms of the main sequence also display a larger scatter and deviation at $z=0$.
The discussion about the physical processes that shape the SFMS is subject to ongoing research.
\citet{fortune:2025} show that a consistent evolution along the main sequence is a rare scenario, demonstrating that most non-quenched galaxies at $z\approx 0$ have undergone extended quiescent periods and rejuvenations, explaining part of the scatter. \citet{teklu:2023} show that the overall decline of the SFMS is not a problem of a lack in gas reservoir, but rather that gas is deposited at larger radii in active galaxies, which systematically decreases surface density and thus conversion efficiency, which explains the trend of increasing depletion time scales towards lower redshifts as described in Sec.~\ref{sec:depl}. However, as discussed by \citet{kelson:2014}, the main sequence can also build up through hierarchical growth following the central limit theorem, and in such a picture, a more detailed investigation of the quenching and rejuvenation cycle of galaxies is needed in the future to shed light on the interplay between gas accretion, star formation, and feedback cycles of galaxies.

\subsection{Kennicutt--Schmidt Relation}
\label{sec:ks-relation}

\begin{figure*}
    \includegraphics[width=0.99\textwidth]{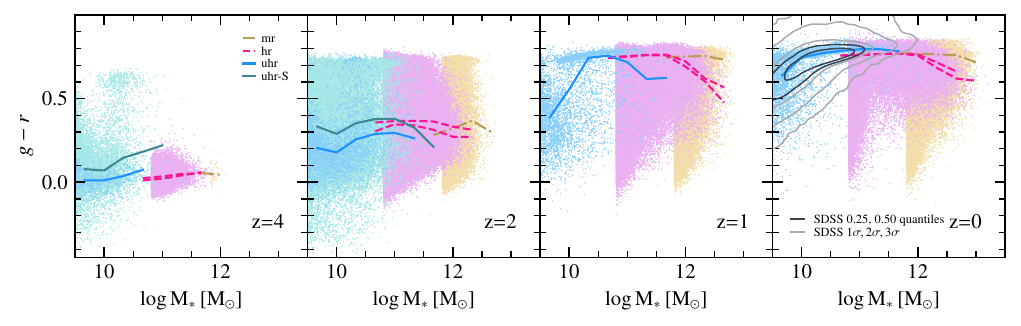}
    \caption{
    Color-mass relation of simulated galaxies at multiple redshifts in comparison with the SDSS sample at $z=0$. The colors of the simulated galaxies are measured from the rest-frame, therefore allowing a direct comparison between redshifts. The SDSS galaxies were obtained from the SDSS DR18 database \citep{almeida:2023}, with a sample of 326\,027 galaxies at redshifts $z=0$--0.1. Their dark contours encircle 25\% and 50\% of the galaxies, and the light contours are the $1\sigma$, $2\sigma$, and $3\sigma$ contours.
    }
    \label{fig:colormass}
\end{figure*}

The star formation rates of galaxies are directly coupled to the gas densities in galaxies, as stars can only form if the gas is dense enough to cool and thus provide the conditions for star formation to occur. Therefore, one of the central relations governing star formation is that between gas density and SFR density \citep{schmidt:1959} and its observational counterpart between their respective projected surface densities, the Kennicutt--Schmidt relation \citep{kennicutt:1989, kennicutt:1998}.
The relation has been found to be approximately a power law, where the classically assumed relation from \citet{kennicutt:1998} is the following:
\begin{equation}
    \Sigma_\mathrm{SFR} = 2.5 \times 10^{-4} \left(\frac{\Sigma_\mathrm{gas}}{1\,M_\odot\,\mathrm{pc}^{-2}}\right)^{1.4}\,M_\odot\,\mathrm{yr}^{-1}\,\mathrm{kpc}^{-2}.
\end{equation}

For the simulations, we determined the surface densities analogously to the observations using
\begin{align}
    \Sigma_\mathrm{cg} &= f_\mathrm{H\textsc{i},H_2} \times 0.5 \frac{M^\mathrm{cg}_{R_{0.1}}}{2\pi R^\mathrm{2D}_{1/2}}, \\
    \Sigma_\mathrm{SFR} &= f_\mathrm{H\textsc{i},H_2} \times 0.5 \frac{\mathrm{SFR}_{R_{0.1}}}{2\pi R^\mathrm{2D}_{1/2}},
\end{align}
where we assumed $f_\mathrm{H\textsc{i},H_2} = 0.25$ as the approximate H\textsc{i}/$\mathrm{H}_2$ mass fraction of the cold gas \citep[e.g.][]{valentini:2023}. Here we used the measurements and aperture as defined in section \ref{sec:localprops} to be consistent in our analysis. Since the observational studies use different tracers and measures for the cold gas, this value may very well be taken to be lower, thus further shifting the simulated points to the left, to lower surface densities.

In Fig.~\ref{fig:ks-relation} we show the SFR surface density--cold gas surface density relation for the \magpath simulations at four redshifts in comparison with several observational studies of \HI and molecular gas.
At all redshifts, the median lines of the simulations are nearly log-linear, as expected and also well converged between resolutions.
From the here presented data, it becomes apparent that the global SFR densities of the galaxy population as a whole declines significantly over time, though the log-linear relation does not change as much, dropping by ${\sim}0.5\,\mathrm{dex}$ for $\log\Sigma_\mathrm{SFR}$ from $z=4$ until today.
For a more detailed analysis of the star formation relations and their evolution with time in \Bfuhr, see \citet{teklu:2023}. Furthermore, the evolution of the Kennicutt--Schmidt relation with time was recently investigated by \citet{kraljic:2024} for the NewHorizon simulation.

We also over-plotted the empirical log-linear relations from \citet{kennicutt:1998} as written above and \citet{daddi:2010} in \cref{fig:ks-relation} at $z=0$, as well as data points from a number of different studies. 
Overall, the galaxies in the \magpath simulations agree well with the observations at all considered redshifts, with only some systematic offsets seen at $z=0$ and 1, most likely due to the different methodologies revolving around determining cold or molecular gas masses and the tracers used for the star formation rate.

The data points from the observational studies are from the following works: \citet{kennicutt:1998} with local normal spiral galaxies and starbursting galaxies at $z=0$, which were recently revisited by \citet{de_los_reyes:2019} and \citet{kennicutt:2021}, respectively, \citet{wyder:2009} with low-surface-brightness galaxies at $z=0$, \citet{daddi:2010} with normal spiral and starburst galaxies at $z=0$, \citet{roychowdhury:2017} with faint dwarf irregular galaxies at $z=0$, and \citet{liu:2015} with spiral and (ultra)luminous IR galaxies (ULIRGs) at $z=0$. At higher redshifts, we include multiple studies up to $z=4$: \citet{bothwell:2010} with ULIRGs at $z=2$, \citet{genzel:2010} with galaxies at $z=1$--3.5, \citet{freundlich:2013} at $z=1.2$, \citet{tacconi:2013} at $z=1$--3, \citet{freundlich:2019} at $z=0.5$--0.8, \citet{bethermin:2023} at $z=4.5$, and \citet{nagy:2023} at $z=1$. Despite the wealth of observations, it is still not entirely clear whether there is a continuous linear relation ranging from non-starbursting to starbursting galaxies and whether this changes with redshift.

At $z=0$ most measurements of the cold gas are of \HI for non-starbursting galaxies \citep{wyder:2009, liu:2015, roychowdhury:2017, de_los_reyes:2019} and of $\mathrm{H}_2$ (determined from CO measurements) for more heavily star-forming galaxies \citep{daddi:2010, liu:2015, de_los_reyes:2019, kennicutt:2021}, for which we added up the two masses where available. For the higher-redshift studies, the molecular gas is mostly obtained from CO measurements \citep{bothwell:2010, freundlich:2013, tacconi:2013, freundlich:2019, nagy:2023}, though \citet{bethermin:2023} use \CII and \citet{genzel:2010} collected galaxies from multiple sources with different methods. The SFRs are estimated based on the UV luminosity \citep{wyder:2009, roychowdhury:2017, nagy:2023}, IR luminosity \citep{liu:2015, kennicutt:2021}, both UV and IR \citep{tacconi:2013, de_los_reyes:2019, bethermin:2023}, radio \citep{bothwell:2010, liu:2015}, \OII \citep{freundlich:2013}, H$\alpha$ \citep{tacconi:2013}, or CO \citep{freundlich:2019}.
As a result of the methodological differences, the slight offsets between the observations themselves and with the median simulation lines are not unexpected.

Overall, we find the star formation rate density to decline with time, but in such a way that the Kennicutt-Schmitt relation always holds, from $z=4$ to $z=0$, that is the decline in star formation rate density occurs in concert with the decline of the cold gas density, without ever breaking the relation between the two. This means that the physical principles that lead to star formation stay the same with redshift, and only the circumstances of how gas is distributed in galaxies and how conditions for star formation are generated changes with redshift. Multiple reasons can be imagined for this, from environmental impact to feedback. However, for the \magpath simulations it was shown by \citet{teklu:2023} that this decline is not due to less gas being present in a galaxy, but rather that it is distributed at overall larger radii, which leads to lower gas surface densities and thereby to lower star formation rates, an effect that is actually also observed.

\begin{figure*}
    \includegraphics[width=0.99\textwidth]{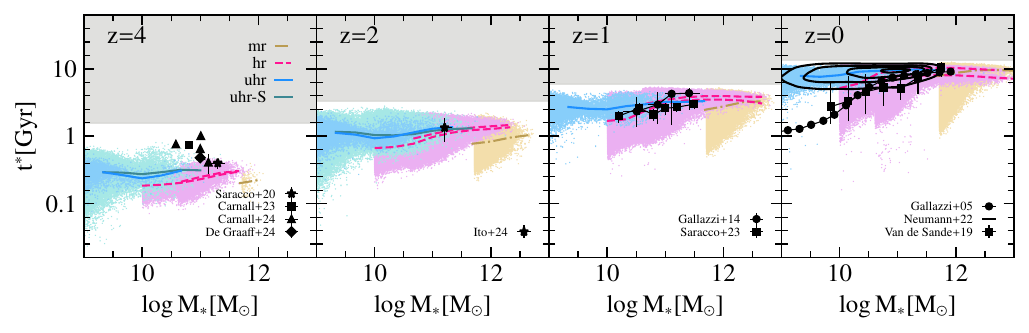}
    \caption{Stellar mass--stellar age relation from $z=4$ down to $z=0$ from left to right, with the \magpath simulations shown in color as indicated by the legend, with the median values shown in darker colors. The gray shaded area represents all ages that are older than the age of the Universe at the respective redshift. The following observations are included for comparison in black: at $z=4$, individual massive quiescent, early-type galaxies are shown from pre-JWST measurements by \citet[][star]{saracco:2020_massive_etg_at_z3}, and JWST measurements by \citet[][square]{carnall:2023}, \citet[][triangles]{carnall:2024}, and \citet[][diamond]{degraaf:2024}. At $z=2$, an individual massive quiescent early-type galaxy observed with JWST from \citet[][star]{ito:2024_onebh} (JWST NIRSpec) is included. At $z=1$, massive galaxies by \citet[][circles]{gallazzi:2014} and passive early-type galaxies from \citet[][squares]{saracco:2023} are shown. At $z=0$, local galaxies from \citet[][circles]{gallazzi:2005}, SAMI galaxies from \citet[][squares]{vandesande:2019}, and results from the MaNGA \textsc{firefly} Value-Added-Catalog \citep[][black contours at the 10th, 32nd, 68th, and 95th quantile]{neumann:2022} are shown.} 
    \label{fig:mar}
\end{figure*}

\subsection{Color--Mass Relation}
\label{sec:colormass}

One of the more direct observables of galaxies from broadband imaging is the color of a galaxy and its relation to either its absolute magnitude in a given band or its total stellar mass \citep[e.g.][]{strateva:2001, baldry:2004, kauffmann:2004}, where a bimodality was discovered, referred to as the red sequence and blue cloud, respectively, with the green valley located between them. This encodes the recent star formation episodes, as more recent star formation leads to bluer colors while older stellar populations result in redder colors. 
At $z=0$, these regions are directly connected to the morphologies of galaxies, where ETGs tend to reside in the red sequence and LTGs in the blue cloud.

In Fig.~\ref{fig:colormass}, we show the color-mass relations for the \magpath simulations at four redshifts with the rest-frame $g-r$ colors, compared to the observed SDSS galaxies at $z=0$ (black and gray contours; \citealp{almeida:2023}). The rest frame was chosen to see the actual physical evolution of the galaxies in a consistent manner in the simulations. At $z=0$, the red sequence can clearly be seen around the median lines at higher masses, which are dominated by ETGs. At lower masses below ${\sim}10^{10.5}\,M_\odot$, the median line drops from the red sequence into the blue cloud, for which we find excellent agreement between \Bfuhr shown in blue and the SDSS sample. For the lower resolution simulations, we see a large scatter at the lower masses of the mass range that these simulations cover, but as shown by the median line of the samples, the blue galaxies at high masses exist in these boxes, but in such low numbers that the overall distribution clearly is following the red sequence as expected. Note that these low number blue galaxies at the high-mass end are those BCGs that live in cool core clusters, where the gas cools from the hot halo as discussed before \citep[see e.g.][for more details]{gonzalez_villalba:2025}.
Thus, we again find very good convergence between the simulations of different volumes and resolutions, with more special objects found in the larger simulation volumes, but also very good agreement with the observations by SDSS for both the red sequence but also the blue cloud. As shown by Khalid et al, 2025 (submitted), the \magpath simulations here perform as good or even better than other simulations like IllustrisTNG or Eagle.

For the SDSS observational sample shown at $z=0$, we queried the publicly available SDSS DR18 database by cross-matching the galaxy data with photometry for the colors and spectroscopy for the redshifts, as well as the star-forming-modeled galaxy stellar mass model table using the approach of \citet{maraston:2009}. We selected only those galaxies with clean photometry with spectroscopic redshifts $z<0.1$ and available stellar mass estimates, resulting in a sample of over 325\,000 galaxies. We plotted their colors determined from the K-corrected magnitudes in the $g$ and $r$ band filters.

At higher redshifts we can see how the red sequence is built up over time, with most galaxies still being very blue at $z=4$, while the star formation is still very high. Subsequently, the galaxies rise to redder colors at $z=2$, where the reddest galaxies already make up the red sequence, and finally most of the red sequence is built up at $z=1$ for all masses. This evolution also reflects the result of the cosmic SFR density with the SFR density declining after $z=2$ (see Sec.~\ref{sec:sfrdens}). The different simulations in the suite are well converged, only having the respective most massive galaxies falling below the red sequence, most likely due to these types of systems tending to be late-forming and thus being a biased sample at the given mass. While the fiducial model simulations again show excellent convergence between resolutions, we also see that the fiducial and the advanced BH models overall agree well. Interestingly, we can see that for the advanced BH model run \Bthuhr shown in turquoise at $z=4$ we already see a small red sequence being built up, which is not present for the fiducial run. While this could now be due to the \Bth being larger than \Bf, we do not see the red sequence being formed at $z=4$ in the \hr runs of \Bt or \Btb, which both are even larger than \Bth. This build-up of an early red sequence is in excellent agreement with the fact that we find quenched galaxies in this simulation already at $z=4$, in good agreement with observations but also different than other simulations, as discussed by \citet{kimmig:2025} and \citet{remus:2025}.

\subsection{Stellar Mass--Age Relation}
\label{sec:ma}

As discussed in the previous section, colors encode the ages of stars in galaxies, with bluer galaxies undergoing recent star formation than redder galaxies, and these colors are correlated with the morphologies of galaxies at $z=0$. As the red sequence is more pronounced for more massive galaxies, a correlation between the actual ages of galaxies and their stellar mass can be well expected at $z=0$. Fig.~\ref{fig:mar} presents the stellar age $t^*_{R_{-1.1}}$ as a function of the stellar mass $\log_{10}(M^*_{R_{0.1}}/M_\odot)$ at redshifts $z = 4, 2, 1,$ and $0$ from the \magpath simulations (colored points), with median lines for each redshift and simulation box. Note that we take the mass-weighted mean of the time since each star formed, not the light-weighted one. The dark shaded areas mark the ages that are older than the age of the Universe at the given redshift and are thus impossible. Overall, we see convergence between the different resolutions and volumes, although here it is not as good as in previous scaling relations, most likely caused by the more efficient high-redshift star formation, which also causes higher metallicities (see Sec.~\ref{sec:mz} and \citealp{tornatore:2007,kimmig:2025}). Generally, we find the average star in the \mr resolution simulations to be slightly younger than in the \hr resolution, with a better agreement between \hr and \uhr overall. At $z=4$, we see the strongest deviations, but as already discussed by \citet{kimmig:2025} this is due to the higher resolution \uhr forming stars earlier than the lower resolution simulations. Note that we do not find a difference between the two BH models in this case.

\begin{figure*}
    \includegraphics[width=0.99\textwidth]{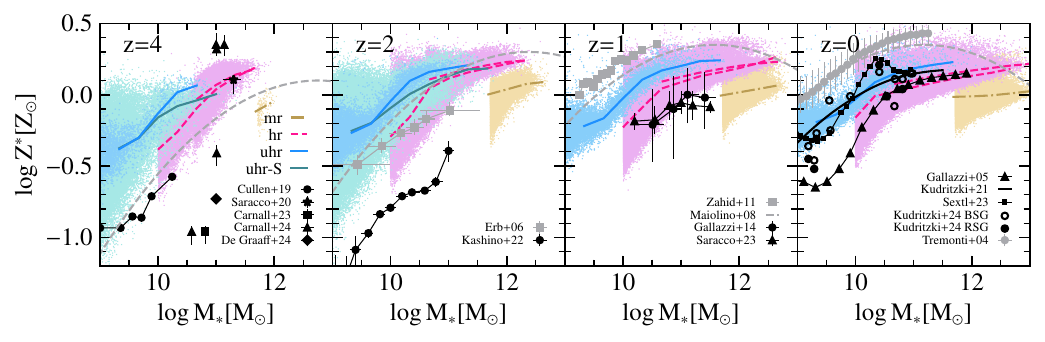}
    \caption{Stellar-mass--stellar-metallicity relation from $z=4$ to $z=0$ from left to right, for the \magpath simulations, with different resolutions shown in colors as indicated in the legend. For comparison, several observations are shown in black: at $z=4$, for star-forming galaxies by \citet[][circles]{cullen:2019}, and for quiescent galaxies from \citet[][star]{saracco:2020_massive_etg_at_z3}, \citet[][square]{carnall:2023_highzbh}, \citet[][diamond]{degraaf:2024} and \citet[][triangles]{carnall:2024}. At $z=2$, observations of stellar metallicities of star-forming galaxies by \citet[][circles]{kashino:2022} are shown. At $z=1$, stellar metallicity measurements by \citet[][circles]{gallazzi:2014} and \citet[][triangles]{saracco:2023} are included, and at $z=0$ stellar metallicities by \citep[][triangles]{gallazzi:2005} and \citet[][squares]{sextl:2023}, together with measurements of individual red and blue supergiants in local galaxies by \citet[][empty circles, filled circles]{kudritzki:2024}, are shown. Furthermore, results from the model by \citet{kudritzki:2021} are shown as solid black line. In addition to the stellar metallicities, gas-phase metallicity measurements are included in gray: at all redshifts, best fits to the data by \citet{maiolino:2008} at the closest matching redshift are shown as gray dashed lines. In addition, the gas-phase metallicity relation from \citet[][gray squares]{erb:2006} at $z=2$, from \citet[][gray squares]{zahid:2011} at $z=1$, and from \citet[][gray circles]{tremonti:2004} at $z=0$ are shown.}
    \label{fig:mzr}
\end{figure*}

At all redshifts, the mass-age relation is mostly flat, with a very mild trend with stellar mass. Less massive galaxies are on average younger, especially visible within the \hr resolution. Ages are especially difficult to obtain from observations, but we overplotted various observational results as further specified in the figure caption of Fig.~\ref{fig:mzr}. At $z=0$, \magpath galaxies agree well with MaNGA results \citep{neumann:2021,neumann:2022} shown as black contours, especially at the \uhr level. They are also consistent with measurements for local galaxies from \citet[][black circles]{gallazzi:2005} and SAMI results from \citet[][black squares]{vandesande:2019} at intermediate masses, and at low masses also for the \hr resolution, although there are deviations from the \uhr galaxies. Interestingly, this is also where the observations disagree between the MaNGA sample and the measurements from SAMI and the local galaxies, with the two different resolutions matching the two different observational results.

At $z=1$, observations by \citet[][black circles]{gallazzi:2014} for massive galaxies and by \citet[][squares]{saracco:2023} for passive early-type galaxies from the VANDALS survey are presented, and we find good agreement with the \magpath simulations. At $z=2$, only one observation is included, an individual massive, quiescent early-type galaxy presented by \citet[][star]{ito:2024_onebh}. The measurement agrees well with the simulations, however, more age measurements would be needed to study the range of observed ages in comparison to the simulations. At $z=4$, several observations are included, all of which are passive massive galaxies \citep{saracco:2020_massive_etg_at_z3,carnall:2023_highzbh,carnall:2024,degraaf:2024}. All of these galaxies are older than the simulated average. However, all but one galaxy from the sample by \citet{carnall:2024} are well within the range of ages and stellar masses predicted from the \Bthuhr simulation. As these galaxies are also suspected to be among the most extreme at their redshift, this fits well with the simulations that also find the quiescent massive galaxies at $z=4$ to be on the extreme end of the galaxy population \citep{kimmig:2025}. More measurements, especially for star-forming galaxies, are required to test the age distribution at high redshifts against the full simulation sample in the future.

\subsection{Stellar Mass--Metallicity Relation}
\label{sec:mz}

Closely related to the ages of the stellar population is its metallicity. Old stars are typically less metal-rich compared to young stars, as the gas that the latter formed from had more time to be chemically enriched.
First detected by \citet{lequeux:1979}, the gas-phase metallicity of star-forming galaxies rises with increasing gas mass. This is the so-called \textit{mass-metallicity relation}. \citet{tremonti:2004} established a tight ($\pm 0.1\,\mathrm{dex}$) correlation between stellar mass and the gas-phase metallicity. As the stars are formed from the gas, a similar relation was expected for the stellar metallicity, and was confirmed by \citet{gallazzi:2005}, which has since been firmly established \citep[e.g.][]{zahid:2017, kudritzki:2021, sextl:2023, kudritzki:2024}. 
We calculate the stellar metallicity as 
\begin{equation}
[\mathrm{Z}] = \log_{10}(Z^*_{R_{0.1}}/Z_\odot),
\end{equation}
using $Z_\odot=0.0142$ \citep{asplund:2009}. Fig.~\ref{fig:mzr} shows the stellar metallicity as a function of the stellar mass $\log_{10}(M^*_{R_{0.1}}/M_\odot)$ at redshifts $z = 4, 2, 1,$ and $0$ for the \magpath simulations (colored points), with median lines for each redshift and simulation box. Overplotted are various observational results, where gas-phase metallicities are converted from $[\mathrm{O/H}] = 12+\log{(\mathrm{O/H})}$ to $[\mathrm{Z}]$ using $[\mathrm{Z}] = [\mathrm{O/H}] - [\mathrm{O/H}]_\odot$, and $[\mathrm{Z/H}]$ is converted to $[\mathrm{Z}]$ using $[\mathrm{Z/H}] = [\mathrm{Z}] - \log_{10}(X)$, with solar values of $[\mathrm{O/H}]_\odot = 8.69$ and $X = m_\mathrm{hydrogen}/m_\mathrm{total} = 0.7154$ as given by \citet{asplund:2009}. Generally, direct measurements from the stellar components are shown in black, while measurements obtained from the gas are shown in gray.

The different resolution levels are largely consistent with each other at all redshifts, with slightly lower metallicity for the lower resolutions. This is a known effect for gas-phase iron abundances \citep{tornatore:2007} which also translates to stars, although \citet{tornatore:2007} mention that the effect is less relevant at the center of clusters. It is likely a consequence of both increased mixing with higher resolution, as well as increased high-redshift star formation in higher-resolution simulations \citep{tornatore:2007}, which causes more chemical enrichment over the span of stellar evolution. 

At each redshift, the metallicity increases with stellar mass, consistent with observations. For all redshifts, the best-fits from the ESO-VLT large program AMAZE by \citet[][gray dashed line]{maiolino:2008} are shown at the closest matching redshift. In addition, at $z=0$ we show the stellar mass--gas-phase metallicity relation for star-forming galaxies by \citet[][gray circles]{tremonti:2004} obtained from stellar evolutionary synthesis and photoionization models, and results for stellar metallicities from \citet[][black triangles]{gallazzi:2005} using stellar population synthesis models. Furthermore, results from a galaxy look-back evolution model by \citet[][solid black line]{kudritzki:2021} are shown together with measurements of individual red and blue supergiant stars in local galaxies from \citet[][empty circles, filled circles]{kudritzki:2024}, and finally metallicities for the young population of star-forming galaxies by \citet[][black squares]{sextl:2023} obtained by using stellar population synthesis are included. 
We find good agreement with \citet{sextl:2023}, \citet{kudritzki:2021}, and \citet{kudritzki:2024} at $z=0$ at low to intermediate stellar masses, and also with \citet{gallazzi:2005} at intermediate to large masses. As expected from the diverse nature of the methods used to infer the stellar metallicities, the observations are not in agreement with one another but show a rather large scatter. This could also be related to the issue with measuring metallicities as discussed in detail by \citet{kewley:2008}.

At $z=1$, the \hr and \mr resolution simulations of \magpath are broadly consistent with observation of the stellar mass--stellar metallicity relation by \citet[][black circles]{gallazzi:2014} for massive galaxies using rest-frame optical spectroscopy from the Magellan telescope, and by \citet[][black triangles]{saracco:2023} for passive early-type galaxies from the VANDALS survey at $1.0 \leq z \leq 1.4$. 
For \Bfuhr shown in blue, the observations lie below the simulated galaxies. However, for comparison we also show the stellar mass--gas-phase metallicity from the DEEP2 survey at $z\sim 0.8$ obtained using strong-line diagnostics from \citet[][gray squares]{zahid:2011}, which are in agreement with the measurements for the gas metallicity by \citet{maiolino:2008}. Both gas measurements are above the metallicities found for the \magpath simulations, which are basically framed in by the stellar and the gas measured metallicity relations found from observations, similar to the results seen at $z=0$.

At $z>1$, most observational results lie below the stellar metallicities from the \magpath simulations, although the best-fit line to the stellar mass--gas-phase metallicity relation by \citet{maiolino:2008} (light gray dashed line) is consistent with our simulated stellar metallicities at $z=2$, and slightly below at $z=4$. However, systematic observations are becoming sparse. 
At $z=2$, the stellar mass--gas-phase metallicity relation for star-forming galaxies obtained using [\NII]/H$\alpha$ ratios by \citet[][light gray squares]{erb:2006} are shown, and stellar metallicity measurements for star-forming galaxies from the zCOSMOS-deep survey by \citet[][black circles]{kashino:2022}. Again, gas-phase metallicities are higher than those found for stars. However, since the measurements by \citet{kashino:2022} are only obtained for star-forming galaxies, which usually have lower metallicities than quiescent galaxies, we do not expect the observations to actually cover the full range of metallicities present in galaxies in general.

Interestingly, we get a glimpse at this at $z=4$. First, observations for star-forming galaxies from the VANDALS survey using rest-frame FUV spectra obtained with VIMOS by \citep[][black circles]{cullen:2019} are shown to be at the lower end of what is found for the \magpath simulations. However, there also exist measurements of the stellar metallicities for quiescent massive galaxies, mostly obtained recently thanks to JWST NIRSpec and presented by \citet[][black diamond]{degraaf:2024}, \citet[][black square]{carnall:2023_highzbh}, and from the EXCELS survey by \citet[][black triangles]{carnall:2024}. Metallicities for one pre-JWST massive quiescent galaxy from \citet[][black star]{saracco:2020_massive_etg_at_z3} obtained using the Large Binocular Telescope was also included. These quiescent galaxies span the whole range of metallicities, from even below what was measured for the star-forming galaxies, to significantly super-solar metallicities even above the most metal-rich galaxies found in the \magpath simulations. This clearly demonstrates the issue of understanding metal enrichment in stars at high redshifts, but also at lower redshifts, from both models and observations, and further studies into this matter are required in the future.

\begin{figure*}
    \includegraphics[width=0.99\textwidth]{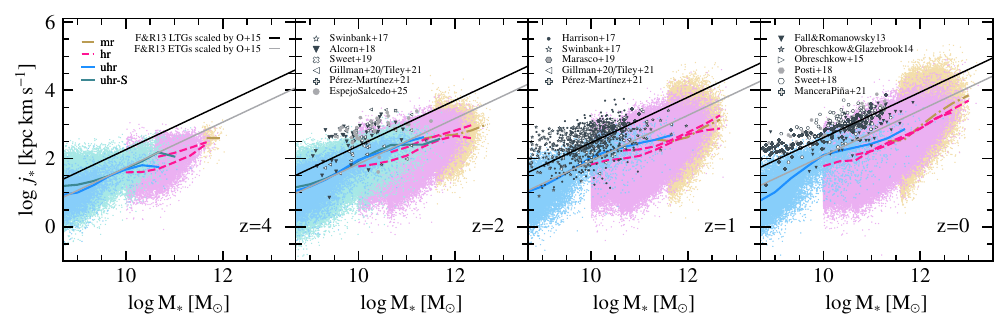}
    \caption{
    Stellar specific angular momentum-stellar mass relation of the simulated galaxies compared to observations at different redshifts.
    The simulated galaxies consist of all types of morphologies, whereas on the observational side all galaxies are disk galaxies at $z=0$ and/or star-forming at higher redshifts, except for the data and lines from \citet{fall:2013} at $z=0$, where both LTGs and ETGs make up the sample.
    For the simulation we show $J^*_{3R_{1/2}}$ and $M^*_{R_{0.1}}$, which are the two quantities which were determined as the best morphological indicators through the $b$-value by \citet{teklu:2015}.
    From the observations, we show the LTG and ETG log-linear relations from \citet{fall:2013} with slopes of 2/3 at $z=0$ in dark gray and light gray, respectively, as well as their evolution with redshift according to the model by \citet{obreschkow:2015}, shown in the other three panels.
    The observational data points included in the panels are from \citet{fall:2013}, \citet{obreschkow:2014}, \citet{obreschkow:2015}, \citet{harrison:2017}, \citet{swinbank:2017}, \citet{alcorn:2018}, \citet{posti:2018}, \citet{sweet:2018}, \citet{sweet:2019}, \citet{gillman:2020}, \citet{mancera_pina:2021}, \citet{tiley:2021}, \citet{perez_martinez:2021}, and \citet{espejo_salcedo:2025}.
    }
    \label{fig:jm-relation}
\end{figure*}

Generally, it is difficult to draw any definitive conclusions on what metallicities should be found at higher redshifts given the beyond $1\,\mathrm{dex}$ disagreements found between individual observations of galaxies \citep{saracco:2020_massive_etg_at_z3,carnall:2023_highzbh,degraaf:2024,carnall:2024}. We find the \magpath simulations to best agree with observations at $z=0$, which is also the redshift with the smallest overall spread between different observations -- even though it is still not insignificant \citep{kewley:2008}. Unless systematics are accounted for correctly, and it is understood where such differences come from, we conclude here only that the \magpath simulations generally well reproduce the expected trend of increasing metallicity with increasing stellar mass, converge among the differing resolution levels within the bounds of what is to be expected \citep{tornatore:2007}, as well as nicely match the lower redshift observations. At higher redshifts, \citet{kimmig:2025} find that metal ratios such as the alpha abundance may be more powerful indicators of other galaxy properties than calibrations involving the total metallicity because the main channel of enrichment for individual elements changes throughout cosmic time \citep{kobayashi:2020}. Early iron abundance is driven exclusively by supernovae type-II, while over longer and thus later times supernovae of type-Ia become increasingly significant \citep{kobayashi:2020}, eventually becoming the dominant enrichment channel. Further details about metallicity and metal distributions in \magpath will be presented in two upcoming studies by Stoiber et al.\ (in prep.), and Kimmig et al.\ (in prep.).

\subsection{Specific Angular Momentum Relation}
\label{sec:jm-relation}

An important dynamical relation of galaxies has been found between the stellar mass and specific angular momentum of galaxies as a consequence of a log-linear relation between halo mass and specific angular momentum \citep[e.g.][]{fall:1983, romanowsky:2012}, which can be used to test models of galaxy formation. For the stellar component, it shows two offset relations for ETGs and LTGs. For this reason, the location in the stellar specific angular momentum-stellar mass plane is a robust measure of morphology in simulation studies, where LTGs have a higher specific angular momentum at a given mass \citep[e.g.][]{teklu:2015, teklu:2017}.
The quantification thereof is performed with the $b$-value \citep{teklu:2015}, calculated at $z=0$ as
\begin{equation}
    b = \log\frac{j}{\mathrm{km}\,\mathrm{s}^{-1}}-\frac{2}{3}\log\frac{M_*}{M_\odot},
\end{equation}
where typical thresholds of $b = -4.73$ and -4.35 have been used to distinguish between ETGs, intermediate-type galaxies, and LTGs, respectively \citep[e.g.][]{emami:2021, valenzuela:2024_streams}.
In particular, \citet{teklu:2015} showed that the $b$-value of a galaxy strongly correlates with the circularity measure of a galaxy, which directly reflects its morphology. 

In Fig.~\ref{fig:jm-relation} we show this relation for the galaxies of the \magpath simulations at multiple redshifts in comparison with the results of several observational studies. As seen from the median lines corresponding to the simulated galaxies, the median $j_*$--$M_*$ relation in \magpath is roughly log-linear with similar slopes at all redshifts and well-converged for the different resolutions. Over time, the specific angular momenta increase as the redshift drops, which is consistent with the theoretical scaling of $j_*$ with redshift ($j_* \propto (1+z)^{-1/2}$), as discussed by \citet{obreschkow:2015}.
The dark gray line at $z=0$ indicates the best fit with the theoretical slope of 2/3 from \citet{fall:2013} to LTGs and the light gray line to ETGs. At higher redshifts, we apply the theoretical scaling of specific angular momentum with redshift to the two lines according to \citet{obreschkow:2015}. We can see that the specific angular momenta of the simulated galaxies follow this theoretical evolution of the scaling relation.

\begin{figure*}
    \includegraphics[width=0.99\textwidth]{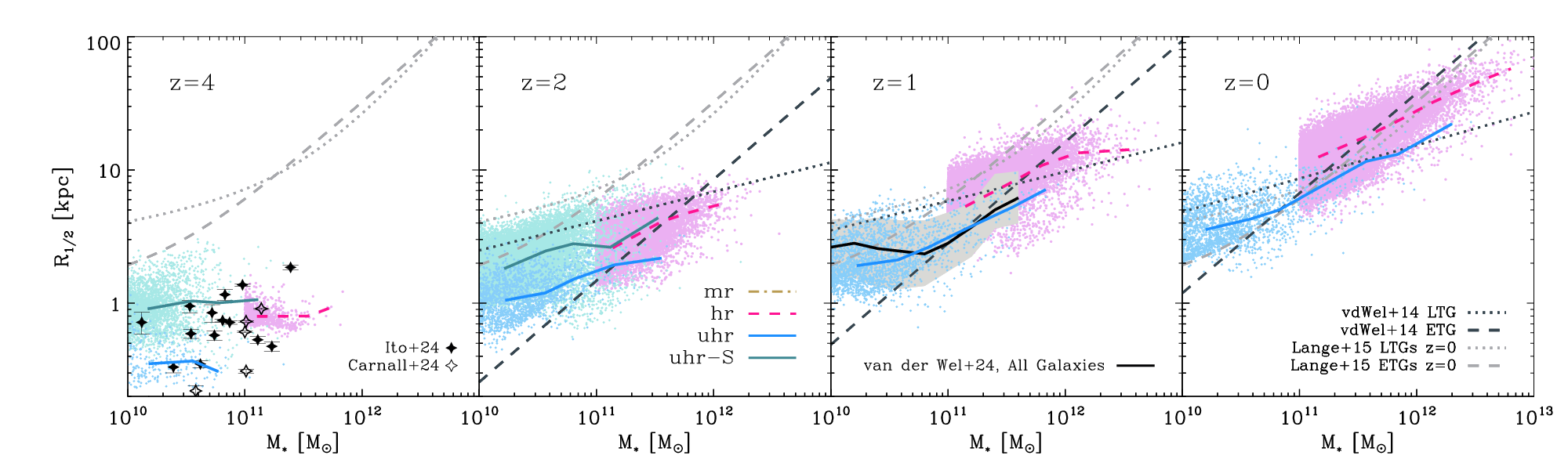}
    \caption{Stellar half-mass radii versus stellar mass for the \magpath simulations, with different resolutions and volumes shown in color as indicated by the legend, at $z=4$, $2$, $1$, $0$ from left to right. Solid colored lines mark the median values for the simulations of different resolutions. Gray dashed and dotted lines included at all redshifts mark the relations found at $z=0$ from the GAMA survey by \citet{lange:2015} for ETGs and LTGs, respectively. The relations found for ETGs and LTGs from HST by \citet{vanderwel:2014} are shown as black dashed and dotted lines, respectively, at the redshifts closest to the depicted redshift. At $z=1$, new measurements based on JWST data by \citet{vanderwel:2024} are shown in black, with the scatter as gray shaded are, for the full galaxy sample and not split into ETGs and LTGs. At $z=4$, only individual measurements for ETGs are available, from \citet[][filled stars]{ito:2024} and \citet[][open stars]{carnall:2024}.}
    \label{fig:mass_size}
\end{figure*}

For the simulations, we selected all the galaxies irrespective of their morphology.
On the observational side, most studies have primarily addressed disk galaxies, with the notable exception of \citet{romanowsky:2012} and \citet{fall:2013}, who considered both ETGs and LTGs and fit log-linear relations to the two groups. Our simulated galaxies agree well with the observed specific angular momenta that are spanned by the different morphologies. The relation between halo spin, morphology and position in the stellar mass vs. specific angular momentum space of galaxies is discussed in detail in \citet{teklu:2015}. The evolution of the simulated specific angular momenta are consistent with the LTG measurements across the three redshifts with available observations ($z=0$--2) and consistent with the theoretically expected evolution \citep[see][]{2016ilgp.confE..41T}. It should be noted that the observations attempt to infer the total 3D specific angular momentum by measuring or extrapolating the rotation curve as far out as possible from projected 2D measurements. For a detailed study and comparison of the \magpath \Bf galaxies and observations with respect to their dynamics and morphology, see \citet{teklu:2015}, \citet{2016ilgp.confE..41T} and \citet{schulze:2018}.

On the observational side, we have collected the data of galaxies at $z=0$ from \citet{fall:2013} for disk and spheroid galaxies, from \citet{obreschkow:2014} for disk galaxies, \citet{obreschkow:2015} for clumpy disk galaxies at $z=0.1$, \citet{posti:2018} for a large range of galaxies from irregular dwarfs up to massive spiral galaxies, \citet{sweet:2018} for disk galaxies extending to higher bulge fractions, and \citet{mancera_pina:2021} for nearby disk galaxies.
While the exact slopes found for the relation differ between the studies, they have generally been consistent with the theoretical value of 2/3.
At higher redshifts, we have used the observational data from \citet{harrison:2017} for star-forming galaxies at $z=0.6$--1 from KMOS for the line-of-sight kinematics, \citet{swinbank:2017} for star-forming galaxies at $z=0.3$--1.7 from KMOS and MUSE, \citet{alcorn:2018} for star-forming galaxies at $z=2$--2.5 from Keck/MOSFIRE, \citet{marasco:2019} for isolated disk galaxies at $z=1$ from KMOS, \citet{sweet:2019} for a bright galaxy at $z=1.62$ from KMOS (we do not show the other $z=1.47$ galaxy in any of the panels as it lies right between $z=1$ and 2), \citet{gillman:2020} and \citet{tiley:2021} for star-forming galaxies at $z=1.2$--1.8 from KMOS, \citet{perez_martinez:2021} for cluster disk galaxies at $z=0.4$--1.4 from different 2D spectroscopic measurements, and \citet{espejo_salcedo:2025} for star-forming galaxies at $z=1.5$--2.5 from KMOS.

Generally, we see a decline in angular momentum with decreasing redshift as predicted by \citet{obreschkow:2015} for all resolutions and volumes of the \magpath simulations, which show again excellent convergence. Compared to the observed disk galaxies, we find slightly lower overall angular momentum at all redshifts, indicating that more angular momentum is captured in observed disks than in the simulations. This could partially be due to resolution limitations, but could also reflect the BH treatment, since \Bthuhr with the advanced BH model shows a larger scatter towards high angular momenta in galaxies at $z=4$ and $z=2$, but we cannot confirm this to hold true to $z=0$ since the simulation only ran to $z=2$ on this resolution level. We will investigate this in more detail in the future. However, as introduced by \citet{teklu:2015} and supported by findings from \citet{genel:2015}, the b-value is an extremely suited quantity to distinguish morphologies of galaxies even if the individual resolution is not as high, as demonstrated by the resolution convergence here.

\subsection{Mass--Size Relation}
\label{sec:masssize}

So far, we have studied mostly general properties of galaxies, but have not yet looked at the spatial distribution of matter in galaxies. However, it is well known that at a given stellar mass, the stars can be distributed rather differently spatially, and thus the galaxies have very different properties. The most direct way in describing the stellar matter distribution of galaxies is through the radius that contains half of the stars in a given galaxy, called the half-mass radius. Assuming a constant mass to light ratio, this should be an excellent approximation of the half-light radius measured from observations, as was also discussed by \citet{genel:2018}. Furthermore, \citet{vandeven:2021} showed that values obtained from 2D and 3D measurements are nearly identical, with a slightly larger scatter for ETGs as shown by \citet{vanderwel:2024}, boosting confidence in comparisons between observations and simulations.

The stellar-mass--size relation of galaxies from the \magpath simulations was already presented in several previous works from different angles, in the context of ETG evolution from $z=2$ to present day \citep{remus:2017}, with respect to kinematic properties \citep{schulze:2018}, from BCGs in clusters to galaxies for dark matter-baryon interactions \citep{harris:2020}, or the in-situ to accreted properties of galaxies \citep{remus:2022}. Focusing on high redshifts, the mass--size relation was studied by \citet{remus:2025}. However, it was never presented in a consistent way for all galaxies from different simulation resolutions and volumes over a broad redshift range so far.

Fig.~\ref{fig:mass_size} shows the mass--size relation for the \magpath simulations, at $z=4$, $2$, $1$, $0$ from left to right, for the \hr and \uhr resolution level simulations. Results for the \mr resolution are not included as the softening of that simulation is so large that it artificially smears out the centers of the galaxies, and that resolution level is also not intended to be used for radial distribution studies of galaxy properties. The median lines of the simulations are marked by darker colored lines. The convergence between the resolutions is not as excellent as in many previously discussed scaling relations, caused by the impact of the larger softening lengths for the lower-resolution simulations. The mass--size relation for ETGs and LTGs at $z=0$ from the GAMA survey from \citet{lange:2015} are included at all redshifts for reference in gray, and the mass--size relations for ETGs and LTGs from \citet{vanderwel:2014} at the respective redshift are shown in black as dashed and dotted lines, respectively. As can be seen clearly, the \uhr resolution simulations shown in blue and turquoise agree excellently with the observations, while the \hr resolution simulation already tends towards too extended smaller galaxies and too compact larger galaxies. However, the overall agreement is still good, showing that for the higher-mass end radial distributions can still be studied.

We do not distinguish our galaxy populations according to morphology here, as it has been shown in previous works that the morphological differences in the mass--size relation are captured nicely by the \magpath simulations by \citet{schulze:2018}, \citet{remus:2022}, and \citet{remus:2025}. Instead, we are interested in the overall distribution. We can see that the population of \magpath galaxies covers the whole size range at a given mass from ETGs to LTGs, and thus is a good representation of the overall galaxy populations. Most interestingly, we can thus directly compare to the newest size measurements from JWST data by \citet{vanderwel:2024}, who also do not focus solely on splitting the galaxy populations by morphology but also give a median to the whole distribution, which is included at $z=1$ in Fig.~\ref{fig:mass_size} as black solid line, with the $1\sigma$ scatter marked as shaded gray area. Interestingly, the median of the \Bfuhr galaxies is in excellent agreement with that observation, even capturing the upturn to larger radii at the low-mass end, although it is not as strong for the \magpath simulations as it is for the observations, indicating that we have slightly too many compact galaxies at the low-mass end compared to the observations. However, the overall agreement is striking and very promising in terms that radial mass distributions are captured well by the simulations.

At the highest redshift of $z=4$, only individual observations from JWST measurements are available for quiescent galaxies by \citet{ito:2024} and \citet{carnall:2024}, and the morphology is not completely clear, as quiescent galaxies do not necessarily need to be ETGs. The spread in radii from observations in fact is rather large, covering the full spread of what is shown from simulations. As discussed by \citet{remus:2025}, the simulation \Bthuhr with the advanced BH model shows a very good agreement with these observations in terms of the mass--size spread, and as also shown here in the left two panels of Fig.~\ref{fig:mass_size}, at both redshifts the radii at fixed stellar mass of for the advanced BH model simulation shown in turquoise compared to the fiducial run in blue are generally larger, indicative of more extended disks in agreement with observations, while at the same time containing very compact galaxies resembling the quiescent galaxy observations. How much of this is due to the larger box volume of that simulation cannot be tested here and needs to be understood in future analyses. 

Overall, the simulation shows a general trend that galaxies at lower redshifts are larger than at higher redshifts, with a stronger evolution for the quiescent galaxies in terms of radial growth than for disk galaxies, in excellent agreement with observations \citep[e.g.][]{vanderwel:2014,vandesande:2013}. We find a tendency for more extended galaxies being more common at the low-mass end, but the trend is not as pronounced as in the observations by \citet{vanderwel:2024}. This evolution indicates that quiescent galaxies are formed in shorter timescale massive starbursts at high redshifts as suggested by \citet{kimmig:2025} and confirmed by observations \citep[e.g.][]{forrest:2020,kakimoto:2024,nanayakkara:2024}, while the lower-redshift evolution is dominated by merger events, especially also minor merger growth, as suggested for example by \citet{naab:2009} and \citet{schulze:2020}. For the disk galaxies, many pathways of growth through redshift exist, and whether this is encoded in their sizes yet needs to be determined in the future.

\subsection{The Fundamental Plane of Galaxies}
\label{sec:funplan}

\begin{figure*}
    \includegraphics[width=0.99\textwidth]{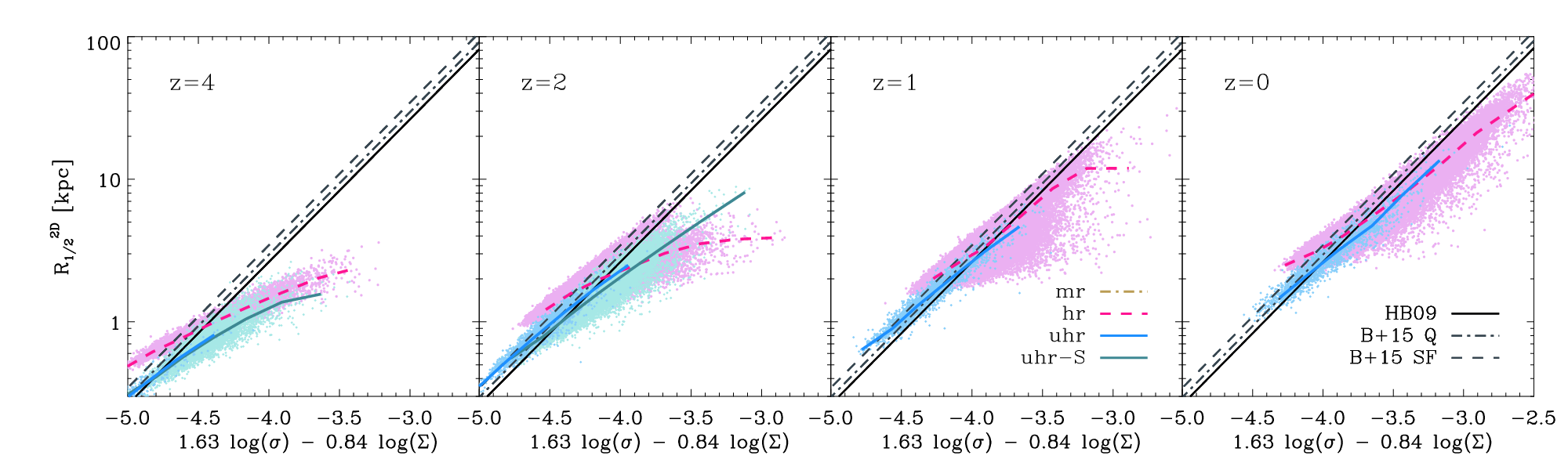}
    \caption{Fundamental plane of galaxies for the \magpath simulations with colors indicating different resolutions and volumes, at $z=4$, $2$, $1$, and $0$ from left to right. For comparison, the observed relation obtained at $z=0$ from all galaxies from \citet{hyde:2009} is shown as solid lines at all redshifts, and the lines obtained for quiescent and star-forming galaxies separately at $z=0$ by \citet{bezanson:2015} are shown as dash-dotted and dashed lines, respectively. As studies indicate that there is no evolution in redshift up to $z=2$ for this relation \citep[e.g.][]{bezanson:2013}, we include these three relations at all redshift for comparison.}
    \label{fig:funplan}
\end{figure*}

The Fundamental Plane of galaxies describes the relation between three characteristic properties of galaxies that span up a plane, namely the size, the central velocity dispersion, and the mean central surface brightness. This was already described and discussed by \citet{hyde:2009}. For comparison with simulations, the mass fundamental plane is more suitable, which simply exchanges the central surface brightness by the stellar mass surface density as described by \citet{bezanson:2015}.
Following \citet{bezanson:2015}, the mass fundamental plane is described as
\begin{equation}
     \log(R_{1/2}) = \alpha \log(\sigma) - \beta \log(\Sigma_*) +\gamma,
\end{equation}
with the classical values from \citet{hyde:2009} being $\alpha=1.629$ and $\beta=-0.840$, and the normalization $\gamma$ usually kept free.
This scaling relation basically combines the mass--size relation that we discussed in the previous section, and the correlation between the mass of a galaxy and its kinematics. For spheroidal galaxies, or ETGs, the observed relation between luminosity and the galaxy velocity dispersion is called Faber-Jackson relation \citep{faber:1976}, which, if a constant mass-to-light ratio is assumed, describes the correlation between the stellar mass and the velocity dispersion. For disk galaxies, or LTGs, the observed relation between the luminosity and the maximum rotational velocity is called the Tully-Fisher relation \citep{tully:1977}, which for the mass instead the luminosity transcribes into the Baryonic Tully-Fisher relation \citep{mcgaugh:2000,ristea:2024}. For the \magpath simulations \Bfuhr, it was shown by \citet{mayer:2023} that the simulated disk galaxies excellently reproduce the observed baryonic Tully-Fisher relation from $z=2$ to $z=0$. 

While both the Faber-Jackson and the Tully-Fisher relation are valid for either spheroidal or disky galaxies, the fundamental plane has been shown to be one plane for all types of galaxies \citep{bezanson:2015}, with only a slight offset for disk galaxies in terms of the normalization. Fig.~\ref{fig:funplan} shows the fundamental plane of galaxies for the \magpath simulations, with colors indicating the different resolutions as given in the legend, for redshifts $z=4,2,1,0$ from left to right, respectively. Overall, we again find convergence between the resolutions at all redshifts, albeit at $z=4$ the \hr resolutions shows an offset in normalization to the \uhr resolution simulations. The \hr resolution simulations also show a larger scatter, albeit this is expected given the much larger number of galaxies in the larger yet less resolved simulations.

For comparison, the observed relation from \citet{hyde:2009} is shown as solid line, and the relations for quiescent and star forming galaxies reported by \citet{bezanson:2015} are shown as dash-dotted and dashed lines, respectively. At $z=1$ and $z=0$ we find the simulations to closely match the findings of the observations, with the simulated fundamental plane being slightly tilted compared to the observations, with the same tilt at both redshifts. From the observational side, most observations report no evolution of the fundamental plane with redshift: \citet{bezanson:2015} report no evolution for both star forming and quiescent galaxies up to $z=1$, and for ETGs no evolution even up to $z=2$, with the caveat for the LTGs that they could not properly be measured out to such high redshifts. Similarly, \citet{zahid:2016} find no evolution up to $z=0.6$, while \citet{degraaff:2021} report no indications of evolution for neither star forming nor quiescent galaxies up to $z=0.8$ from the LEGA-C survey, and report both types of galaxies to reside in the same plane. The only deviations from this behavior are reported for galaxies inside a galaxy cluster environment, where \citet{saracco:2020} report a tilt for the fundamental plane at $z=0$ compared to the field. However, \citet{holden:2010} do not find a tilt for galaxies in galaxy cluster environments at $z=0.8$, and even reporting the tilt to be identical to that reported for $z=0$ field galaxies.

The only change in redshift behavior reported in observations reflect the change with redshift seen in the simulations: galaxies at lower redshifts have larger radii at fixed mass, and thus the whole galaxy population moves from the lower left to the upper right on the fundamental plane, however, the tilt if found to be constant. This is in agreement with other simulations as well: For IllustrisTNG, \citet{lu:2020} report no change in the plane for their ETGs up to $z=2$, while \citet{degraaff:2023} report for Eagle that the star forming and quiescent galaxies live on the same plane as well. We find the \magpath simulations to agree with this, and at $z=0$ \citet{remus:2016} showed the fundamental plane for ETGs and LTGs separately to agree with observations as well.

At high redshifts of $z=4$ and $z=2$, we find an interesting deviation from the plane for the \magpath simulations, however: the tilt of the overall fundamental plane becomes steeper for all simulations. We investigated the origin of this split, and found the source of this to be extremely gas rich galaxies. If we plot quiescent galaxies at these redshifts only, the slope of the resulting fundamental plane is the same as at lower redshifts. If, however, we include star forming gas rich galaxies at redshifts before cosmic noon, they still form a fundamental plane that is relatively thin, however, their slope is tilted compared to the plane of the quiescent galaxies. On the one hand, this explains the behavior of the fundamental plane from high to low redshifts in the \magpath simulations, since the amount of very gas rich, highly star forming galaxies decreases strongly after the peak of star formation around $z=2$. On the other hand, the origin of this second plane for gas rich star forming galaxies at high redshifts is puzzling, especially since it is not found for the star forming galaxies at $z=1$ or $z=0$, and will be subject to an additional study in the future.

\begin{figure*}
    \includegraphics[width=0.99\textwidth]{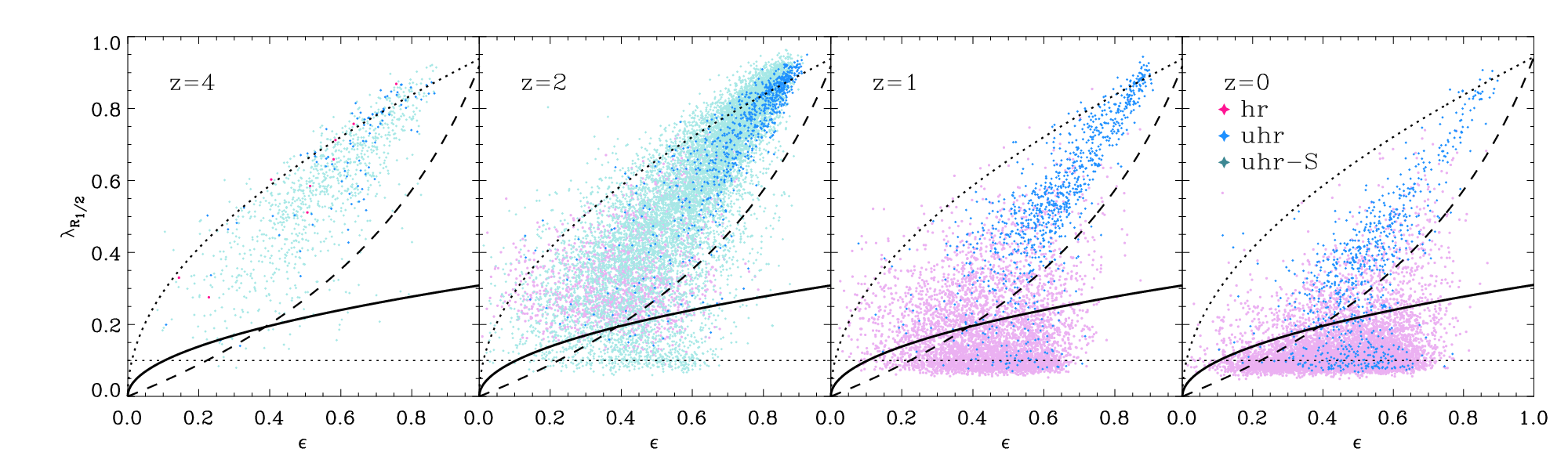}
    \caption{Rotational support--ellipticity plane from $\lambda_{R_e}$ and $\epsilon_e$ of the simulated galaxies at four redshifts. The properties were determined in a light-weighted manner from the rest-frame r-band fluxes and from the edge-on perspective. The solid black line indicates the threshold $\lambda_R = 0.31\sqrt{\epsilon}$ between fast (above) and slow (below) rotators from \citet{emsellem:2011}. The dashed line indicates the theoretical prediction for an edge-on ellipsoidal galaxy with anisotropy $\delta = 0.8 \epsilon$ following \citet{emsellem:2011} and \citet{schulze:2018}. The curved dotted line shows the theoretical model for a galaxy with anisotropy $\delta=0$ following \citet{binney:2005} and \citet{schulze:2018}. The horizontal dotted line at $\lambda_R = 0.1$ indicates an approximate lower bound of $\lambda_R$ due to the resolution limit of the simulations \citep[e.g.][]{bois:2010, naab:2014, schulze:2018}.}
    \label{fig:l_e}
\end{figure*}

\subsection{Resolved Spatial Distributions: The $\lambda_\mathrm{R}$--$\epsilon$ Plane}

Finally, we place our focus on the inner resolved kinematics of galaxies. While many scaling relations have implications for larger-scale structure formation of halos and their baryonic components, the small-scale kinematics in the inner regions of galaxies are not covered by these. One way of quantifying the inner kinematics and spatial distribution from resolved IFU observations is through the velocity-velocity dispersion ratio $V/\sigma$ and the ellipticity \citep[e.g.][]{binney:2005, cappellari:2007}, both of which can be determined from projection. A new normalized measure for how rotationally-dominated or dispersion-dominated a galaxy is was later introduced by \citet{emsellem:2007, emsellem:2011} through the $\lambda_R$ parameter (see \cref{sec:localprops}).

Simulated galaxies in cosmological simulations have generally had problems fully reproducing the observed galaxy distribution in the $\lambda_R$--$\epsilon$ plane \citep[e.g.][]{vandesande:2019, pulsoni:2020, vandesande:2021}, where for instance overly many elongated slow-rotating galaxies are found in the simulations, whereas these types of objects hardly exist in observations. Because of these difficulties, this relation is an ideal way for discriminating between galaxy formation models used in cosmological simulations. In previous studies it has been found that the galaxies of \magpath align very well with the observed $V/\sigma$ and $\lambda_R$--$\epsilon$ properties of multiple IFU surveys, where \citet{schulze:2018} compared them to ATLAS\textsuperscript{3D}, CALIFA, SAMI, and SLUGGS, and \citet{vandesande:2019} to ATLAS\textsuperscript{3D}, CALIFA, MASSIVE, and SAMI, finding overall good agreement between simulation and observation. Similarly, \citet{valenzuela:2024_streams} compared the inner kinematics with respect to the outer tidal feature structures of the simulated galaxies to MATLAS galaxies, again finding consistent behavior between simulation and observation, and \citet{valenzuela:2024} found consistent kinematic behavior related to the intrinsic shapes of galaxies compared to ATLAS\textsuperscript{3D}. Finally, further predictive and comparative studies on the kinematics in the outer regions and on higher-order moments have been performed by \citet{schulze:2020} and Remus et al.~(in prep.), respectively.
In contrast, other simulation suites struggle to fully reproduce these kinematic relations: \citet{vandesande:2019} found that EAGLE, HYDRANGEA, and HORIZON-AGN galaxies are generally too round at a given $V/\sigma$ value, whereas \magpath galaxies cover the entire range of ellipticities, but have slightly too low rotational support at a given ellipticity. At the other extreme, \citet{pulsoni:2020} found that a large fraction of TNG50 and TNG100 galaxies are overly elongated with slow-rotating galaxies, which they excluded from their analysis due to there being no observational counterparts to those systems.

At higher redshifts, the kinematic analyses are currently rather limited from both the observational and simulated point of view. For \magpath, \citet{schulze:2018} and \citet{kimmig:2025} have made predictions at redshifts up to $z=2$ and $z=5$, respectively, where \citet{kimmig:2025} also connected the inner kinematics to the quenching of massive galaxies at early times up to $z=5.3$, finding that quenched galaxies tend to be faster rotating systems at time of quenching. On the observational side, the $\lambda_R$--$\epsilon$ kinematic plane is difficult to produce in a statistically complete way due to the necessity of highly resolved and low signal to noise IFU observations. For instance, it has been shown for the MAGPI galaxies at $z=0.3$ by \citet{derkenne:2024} for a small sample of about 50 galaxies.

While these previous studies of \magpath galaxy kinematics have addressed different questions with appropriate methodologies, where for instance mass-weighted kinematics were determined to understand the physical motion of the stars, and projections were chosen to be edge-on or random as needed for the comparisons. For this paper we decided on a balance between observationally comparable and physical meaningfulness: in \cref{fig:l_e} we show the $\lambda_R$--$\epsilon$ plane for our simulated galaxies at four redshifts, where we determined the projected quantities in a light-weighted manner (by the r-band filter flux) to better correspond to the observations, but from an edge-on projection, which presents the more physical description of a galaxy's kinematics, despite not being directly comparable to observations (for this we refer the reader to the previously mentioned studies on \magpath kinematics). It is immediately clear from the figure that at higher redshifts the galaxies tend to be very rotationally dominated (high values of $\lambda_R$), whereas at later times there is a large number of slowly-rotating galaxies. The mass trend of more massive galaxies being more slowly-rotating can also be seen from the \hr-resolution galaxies (pink data points) being located further down than the \uhr-resolution galaxies (blue and green). The solid line is the commonly used threshold of $\lambda_R = 0.31\sqrt{\epsilon}$ between fast and slow rotators from \citet{emsellem:2011}. Due to the light-weighted nature of the kinematic properties and the commonly more rotationally supported younger stars formed in gas disks, the $\lambda_R$ values shown here in \cref{fig:l_e} are higher than the mass-weighted counterparts found in previous studies. These higher values are in fact more in line with what observers find, who also obtain the kinematics from light-weighted measurements \citep[e.g.][]{schulze:2018, vandesande:2019}.

\begin{figure*}
    \includegraphics[width=0.99\textwidth]{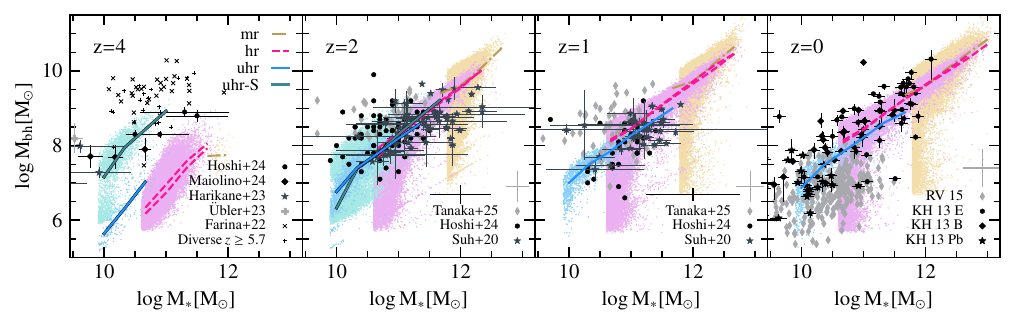}
    \caption{Black hole mass versus stellar mass from $z=4$ to $z=0$ of the \magpath simulations (colored points, with median lines). Shown are also observations at $z=4$ from \citet{farina:2022,uebler:2023,harikane:2023,maiolino:2024,hoshi:2024}, at $z=2$ and $z=1$ from \citet{suh:2020,hoshi:2024,tanaka:2025}, and locally at $z=0$ from \citet{kormendy:2013,reines:2015}. The gray crosses in the lower right show the average measurement error given by \citet{tanaka:2025} at $z=2$ and $z=1$, and by \citet{reines:2015} at $z=0$, while black is given by \citet{hoshi:2024} at $z=2$ and $z=1$.}
    \label{fig:magorrian}
\end{figure*}

\section{Scaling Relations Through Cosmic Time~V: The~Black~Hole~Relations}
\label{sec:scale3}

Supermassive black holes reside at the center of galaxies, and contribute significantly to the evolution of their hosts by heating as well as directly expelling vast amounts of gas, thereby regulating the star formation \citep{kormendy:2013,dubois:2014,steinborn:2015,zinger:2020}, potentially even down to dwarf galaxies \citep{silk:2017,greene:2020}. We present here briefly the established scaling relations from the literature. Throughout we include only central galaxies, with galaxy stellar mass cuts as $M_{R_{0.1}}^*\geq 5\times 10^{9}\,M_\odot$, $\geq 4\times 10^{10}\,M_\odot$ and $\geq 4\times 10^{11}\,M_\odot$ for the \uhr, \hr, and \mr resolutions, respectively. By necessity we only include galaxies that host black holes.

\subsection{The Magorrian Relation}
\label{sec:maggo}

A link between the stellar mass in the central bulge of galaxies and their SMBH has been established for a long time in the local Universe \citep{magorrian:1998,haering:2004,kormendy:2013}, but has recently been extended up to redshifts of $z>4$ \citep{maiolino:2024}. Observationally, black hole masses of broad-line AGNs are typically determined via single-epoch virial estimations \citep{greene:2005,vestergaard:2009}. These rely on the assumption that the central gas kinematics are dominated by the black hole, such that the mass is given by the physical size of the broad line region and the gas velocity. The former is found to correlate with the AGN luminosity \citep{bentz:2009,gravity:2020}, while the latter is determined via the width of the chosen broad emission line \citep{reines:2015}. Typically used are H$\alpha$ and H$\beta$, which at high redshifts however move out of the K-band and are thus inaccessible from the ground \citep{farina:2022}, where instead \MgII and \CIV are employed instead, though \citet{shen:2012} find \MgII to be more reliable. However, the measurements of both the black hole mass and galaxy stellar mass are known to suffer from biases which can be significant, as discussed in great depth by \citet{farrah:2023}.

Fig.~\ref{fig:magorrian} shows the central SMBH mass versus the galaxy stellar mass $M_{R_{0.1}}^*$ from $z=4$ to $z=0$ of the \magpath simulations (colored points), with median lines for each simulation box. We find excellent agreement between the different resolution levels of the boxes, as well as with the observations from $z=2$ down to $z=0$. For observations of galaxies with a range of redshifts, we split them into the panels with redshift cuts as $z>3$, $3>z>1.5$, $1.5>z>0.5$, and $0.5>z$, going from left to right. 

At $z=2$ and $z=1$, the observations by \citet{suh:2020} are of 100 X-ray-selected broad-line AGNs from the Chandra-COSMOS Legacy Survey \citep{civano:2016}, with galaxy stellar mass determined via multi-component SED fitting and the black hole mass via the single-epoch virial method on the H$\alpha$, H$\beta$, or \MgII emission lines from the Keck/DEIMOS optical and Subaru/FMOS near-IR (NIR) spectroscopy. \citet{hoshi:2024} take the optical variability-selected type-I AGN sample by \citet{kimura:2020} at $0.3<z<3.5$, estimating black hole masses via \CIV, \MgII, H$\beta$, and H$\alpha$ broad line emission and the single-epoch virial method, while the host galaxy mass is determined from SED fitting on images from COSMOS2020 \citep{weaver:2022} the Chandra-COSMOS Legacy Survey \citep{marchesi:2016}. \citet{tanaka:2025} observe 107 X-ray-selected type-I active galactic nuclei (AGNs), determining the stellar masses from a decomposition on JWST images from COSMOS-Web \citep{casey:2023} and PRIMER, while the black hole masses are drawn from \citet{schulze:2014} and \citet{schulze:2018_bhs}, who use again broad line measurements of H$\alpha$, H$\beta$, or \MgII, and a virial mass estimation relation. 

At $z=0$, \citet{kormendy:2013} catalog black hole and galaxy stellar masses from various sources, while \citet{reines:2015} select 244 broad-line AGNs from the NASA Sloan Atlas (NSA), based on the SDSS Data Release 8 spectroscopic catalog \citep{aihara:2011}. Stellar masses are calculated from the i-band, with stellar mass-to-light determined via the $g-i$ color following \citep{zibetti:2009}, while black hole masses come from single-epoch virial estimation via H$\alpha$ broad line emission. 

At $z=4$, we find noticeable differences between the different \magpath simulations, where \Bth galaxies of the same stellar mass host black holes that are around an order of magnitude more massive compared to those found in \Bz, \Btb, \Bt, and \Bf, which agree more closely with each other. This is because of the different black hole model implemented for \Bt, given by \citet{steinborn:2015} and also described in Sec.~\ref{sss:abhm}. This newer model allows for a faster growth of the black holes at high redshift (while still remaining Eddington-limited), which have recently been revealed by JWST observations to indeed be more massive compared to their low-redshift counterparts \citep{pacucci:2023,maiolino:2024}.

We find that \Bth matches the observed black hole masses at $z=4$ by \citet{uebler:2023,harikane:2023,maiolino:2024,hoshi:2024}, although those by \citet{farina:2022} tend to be higher by around half an order of magnitude. Nonetheless, \Bth well reproduces the observed relations both at $z=4$ and $z=2$, implying that it is crucial to include a two-channel growth for the SMBH in simulations, to differentiate the efficient accretion of turbulent cold gas flows \citep{gaspari:2013} as compared to hot gas, which more resembles the assumptions of a hot diffuse medium for Bondi-Hoyle-Lyttleton accretion \citep{hoyle:1939,bondi:1944,bondi:1952}.

The observation by \citet{uebler:2023} is for an individual type-1.8 AGN at $z=5.55$ performed with JWST/NIRSpec, with the host stellar mass determined by a fit to the continuum emission, which they assume to be dominated by the galaxy (where it is worth noting that \citealp{barchiesi:2023} find a higher stellar mass when using broadband photometry), while the black hole mass is estimated via the low-redshift calibration of the single-epoch virial estimate using the H$\alpha$ broad line width as given by \citet{reines:2013}, with little variation when instead using H$\alpha$ or H$\beta$ and the calibrations by \citet{greene:2005}. \citet{harikane:2023} meanwhile observe 10~type-I AGN at $4<z<7$ with JWST/NIRSpec and determine the black hole mass via the H$\alpha$ width following \citet{greene:2005}, while stellar mass is determined via SED fitting with a model of the PSF as the AGN and S\'ersic profile \citep{sersic:1963} for the host (at least in the case of the two AGNs plotted within the range of Fig.~\ref{fig:magorrian}). \citet{maiolino:2024} observe thirteen galaxies as part of the JADES Survey \citep{eisenstein:2023}, where we plot here only their BLR-1 components for candidate merging black holes, with galaxy stellar masses determined via SED fitting with a power-law continuum for the AGN, while the black hole masses are estimated via H$\alpha$ line widths following \citet{reines:2013}.

\begin{figure*}
    \includegraphics[width=0.99\textwidth]{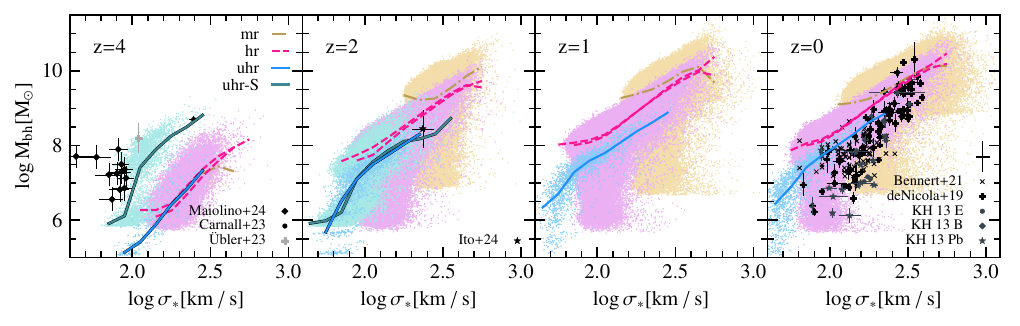}
    \caption{Black hole mass versus central stellar velocity dispersion $\sigma_*$ from $z=4$ to $z=0$ of the \magpath simulations (colored points, with median lines). Shown are also observations of individual objects at $z=5.5$ \citep{uebler:2023}, $z=4.6$ \citep{carnall:2023_highzbh} and $z=2.1$ \citep{ito:2024_onebh}, the JADES sample by \citet{maiolino:2024} at $z>4$ and compilations of observations at $z=0$ by \citet{kormendy:2013,denicola:2019,bennert:2021}. For \citet{bennert:2021}, we show their observational error as a black cross on the right.}
    \label{fig:sig_bh}
\end{figure*}

Finally, we also show the Near-IR observations by \citet{farina:2022} performed with X-shooter \citep{vernet:2011}, with most of their black hole masses determined via the \MgII broad line and the low-redshift calibration by \citet{shen:2011} with a smaller number using the \CIV line via \citet{vestergaard:2006}. We note that their higher values in comparison to \magpath may, at least in part, be due to the chosen calibration, with \citet{farina:2022} reporting that when they instead follow \citet{vestergaard:2009} for the \MgII lines they find an average reduction in BH mass of $0.2\,\mathrm{dex}$, and for the highest outlier (J0100+2802) the difference is $0.5\,\mathrm{dex}$. The other BH with mass above $10^{10}\,M_\odot$ (P009-10) is calibrated via \CIV, where applying the correction by \citet{coatman:2017} would result in a nearly $1\,\mathrm{dex}$ reduction in mass. \citet{farina:2022} additionally compile data for quasars at $z\geq5.7$ \citep{willot:2015,willot:2017,izumi:2018,izumi:2019,izumi:2021,pensabene:2020}, which we also plot in Fig.~\ref{fig:magorrian} (black pluses). 

\subsection{Black-Hole-Mass--$\sigma$ Relation}
\label{sec:bhsig}

Given the relation between host stellar masses $M_*$ and their central stellar velocity dispersion $\sigma_*$, it is not unsurprising that observations also find a black-hole-mass--sigma relation \citep{ferrarese:2000,kormendy:2013,denicola:2019}, which has over time been argued to be more fundamental than the Magorrian relation \citep{maiolino:2024}, although \citet{kormendy:2013} find the scatter for both relations in the local Universe ($z=0$) to be similar. Gas flow toward the center is regulated by the balance between the strength of the black hole outflow (which depends on the black hole mass as $\dot{E}\propto\dot{M}_\mathrm{bh}\propto M_\mathrm{bh}^2$, \citealp{steinborn:2015}) and the depth of the potential well, which works to draw in more gas, where such a balance for a black hole accreting at Eddington rate gives a maximum scaling mass as $M_\mathrm{bh}\propto \sigma_*^5$ \citep{silk:1998,ferrarese:2000}.

In Fig.~\ref{fig:sig_bh} we plot the central stellar velocity dispersion for the simulations (determined in projection within 1~stellar half-mass radius) versus the central SMBH mass from $z=4$ to $z=0$, plotting also observations for comparison. We find that the differing resolutions of the \magpath simulation suite converge well among each other, aside from \Bth at $z=4$ where the differing black hole implementation allows earlier growth of the black holes \citep{steinborn:2015,kimmig:2025}. Furthermore, the high and medium resolutions (\hr and \mr) tend toward lower dispersion at the same black hole mass by around $0.2\,\mathrm{dex}$ at later times as compared to the highest resolution (uhr). 

Comparing to the observations at $z=0$ by \citet{kormendy:2013}, \citet{denicola:2019}, and \citet{bennert:2021} in the left panel of Fig.~\ref{fig:sig_bh}, we find generally good agreement, in particular for the more massive galaxies. We find also that at the low black hole-mass end, the observations tend toward higher velocity dispersions compared to the simulations. We note that whenever there are observations by \citet{kormendy:2013} and \citet{denicola:2019} for the same object, we plot only the latter. For $M_\mathrm{bh}\propto \sigma_*^\beta$, the highest resolution simulation \Bf finds a slope of $\beta=3.4\pm0.2$, somewhat more shallow than the values of $\beta=4.8\pm0.5$ and $\beta=4.377\pm0.29$ found by \citet{ferrarese:2000} and \citet{kormendy:2013} but in closer agreement with $\beta=4.02\pm0.32$ determined by \citet{tremaine:2002} as well as the more recent value by \citet{bennert:2021}, who find $\beta=4\pm0.25$ for their sample of $50$~AGNs and $51$~quiescent galaxies.

We find that at $z=0$ the observations by \citet{bennert:2021} lie closest to the simulations, but still generally tend toward higher velocity dispersions, in particular at lower black hole masses. This is interesting when considering that we found overall better agreement when plotting the black hole mass against the bulge mass (see Fig.~\ref{fig:magorrian}), which may indicate that the simulations find the correct amount of stellar mass, but should be distributing it differently kinematically. This matches the general behavior that simulations of galaxies in clusters tend to find lower velocity dispersions compared to observations \citep{meneghetti:2020}, except at the higher-mass end \citep{bahe:2021}. This is not only seen for galaxies in clusters, but also found for galaxies measured with integral field units like the SAMI galaxies, or ATLAS\textsuperscript{3D}, for which \citet{vandesande:2019} showed that the measured velocity dispersions from observations are larger than what is found in simulations. This mismatch between simulations and observations is as of yet an unsolved issue.

\begin{figure*}
    \includegraphics[width=0.99\textwidth]{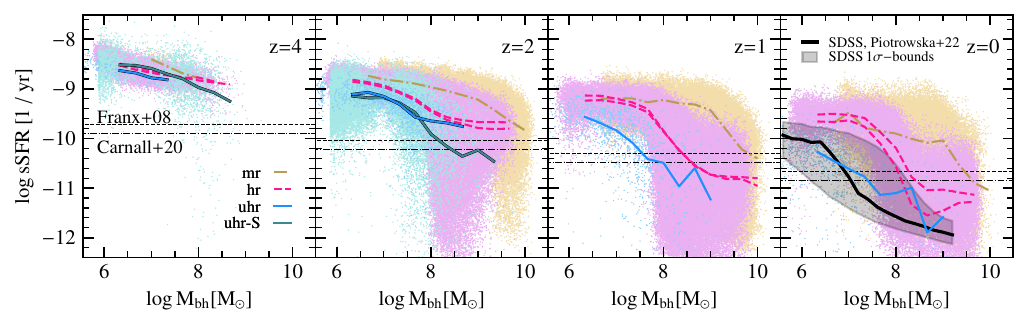}
    \caption{The specific star formation rate as a function of black hole mass. Horizontal black dashed and dash-dotted lines are typical cuts employed to separate the star-forming and quiescent populations by \citet{franx:2008} and \citet{carnall:2020}, while the black solid line shows the SDSS median trend as given by \citet{piotrowska:2022}.}
    \label{fig:bh_ssfr} 
\end{figure*}

At $z=2$ in Fig.~\ref{fig:sig_bh}, we find that the individual observation by \citet{ito:2024_onebh} well agrees with all our simulations. At $z=4$, \Bth best reproduces the observations by \citet{uebler:2023}, \citet{carnall:2023_highzbh}, and \citet{maiolino:2024}, while the other boxes' black hole masses are too small comparatively (as already noted for Fig.~\ref{fig:magorrian}). We note that the black hole masses measured by \citet{uebler:2023} and \citet{maiolino:2024} lie toward the upper end of our range of simulated galaxies even for \Bth, presenting the opposite challenge to $z=0$ of requiring lower dispersions in the simulation for the given black hole masses. Given that, again, the stellar masses are in better agreement, and also that the more massive end as represented with the observation by \citet{carnall:2023_highzbh} is in much better agreement, it is not unlikely that the limiting factor here is numerical in nature. Earlier seeding of black holes at lower masses, as well as a better resolution may be sufficient to better reproduce the observations at this low-mass, compact end. For example, the gravitational smoothing length of \Bth is around $1~\mathrm{kpc}$, limiting the compactness (and therefore minimum dispersion) that can be achieved at high redshifts. Other possible improvements include self-consistently tracing the BH spin through time \citep{sala:2024}.

We note that, for the JWST observations by \citet{maiolino:2024} at $z>4$, the black hole masses are calculated as described above, while the central stellar velocity dispersion is determined via the narrow component of H$\alpha$ where available, and otherwise of [\OIII]. This is therefore an ionized gas velocity dispersion, which may have some differences to that of the stars \citep{bezanson:2018}, such that \citet{maiolino:2024} correct their values upwards by $0.12$--$0.18\,\mathrm{dex}$. The same is true for the galaxy observed at $z\approx5.5$ by \citet{uebler:2023}, with the black hole mass determined as above while the velocity dispersion is given as the narrow line dispersion H$\alpha$, with an upwards correction factor of $+0.1\,\mathrm{dex}$ \citep{bezanson:2018}. \citet{carnall:2023_highzbh} similarly determine the black hole mass via the flux and broad-line width of H$\alpha$ following \citet{greene:2005}, but calculate the velocity dispersion from their Bagpipes \citep{carnall:2018} full spectral fit \citep{carnall:2023_highzbh}. Similarly, \citet{ito:2024} calculate the velocity dispersion via fit to the NIRSpec spectrum using \textsc{pPXF} \citep{cappellari:2017}, while the black hole mass comes from the broad H$\alpha$ emission line following \citet{reines:2013}.

\subsection{Specific Star Formation Rate versus BH Mass}
\label{sec:ssfrbhm}

\begin{figure*}
    \includegraphics[width=0.99\textwidth]{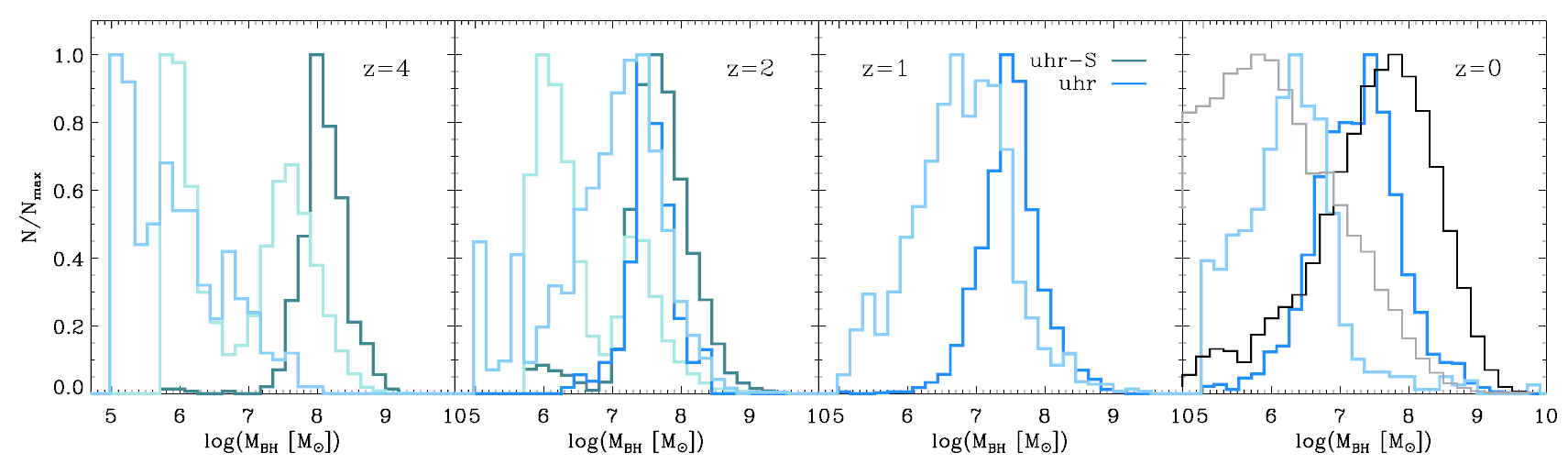}
    \caption{Histograms of the black hole mass distributions for both star-forming (lighter colors) as well as quiescent galaxies (darker colors) for the \uhr resolution simulations \Bth and \Bf, normalized by the peak value of all bins to be comparable to the SDSS data from \citet{piotrowska:2022} for star-forming (light gray) and quiescent (dark gray) galaxies at $z=0$.}
    \label{fig:bh_fq}
\end{figure*}

Finally, within the space of black hole scaling relations, we consider the integrated impact of the black hole (given by its current mass, see \citealp{bluck:2020,piotrowska:2022}) on the host galaxy's current specific star formation rate (sSFR) in Fig.~\ref{fig:bh_ssfr}. To be comparable to the results for SDSS found by \citet{piotrowska:2022}, in addition to the general selection cuts for this section, here we apply also their cuts as $M_\mathrm{vir}>10^{11}\,M_\odot$, $10^{12}\,M_\odot>M_*>10^{9}\,M_\odot$, and $\mathrm{SFR}_\mathrm{tot}>0$. For reference, we also plot as black horizontal lines some commonly employed thresholds on the sSFR of a galaxy to define quiescence, with the threshold of sSFR$<0.3/t_\mathrm{hubble}$ (dashed) by \citet{franx:2008} and sSFR$<0.2/t_\mathrm{hubble}$ (dash-dotted) as given by \citet{carnall:2020}, and $t_\mathrm{hubble}$ the age of the Universe at the given redshift in units of years. We find that, generally, higher black hole masses result in lower star formation rates for all simulations, as expected, with a shallow decline up until around $M_\mathrm{bh}=10^{7}\,M_\odot$ followed by a steeper drop toward $M_\mathrm{bh}=10^{8}\,M_\odot$, beyond which most galaxies are quiescent. This is in agreement with the location of this dip found at $z=0$ for SDSS galaxies by \citet{piotrowska:2022} at around $M_\mathrm{bh}\approx10^{7}\,M_\odot$. We note that the lower resolution simulations are offset toward consistently higher sSFR compared to \uhr because of their much higher minimum stellar mass cut. When we set the minimum stellar mass of \Bf to $M_*\geq4\times10^{10}\,M_\odot$, as for \Bt and \Btb, the median trends among the simulations agree. 

We also find a significant scatter at high black hole masses, where within each simulation some galaxies are able to retain high star formation rates, while others are quenched. This is true in particular at $z=4$ for \Bth, which already has a noticeable number of galaxies scattering down to quiescence, resulting in better agreement with observed quiescent number densities as discussed for Fig.~\ref{fig:numb_dens}, even while the median trends of all simulations agree remarkably well at this redshift. The necessity of this scatter in reproducing observations is discussed in more detail by \citet{kimmig:2025}. At low redshifts, we find that the higher-resolution simulation, which reaches lower stellar masses more comparable with SDSS, also best matches the median behavior found by \citet{piotrowska:2022}.

Consequently, for the higher-resolution simulations, we can compare the distribution of black hole masses for star-forming and quiescent galaxies with that found in SDSS at $z=0$ by \citet{piotrowska:2022} in Fig.~\ref{fig:bh_fq}. We find that at all redshifts quiescent galaxies consistently host more massive black holes than their star-forming counterparts. The transition between the two populations occurs for \magpath at around $M_\mathrm{bh}=10^{7}\,M_\odot$, which is in remarkable agreement with the distribution of SDSS galaxies. Additionally, \Bf quiescent galaxies also host a tail end of AGNs with lower masses, matching the behavior in SDSS, although there tends to be a slightly lower frequency of star-forming galaxies with more massive black holes compared to observations.

\section{Data Availability}
\label{sec:DataAvail}

\subsection{Flat Light-cones}
One of the data products readily available from the \magpath suite is a set of light-ones at different depths and wavelengths. 

\begin{figure*}
    \includegraphics[width=0.49\textwidth]{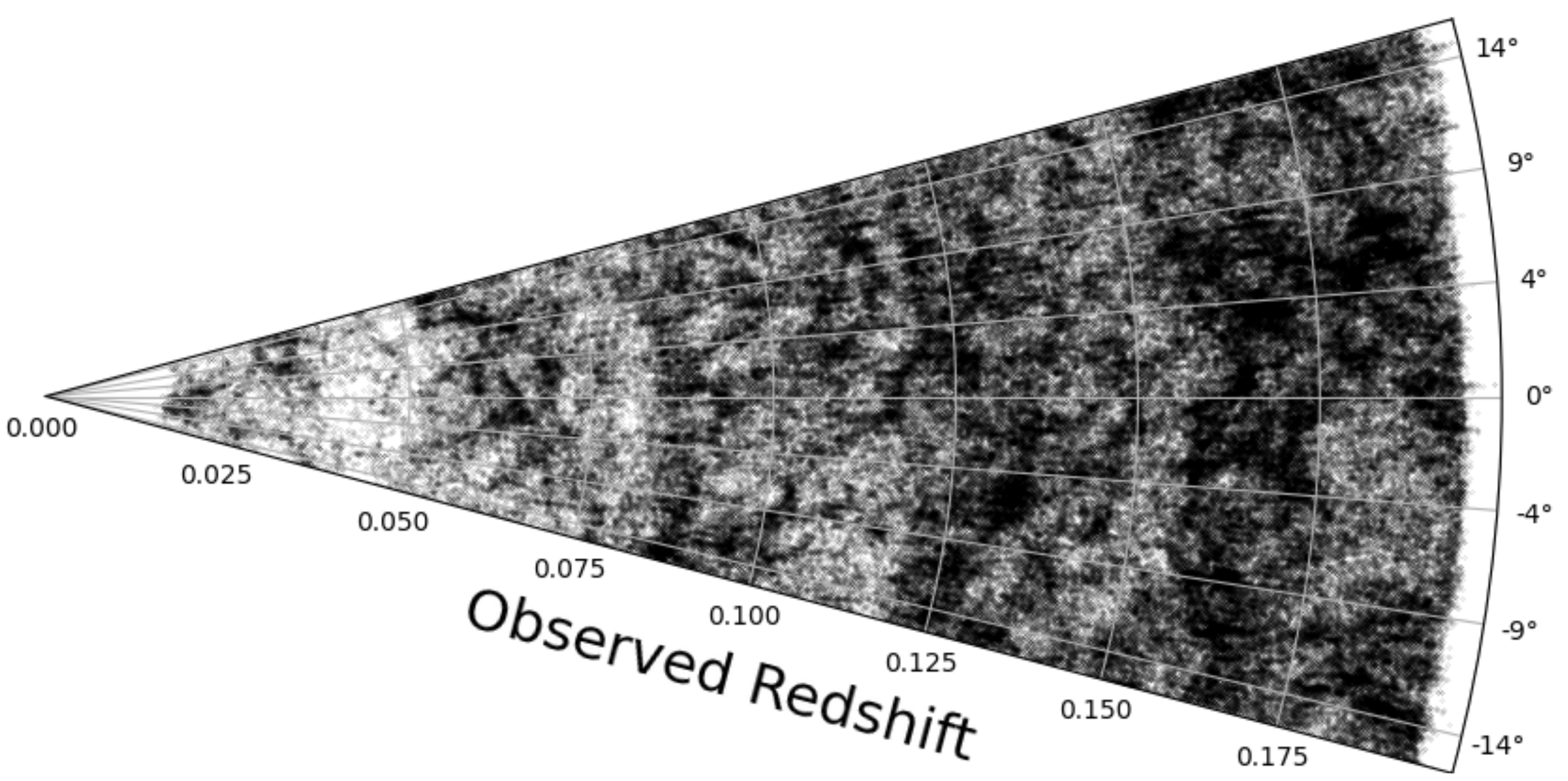}
    \includegraphics[width=0.49\textwidth]{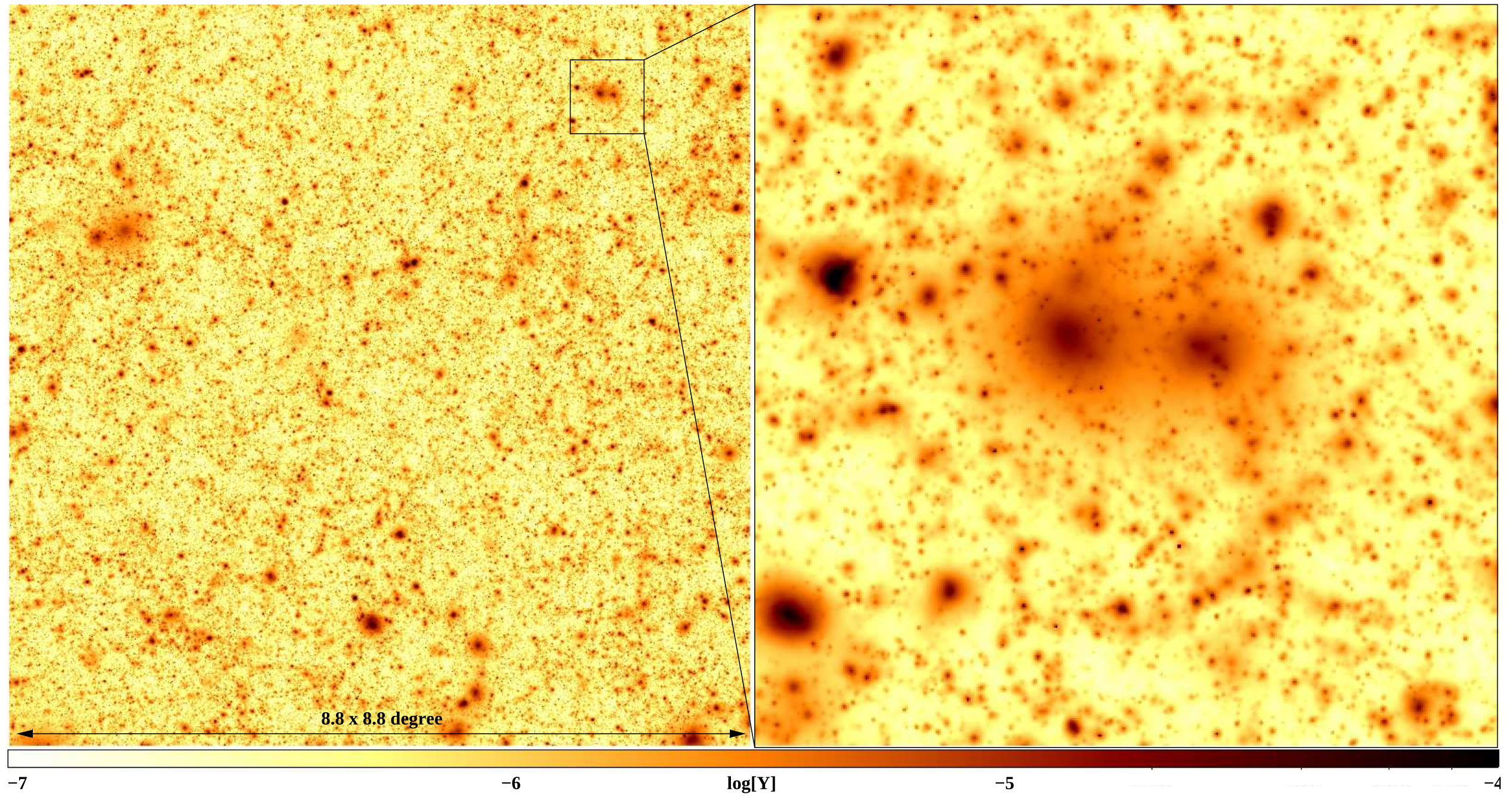}
    \caption{Shown on the left is the distribution of the galaxies within the LC30 \citep{Marini2024}, illustrating the geometry of a flat light-cone. On the right, the $8.8\times8.8$ square degree thermal SZ light-cone \citep{2016MNRAS.463.1797D} is shown with an additional zoom onto a galaxy cluster within it.}
    \label{fig:flat_lightcone}
\end{figure*}

There are various realizations of SZ light-cones derived from the different simulations, including five $13x13$ square degree light-cones out to $z=1.98$ \citep{2014MNRAS.440.2610S} and one $8.8\times8.8$ square degree deep light-cone out to $z=5.2$ based on \Bomr \cite{2016MNRAS.463.1797D} as well as an $35\times35$ square degree light-cone for thermal and kinetic SZ, based on \Bomr out to $z=2.15$ \citep{2018MNRAS.478.5320S}. They also provide halo catalogs for the individual slices they constitute of. In Fig.~\ref{fig:flat_lightcone} two images of light-cones are shown exemplarily to illustrate the geometry as well as the richness of structures within them. A $30\times30$ square degree light-cone obtained from \Bthr was used to study X-ray properties of galaxy clusters and to test for biases in temperature measurements \citep{2023A&A...675A.150Z} well as to demonstrate that the average energy of the observed cluster can be used as a proxy of its temperature \citep{2024arXiv240812026K}.  

A set of multi-wavelength light-cones are extracted from the parent cosmological \Bthr. We produced one light-cone with an aperture equivalent to $30\times30$ square degrees down to $z=0.2$ and four with $5\times5$ square degrees down to $z=1.9$. The data available comprises the X-ray emission (gas particles, black holes, and X-ray binaries), the integrated mean Compton $Y$ parameter in the SZ emission, the galaxy optical magnitude in the SDSS filters, and a list of properties for the galaxies and halos, as identified by \subfind. The geometric structure of the light-cones affects intrinsic properties that are distorted due to redshift dimming, peculiar velocities, and projection effects. We extensively describe the design of the light-cone in \cite{Marini2024, Marini2025}.

In the following, we give an overview of the data available for each wavelength.
\begin{itemize}
    \item The X-ray emission is calculated with {\sc Phox}, as described in Sect. \ref{sec:lxm}. Additionally, we post-process the synthetic photon list with SIXTE \citep{Dauser2019} to extract a mock observation with eROSITA in scanning mode, for exposure times equivalent to eFEDS and eRASS:4 \citep{eROSITA_ScienceBook}. The resulting mock observation, including all eROSITA instrumental effects and calibrations, are processed with eROSITA Science Analysis Software System (eSASS) to extract extended and point source detections as done in \cite{Merloni2024} and \cite{Bulbul2024} for eRASS:4. A full description on this data product is provided in \cite{Marini2024}.
    \item An extensive description of the SZ emission and comparison with the observational data of the \magpath predictions is provided in \cite{2016MNRAS.463.1797D}. A full description of this data product is provided in \cite{2016MNRAS.463.1797D}.
    \item Optical properties of the galaxy population are provided in the intrinsic and observed magnitudes. These are based on the single stellar population model, described in \cite{Saro2006}. The absolute rest-frame magnitude completeness is at $r=-21$. To accurately account for observational uncertainties and incorporate the K correction, the rest-frame magnitudes given by \magpath have been fitted through a standard SED fitting technique with CIGALE (\citealt{Boquien2019} and references therein). The best-fit template is then shifted to the observed frame at the galaxy's redshift, and the magnitudes in the desired filters are recalculated from this. A full description of this data product is provided in \cite{Marini2025}.
\end{itemize}
The data here described are available from the corresponding author upon reasonable request.

\subsection{Full-Sky Light-cones}

To produce full sky light-cones we use {\sc SMAC} \citep{dolag:2005} which allows us to project a spherical shell within a simulation onto a healpix map. This was first applied to \Bomr of \magpath to produce thermal and kinetic SZ maps out to a limiting redshift range of $z=0.17$ to study the large modes within the observed SZ power spectrum. This full-sky map was extended using \Bomr to reach a redshift of $0.5$ by this simple technique. Duplicating the box twice in each spacial direction allowed us to extend this even up to a redshift of $z=1.2$ \citep{2022MNRAS.513.2252C}. This full-sky light-cone is in qualitative agreement with the analog created by websky. To contrast the imprint of the local super clusters onto the SZ power spectrum, \citet{2024A&A...692A.180J} use 27 realizations of full sky maps produced from \Bthr up to 110 Mpc distance.

\begin{figure*}
    \centering
    \includegraphics[width=0.99\textwidth]{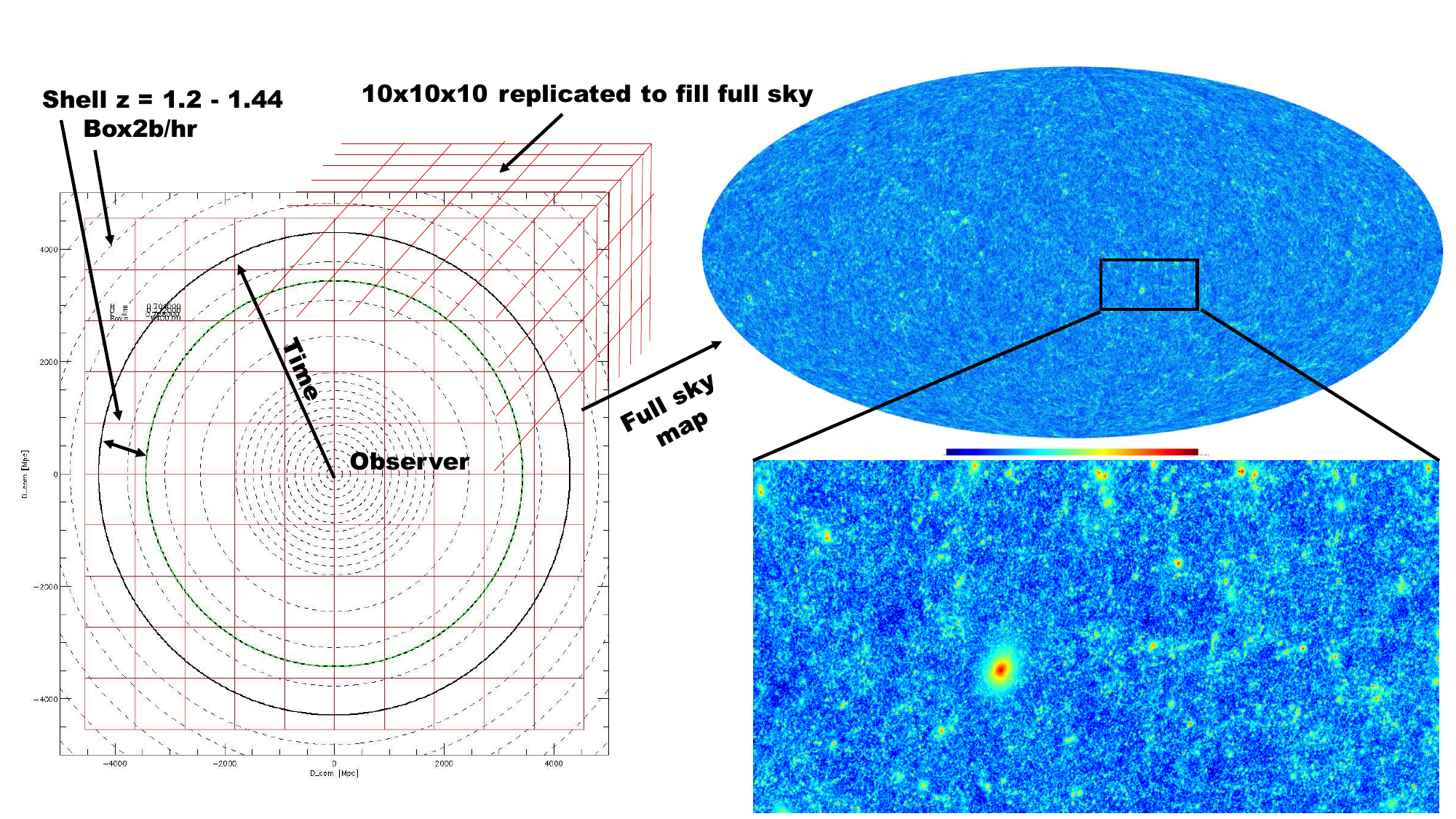}
    \caption{Shown is the geometry of one slice for the full tSZ sky light-cone from \Btbhr on the left. Although the simulation is almost one 1Gpc on each side, 1000 replicas are needed to fill the full sky. The right panel shows the full sky map of the whole light-cone with a zoom onto one galaxy cluster. One snapshot of this simulation covers 4.8 TB on disk and to produce the whole light-cone more than 10 PB of data has to be processed.}
    \label{fig:fullsky_lightcone}
\end{figure*}

Meanwhile, a high-resolution full-sky map of thermal SZ out to redshift $z=1.44$ has been generated combing \Bthr and \Btbhr, as visualized in Fig.~\ref{fig:fullsky_lightcone}. Despite the Gpc size of \Btbhr the box has to be replicated 1000 times to fill the full sky in the final redshift slice. Although this means that individual clusters appear several times in the individual slices, the different geometry of the contribution on the replicated boxes to the individual slices avoids replications of the surrounding structures and different projection directions and therefore effectively leads to quite different appearances of the same cluster, as can be seen from the sketch of full-sky construction in the left part of Fig.~\ref{fig:fullsky_lightcone}. Note that for this task, very effectively parallelized post-processing routines have to be used. By the need for such a large number of replicas the final full-sky light-cone as shown in the right of Fig.~\ref{fig:fullsky_lightcone} was created by the accumulation of the results of processing almost 10 PB of simulation data.  

\subsection{Web Portal}
\label{sec:web}

The data of most of the described \magpath simulations are made available to the community within a first test operation of the cosmological simulation web portal.\footnote{c2papcosmosim1.srv.lrz.de} More details can be found in \citet{ragagnin:2017webportal}. Users can access data products extracted from the simulations via a user-friendly web interface, browsing through visualizations of cosmological structures while guided by metadata queries helping to select galaxy clusters and galaxy groups of interest. At the moment, {\sc Phox} is the first enabled service on this platform and allows to perform virtual X-ray observations where FITS files with photon lists are returned in the so-called {\sc simput} format, taking the specifications of various, existing and future X-ray telescopes into account. In addition, the {\sc Smac} service \citep{dolag:2005} allows the making of synthetic maps of various, fundamental, and idealized quantities (e.g. density, temperature, thermal SZ, kinetic SZ, and many more). For more complex analysis, the {\sc SimCut} service allows one to get the full simulation data of a region cut out around halos for download, which then can be analyzed in detail. 
 
\begin{figure*}
    \centering
    \includegraphics[width=0.99\textwidth]{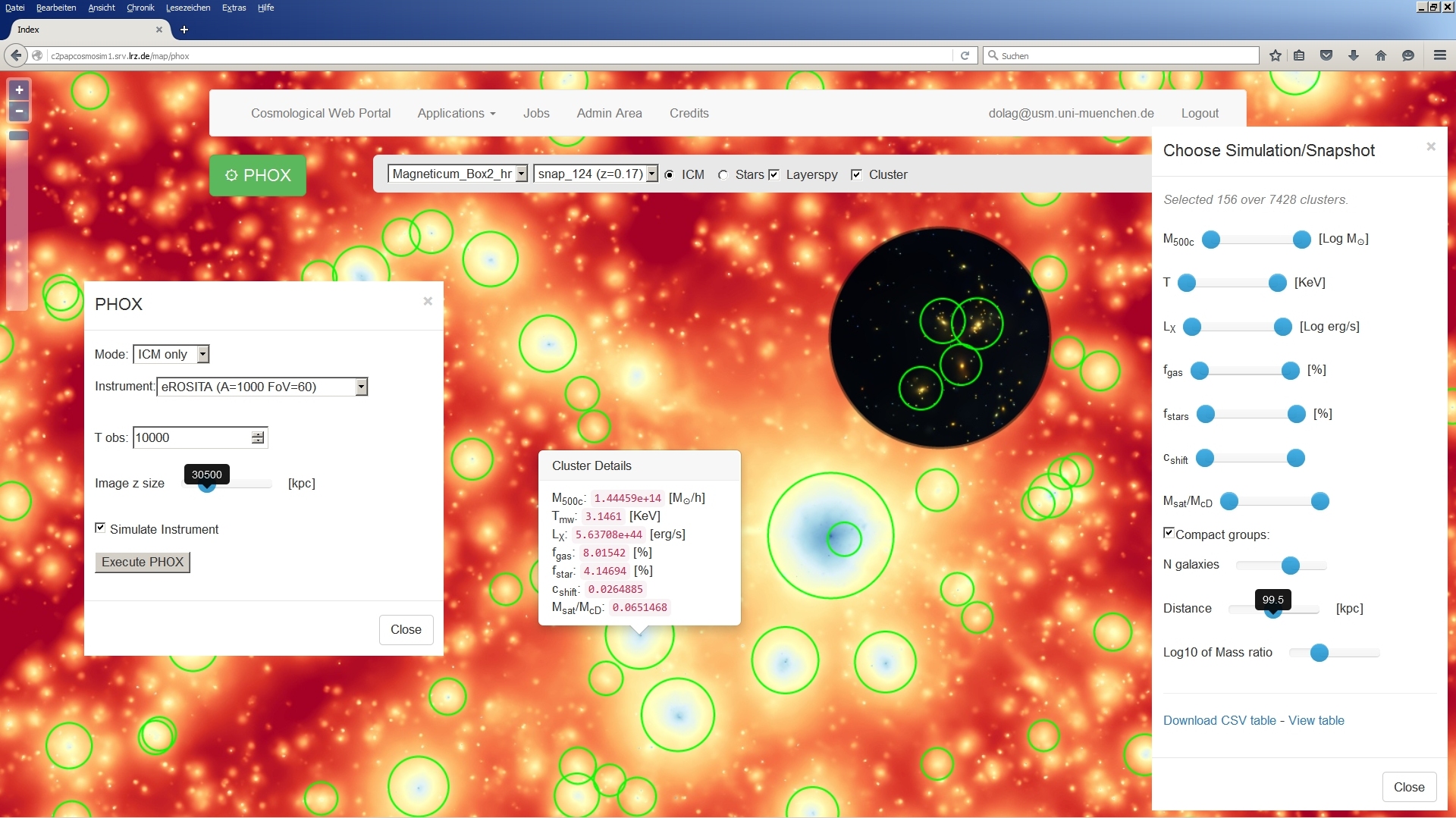}
    \caption{Shown is the graphical interface of the web portal, selecting \Bthr from the \magpath simulation at $z=0.17$, visualizing the diffuse media with the layer-spy option for the stellar component on and clusters and groups overlayed as circles. The pop-up shows the properties of the currently chosen cluster. On the right, there is the cluster restriction interface shown which also allows to download all metadata of the selected clusters as CSV table. The left panel shows the {\sc Phox} service interface.}
    \label{fig:web_vizual}
\end{figure*}

The visual front end allows exploring the cosmological structures within the simulation based on spanning through and zooming into high resolution, 4096-megapixel size images available for 40 outputs of the simulation at various redshifts, as shown in Gig.~\ref{fig:web_vizual}. Generally, two different visuals can be used at the same time. They represent the diffuse baryonic medium, visualized, color coded according to its X-ray emission, thermal SZ, or pressure jumps to find cosmic shocks. Alternatively, the stellar component is visualized according to the density of the stars and color-coded by the mean age of the stellar population using {\sc Splotch} \citep{dolag:2008}. Additionally, the position of galaxy clusters and groups can be overlayed as circles and an information panel on the cluster properties gets visible as soon as a galaxy cluster or group is selected. In addition, the user is allowed to perform complex queries of the metadata of the galaxy clusters and groups. This can be done interactively by using an offered interface and allows to select clusters based on different, physical properties. The available metadata allows not only to select clusters by their mass or temperature but also by their gas and star fraction and even by some dynamical state indicators like center-shift or stellar mass fraction between the central galaxy and satellite galaxies. The metadata can also be always downloaded as CSV file. The {\sc ClusterInspect} service allows to produce some simple scatter-plots directly on the web portal. The {\sc HaloConversion} interface allows to convert halo properties between different overdensity definitions and cosmologies, following \citet{2021MNRAS.500.5056R}. Several more services are currently in preparation and will be made available to the user community in the near future.

The {\sc Phox} service also \citep{biffi:2012} allows to perform synthetic X-ray observations of the ICM component of the selected galaxy cluster. Here the user can choose the size of the region of the simulation to be included along the line of sight (currently up to 100~Mpc) as well as to also include photons from the AGN \citep{2018MNRAS.481.2213B}. The service then, as soon as executed, returns the idealized list of observed photon events according to the instrument specification (effective area and field of view) and the chosen observing time. For some specific X-ray instruments, among them Athena+, Chandra, XMM, eROSITA, the user can request additional results of an instrument simulation based on several publicly available simulators such as {\sc SIXTE}, {\sc pyXSIM} and others, which then returns event files which takes the actual instrument specifications (like energy-dependent effective area and beam smearing) into account.

Future services will include accurate lensing maps and optical maps in various filters, enabling us to fully explore the potential of the underlying simulations. All such services have an optimized reading routine that allows to efficiently extract a spacial region around any position within the full simulation data, which can exceed 20TB per snapshot as in the case of \Bzmr. The technical details are explained in \citet{ragagnin:2017webportal} and also summarized in Appendix A1, demonstrating that the reading of a spatial region of a couple of virial radii around the most massive clusters is performed in less than one second, which is a key feature which allows to provide services on a web portal which operate directly on the fill simulation data. 

\section{Summary, Discussion, and Conclusion}
\label{sec:end}

In this study we have introduced the full set of simulations belonging to the \magpath hydrodynamical cosmological suite of simulation (www.magneticum.org), the simulation set that self-consistently still covers the largest range in box simulation volumes and resolutions of all cosmological hydrodynamical simulation sets currently on the market. The dynamical range of these simulations covers 7 orders of magnitude in mass range. The box volumes reach from $56.35~\mathrm{Gpc}^3$ for the largest simulation \Bz, down to $1.7\times10^{-5}\mathrm{Gpc}^3$ for the smallest volume \Bfi, with dark matter resolutions ranging from $m_\mathrm{DM}=1.3\times10^{10}\,M_\odot$ for the lowest resolution \mr to $m_\mathrm{DM}=1.9\times10^{6}\,M_\odot$ for the highest resolution \xhr, with according approximate stellar particle masses of $m_*=6.5\times10^{8}\,M_\odot$ for \hr to $m_\mathrm{DM}=1\times10^{5}\,M_\odot$ for \xhr. The flagship simulations for each resolution are \Bzmr, \Bthr, and \Bfuhr for runs down to $z=0$, and \Btbhr and \Bthuhr for higher redshift studies. With this setup, the \magpath simulations are unprecedentedly well suited for studying the formation and evolution of structures from galaxies to galaxy clusters in a self-consistent manner, and also allow for analyses of the connection of these structures to cosmological studies, especially since the \magpath suite of simulations is accompanied by 15 simulations of \Bomr probing a large range of cosmologies as introduced by \citet{singh:2020}. These simulations are not part of this study but belong to the \magpath simulation family.

The details of the physics included in the \magpath simulations were discussed in detail, highlighting also the extreme technical achievement with respect to code performance, which allowed \Bzmr and \Btbhr to be performed despite the extremely large number of resolution elements 10 years before any other simulation of that kind could be performed otherwise. Furthermore, the \magpath simulation includes two different treatments of BH physics, the fiducial model that was introduced by \citet{hirschmann:2014} and is discussed in detail here as well, and the advanced BH model that was introduced by \citet{steinborn:2015} and has proven to be especially good at reproducing high redshift observations that other simulations struggle to capture \citep[e.g.][]{kimmig:2025,remus:2025,weller:2025}. One specialty of the \magpath simulations is that the BHs are not fixed to the potential minimum but can wobble during merger events, which is a more realistic treatment of BHs during merger events as also shown by \citet{steinborn:2016}, leading to BH feedback not immediately killing the galaxy before the merger event between the stellar components could even begin properly. A comparison of the feedback implementations within the different simulations currently often used is presented by \citet{2025arXiv250206954V}. The \magpath simulations were all performed with an advanced version of the GADGET-3 code \citep{springel:2005}, including code performance improvements as introduced by \citet{beck:2015}, and artificial viscosity and conductivity to deal with the hydrodynamic issues in standard Gadget-2 \citep{dolag:2005viscosity,dolag:2004,arth:2014}, among others, with more improvements being actively worked on \citep{marin-gilaber:2022,marin-gilabert:2024}.
Another specialty of the \magpath simulations compared to other hydrodynamical cosmological simulations on the market is that their subgrid physics was not tuned to reproduce the stellar mass functions, but instead, the physical modules were chosen to capture the hot gas component in galaxy clusters at $z=0$. Thus, scaling relations for the stellar components are not tunwed for but rather evolve naturally from this simulation, perfectly complementing other simulations currently available, as discussed by \citet{popesso:2024}.

We present 28 scaling relations obtained from the \magpath simulations for the full range of resolutions and simulation volumes, in comparison to observations. The 28 relations are split into 4 mass functions, 4 global evolutions, 7 scaling relations for global halo properties, 9 scaling relations for internal or dynamical properties of galaxies, and 4 BH scaling relations. The 4 mass functions are for dark matter, stars, gas, and BHs. The 4 global evolutions are the cosmic star formation rate density, the depletion timescale through cosmic time, the all galaxy and quenched galaxy number densities through redshift, as well as the most massive halo through cosmic time. The 7 global scaling relations are the stellar-mass--halo-mass relations, the baryon conversion efficiency, the gas-mass--halo-mass relation, the temperature-mass relation, the Compton Y-mass relation, the Lx-mass and the Iron-metalicity--mass relation. The 9 local scaling relations are: the galaxy star formation main sequence, the Kennicutt-Schmitt relation, the color-mass diagram and the red sequence and blue cloud, the stellar mass-age relation, the mass-metalicity relation, the specific angular momentum relation and the b-value, the mass-size relation, the fundamental plane of galaxies, and the resolved stellar kinematics relation. Finally, the four BH scaling relations investigated in this work are the Magorrian relation, the black-hole--$\sigma$ relation, the specific star formation rate-BH mass relation, and the BH mass distribution split by galaxy type.
For all scaling relations we find excellent convergence between the simulations of different resolutions and volumes, without adding additional fudge factors, clearly showing that the \magpath simulations are indeed perfectly suited to study structure formation over such a broad range in masses. For some scaling relations, where internal galaxy properties were investigated, the \mr resolution reaches its resolution limit imposed by the softening, and thus cannot be used for such studies. On the other hand, the small box volumes are too small to be usable for statistics on the high mass end of galaxy clusters. Thus, a careful choice of the simulations to use for a given scientific question is important, not due to convergence issues but rather cosmic variance and resolution.

The comparison to observations revealed a surprisingly good agreement for all scaling relations at all redshifts, from $z=4$ to the present day. The best agreements were found for those scaling relations involving the halo mass or gas properties, where the agreement with observations and models is excellent. The strongest deviations were found for the stellar mass function and the stellar-mass--halo-mass relations, but both relations also show large uncertainties from observations due to very different measurement methods, but also possibly driven by the local Universe not being perfectly representative of the median behavior of structures in the universe \citep[e.g.][]{dolag:2023}. Generally, deviations are strongest for the low mass end, indicating that our stellar feedback is not as strong as suggested to be necessary to capture dwarf galaxies properly \citep[e.g.][]{wang:2015,tollet:2019}. These deviations vanish once feedback from BHs sets in, and the agreement for masses from MW mass galaxies to galaxy groups with observations is excellent. At the high mass end, we seem to have too many stars concentrated at the BCG, which could be either due to missing physics, AGN feedback, or observational uncertainties as discussed in detail. Interestingly, the cosmic star formation rate density and the number densities and quenched number densities are captured extremely well, indicating that overall we do not produce too many stars in the simulations. 

Kinematic properties are captured very well by the simulation, visible in a large variety of galaxy scaling relations captured excellently, like the fundamental plane, the Kennicutt-Schmitt relation, the $\lambda_R--\epsilon$ relation, or the mass-size relation. However, as reported for other simulations as well, we find our simulated velocity dispersion to be too small compared to observations, showing in all relations including the velocity dispersion as already discussed by \citet{vandesande:2019} and others. This is still an open riddle to solve for the simulation community, which clearly does not depend on the parameters fitted for the subgrid models, as it is nearly identical for all large cosmological simulation suites, seen from field galaxies to galaxies in the most massive clusters.

The fact that the \magpath simulations are equally efficient in reproducing scaling relations as simulations like IllustrisTNG and Eagle, which are scaled to the stellar properties directly, supports one important point with regards to reproducing observed relations in simulations and models: there are multiple ways to obtain the same scaling relation. This is due to the fact that structures in our Universe grow hierarchically, and hierarchical growth imposes the central limit theorem, thus producing scaling relations and tightening them with time (e.g., \citealp{donofrio:2021} for a review, or \citealp{remus:2013} for the specific case of total density slopes). However, this clearly shows that the important physics is actually not encoded in the tightness of the scaling relation itself, but rather in the origin of the scatter of a given scaling relation, which needs to also be reproduced correctly within simulations to understand the formation of structures from cosmic dawn to present day self-consistently. In particular the recent results are especially enlightening when looking at the high redshift Universe and the challenges imposed by recent JWST observations. Where other simulations struggle, the high redshift performance is excellent for the \magpath \uhr simulations, with the advanced BH feedback model overall performing the best. As shown in the section of the BH scaling relations, this comes together with a large scatter in the BH scaling relations, that actually enables a broad spread in galaxy and BH properties to be reproduced at high redshifts, which is exactly what is observed. For more details on this, as well as the other scaling relations presented here, we provide an overview of all publications using the \magpath simulation suite, with comparisons to various observational datasets, on the project webpage\footnote{http://www.magneticum.org/publications.html}.

\subsection{Lessons learned}

One of the main lessons learned from the \magpath simulations is that anchoring the cosmological simulation on galaxy clusters can be a very powerful tool to constrain galaxy evolution. Especially the ICM with its well-measured metal content, distribution and evolution gives a record on galaxy formation from early times till now and also strongly displays how AGN feedback works in such environment. This gives a very powerful leverage on calibrating numerical models for galaxy formation. Having more than one hundred studies that used results from the \magpath simulations, it is difficult to give a full summary of the obtained insights. However, there are some main results which are worth repeating in the following.

As the \magpath simulations very well reproduce the observed metal content and composition within the ICM as well as within galaxies simultaneously \citep{dolag:2017}, they can be used to investigate long-standing questions regarding the metal distribution in the universe. One of those long-lasting and unsolved questions is the so-called iron conundrum observed in galaxy clusters, e.g. meaning that the amount of iron observed within the ICM exceeds the amount of iron which can be expected to be produced by the observed stellar population of the galaxies within the cluster. Interestingly, this conundrum is not present in the simulations. Here, the iron share is, as expected, of order one, confirming that the assembly process and evolution of galaxy clusters, including merging events, stellar and AGN energy feedback, accretion of pristine gas along with pre-enriched gas, and the accretion of galaxies or galaxy groups, do not significantly alter the final baryonic and metal budget in massive clusters in the respect to the expectations \citep{2025A&A...698A.238B}. In the simulations this comes with the larger stellar mass formed within massive galaxy clusters, especially stellar mass within the BCG and ICL, which is only rarely inferred with such high values in observations. However, there are still large uncertainties involved when comparing the ICL inferred directly from the simulations with the way it is measured in observations, which might bring them both closer together, as discussed in \citet{brough:2024}. Note that despite these complications in comparison with observations, the fraction of stellar mass in the BCG and ICL i the simulations relates very closely to the galaxy cluster mass assembly history \citep{kimmig:2025b}.

While reproducing the Magorrian relation within cosmological simulations seems to be accomplished by almost all models, the luminosity function of AGNs seems to be a difficult property to be reproduced by most of the different implementations of AGN feedback \citep{2022MNRAS.509.3015H}. However, the implementation within the \magpath simulations seems to obtain reasonable good match over the evolution of the AGN luminosity function \citep{hirschmann:2014,2018MNRAS.481.2213B}, mostly by being more bursty, but at the same time very gentle and also setting in at more early times. Although the numerical implementation in general typically leads to an inverse trend with respect to the mass of galaxy clusters for the scale on which the AGN injects its power as well as for the amount of AGN feedback when operating in the so-called radio feedback mode, overall it reproduces the appearance of cool-core and non cool-core systems and their mass dependence surprisingly accurately, mostly thanks to the inclusion of thermal conduction \citep{gonzalez_villalba:2025}.     

On galaxy scales, the simulations not only reproduce the global trend that the fraction of star-forming gas in galaxies is decreasing with time, they also show that a major fraction of the galaxies undergoes long-term cycles of quenching and rejuvenation on gigayear timescales \citep{fortune:2025}. While this on-off cycle is related to the geometry of gas accretion at the halo outskirts, the simulations also show that the overall declining trend is not a matter of the total available cold gas but how the distribution of this gas within the galaxies is changing with time \citep{teklu:2023}. When investigating the stellar dynamics of early-type galaxies we found that the fast-rotating population is already in place at $z=2$ and could demonstrate that at least 30\% of the population of the slow-rotating early-type galaxies evolved from the fast-rotating population by a mayor merger event \citep{schulze:2018}. We also find a significant fraction of quenched galaxies at high redshift (e.g. $z\approx3.5$), which are quenched through a rapid burst of star formation and subsequent AGN feedback caused by a particularly isotropic collapse of surrounding gas, occurring on timescales of around 200 Myr or shorter. The resulting quenched galaxies host stellar components that are kinematically fast-rotating and alpha-enhanced \citep{kimmig:2025}.

\subsection{Outlook}
Therefore, for the future, it is important to simulate big volumes at high resolutions needed to capture the full spread of galaxy formation at high redshifts fully, which will be done in the future with the \magpath spinoff DAWN (Remus et al., in prep). Furthermore understanding how far the local Universe is representative of the scaling relations and their scatter for the general Universe also poses an important challenge, as all our models are focussed on what we have studied the best, namely our local environment. If that environment for some reason is not representative, then our models are erroneous by design. These questions will be addressed by further future spinoffs of the \magpath simulations, namely the local Universe constrained simulations SLOW \citep[e.g.][]{dolag:2023,seidel:2024} and high-resolution zoom suites of general galaxy cluster from the COMPASS simulations (Kimmig et al, in prep) and the Local Universe Cluster analogs simulations LowerDecks (Seidel at al., in prep). This will be exciting research to come in the near future, to boost our understanding of structure formation in the Universe from cosmic dawn to the present day.

\section*{Acknowledgments}

The \magpath simulations were performed at the Leibniz-Rechenzentrum with CPU time assigned to the Project \textit{pr83li}. This work was supported by the Deutsche Forschungsgemeinschaft (DFG, German Research Foundation) under Germany's Excellence Strategy EXC-2094\,--\,390783311. We are especially grateful for the support by M.~Petkova through the Computational Center for Particle and Astrophysics (C2PAP).
LMV and KD acknowledge support by the COMPLEX project from the European Research Council (ERC) under the European Union’s Horizon 2020 research and innovation program grant agreement ERC-2019-AdG 882679.
LCK acknowledges support by the DFG project nr. 516355818. 
IM and PP acknowledge support from the European Research Council (ERC) under the European Union’s Horizon Europe research and innovation program ERC CoG (Grant agreement No.~101045437, PI P.~Popesso). SVZ and VB acknowledge support by the DFG project nr. 415510302.

Funding for the Sloan Digital Sky Survey V has been provided by the Alfred P. Sloan Foundation, the Heising-Simons Foundation, the National Science Foundation, and the Participating Institutions. SDSS acknowledges support and resources from the Center for High-Performance Computing at the University of Utah. SDSS telescopes are located at Apache Point Observatory, funded by the Astrophysical Research Consortium and operated by New Mexico State University, and at Las Campanas Observatory, operated by the Carnegie Institution for Science. The SDSS web site is \url{www.sdss.org}. 

SDSS is managed by the Astrophysical Research Consortium for the Participating Institutions of the SDSS Collaboration, including Caltech, The Carnegie Institution for Science, Chilean National Time Allocation Committee (CNTAC) ratified researchers, The Flatiron Institute, the Gotham Participation Group, Harvard University, Heidelberg University, The Johns Hopkins University, L’Ecole polytechnique fédérale de Lausanne (EPFL), Leibniz-Institut für Astrophysik Potsdam (AIP), Max-Planck-Institut für Astronomie (MPIA Heidelberg), Max-Planck-Institut für Extraterrestrische Physik (MPE), Nanjing University, National Astronomical Observatories of China (NAOC), New Mexico State University, The Ohio State University, Pennsylvania State University, Smithsonian Astrophysical Observatory, Space Telescope Science Institute (STScI), the Stellar Astrophysics Participation Group, Universidad Nacional Autónoma de México, University of Arizona, University of Colorado Boulder, University of Illinois at Urbana-Champaign, University of Toronto, University of Utah, University of Virginia, Yale University, and Yunnan University.

The following software was used for this work: Julia \citep{bezanson17:julia}, Matplotlib \citep{hunter07:matplotlib}, IDL.
\bibliographystyle{style/aa}
\bibliography{scaling}

\appendix

\section{Data-IO}
\begin{figure}
\includegraphics[width=0.5\textwidth]{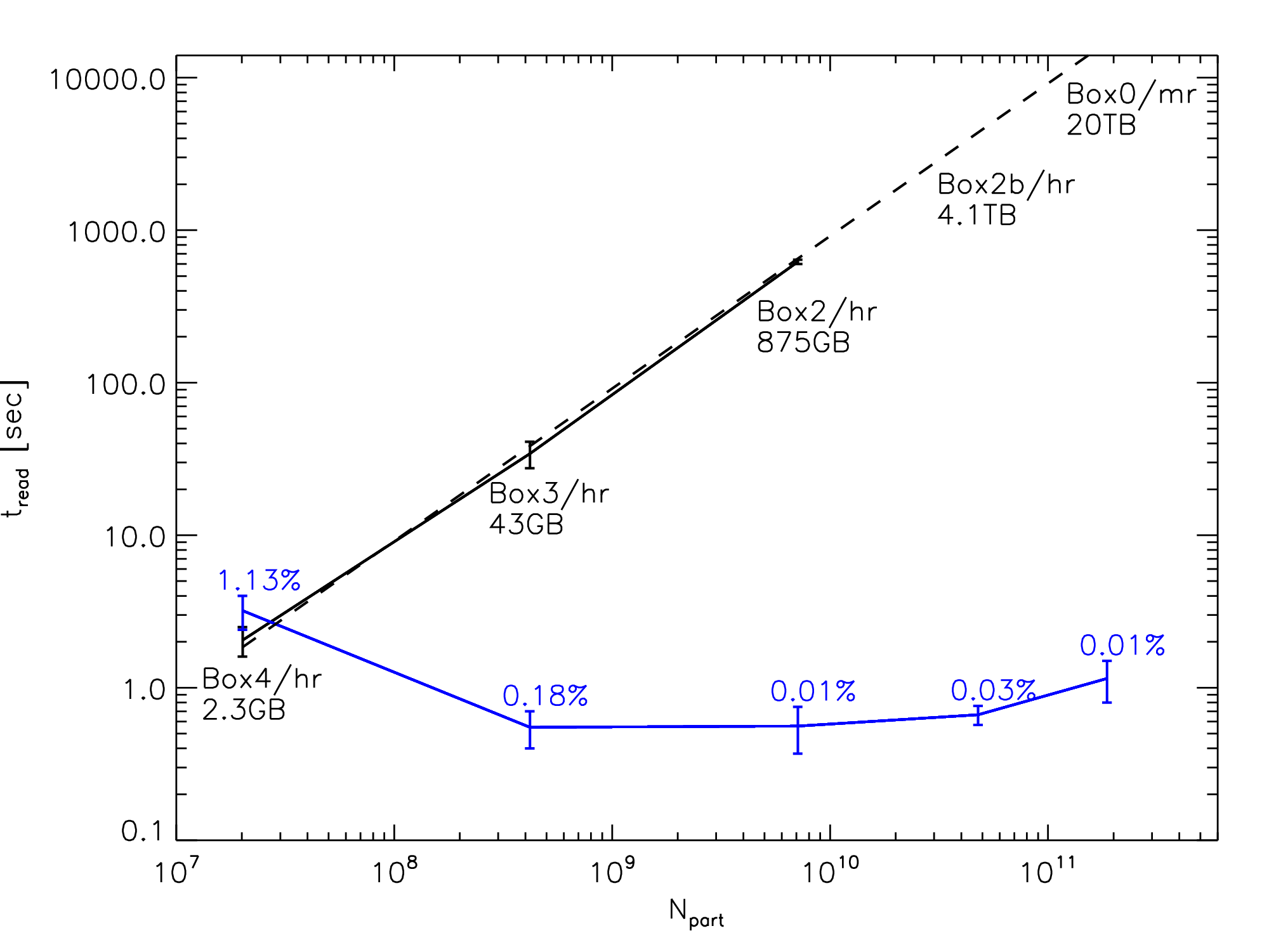}
\caption{Wall clock time to read all data for the stellar component from most massive galaxy cluster in the different simulation volumes, there the x-axis denote the total number of particles in each simulation. The black line shows the brute force approach by reading all stellar data and doing the spatial selection afterwards. The black labels give the according simulation and the total size of the associated snapshot. The blue lines show the result when using the key-index files, which allows to read out only the relevant part of the snapshot files, which is given as the labeled fraction in percent.}
\label{fig:read_keys}
\end{figure}

For the \magpath project the output is structured in a special way to allow a general, very fast access to a spatially defined regions. While many other simulation campaigns order the output according to the particles within halos, which allows to access the particles belonging to an individual halo in a very fast way, we developed for the \magpath simulations  an especially structured output. For details, see \citet{ragagnin:2017webportal}. In short, before writing the data to disk, a domain decomposition based on a shortened Peano-Hilbert key is performed, which is optimized to have one continuous segment in each snapshot file. Note that in contrast to the standard domain decomposition in Gadget which is using a 64 bit long key, we are using a shorter, 32 bit long key which strongly reduces the later reading overhead. Afterwards, index files are created holding the mapping from key index to offset and length information of the particle data stored in the snapshot files. For further optimization, these indexing is grouped to allow for continuous blocks of pixels.     

This procedure has been shown to work extremely efficiently. The creation, reading, and comparing operations needed for selection of the {\it pixel} lists introduce only a minimal overhead. Fig.~\ref{fig:read_keys} demonstrates that the reading of all stellar particles within the virial radius of the most massive galaxy cluster in the simulations takes significantly less than 1 second in all simulations, including \Bzmr, which contains more than $10^{11}$ particles and the information has to be filtered out of individual snapshots, which in total occupy 20TB on disk.

\section{Web-portal usage statistics}

\begin{figure}
\includegraphics[width=0.5\textwidth]{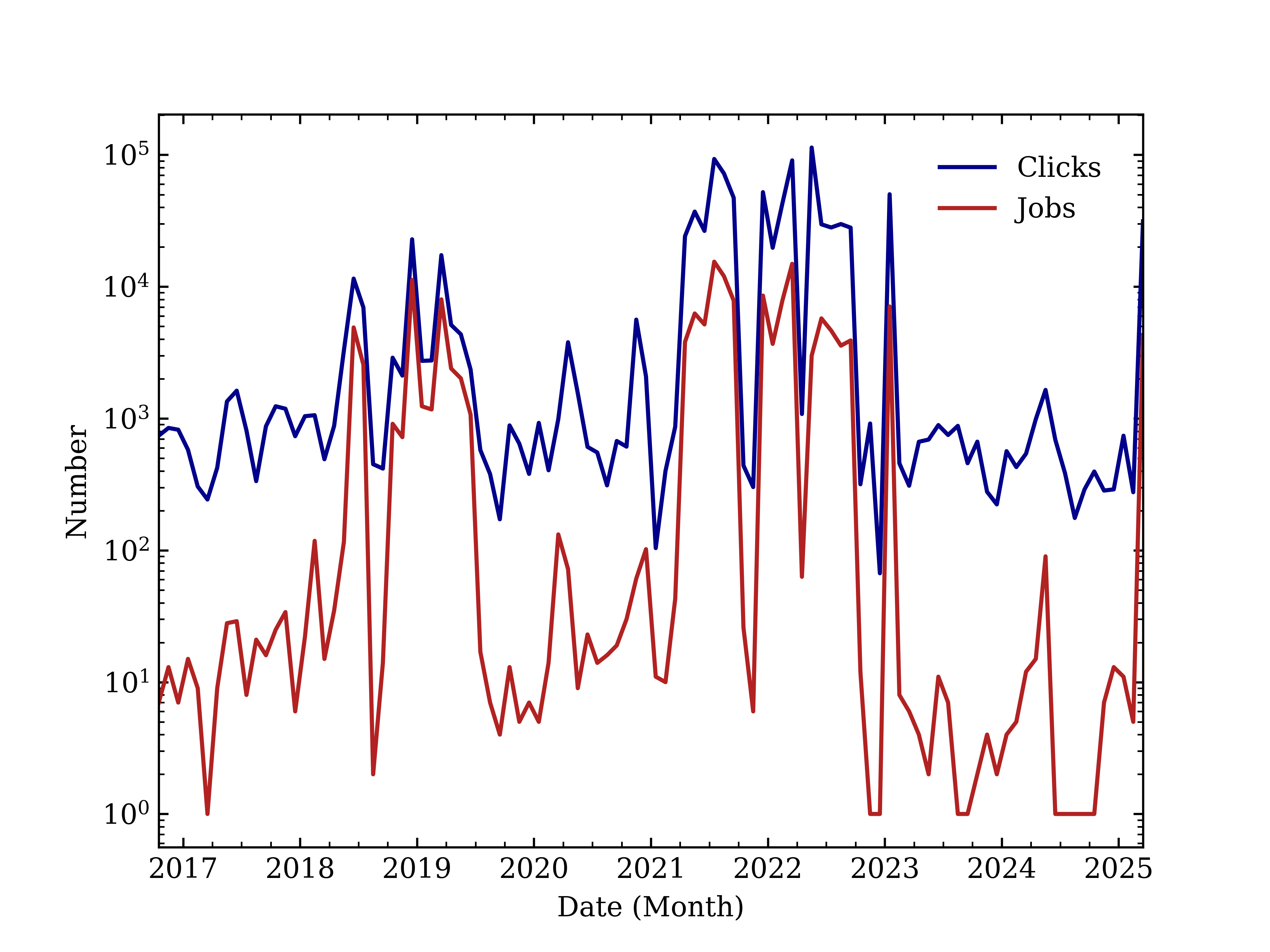}
\caption{Usage statistics from the web portal. Shown is the activity per month since 2017. The blue line marks the general clicks on the front-end, whereas the red line shows the number of executed post-processing jobs. On average, there are between ten and hundreds of post-processing jobs executed per month, while in some cases these numbers slightly exceed tens of thousands of jobs processed in one month.}
\label{fig:portal_statistics}
\end{figure}

The above described output strategy allowed to make not only already post-processed data available but also to extend out web portal (as described in \ref{sec:web}) to allow the community to apply standard post-processing workflows interactively onto the full simulation data. Fig.~\ref{fig:portal_statistics} shows that for almost 10 years users are processing the simulation data directly from the web portal, sometimes reaching almost ten thousand postprocessing jobs per month. In total, the web portal sofar received more than one million clicks and processed more than 150 thousand user defined post-processing jobs. 

\section{Aperture Cuts}
\begin{figure*}
\includegraphics[width=0.98\textwidth]{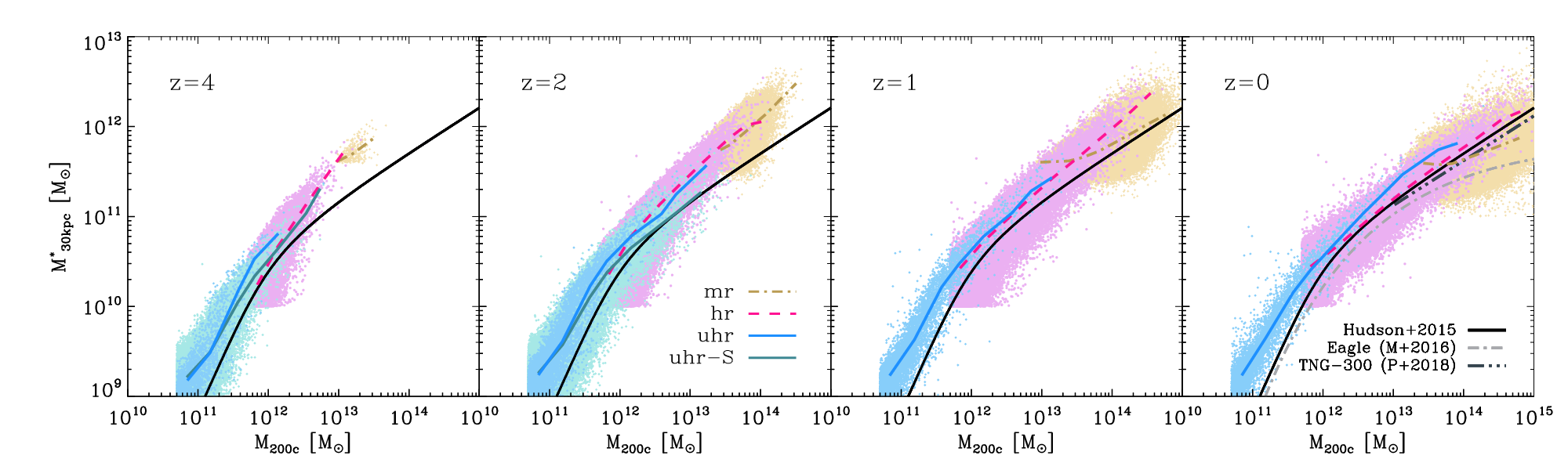}
\caption{Stellar-mass--halo-mass relations for the \magpath simulations, with colors indicating the different resolutions as indicated in the legend. Here, the halo mass is given as $M_\mathrm{200,c}$, and for the stellar mass all stars within a fixed aperture of $30~\mathrm{kpc}$ that are allocated to the central galaxy by \subfind are used, denoted $M_\mathrm{30kpc}^*$. The black solid line shows the SMHM relation found from lensing at $z\approx0$ by \citet{hudson:2015}, as a reference at all redshifts. In addition, stellar-mass--halo-mass relations found using the exact same definitions from the IllustrisTNG simulations from \citet[][dark gray dash-dot-dotted line]{pillepich:2018} and the EAGLE simulations from \citet[][light gray dash-dotted line]{matthee:2017} are shown.}
\label{fig:SMHM_ap30}
\end{figure*}

In Fig.~\ref{fig:SMHM_ap30} we show the measured SMHM relation similar to Fig.~\ref{fig:smhm200}, but for an aperture cut of $30~\mathrm{kpc}$, to compare to results from the IllustrisTNG simulations from \citet[][dark gray dash-dot-dotted line]{pillepich:2018} and the EAGLE simulations from \citet[][light gray dash-dotted line]{matthee:2017}. In addition, observations from the weak lensing survey CFHTLenS by \citet{hudson:2015} at $z=0$ are shown as a solid black line, and we include this line for reference in all redshift panels. For further details, see discussion in section \ref{sec:smhm}.

\label{lastpage}
\end{document}